\pdfoutput=1
\documentclass[cernpreprint,texlive=2023,atlasdraft=false,UKenglish, texmf, orcidlogo]{atlasdoc}
 
\usepackage{atlaspackage}
\usepackage{atlasphysics}
\usepackage{atlasbiblatex}
 
\newcommand{\AtlasCoordFootnote}{
ATLAS uses a right-handed coordinate system with its origin at the
nominal interaction point (IP) in the centre of the detector and the
\(z\)-axis along the beam pipe. The \(x\)-axis points from the IP to
the centre of the LHC ring, and the \(y\)-axis points upwards. Polar
coordinates \((r,\phi)\) are used in the transverse plane, \(\phi\)
being the azimuthal angle around the \(z\)-axis. The pseudorapidity is
defined in terms of the polar angle \(\theta\) as \(\eta = -\ln
\tan(\theta/2)\) and is equal to the rapidity $ y = \frac{1}{2} \ln
\left( \frac{E + p_z c}{E - p_z c} \right) $ in the relativistic
limit. Angular distance is measured in units of \(\Delta R \equiv
\sqrt{(\Delta y)^{2} + (\Delta\phi)^{2}}\).
}
 
\graphicspath{{logos/}}
 
\usepackage{ANA-TOPQ-2020-09-PAPER-defs}

\AtlasTitle{Measurement of differential cross-sections in $t\bar t$ and $t\bar t+$jets production in the lepton+jets final state in $pp$ collisions at $\sqrt s=13$ TeV using 140 fb$^{-1}$ of ATLAS data}
 
\author{The ATLAS Collaboration}
 
\AtlasJournalRef{JHEP 08 (2024) 182}
\AtlasDOI{10.1007/JHEP08(2024)182}
\PreprintIdNumber{CERN-EP-2024-163}
 
\AtlasAbstract{
Differential cross-sections for top-quark pair production, inclusively
and in association with jets, are measured in $pp$ collisions at a
centre-of-mass energy of 13 TeV with the ATLAS detector at the LHC
using an integrated luminosity of 140 fb$^{-1}$. The events are selected
with one charged lepton (electron or muon) and at least four jets. The
differential cross-sections are presented at particle level as
functions of several jet observables, including angular correlations,
jet transverse momenta and invariant masses of the jets in the final
state, which characterise the kinematics and dynamics of the
top--antitop system and the hard QCD radiation in the system with
associated jets. The typical precision is 5\%--15\% for the absolute
differential cross-sections and 2\%--4\% for the normalised
differential cross-sections. Next-to-leading-order and
next-to-next-to-leading-order QCD predictions are found to provide an
adequate description of the rate and shape of the jet-angular
observables. The description of the transverse momentum and invariant
mass observables is improved when next-to-next-to-leading-order QCD
corrections are included.
}

\hypersetup{pdftitle={ATLAS document},pdfauthor={The ATLAS Collaboration}}
 
\begin{document}
 
\maketitle
 
\tableofcontents
 
\section{Introduction}
The production of top-quark pairs ($\tt$) in proton--proton ($pp$)
collisions provides a testing ground for the Standard Model (SM) at the \TeV\
scale. Differential cross-sections can be used to extract fundamental
parameters of the SM, such as the top-quark
mass~\cite{jhep:06:019,epj:c83:963} and the strong coupling contant
$\as$~\cite{epj:c80:658}, and to derive constraints~\cite{epj:c80:658}
on the parton distribution functions (PDFs). These measurements are
also sensitive to the existence of new physics beyond the
SM~\cite{jhep:06:063,jhep:12:068}, which would appear as deviations of
the measurements from the SM predictions. In addition, for all
searches for new physics in which $\tt$ production constitutes a
background, the $\tt$ differential cross-sections are particularly
useful to highlight the differences between Monte Carlo (MC)
generators and can be used to tune the parameters of those
generators.
 
The
ATLAS~\cite{epj:c73:2261,pr:d90:072004,jhep:06:100,pr:d93:032009,epj:c76:538,pr:d94:092003,epj:c77:292,jhep:11:191,jhep:10:159,jhep:04:046,jhep:01:033,epj:c79:1028}
and
CMS~\cite{epj:c73:2339,epj:c75:542,pr:d94:052006,pr:d94:072002,epj:c77:459,jhep:04:060,pr:d95:092001,jhep:06:002,pr:d97:112003,jhep:02:149,epj:c80:658,pr:d103:052008}
Collaborations have published measurements of $\tt$ differential
cross-sections at centre-of-mass energies $\!\!\sqrt s=7,\ 8$ and
13~\TeV\ in $pp$ collisions at the LHC~\cite{jinst:3:S08001} using
final states containing a pair of leptons or a single lepton plus
jets, both in the full phase space using parton-level variables and in
fiducial phase-space regions using observables constructed from
final-state particles (particle level). These results have been used
extensively to improve the modelling of MC
generators~\cite{ATL-PHYS-PUB-2016-016,ATL-PHYS-PUB-2016-020,ATL-PHYS-PUB-2017-007,ATL-PHYS-PUB-2018-009,epj:c80:4}
and to reduce the uncertainties in the  gluon PDF~\cite{jhep:04:040}.
 
The analysis presented here focuses on the study of jet observables,
some of which have not been explored in previous analyses, in
the $\tt$ and $\tt+$jets systems. Measurements of differential cross
sections in the boosted regime are included in a previous
publication~\cite{jhep:04:80}. In the SM, the top quark decays
almost exclusively into a $W$ boson and a $b$-quark. The signature of
a $\tt$ decay is therefore determined by the $W$ boson decay
modes. This analysis makes use of the \ljets\ ($\ell=e,\ \mu$) $\tt$ decay mode,
also called the semileptonic channel, where one $W$ boson decays into
an electron or a muon and a neutrino, and the other $W$ boson decays
into a quark--antiquark pair\footnote{Events in which the $W$ boson
decays into an electron or muon through a $\tau$-lepton decay may
also meet the selection criteria.}. Since the reconstruction of the top
quark depends on its decay products, in the following the top
quark is referred to as `hadronically or leptonically decaying top
quark', depending on the $W$ boson decay mode.
 
Absolute and normalised differential cross-sections for the $\tt$ and
$\tt+$jets systems are measured as functions of various jet
observables to characterise the kinematics, topology and dynamics of
the $\tt$ system and are also sensitive to initial- and final-state
quantum chromodynamics (QCD) radiation effects, represented by the
production of additional jets. Some of these observables are measured
for the first time. State-of-the-art next-to-leading-order (NLO) and
next-to-next-to-leading-order (NNLO) QCD predictions interfaced to
parton-shower (PS) algorithms are compared with the measurements.
The measured particle-level differential cross-sections as functions
of the jet observables presented here benefit from the increase in
integrated luminosity and the reduced detector-related uncertainties
compared to previous publications. In addition, the comparison of
NNLO+PS differential cross-sections for jet observables at particle
level with the measurements is performed for the first
time. Therefore, more stringent tests of perturbative QCD (pQCD) are
achieved than in the previous analyses.
 
\section{ATLAS detector}
The ATLAS detector~\cite{PERF-2007-01} at the LHC covers nearly the
entire solid angle around the collision
point.\footnote{\AtlasCoordFootnote} It consists of an inner tracking
detector surrounded by a thin superconducting solenoid,
electromagnetic (EM) and hadron calorimeters, and a muon spectrometer
incorporating three large superconducting air-core toroidal magnets.
 
The inner-detector system (ID) is immersed in a \qty{2}{\tesla} axial
magnetic field and provides charged-particle tracking in the range of
\(|\eta| < 2.5\). The high-granularity silicon pixel detector covers
the vertex region and typically provides four measurements per track,
the first hit normally being in the insertable B-layer (IBL) installed
before Run~2~\cite{ATLAS-TDR-19,jinst:13:T05008}. It is followed by the
silicon microstrip tracker (SCT), which usually provides eight
measurements per track. These silicon detectors are complemented by
the transition radiation tracker (TRT), which enables radially
extended track reconstruction up to \(|\eta| = 2.0\). The TRT also
provides electron identification information based on the fraction of
hits (typically 30 in total) above a higher energy-deposit threshold
corresponding to transition radiation.
 
The calorimeter system covers the pseudorapidity range
\(|\eta| < 4.9\). Within the region \(|\eta|< 3.2\), electromagnetic calorimetry
is provided by barrel and  endcap high-granularity lead/liquid-argon
(LAr) calorimeters, with an additional thin LAr presampler covering
\(|\eta| < 1.8\) to correct for energy loss in material upstream of
the calorimeters. Hadron calorimetry is provided by the
steel/scintillator-tile calorimeter, segmented into three barrel
structures with \(|\eta| < 1.7\), and two copper/LAr hadron endcap
calorimeters. The solid angle coverage is completed with forward
copper/LAr and tungsten/LAr calorimeter modules optimised for
electromagnetic and hadronic energy measurements respectively.
 
The muon spectrometer (MS) comprises separate trigger and
high-precision tracking chambers measuring the deflection of muons in
a magnetic field generated by the superconducting air-core toroidal
magnets. The field integral of the toroids ranges between \num{2.0}
and \qty{6.0}{\tesla\metre} across most of the detector. Three layers
of precision chambers, each consisting of layers of monitored drift
tubes, cover the region \(|\eta| < 2.7\), complemented by
cathode-strip chambers in the forward region, where the background is
highest. The muon trigger system covers the range \(|\eta| < 2.4\)
with resistive-plate chambers in the barrel, and thin-gap chambers in
the endcap regions.
 
Interesting events are selected by the first-level trigger system
implemented in custom hardware, followed by selections made by
algorithms implemented in software in the high-level
trigger~\cite{epj:c77:317}. The first-level trigger accepts events
from the \qty{40}{\MHz} bunch crossings at a rate below
\qty{100}{\kHz}, which the high-level trigger further reduces in order
to record events to disk at about \qty{1}{\kHz}.
 
An extensive software suite~\cite{2404.06335} is used in
data simulation, in the reconstruction and analysis of real and
simulated data, in detector operations, and in the trigger and data
acquisition systems of the experiment.
 
\section{Data sample}
The data used in this analysis were collected with the ATLAS detector
during the $pp$ collision running periods of 2015--2018, when the LHC
operated at a centre-of-mass energy of $\sqrt s=13$~\TeV. Only events
taken during stable beam conditions and satisfying detector- and
data-quality requirements~\cite{jinst:15:p04003}, which include the
calorimeters, muon system and inner tracking detectors operating
normally, are considered. The total integrated luminosity of the
collected sample is {\mbox{$140$~\fb1}}. The uncertainty in the
combined 2015--2018 integrated luminosity is
$0.83\%$~\cite{epj:c83:982}, obtained using the LUCID-2
detector~\cite{jinst:13:p07017} for the primary luminosity
measurements, complemented by measurements using the inner detector
and calorimeters.
 
\section{Monte Carlo samples}
Samples of MC generated events are used to study the characteristics
of the signal events. These samples are also used to perform the
unfolding procedure to obtain the measured cross-sections. In
addition, several MC samples are used to determine the background to
the signal process, except for the background from misidentified or
non-prompt leptons (called `fakes' background), which relies instead
on a data-driven technique (see Section~\ref{backcon}). All the MC
samples are described below and summarised in Table~\ref{tab:MC}.
 
Most of the samples of generated events were passed through a full detector
simulation~\cite{epj:c70:823} based on the \GEANT~\cite{nim:a506:250}
program. Other samples, used for estimating some of the systematic
uncertainties, were produced using the fast simulation software
Atlfast-II. All MC samples were reconstructed and analysed by the same
algorithms as for the data.
 
Pile-up from additional $pp$ collisions in the same and neighbouring
bunch crossings was simulated by overlaying each MC event with a
variable number of simulated inelastic $pp$ collisions generated using
\pyt~8.186~\cite{cpc:178:852} with the ATLAS set of tuned parameters
for minimum-bias events (A3 tune)~\cite{ATL-PHYS-PUB-2016-017}. The MC
events were weighted  (`pile-up reweighting') so that the distribution of the
average number of interactions per bunch crossing matches that
observed in the data.
 
\subsection{Signal simulation samples}
\label{sec:signalMC}
In this section the MC generators used for the simulation of the $\tt$
event samples are described for the nominal sample, the alternative
samples used to estimate systematic uncertainties and the other
samples used in the comparisons of the measured differential
cross-sections. The top-quark mass ($\mt$) and width were set to
$172.5$~\GeV\ and $1.32$~\GeV~\cite{prl:110:042001},
respectively, in all samples; additional samples, in which $\mt$ and
the width of the top quark were varied, are used to study the effect of
these parameters (see below).
 
The nominal signal $\tt$ events were
generated using the NLO matrix elements (MEs)
implemented in the
\PowhegBox~\cite{jhep:09:126,jhep:11:040,jhep:11:070,jhep:06:043}
generator with the NNPDF3.0NLO PDF set~\cite{jhep:04:040}. Events where both
top quarks decayed hadronically were not included. The PS,
fragmentation, and the underlying events were simulated using
\pyt~8.230~\cite{cpc:191:159} with the NNPDF2.3LO PDF
set~\cite{np:b867:244} and the A14
tune~\cite{ATL-PHYS-PUB-2014-021}. The $\hdamp$ parameter, which
controls the $\pt$ of the first gluon emission beyond the Born
configuration in \PowhegBox, was set to
$1.5\,\mt$~\cite{ATL-PHYS-PUB-2016-020}. The main effect of this
parameter is to regulate the high-$\pt$ emission against which the
$\tt$ system recoils. Signal $\tt$ events generated with these
settings are referred to as the nominal signal sample. In all the
following figures and tables the predictions based on this MC sample
are referred to as `\pwp'.
 
Additional signal samples were generated for estimating modelling
uncertainties (see Section~\ref{syst}). These samples are:
\begin{itemize}
\item Events were generated using the same \PowhegBox\
set-up but interfaced to
\her~7.1.3~\cite{epj:c76:196,epj:c58:639}, which includes an
angular-ordered PS model. This sample was generated using
the H7.1-Default tune~\cite{jhep:10:113} and the MMHT2014LO PDF
set~\cite{epj:c75:204} in the PS. This MC sample is
referred to as `\pwh'.
\item Events were generated with \mgamcatnlo~2.3.3~\cite{jhep:07:079}
and the NNPDF3.0NLO PDF set. The PS, fragmentation, and the
underlying events were simulated using \her~7.1.3 with the MMHT2014LO
PDF set and the H7.1-Default tune.
This MC sample is referred to as `\amcnloh'.
\item Events were generated with \sher~2.2.12~\cite{sp:7:034}, which models
the zero and one additional-parton processes at NLO accuracy and up to
four additional partons at leading-order (LO) accuracy using the \textsc{MEPS@NLO}\
prescription~\cite{jhep:04:027}, with the NNPDF3.0NNLO PDF
set~\cite{jhep:04:040}. The matrix elements were calculated with the
\textsc{Comix}~\cite{jhep:12:039} and
\textsc{OpenLoops}~\cite{epj:c79:866,prl:108:111601,cpc:212:220}
libraries. The calculation uses its own parton-shower tune and
hadronisation model~\cite{jhep:03:038}. These MC samples are
referred to as `\sher~2.2.12'.
\item A sample was produced with the same settings as the nominal
\pwp\ sample but with the $\hdamp$ parameter set to $3\,\mt$. This MC
sample is referred to as `\pwp\ $\hdamp$'.
\item Two other \pwp\ samples with $\mt=169$~\GeV\ or $\mt=176$~\GeV\ were
also generated. All other parameters remained the same as in the
nominal \pwp\ sample, except for the width of the top quark, which
was set to $1.23$~\GeV\ and $1.42$~\GeV\ for $\mt=169$~\GeV\ and
$\mt=176$~\GeV, respectively. These MC samples are referred
to as `\pwp\ $\mt$'.
\end{itemize}
 
All the samples mentioned above, except for that used to unfold the
data (see Section \ref{xcross}), were processed using fast
simulation. All these MC predictions were normalised to NNLO plus
next-to-next-to-leading-log (NNLL) calculations in pQCD using a
$k$-factor. The cross-section used to evaluate the $k$-factor is
$\sigma_{\tt}=832^{+20}_{-29}\ (\mathrm{scale})\pm
35\ (\mathrm{PDF},\as)\pm 23\ (\mt)$~pb, as calculated with the
Top++2.0~\cite{cpc:185:2930,np:b855:695,pl:b710:612,prl:109:132001,jhep:12:054,jhep:01:080,prl:110:252004}
program at NNLO in pQCD, including soft-gluon resummation at
NNLL, and assuming $\mt=172.5$~\GeV. This calculation does not include
electroweak corrections. The first uncertainty comes from
the independent variation of the factorisation ($\muF$) and
renormalisation ($\muR$) scales, the second one is associated with
variations in the PDFs and $\as$, following the
PDF4LHC prescription~\cite{jp:g43:023001} with the MSTW2008 NNLO at $68\%$
confidence-level error~\cite{epj:c64:653}, CT10
NNLO~\cite{pr:d82:074024} and NNPDF2.3 5f FFN~\cite{np:b867:244} PDF
sets, and the third uncertainty arises from the assumed value of $\mt$.
 
\subsection{NNLO signal simulated samples}
The first matched computation of top-quark pair production at
NNLO in QCD, including all-order radiative corrections as implemented via
PS simulations, is reported in Ref.~\cite{prl:127:062001}
and referred to as the {\textsc{MiNNLOPS}} procedure. This procedure
involves three steps. In the first step, $\tt$ pairs plus one light
parton were generated using the \PowhegBox\ program, which provides
a NLO calculation and includes a second light parton. The second
step involves the calculation of the limit in which the light partons
generated in the previous step become collinear to the $\tt$ pair;
appropriate Sudakov form factors and higher-order terms are used to
guarantee that the calculation remains finite and accurate at NNLO for
inclusive $\tt$ production. The third step includes the generation of
the kinematics of the second radiated parton using \PowhegBox,
which was accounted for only inclusively in the first step; in this
way, the NLO accuracy of the $\tt$+jet cross-section is
preserved. After this third step, subsequent radiation was generated
using the PS method of \pyt. For the generation of this
sample, the NNPDF3.0NNLO PDF set was used. In addition to the PS step,
\pyt~8.245 was also used for the fragmentation and underlying-event
simulation with the NNPDF2.3LO PDF set and the A14 tune.
 
This prediction was also normalised to ${\cal O}$(NNLO+NNLL) in pQCD using
the cross-section $\sigma_{\tt}=832$~pb calculated with the
Top++2.0 program, as mentioned above. In all the following figures and
tables the predictions based on this sample are referred to as `\pwpm'.
 
\begin{table*}[t]
\caption{Summary of signal and background MC samples, showing the
event generator for the hard-scattering process, the order in pQCD
of the cross-section used for normalisation together with its value
and total uncertainty, the PDF choice, as well as
the parton-shower generator and the corresponding tune used.}
\scriptsize
\setlength{\unitlength}{1.0cm}
\begin{picture} (18.0,7.5)
\put (0.1,3.3){\centerline{\scalebox{0.90}{
\begin{tabular}{llllll} \toprule
Physics process                  & Generator             & PDF set for
& Parton shower   & Tune         & Cross-section   \\
& &                        hard process &                 &              & normalisation   \\
& &                         &                    &  & $\sigma$ (pQCD order)     \\ \midrule
Signal samples: & & & & & $832\pm 51$ pb (NNLO+NNLL) \\
$-$ \pwp &  \PowhegBox & NNPDF3.0NLO & \pyt~8.230 & A14          &  \\
$-$ \pwh &  \PowhegBox &  NNPDF3.0NLO & \her~7.1.3  & H7.1-Default  &     \\
$-$ \amcnloh & \mgamcatnlo~2.3.3 &  NNPDF3.0NLO &  \her~7.1.3  & H7.1-Default  &  \\
$-$ \sher~2.2.12 & \sher~2.2.12 & NNPDF3.0NNLO  &  \sher  & \sher  &     \\
$-$ \pwpm &   \PowhegBox &   NNPDF3.0NNLO  &  \pyt~8.245 &  A14  &       \\
&  &     &  + {\textsc{MiNNLOPS}} &    &       \\
& & & & & \\
Background samples: & & & & & \\
$-$ single-top quark samples:  &  & &   &  & \\
\hspace{0.2cm} $s$-channel (top)  &  & &   &  & $6.4\pm 0.2$ pb (NLO) \\
\hspace{0.2cm} $s$-channel (anti-top)  &  & &   &  & $3.4\pm 0.2$ pb (NLO) \\
\hspace{0.2cm} $t$ channel (top)  &  & &   &   & $136\pm 5$ pb (NLO) \\
\hspace{0.2cm} $t$ channel (anti-top)  &  & &   &    & $81\pm 4$ pb (NLO) \\
\hspace{0.2cm} $tW$ channel  &  & &   &   & $72\pm 4$ pb (NLO+NNLL) \\
\ \ $\bullet$ \pwp & \PowhegBox &   NNPDF3.0NLO & \pyt~8.230 & A14 & \\
\ \ $\bullet$ \pwh & \PowhegBox &   NNPDF3.0NLO & \her~7.0.4  & H7-UE-MMHT  &     \\
$-$ \wjets & \sher~2.2.1 &    NNPDF3.0NNLO         & \sher\ & \sher\      &
$20100\pm 1000$ pb (NNLO) \\
$-$ \zjets & \sher~2.2.1 &    NNPDF3.0NNLO         & \sher\ & \sher\      &
$1906\pm 95$ pb  (NNLO)         \\
$-$ diboson & \sher~2.2.2 &   NNPDF3.0NNLO         & \sher\         & \sher\      &
$211\pm 8$ pb   (NLO)        \\
$-$ $\tt V$  &  \mgamcatnlo~2.3.3 &   NNPDF3.0NLO  & \pyt~8.230   & A14 &
$1.5\pm 0.1$ pb (NLO)          \\
$-$ $\tt H$  &  \PowhegBox &   NNPDF3.0NLO  & \pyt~8.230   & A14          &
$0.51\pm 0.05$ pb (NLO)           \\ \bottomrule
\end{tabular}
}}}
\end{picture}
\label{tab:MC}
\end{table*}
 
\subsection{Theoretical uncertainties in the signal predictions}
\label{thunc}
 
The theoretical uncertainties in the differential cross-section
predictions are estimated in the following way:
 
\begin{itemize}
\item The uncertainties in the QCD predictions from \pwp\ and \pwpm\
due to higher-order terms are estimated by repeating the calculations
using values of $\muR$ and $\muF$ scaled by factors $0.5$ and
$2$; the scales are varied independently.
\item The uncertainty in the QCD predictions from \pwp\ due to the
uncertainty in the proton PDFs is estimated using the $100$ replicas
from the NNPDF3.0 analysis~\cite{jhep:04:040}.
\item The uncertainty in the QCD predictions from \pwp\
due to the uncertainty in $\as$ is estimated by repeating the
calculations using two additional sets of proton PDFs from the
NNPDF3.0 analysis, for which different values of $\as$ at the mass
of the $Z$ boson are assumed in the fits, namely $0.117$ and
$0.119$; in this way, the correlation between $\as$ and the PDFs is
preserved.
\item The uncertainties in the QCD predictions from \pwp\ and \pwpm\
due to the amount of initial- and final-state radiation in the
PS are assessed by varying the corresponding parameter of
the A14 PS tune (Var3c) and by varying the scale
$\muR^{\mathrm{FSR}}$ by factors of $2$ and $0.5$.
\end{itemize}
 
The total theoretical uncertainty in the NLO differential cross-section
predictions of \Pow\ is obtained by adding in quadrature the
uncertainties listed above and it is dominated by the contribution
from the scale variations. The total theoretical uncertainty is
approximately $10\%$ to $15\%$ for the theoretical predictions from
\Pow\ as functions of the angular observables, whereas for the
transverse-momentum and invariant-mass observables the uncertainty
rises to approximately $34\%$ to $42\%$ in the tail of the distributions.
The uncertainty from the scale variations in the \amc\ predictions is
very similar to that of \Pow. For the \sher~2.2.12 predictions, it is
a factor between approximately one and three times larger than for
\Pow. This increase in the uncertainty for the \sher\ calculation is
expected due to the presence of a matrix element calculation with up
to four additional partons at LO. These LO terms lead to larger
uncertainties when varying the scales. For \Pow\ and \amc\ these LO
terms are absent and handled by the PS at lower accuracy and,
when comparing these uncertainties with those from \sher~2.2.12,
it is necessary to take into account the fact that these uncertainties
in \Pow\ and \amc\ are only partly evaluated since not all
contributions to the uncertainty can be estimated.
 
For \pwpm, only scale variations are available; in these calculations
at NNLO, a significant reduction of the scale uncertainty relative
to those for the NLO calculations is obtained. The uncertainties due
to the PDFs and $\as$ are expected to be similar to, or smaller than,
those affecting the NLO prediction~\cite{pr:d109:054027}; the uncertainty
due to the scale variations in the NNLO prediction is still larger
than those arising from the PDFs and $\as$ uncertainties. For the
comparison with the data in Section~\ref{res}, the total uncertainty
associated with the \pwpm\ prediction is taken as the sum in
quadrature of the scale uncertainties, the uncertainties in the
PS and, as an approximation, the uncertainties due to the PDFs and
$\as$ as estimated from \pwp\ at NLO.
 
\subsection{Background simulation samples}
\label{sec:BackgroundMC}
Several processes can produce the same final state as the $\tt$
\ljets\ decay mode. The events produced by these backgrounds need to be
estimated and subtracted from the data to determine the top-quark pair
cross-sections. The MC simulation used for these background processes
are described below:
\begin{itemize}
\item Single top-quark production processes in the $t$- and $s$-channels and in
association with a $W$ boson were modelled at NLO in
QCD using \PowhegBox\ with the NNPDF3.0NLO PDF set. The PS and
hadronisation effects were  simulated with \pyt~8.230,
using the A14 tune and the NNPDF2.3LO PDF set. The $t$-channel
sample was generated with the four-flavour scheme. The
diagram-removal scheme (DR)~\cite{jhep:07:029} was employed in the
$tW$ simulation to handle the interference with $\tt$
production~\cite{ATL-PHYS-PUB-2016-020}. The single-top-quark
cross-section for the $tW$ channel was normalised using its NLO+NNLL
prediction, while the $t$- and $s$-channels were normalised using
their NLO
predictions~\cite{pr:d83:091503,pr:d82:054018,pr:d81:054028,cpc:182:1034,cpc:191:74}.
Events generated using the diagram-subtraction
method~\cite{epj:c71:1547} (DS) instead of the DR technique were
considered as a systematic uncertainty in the modelling of this
background (see Section~\ref{syst}). In addition, events generated
with \PowhegBox\ and interfaced to \her~7.0.4 were also
generated using the H7-UE-MMHT tune~\cite{jhep:10:113} and the
MMHT2014LO PDF.
\item Events containing a $W$ or $Z$ boson in association with jets
(\wjets\ and \zjets) were simulated with the
\sher~2.2.1 generator using NLO-accurate MEs for up
to two jets, and MEs accurate to LO for up to four
jets calculated with the \textsc{Comix} and \textsc{OpenLoops}
libraries. They were matched to the \sher\ PS
using the ME+PS@NLO prescription and the tune developed by the \sher\
authors with the NNPDF3.0NNLO PDF set. The \wjets\ and \zjets\
predictions were normalised to the NNLO
cross-sections~\cite{pr:d69:094008}.
\item Diboson production ($WW$, $ZZ$ and $WZ$),
with one of the bosons decaying hadronically and the other
leptonically, was generated using \sher~2.2.2 with MEs computed at
NLO accuracy in QCD with up to one additional parton and at LO
accuracy with up to three additional partons. The NNPDF3.0NNLO PDF
set was used. These predictions were normalised to the NLO
cross-sections evaluated by the generator.
\item The \mgamcatnlo~2.3.3 generator was used for the production of
$\tt V$, with $V=W,\ Z$. The events were modelled at NLO with the
NNPDF3.0NLO PDF set. \pyt~8.230 with the A14 tune and the NNPDF2.3LO
PDF set was used to simulate the PS. These predictions were
normalised to the NLO cross-sections evaluated by the generator.
\item The production of $\tt H$ events was modelled at NLO
using the \PowhegBox\ generator with the NNPDF3.0NLO PDF
set. \pyt~8.230 with the A14 tune and the NNPDF2.3LO PDF set was used
to simulate the PS. These predictions were normalised to the
fixed-order calculation including NLO QCD and electroweak
corrections~\cite{1610.07922}.
\end{itemize}
 
\section{Object reconstruction and event selection}
The following sections describe the detector- and particle-level
objects used to characterise the final-state event topology and to
define the fiducial phase-space regions for the measurements.
 
\subsection{Detector-level object reconstruction,
identification,
calibration and selection}
\label{sec:objects:reco}
Primary vertices are formed from
reconstructed tracks that are spatially compatible with the
interaction region. The hard-scatter primary vertex is chosen to be
the one with at least two associated tracks and the highest $\sum
\pt^2$, where the sum extends over all tracks with $\pt>0.5$~\GeV\
matched to the given vertex~\cite{ATL-PHYS-PUB-2015-026}.
 
Electron candidates are reconstructed by matching tracks in the ID to
energy deposits in the EM calorimeter~\cite{jinst:14:P12006}. The
electron candidates are required to have a transverse momentum
$\pt>27$~\GeV\ and a pseudorapidity $|\eta|<2.47$, excluding the
transition region between the barrel and endcap calorimeters
($1.37<|\eta|<1.52$). The longitudinal impact parameter $z_0$ of the
associated track is required to satisfy $|z_0\sin\theta|<0.5$~mm,
where $\theta$ is the polar angle of the track, and the transverse
impact parameter significance $|d_0|/\sigma(d_0)<5$, where $d_0$ is
the transverse impact parameter and $\sigma(d_0)$ is its
uncertainty. The impact parameters $z_0$ and $d_0$ are calculated
relative to the beam spot and the beam line, respectively. The
electron candidates must satisfy a `tight' likelihood-based
identification criterion based on shower shape and track selection
requirements~\cite{jinst:14:P12006}. The candidates must also satisfy
$\pt$- and $\eta$-dependent isolation requirements based on tracks and
energy clusters, defined for the `tight' isolation working
point~\cite{jinst:14:P12006}; these isolation requirements are used to
reduce the background from jets misidentified as prompt leptons (fake
leptons) or due to semileptonic decays of heavy-flavour hadrons
(non-prompt real leptons).
 
Muon candidates are reconstructed by combining tracks reconstructed in
the ID and the MS~\cite{epj:c81:578}. The track $\pt$ is determined
through a global fit to the hits, which takes into account the energy
loss in the calorimeters. The track associated with the muon candidate
must have $|z_0\sin\theta|<0.5$~mm and $|d_0|/\sigma(d_0)<3$.
Candidates are required to have $\pt>27$~\GeV\ and
$|\eta|<2.5$. To reduce the background from muons originating from
heavy-flavour decays inside jets, muons are required to be isolated
using track-quality and isolation criteria similar to those applied to
electrons~\cite{epj:c81:578}.
 
Jets are reconstructed using the anti-$\kt$
algorithm~\cite{jhep:04:063} with radius parameter $R=0.4$ as
implemented in the \fj\ package~\cite{epj:c72:1896}. Jet
reconstruction is performed on particle flow objects
(PFlow)~\cite{epc:c77:466}. The PFlow algorithm directly combines
measurements from both the tracker and the calorimeter to form the
input signals for jet reconstruction, which are intended to
approximate individual particles. They are calibrated to be consistent
with electromagnetic clusters using corrections determined in
simulation and inferred from test-beam data. Jet four-momenta are then
corrected for pile-up effects using the jet-area
method~\cite{jhep:04:005}. To reduce the number of jets originating
from pile-up, an additional selection criterion based on a jet-vertex
tagging (JVT) technique is applied. The JVT is a likelihood
discriminant that combines information from several track-based
variables~\cite{epj:c76:581} and the criterion is only applied to jets
with $\pt<60$~\GeV\ and rapidity $|y|< 2.4$. The jet energy and
direction are calibrated using an energy- and $\eta$-dependent
simulation-based calibration scheme with in situ corrections based on
data~\cite{epj:c81:689}. The jets are selected if they have
$\pt>25$~\GeV and $|y|<2.5$.
 
To identify  jets containing $b$-hadrons, a flavour tagging algorithm
called DL1r~\cite{ATL-PHYS-PUB-2017-013} based on a deep neural
network is used. In this approach, the jet flavours are considered
equally during training and multiple output nodes are used, each node
trained to predict a specific jet flavour, which provides a highly
flexible tagger. The input information is a combination of several
discriminants. Jets are considered as $b$-tagged if the value of the
DL1r-algorithm output is larger than a certain threshold that provides
a $70\%$ $b$-jet tagging efficiency obtained in $\tt$ MC events. The
working point of $70\%$ is selected after studying the number of
events obtained in data, contributions from the different types of
background processes and systematic uncertainties.
 
To remove ambiguities in the identification
of the objects mentioned above, a procedure called `overlap removal'
is applied to assign a unique hypothesis to each object. If a muon
shares a track with an electron, it is likely to have undergone
bremsstrahlung and hence the electron is not selected. To prevent
double-counting of electron energy deposits as jets, the jet closest
to a reconstructed electron is discarded if $\Delta R(\mathrm{jet},e)<0.2$.
Subsequently, to reduce the impact of non-prompt electrons, if
$\Delta R\left(\mathrm{jet},e\right) < 0.4$, then that electron is
removed. In cases where a jet is within $\Delta R\left(\mathrm{jet},\mu\right) = 0.4$
of a muon, if the jet has fewer than three tracks the jet is removed
whereas if the jet has at least three tracks the muon is removed.
 
The missing transverse momentum in
the event, whose magnitude is denoted in the following by $\etmiss$,
is defined as the negative vector sum of the $\pt$
of the reconstructed and calibrated objects in the
event~\cite{2402.05858}. This sum also includes the momenta of the ID
tracks that are matched to the primary vertex but are not associated
with any other reconstructed objects.
 
\subsection{Particle-level object definition}
Particle-level objects are defined in simulated events using only
stable particles, which are defined as those with a proper lifetime
$\tau>30$~ps. The fiducial phase space used for the measurements in
this analysis is defined using a series of requirements applied to
particle-level objects analogous to those used in the selection of the
detector-level objects.
 
Electrons and muons are required to not
originate from a generated hadron in the MC event, either directly or
through a $\tau$-lepton decay. This ensures that the lepton comes from
an electroweak decay without requiring a direct match to a
$W$ boson. Events where the $W$ boson decays into a leptonically
decaying  $\tau$-lepton are accepted. The four-momenta of the bare
leptons are then modified by adding the four-momenta of all photons, not
originating from hadron decay, within a cone of size $\Delta R=0.1$, to
take into account final-state photon radiation. Such `dressed leptons'
are then required to have $\pt>27$~\GeV\ and $|\eta|<2.5$.
 
Particle-level jets are reconstructed using the same
anti-$\kt$ algorithm used at the detector level with $R=0.4$. The
jet-reconstruction procedure takes as input all stable particles,
except for charged leptons and neutrinos not from hadron decay as
described above. Particle-level jets are
required to have $\pt>25$~\GeV\ and $|y|<2.5$.
 
A particle-level jet is identified as a $b$-jet
if a hadron with $\pt>5$~\GeV\ containing a $b$-quark is matched to the
jet through the ghost-matching technique described in
Ref.~\cite{jhep:04:005}.
 
Particle-level objects are subject to
different overlap removal criteria than reconstructed objects.
After lepton dressing and jet clustering, events in which the
separation of the selected charged lepton and a jet is $\Delta R<0.4$
are excluded. This removes $6.5\%$ of the events at particle level.
 
Neutrinos from hadron decays,
either directly or via a $\tau$-lepton decay, are not considered. All
other neutrinos are included. The particle-level missing
transverse momentum is calculated from the four-vector sum of these
selected neutrinos.
 
\subsection{Particle- and detector-level event selection}
\label{sec:selection}
The event selection comprises a set of requirements based on the
general event quality and on the reconstructed objects, defined above,
that characterise the final-state event topology. The event selection
follows closely that of the previous analysis~\cite{epj:c79:1028}. In
addition in this analysis, three channels are selected according to
the number of jets in the final state. The selection described
below is applied to the reconstruction-level objects in the data and
the MC samples. A similar event selection is applied to the
particle-level objects in the MC, except for the trigger and primary
vertex requirements.
 
Events were selected online during data taking by
single-electron~\cite{epj:c80:47} or
single-muon~\cite{jinst:15:P09015} triggers. Multiple triggers were
combined in a logical OR to increase the selection efficiency. The
lowest-threshold triggers utilised isolation requirements to reduce
the trigger rate. The isolated-lepton triggers had $\pt$ thresholds
of $20$~\GeV\ for muons and $24$~\GeV\ for electrons in 2015 data, and
$26$~\GeV\ for both lepton types in 2016, 2017 and 2018 data. They
were complemented by other triggers with higher $\pt$ thresholds but
no isolation requirements to increase the trigger efficiency.
 
Offline, events are required to have a reconstructed
primary vertex with two or more associated tracks and to contain exactly
one reconstructed charged-lepton (electron or muon) candidate with
$\pt>27$~\GeV. Each event is
also required to contain at least four jets with $\pt>25$~\GeV\ and
$|y|<2.5$, of which at least two must be tagged as $b$-jets. This is
the first channel studied, called `$\tt$ inclusive'. The second
channel is selected by requiring at least five jets and called
`$\tt+$1jet'. The third channel is selected by requiring the presence
of at least six jets and called `$\tt+$2jets'. In this way, the $\tt+$2jets
sample is a subset of the $\tt+$1jet and the $\tt+$1jet is a subset of
the $\tt$ inclusive sample.
 
\section{Background determination}
\label{backcon}
After the event selection explained in Section~\ref{sec:selection},
various backgrounds contribute to the event yields. Data-driven
techniques are used to estimate the background that derives from
events containing jets mimicking the signature of charged leptons or
leptons from hadron decays, for which precise enough simulations are not
available. Background from other physics processes, which include
leptons in the final state, are taken into account using the MC
samples described in Section~\ref{sec:BackgroundMC}.
 
Dilepton top-quark pair events (including decays into $\tau$-leptons)
can satisfy the event selection and are considered in the analysis as
signal at both the detector and particle levels. All the $\tt$
semileptonic events that satisfy the fiducial selection described
in Section~\ref{sec:selection} are considered as signal. For this
reason, the leptonically decaying top quark decays into $\tau$-leptons are
considered as signal only if the $\tau$-lepton decays
leptonically. Cases where both top quarks decay into a $\tau$-lepton,
which in turn decays into hadrons, are accounted for
in the multijet background.
 
\subsection{Background determination from MC samples}
The single-top-quark background, comprising $t$-channel,
$s$-channel and $tW$ production modes, constitutes the largest
contribution, amounting to $4.3\%$, $3.6\%$ and $3.1\%$ of the
total event yield for the $\tt$ inclusive, $\tt+$1jet and $\tt+$2jets
channels, respectively.
The \wjets\ background represents the second largest
contribution to the background, amounting to $2.3\%$, $1.9\%$ and
$1.8\%$ for the three channels, respectively.
The background contributions from each of the \zjets, diboson,
$\tt V$ and $\tt H$ samples of events to the total yield is
smaller than $1\%$ for all three channels.
 
\subsection{Background from fake leptons}
Multijet production processes, including the production of hadronically
decaying $\tt$ pairs, have a large cross-section and mimic the
\ljets\ signature due to fake leptons or non-prompt real leptons. The
multijet background is estimated directly from data using a matrix
method~\cite{jinst:18:T11004}. The estimate is based on the introduction
of a `loose' lepton definition, obtained by removing the isolation
requirement and loosening the identification criteria given
in Section~\ref{sec:objects:reco}. The number of fake and non-prompt
leptons contained in the signal region is evaluated by inverting the
matrix that relates the number of `loose' and `tight' leptons to the
number of real and fake leptons. This matrix is built using the
efficiencies for fake and real leptons to satisfy the `tight'
selection. The fake-lepton efficiency is measured using data in
control regions dominated by multijet background with the real-lepton
contribution subtracted using MC simulation. The real-lepton efficiency is
extracted by applying a tag-and-probe technique using leptons from
$Z$ boson decays.
The multijet background (called
fakes) contributes $1.4\%$, $1.3\%$ and $1.3\%$ to
the total event yield of each of the three channels, respectively.
 
\section{Pseudo-top reconstruction algorithm}
\label{sec:reconstruction}
The pseudo-top algorithm~\cite{jhep:06:100} is used to reconstruct
the leptonically and hadronically decaying top quarks. This algorithm reconstructs the
four-momenta of the top quarks and their complete decay chain from
final-state objects, namely the charged lepton (electron or muon),
missing transverse momentum, and four jets, two of which are
$b$-tagged, using the same method both at detector and particle levels
which is described below.
 
First, the two $b$-jet candidates are chosen. In events with more than
two $b$-jet candidates, only the two with the highest $\pt$ values are
considered as the $b$-jets from the decay of the top quarks,
and called `jet-b1' and `jet-b2', with
$p_{\textrm{T}}^{\textrm{jet-b1}}>p_{\textrm{T}}^{\textrm{jet-b2}}$.
The rest of the jets are considered as non $b$-jet candidates.
 
The two non-$b$-tagged jets to form the hadronically decaying $W$
boson are
chosen from the available jets in each event (excluding those already
tagged as coming from a $b$-quark) such that the invariant
mass of the pair is closest to $\mw$ and called `jet-W1' and
`jet-W2', with $\Ptjeta>\Ptjetb$.
 
For the neutrino, the $x$ and $y$ components of its transverse
momentum are taken from the corresponding components of the missing
transverse momentum, whereas the $z$ component is calculated by
imposing the $W$ boson mass constraint on the invariant mass of the
charged-lepton--neutrino system. If the resulting quadratic equation
has two real solutions, the one with the smaller value of $|p_z|$ is
chosen. If the discriminant is negative, only the real part is considered.
 
The leptonically decaying $W$ boson is reconstructed from the charged
lepton and the neutrino. The leptonically decaying top quark is reconstructed from
the leptonically decaying $W$ boson and the $b$-tagged jet closest (jet-b1 or jet-b2)
in $\Delta R$ to the charged lepton. The
hadronically decaying $W$ boson is reconstructed from the two
non-$b$-tagged jets (jet-W1 and jet-W2) whose invariant
mass is closest to the mass of the $W$ boson, as mentioned above, and
the hadronically decaying top quark is reconstructed from the
hadronically decaying $W$ boson and the other $b$-tagged jet. This
choice yields the best performance of the algorithm in terms of the
correspondence between the detector and particle levels.
 
\section{Event yields, observables and detector-level distributions}
\label{observables}
\subsection{Event yields}
The individual \ejets\ and \mjets\ channels have very similar
distributions at detector level, therefore they are
combined by summing the distributions. Table~\ref{tabflow} shows the
final yields from the three channels for the combined \ljets\ decay mode
for data, signal and background events. The selection leads to a
sample with an expected background of $9\%$, $8\%$ and $8\%$ for the
three channels, respectively. The overall difference between the data and
the predictions is $2\%$, $2\%$ and $1\%$ for the three channels,
which means that there is a good agreement between the SM prediction
and the data, after taking into account the systematic uncertainties (see
Section~\ref{syst}).
 
\begin{table}[h]
\caption{
Observed and expected number of events in the \ljets\ decay mode after
the full event selection. The systematic uncertainties (see
Section~\ref{syst}) are also included. Due to
correlations, the quadratic sum of the different systematic
uncertainties might not add up to the total systematic uncertainties.
}
\setlength{\unitlength}{1.0cm}
\begin{picture} (18.0,7.5)
\put (0.0,3.5){\centerline{\scalebox{1.0}{
\begin{tabular}{l r @{ $\pm$ } l r @{ $\pm$ } l r @{ $\pm$ } l}
\toprule
& \multicolumn{6}{c}{Yield for channel} \\
Process & \multicolumn{2}{c}{$\tt$ inclusive} &
\multicolumn{2}{c}{$\tt+$1jet} &
\multicolumn{2}{c}{$\tt+$2jets}
\\ \midrule\midrule
$\tt$ &
$4\ 120\ 000$ & $690\ 000$ &
$2\ 110\ 000$ & $420\ 000$ &
$860\ 000$ & $240\ 000$ \\ \midrule
single-top quark &
$194\ 000$ & $33\ 000$ &
$82\ 000$ & $19\ 000$ &
$28\ 900$ & $8\ 300$ \\ \midrule
$W+$jets &
$103\ 000$ & $53\ 000$ &
$44\ 000$ & $23\ 000$ &
$17\ 200$ & $8\ 900$ \\ \midrule
fakes &
$66\ 000$ & $33\ 000$ &
$30\ 000$ & $15\ 000$ &
$12\ 000$ & $6\ 000$ \\ \midrule
$Z+$jets &
$37\ 000$ & $19\ 000$ &
$14\ 600$ & $7\ 600$ &
$5\ 400$ & $2\ 800$ \\ \midrule
$\tt V$ &
$13\ 100$ & $1\ 100$ &
$10\ 300$ & $900$ &
$6\ 470$ & $660$ \\ \midrule
diboson &
$6\ 000$ & $600$ &
$2\ 820$ & $350$ &
$1\ 220$ & $180$ \\ \midrule
$\tt H$ &
$5\ 460$ & $320$ &
$4\ 710$ & $310$ &
$3\ 260$ & $270$ \\ \midrule\midrule
Total SM prediction &
$4\ 540\ 000$ & $710\ 000$ &
$2\ 300\ 000$ & $440\ 000$ &
$940\ 000$ & $240\ 000$ \\ \midrule
Data &
\multicolumn{2}{l}{$4\ 445\ 113$} &
\multicolumn{2}{l}{$2\ 248\ 410$} &
\multicolumn{2}{l}{$924\ 751$} \\
\bottomrule
\end{tabular}
}}}
\end{picture}
\label{tabflow}
\end{table}
 
\subsection{Observables and detector-level distributions}
Measurements of differential cross-sections in the $\tt$ inclusive,
$\tt+$1jet and $\tt+$2jets channels, as defined in
Section~\ref{sec:selection}, are performed as functions of different
jet observables.
 
In the following, the two jets used to reconstruct the
hadronically decaying $W$ boson are called `jet-W1' and `jet-W2, with
$\Ptjeta>\Ptjetb$ (as defined in Section~\ref{sec:reconstruction}).
For the $\tt+$1jet and $\tt+$2jets channels, the additional jets,
called `jet-rad1' and `jet-rad2', are chosen as follows:
`jet-rad1' is the highest $\pt$ jet of those that are not part
of the $\tt$ system; and `jet-rad2' is the second-highest $\pt$
jet of those that are not part of the $\tt$ system.
 
\begin{figure}[h]
\setlength{\unitlength}{1.0cm}
\begin{picture} (18.0,20.5)
\put (0.0,13.3){\includegraphics[width=8cm]{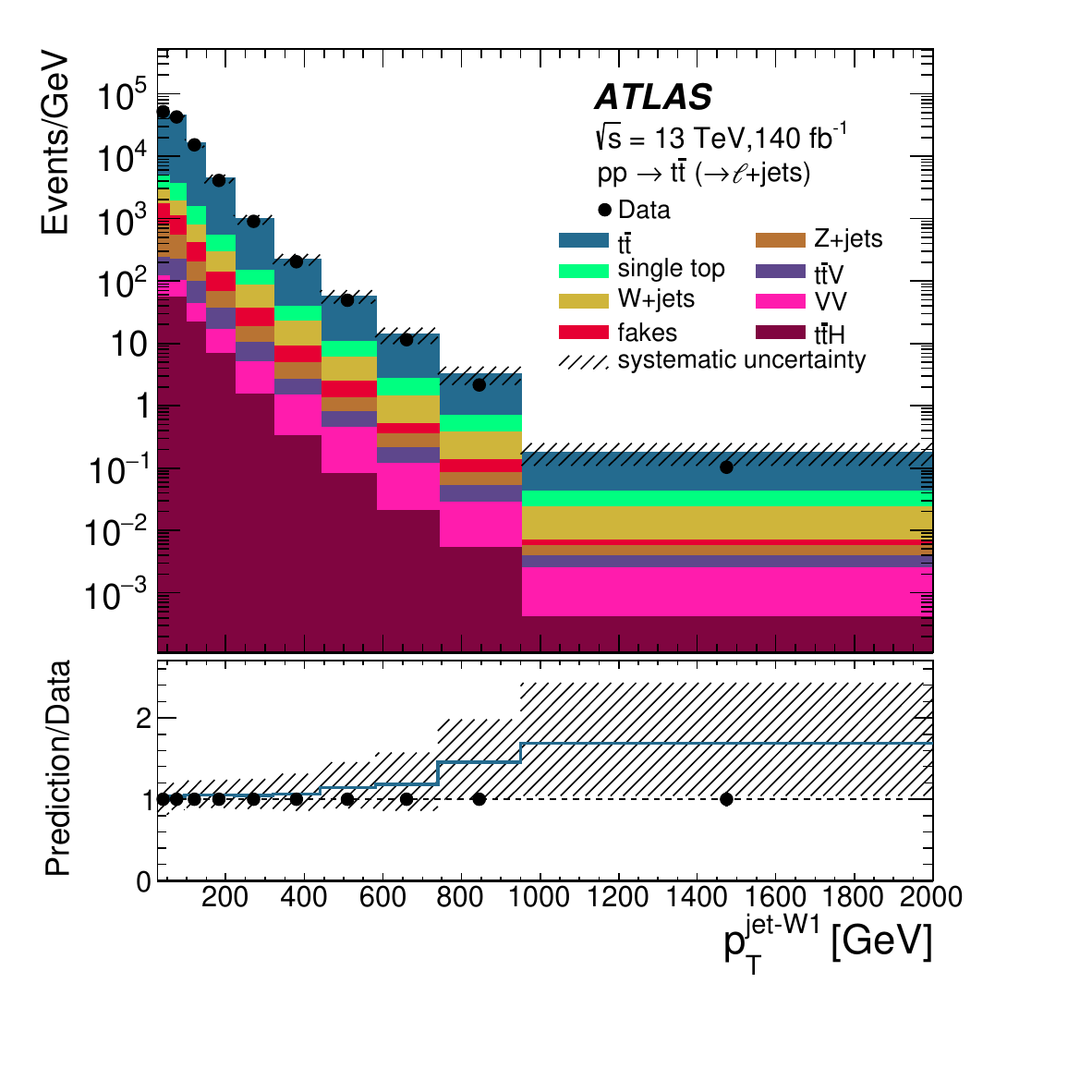}}
\put (8.0,13.3){\includegraphics[width=8cm]{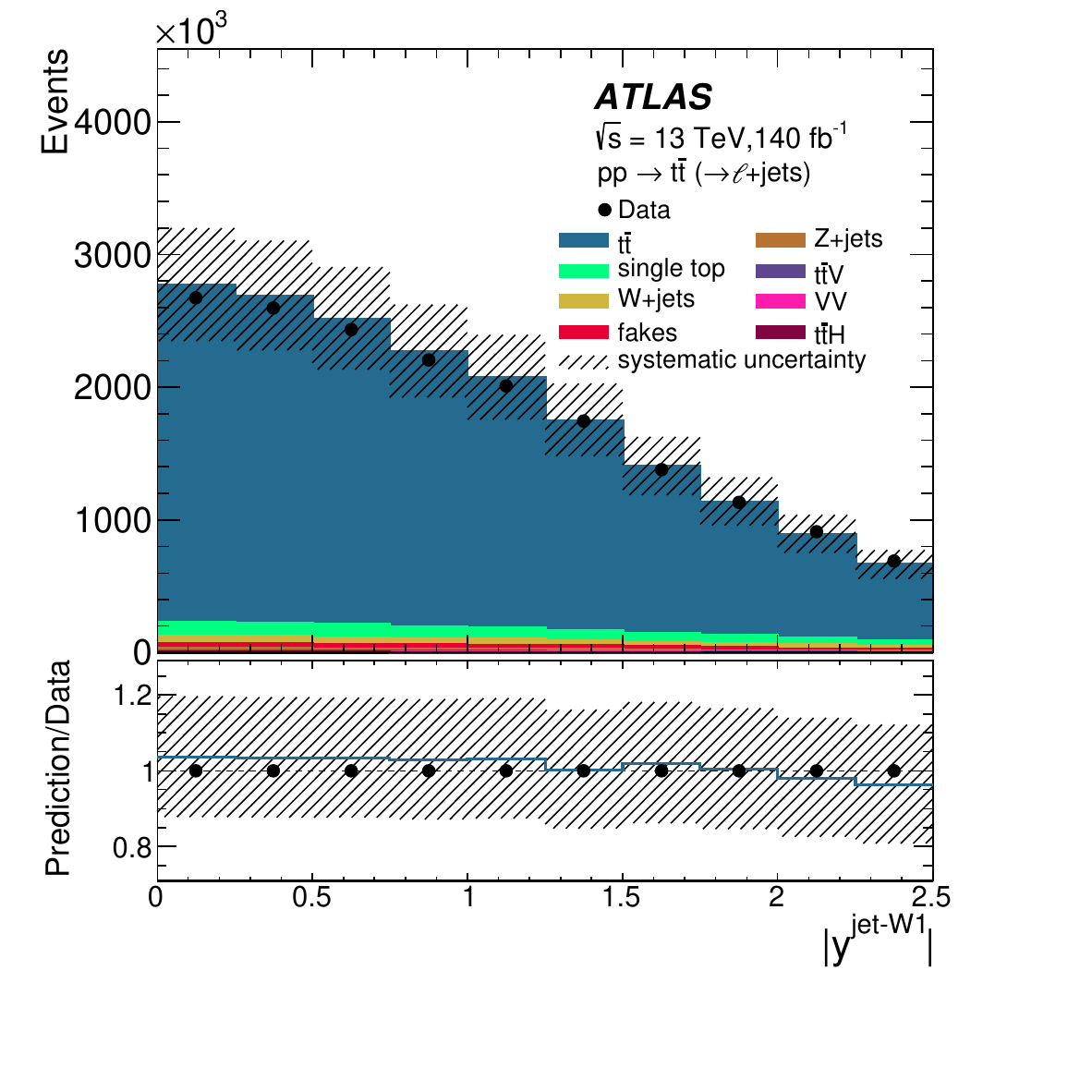}}
\put (0.0,6.3){\includegraphics[width=8cm]{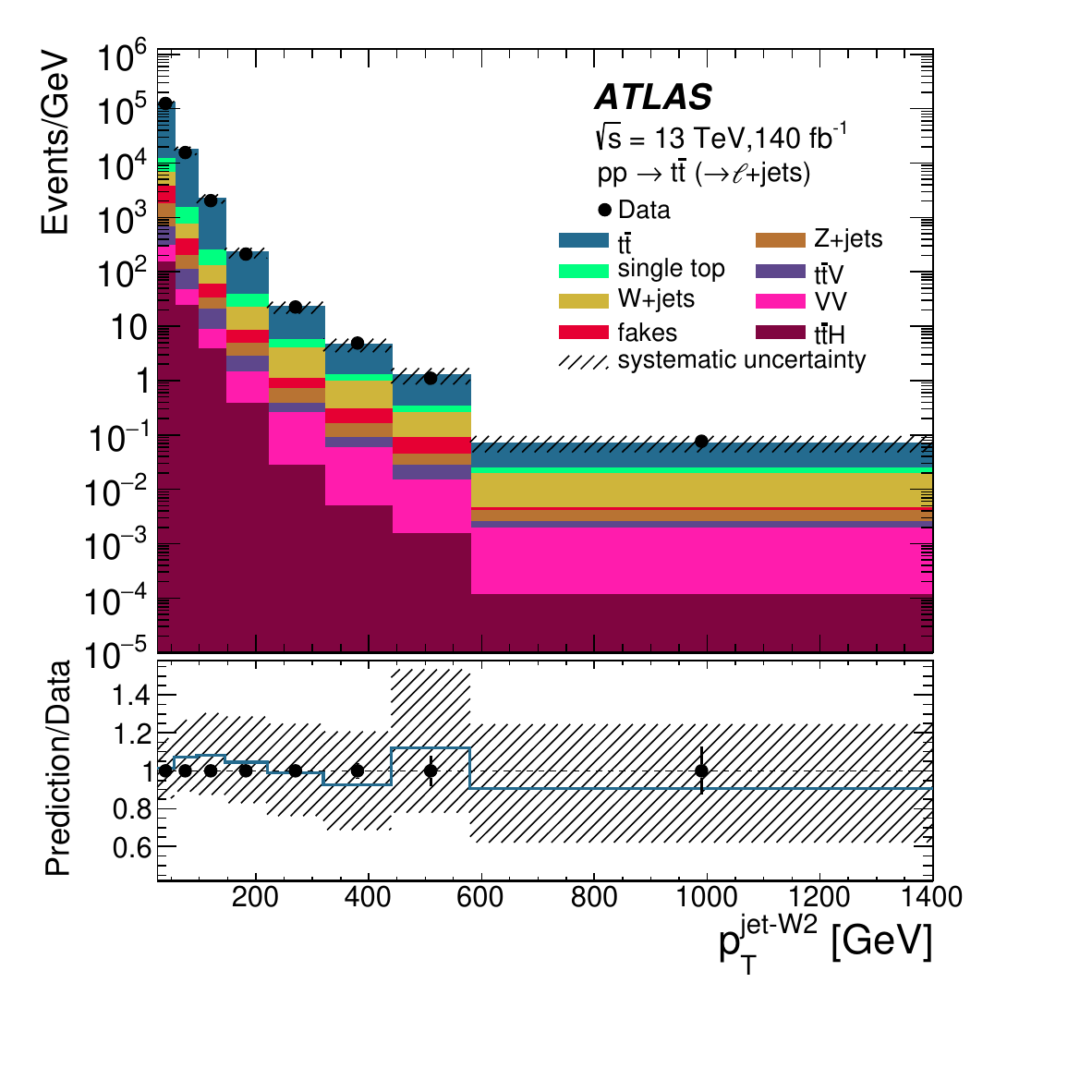}}
\put (8.0,6.3){\includegraphics[width=8cm]{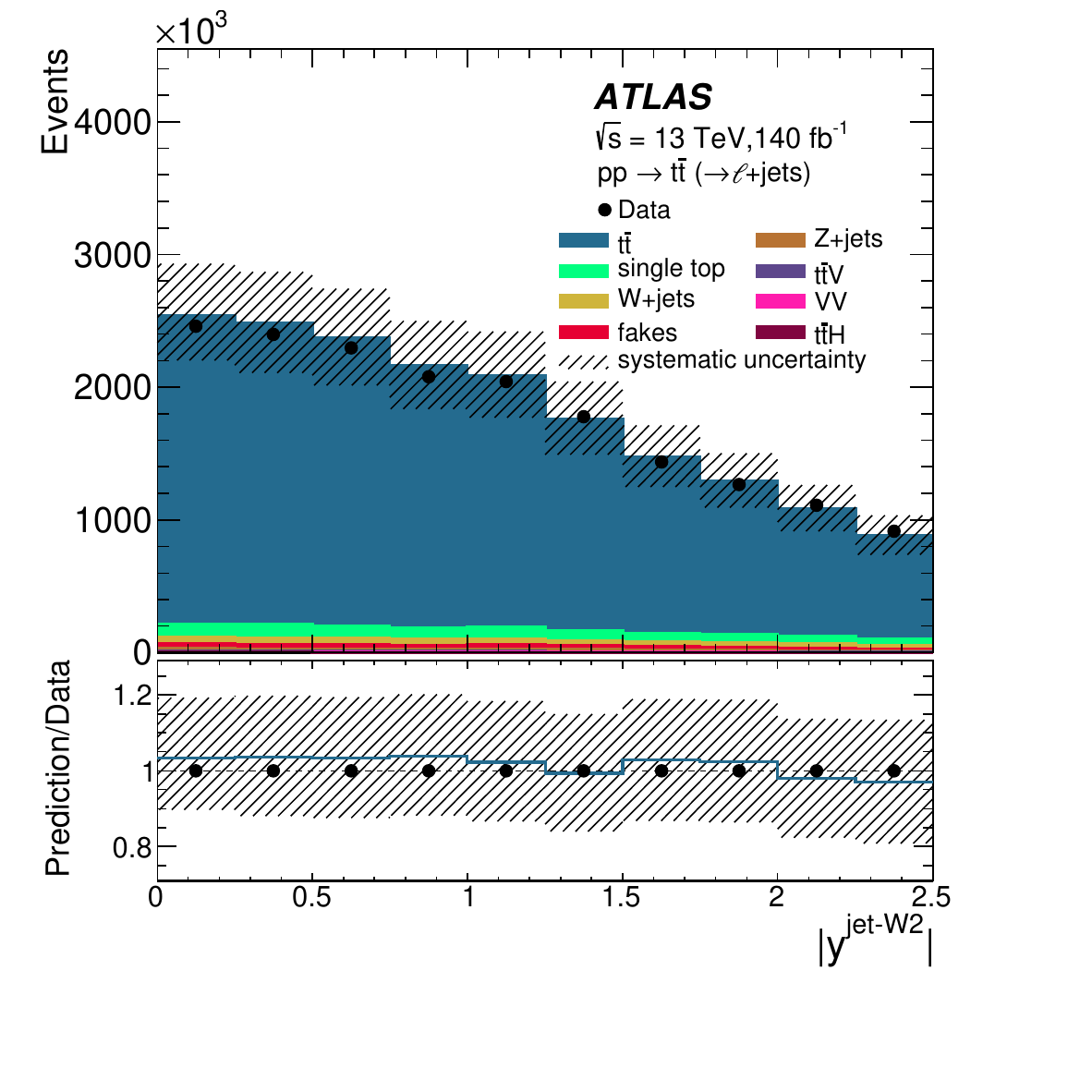}}
\put (0.0,-0.7){\includegraphics[width=8cm]{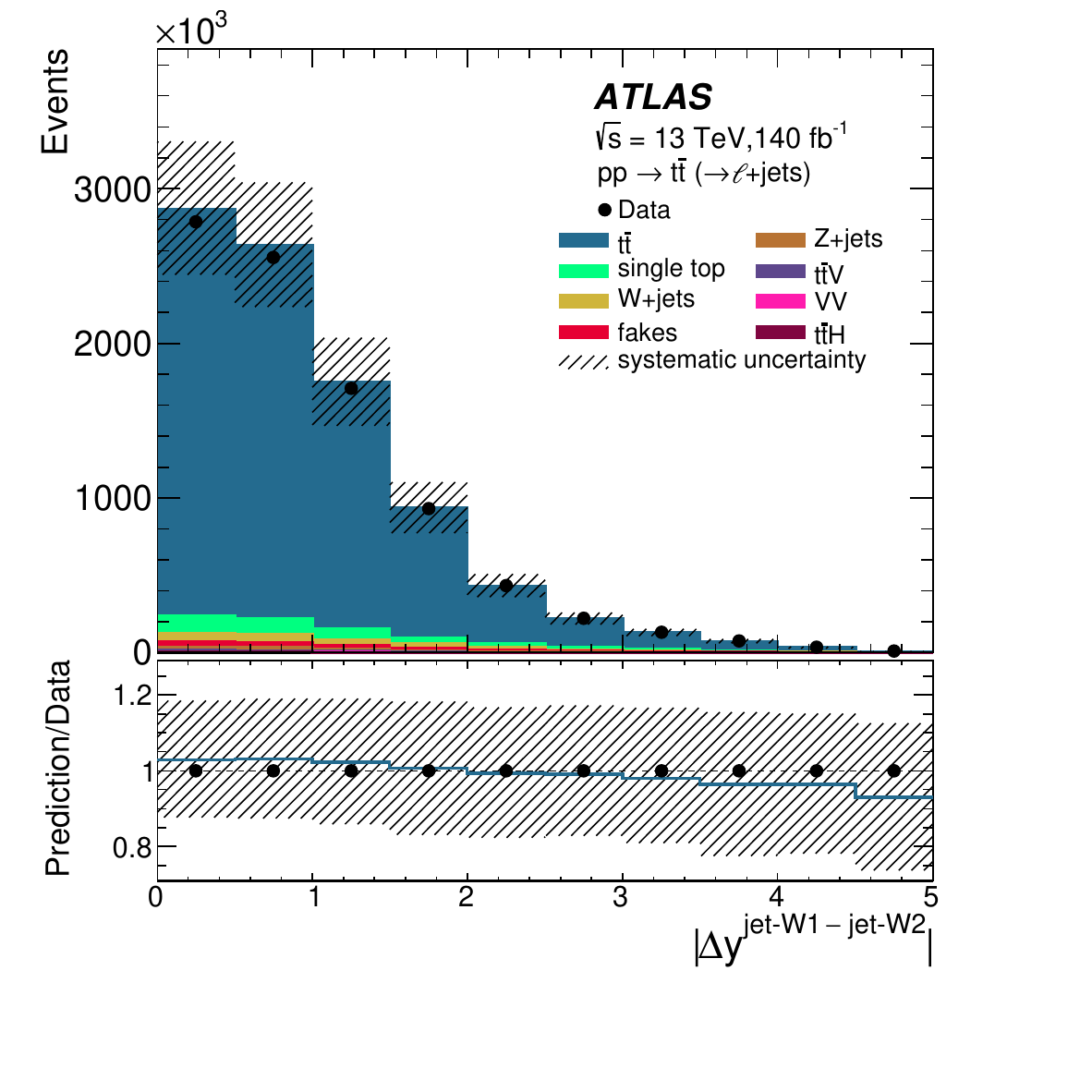}}
\put (8.0,-0.7){\includegraphics[width=8cm]{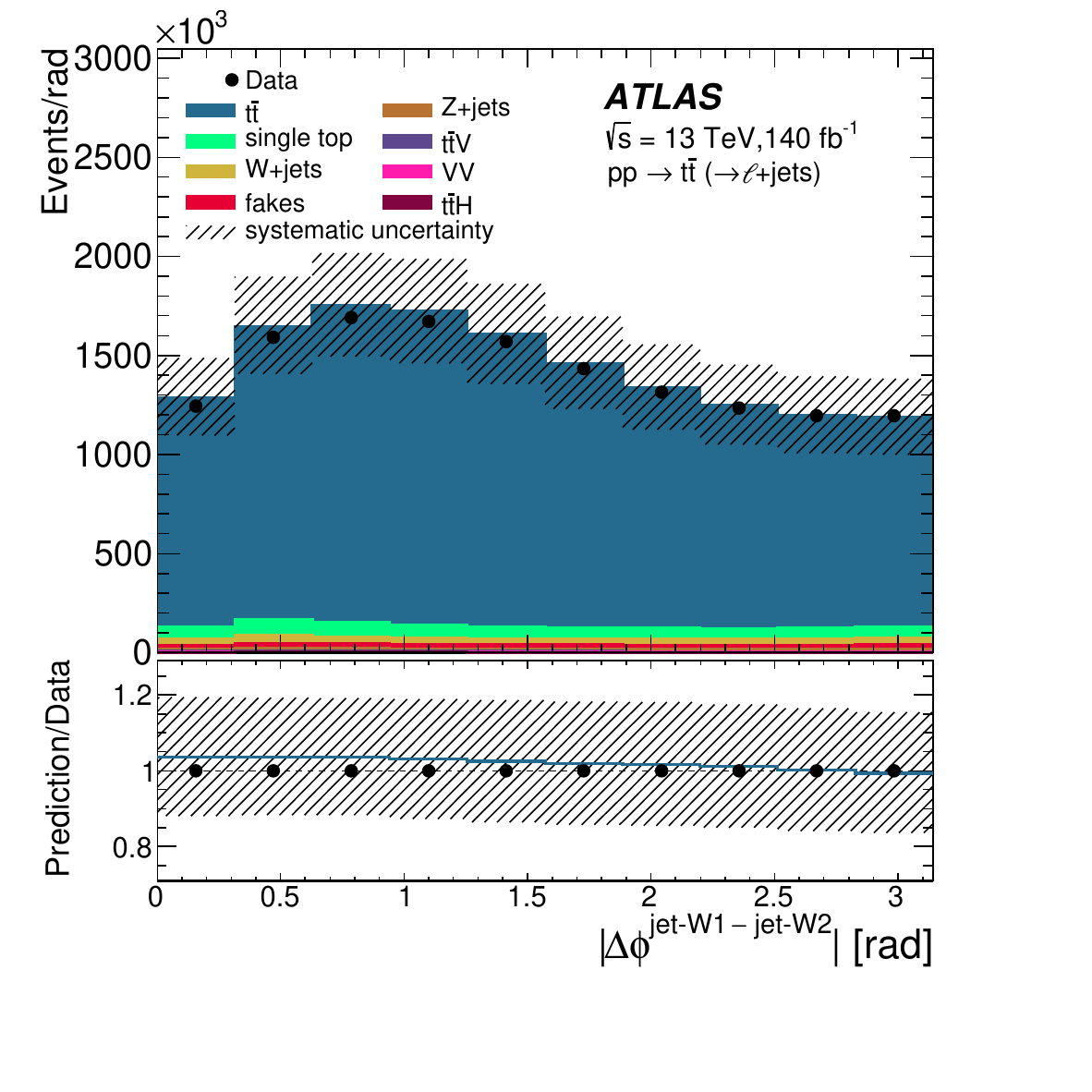}}
\put (3.5,14.2){\small (a)}
\put (11.5,14.2){\small (b)}
\put (3.5,7.2){\small (c)}
\put (11.5,7.2){\small (d)}
\put (3.5,0.2){\small (e)}
\put (11.5,0.2){\small (f)}
\end{picture}
\caption{
The measured (a) $\Ptjeta$, (b) $|\Rapjeta|$,
(c) $\Ptjetb$, (d) $|\Rapjetb|$,
(e) $|\deltayab|$ and (f) $|\deltaphiab|$
distributions divided by the bin width (dots) in the $\tt$ inclusive channel.
The error bars represent the statistical uncertainty in the data; for many
points, the uncertainty is smaller than the marker size and, thus,
not visible.
The hatched bands represent the systematic uncertainties (see
Section~\ref{syst}).
The prediction of the signal and various
backgrounds (histograms) are
also included. The lower panels of each figure show the ratio of the
total prediction and the data distributions.
}
\label{fig10}
\end{figure}

The following observables are measured in the $\tt$ inclusive channel:
\begin{itemize}
\item The kinematics of the two jets forming the $W$ boson are
characterised by measuring the transverse momenta of jet-W1
($\Ptjeta$) and jet-W2 ($\Ptjetb$) and the absolute value of their
rapidities ($|\Rapjeta|$ and $|\Rapjetb|$).
\item The topology of the two jets forming the $W$ boson is
characterised by measuring $|\deltayab|$, defined as the absolute
value of the difference in rapidity between jet-W1 and jet-W2, and
\break $|\deltaphiab|$, defined as the absolute value of the difference in
azimuth between jet-W1 and jet-W2.
\end{itemize}
Figure~\ref{fig10} shows the data, backgrounds and signal MC
distributions at detector level for these observables. Some
differences in shape between the distributions for the data and the SM
predictions are observed, especially at high values of $\Ptjeta$,
which are covered by the systematic uncertainties.
 
\begin{figure}[h]
\setlength{\unitlength}{1.0cm}
\begin{picture} (18.0,20.5)
\put (0.0,13.3){\includegraphics[width=8cm]{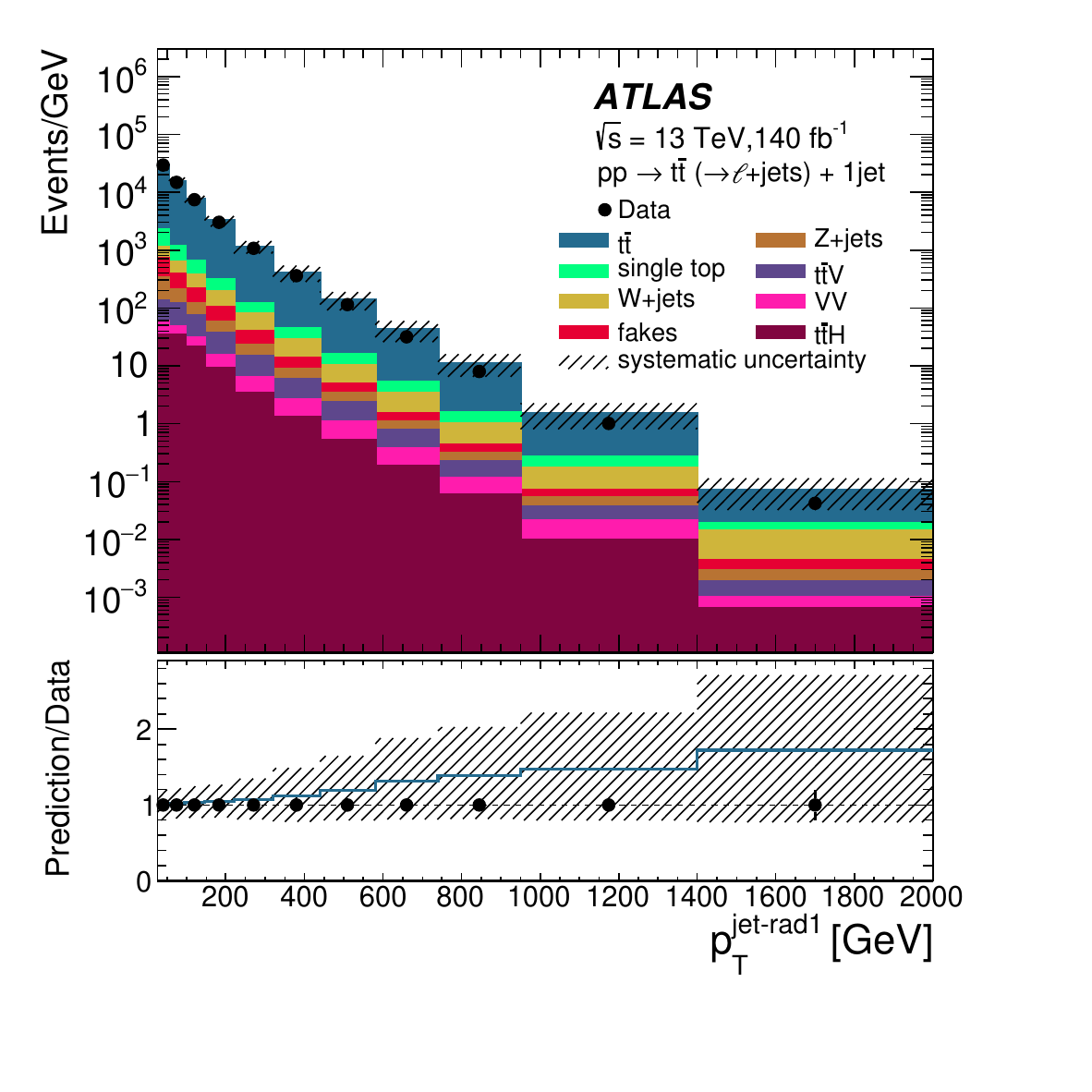}}
\put (8.0,13.3){\includegraphics[width=8cm]{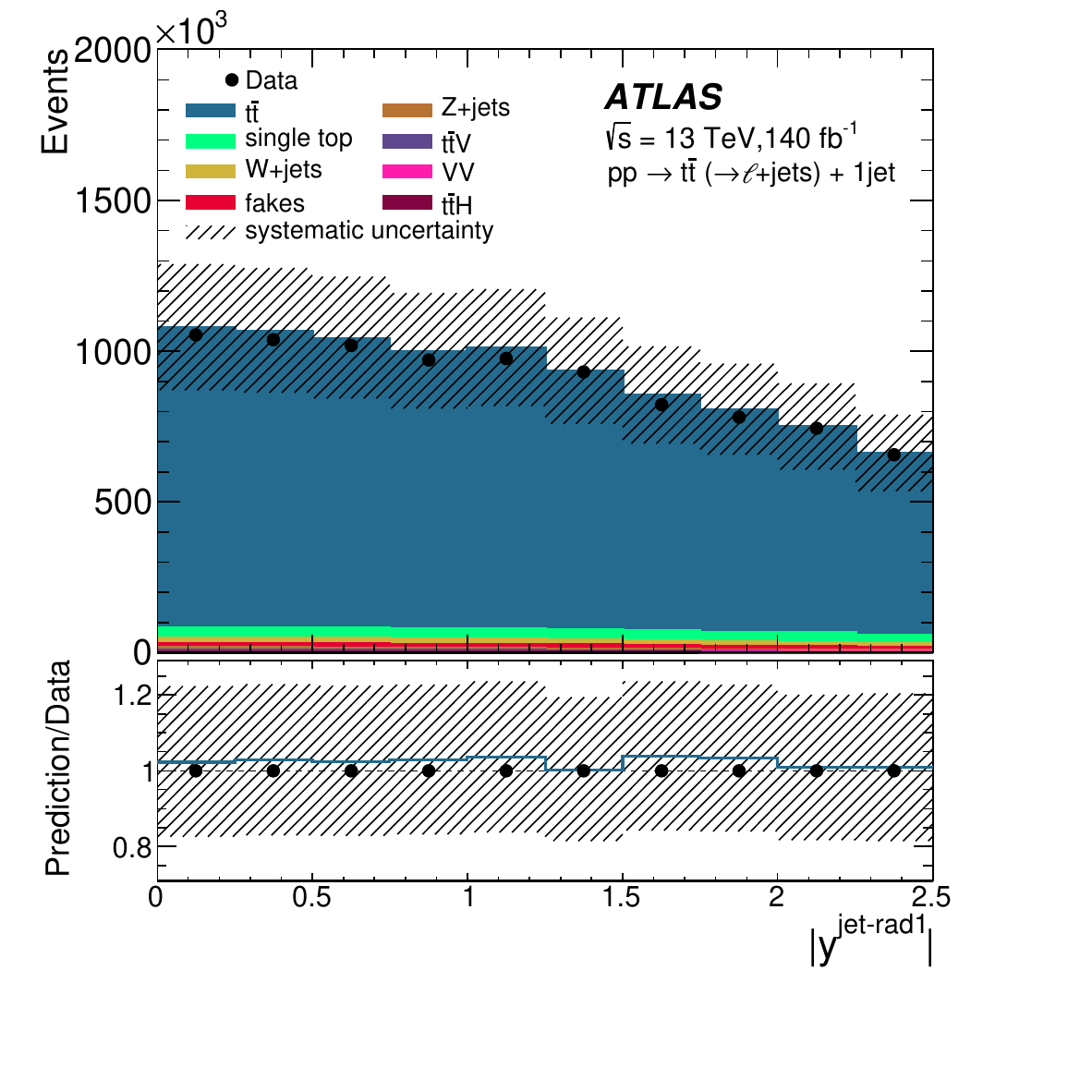}}
\put (0.0,6.3){\includegraphics[width=8cm]{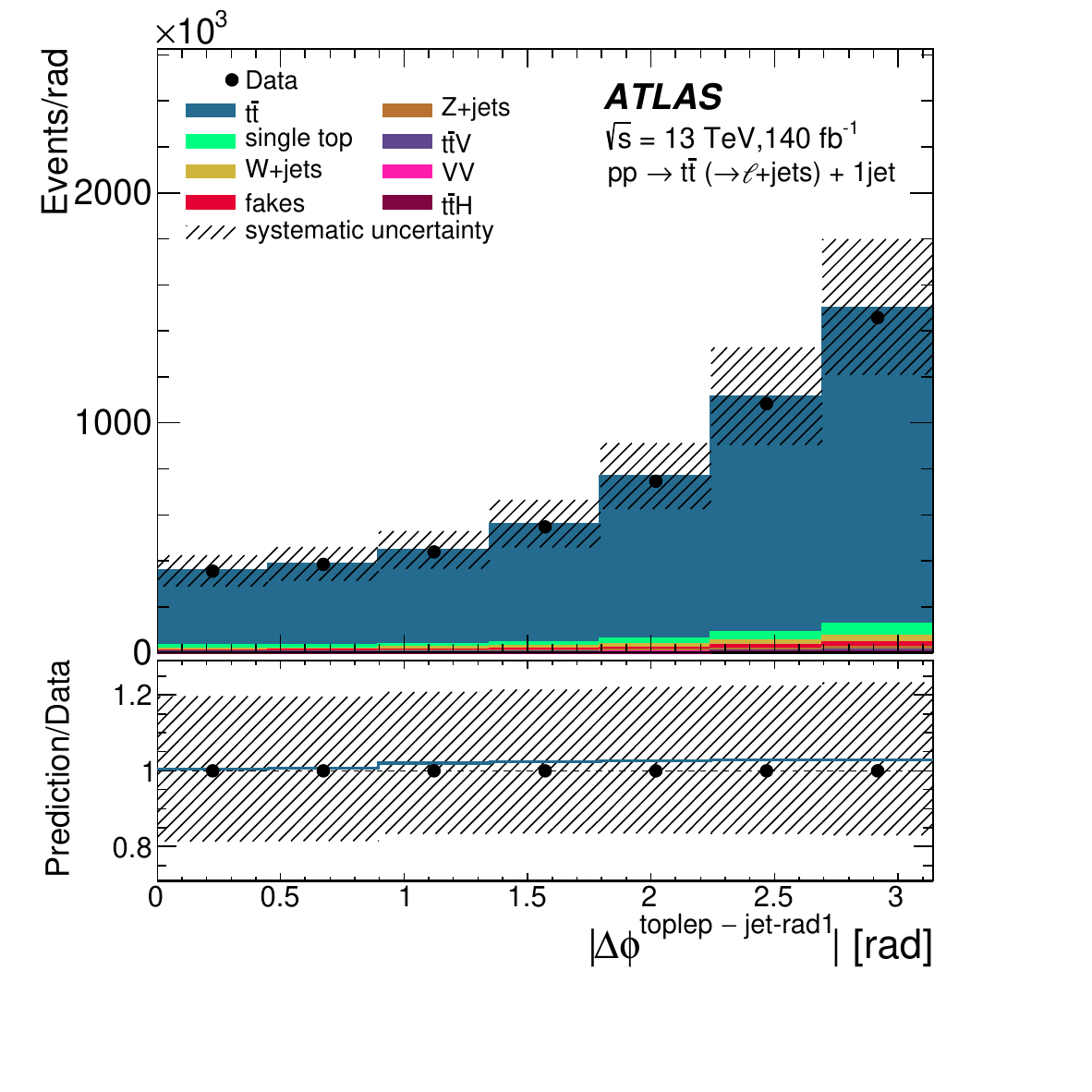}}
\put (8.0,6.3){\includegraphics[width=8cm]{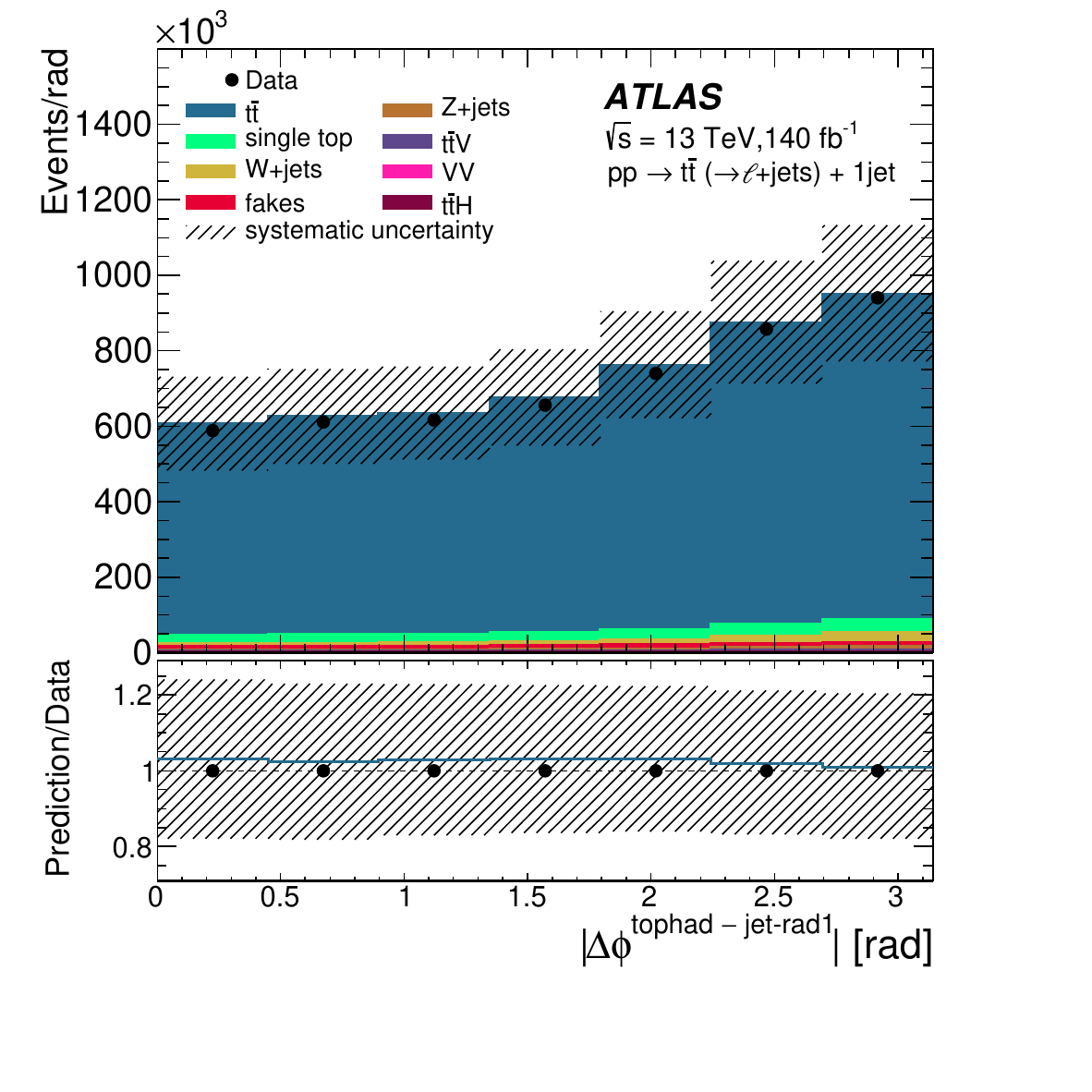}}
\put (0.0,-0.7){\includegraphics[width=8cm]{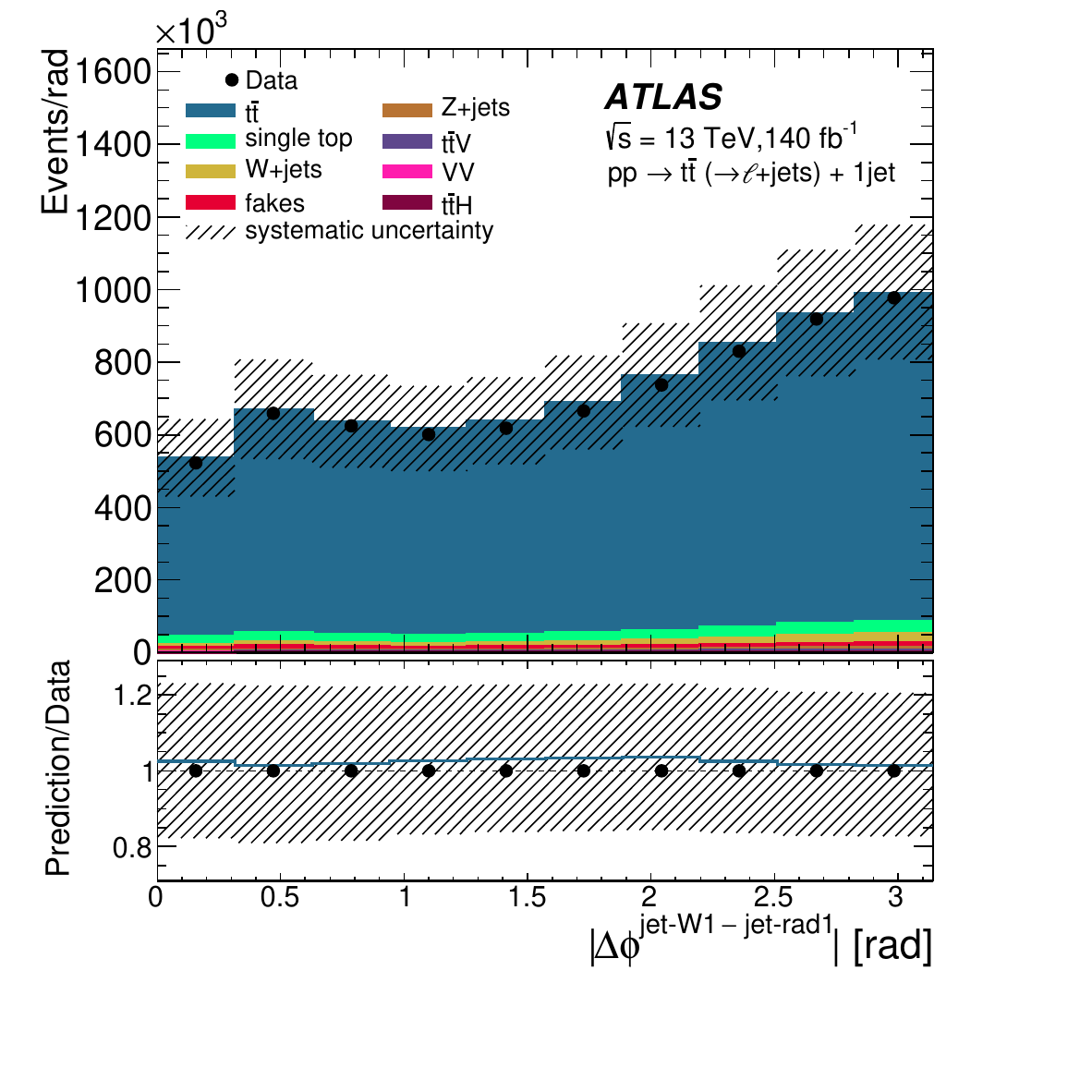}}
\put (8.0,-0.7){\includegraphics[width=8cm]{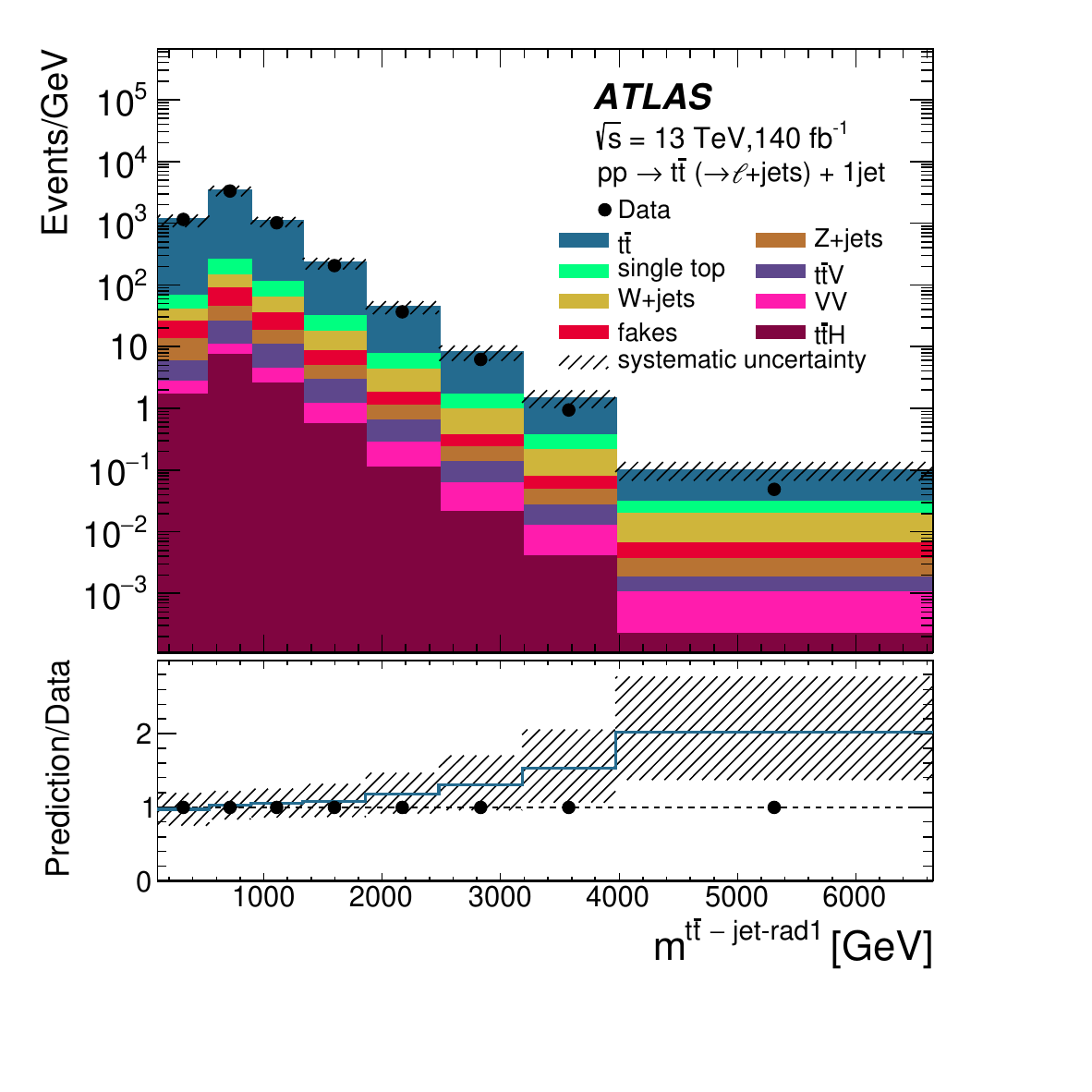}}
\put (3.5,14.2){\small (a)}
\put (11.5,14.2){\small (b)}
\put (3.5,7.2){\small (c)}
\put (11.5,7.2){\small (d)}
\put (3.5,0.2){\small (e)}
\put (11.5,0.2){\small (f)}
\end{picture}
\caption{
The measured (a) $\Ptjete$, (b) $|\Rapjete|$,
(c) $|\deltaphitle|$, (d) $|\deltaphithe|$,
(e) $|\deltaphiae|$ and (f) $\mtte$
distributions divided by the bin width (dots) in the $\tt+$1jet channel.
The error bars represent the statistical uncertainty in the data; for many
points, the uncertainty is smaller than the marker size and, thus,
not visible.
The hatched bands represent the systematic uncertainties (see
Section~\ref{syst}).
The prediction of the signal and various
backgrounds (histograms) are
also included. The lower panels of each figure show the ratio of the
total prediction and the data distributions.
}
\label{fig11}
\end{figure}

The following observables are measured in the $\tt+$1jet channel:
\begin{itemize}
\item The kinematics of the first gluon emission is characterised by
measuring the transverse momentum of jet-rad1 ($\Ptjete$) and the
absolute value of its rapidity ($|\Rapjete|$).
\item The topology of the first gluon emission is characterised by
measuring $|\deltaphitle|$, defined as the absolute value of the
difference in azimuth between the leptonically decaying top quark (toplep) and
jet-rad1, and   $|\deltaphithe|$, defined as the absolute value of
the difference in azimuth between the hadronically decaying top
quark (tophad) and
jet-rad1. In addition, the absolute value of the difference in
azimuth between jet-W1 and jet-rad1 ($|\deltaphiae|$) provides additional
insight into the topology of these events.
\item The dynamics of the $\tt+$1jet events is characterised by
measuring the invariant mass of the $\tt-$jet-rad1 system ($\mtte$).
\end{itemize}
Figure~\ref{fig11} shows the data, backgrounds
and signal MC distributions at detector level for these
observables. There is an adequate description of the measured distributions
by the SM prediction; the $\mtte$ measured distribution is described
by the SM prediction, except for the $\mtte>4$~\TeV\ region.
 
\begin{figure}[h]
\setlength{\unitlength}{1.0cm}
\begin{picture} (18.0,17.5)
\put (0.0,6.3){\includegraphics[width=8cm]{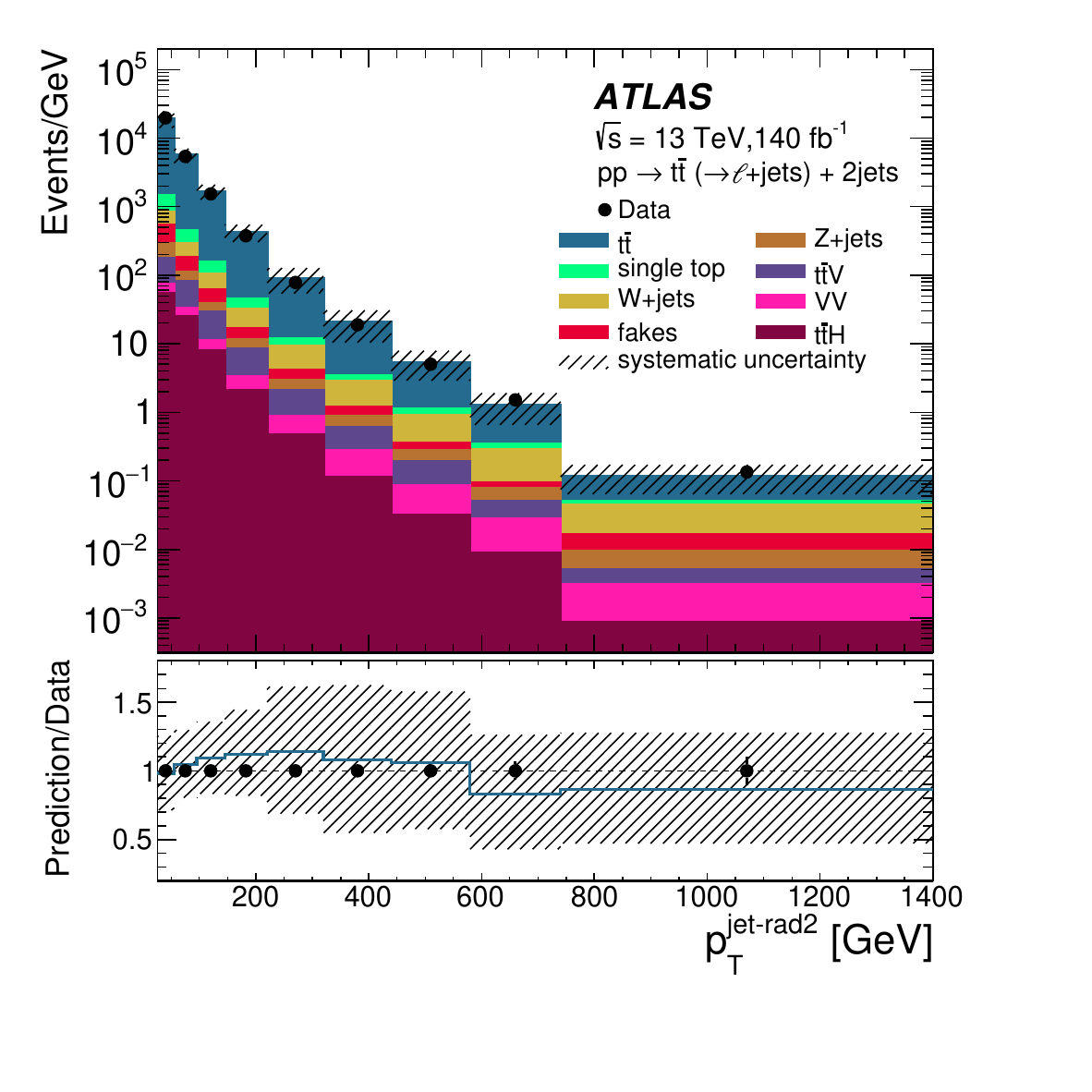}}
\put (8.0,6.3){\includegraphics[width=8cm]{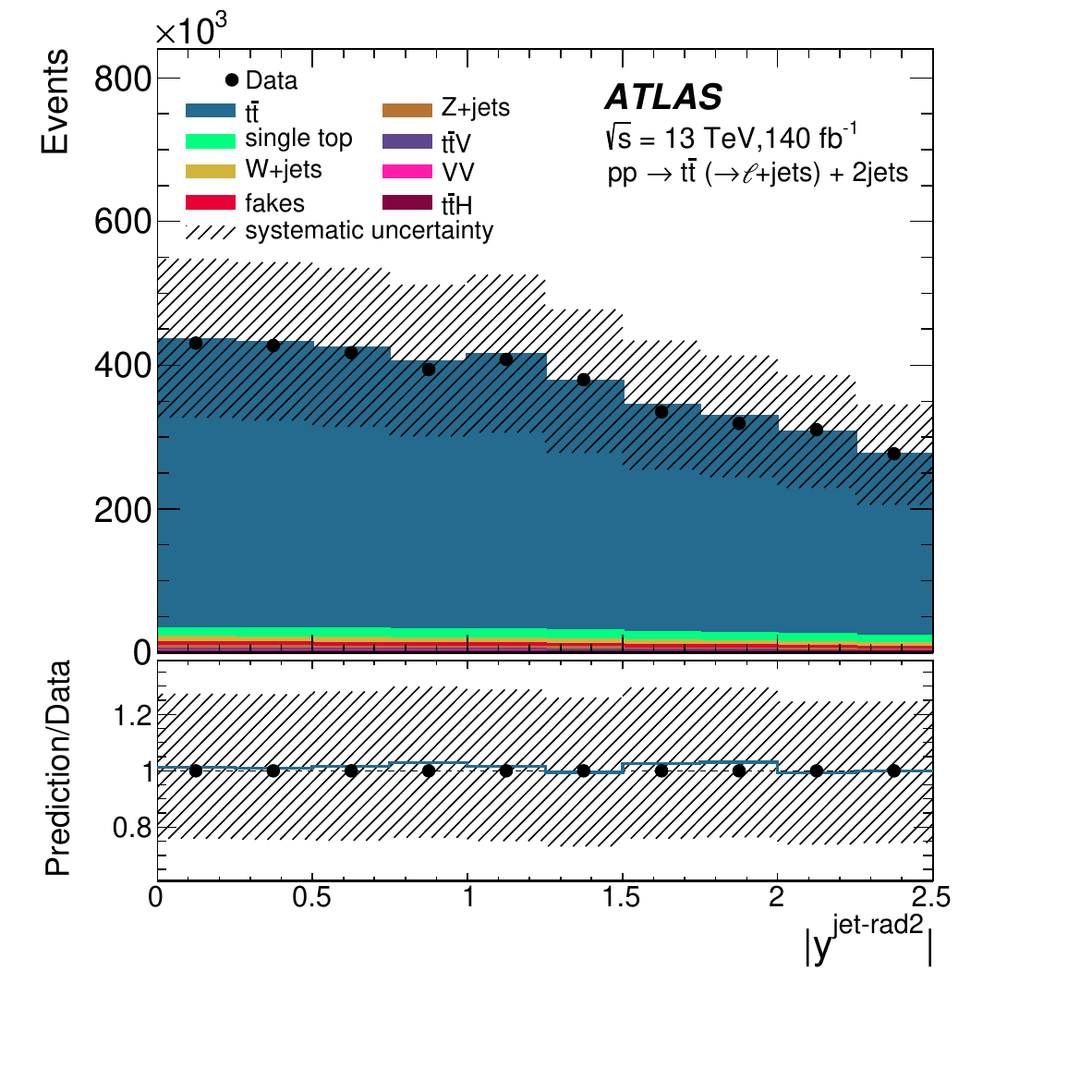}}
\put (0.0,-0.7){\includegraphics[width=8cm]{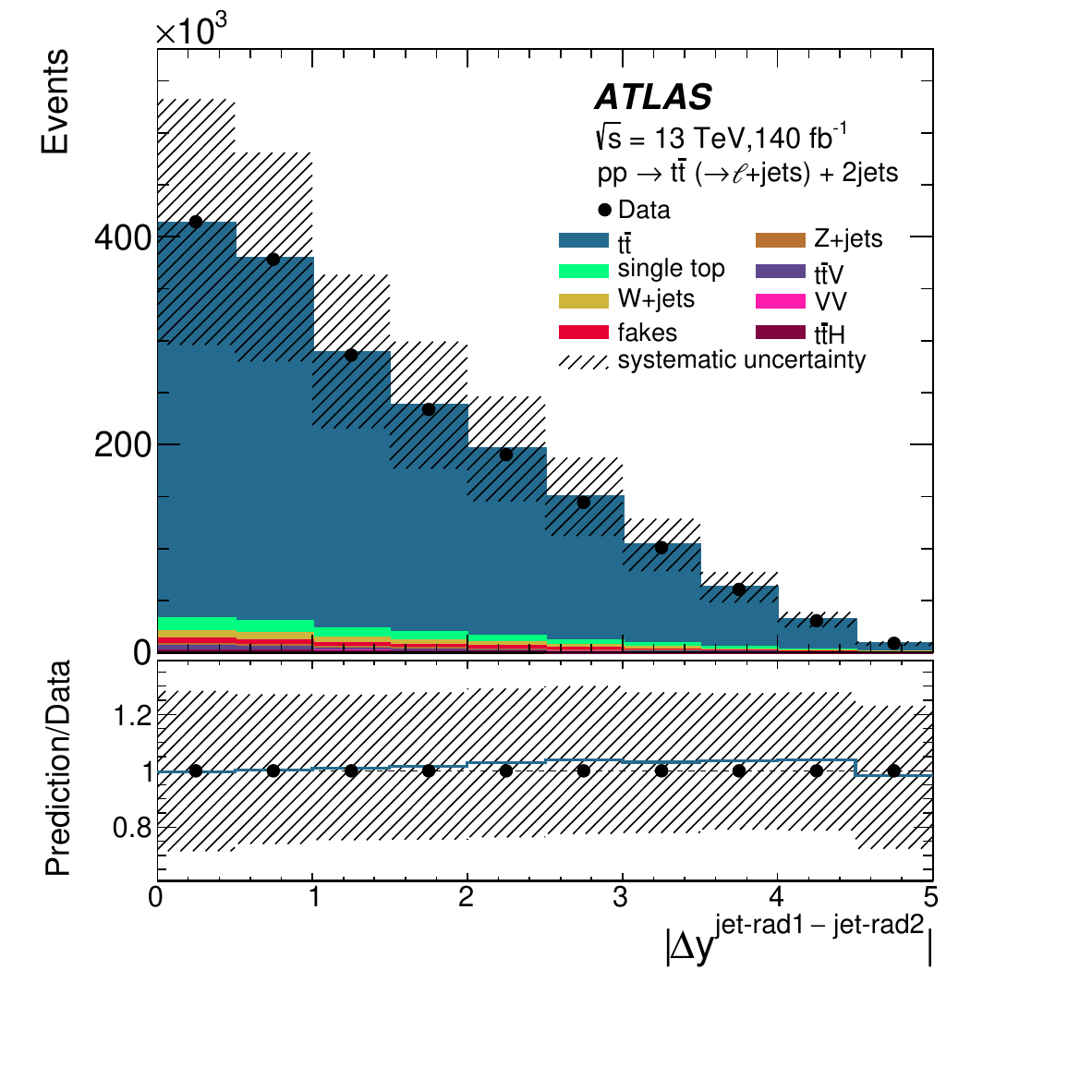}}
\put (8.0,-0.7){\includegraphics[width=8cm]{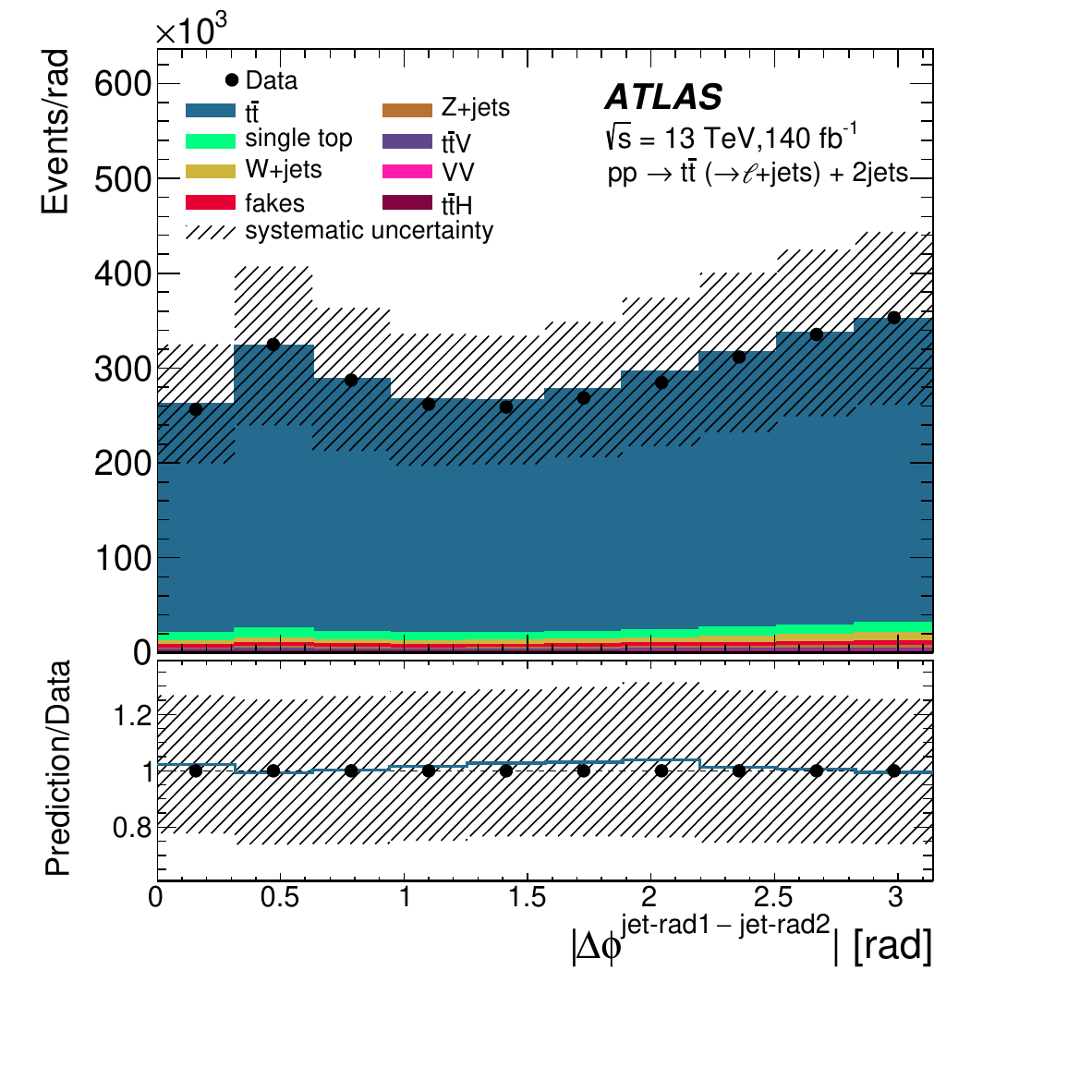}}
\put (3.5,7.2){\small (a)}
\put (11.5,7.2){\small (b)}
\put (3.5,0.2){\small (c)}
\put (11.5,0.2){\small (d)}
\end{picture}
\caption{
The measured (a) $\Ptjetf$, (b) $|\Rapjetf|$,
(c) $|\deltayef|$ and (d) $|\deltaphief|$
distributions divided by the bin width (dots) in the $\tt+$2jets channel.
The error bars represent the statistical uncertainty in the data; for many
points, the uncertainty is smaller than the marker size and, thus,
not visible.
The hatched bands represent the systematic uncertainties (see
Section~\ref{syst}).
The prediction of the signal and various
backgrounds (histograms) are
also included. The lower panels of each figure show the ratio of the
total prediction and the data distributions.
}
\label{fig12}
\end{figure}

\begin{figure}[h]
\setlength{\unitlength}{1.0cm}
\begin{picture} (18.0,17.5)
\put (0.0,6.3){\includegraphics[width=8cm]{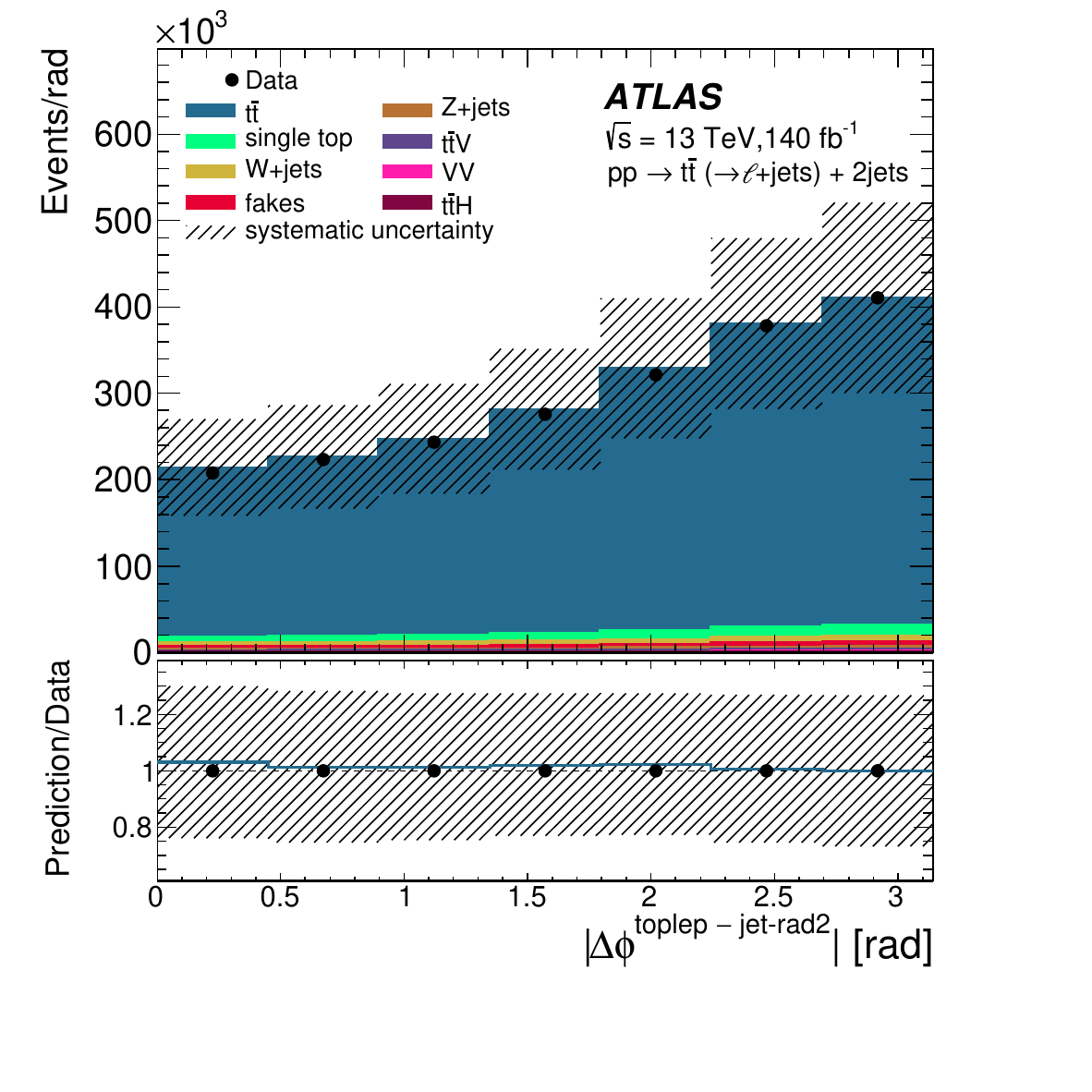}}
\put (8.0,6.3){\includegraphics[width=8cm]{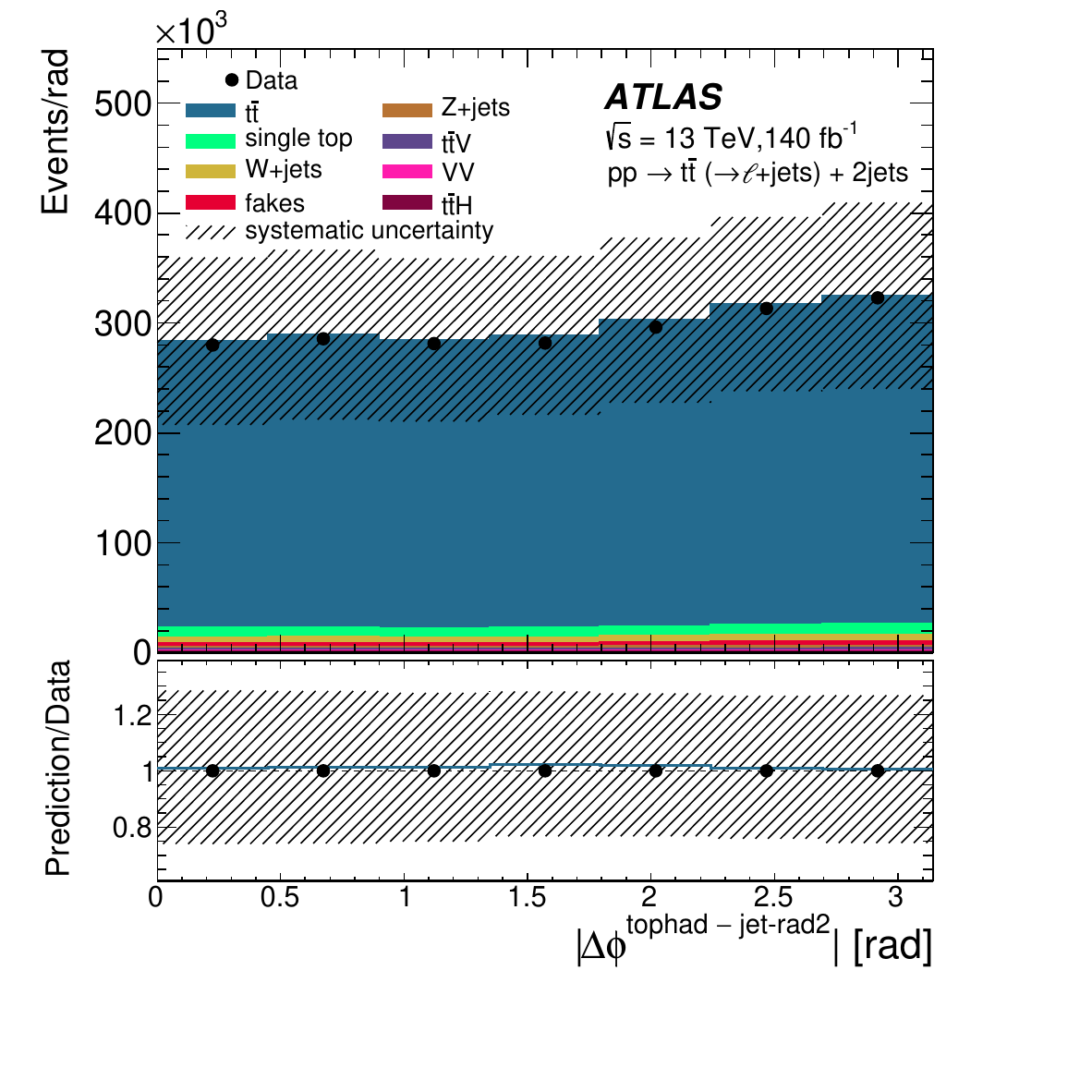}}
\put (0.0,-0.7){\includegraphics[width=8cm]{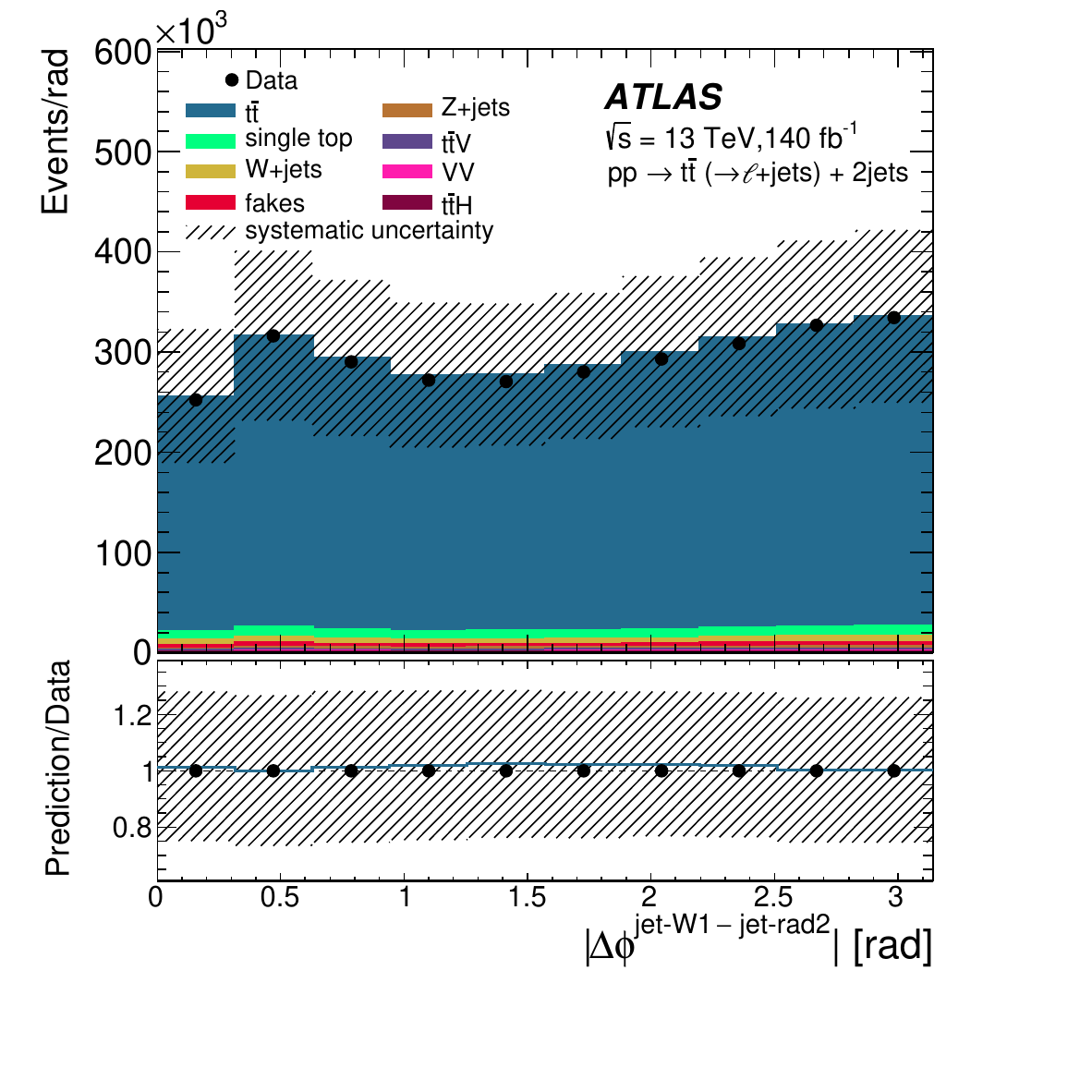}}
\put (8.0,-0.7){\includegraphics[width=8cm]{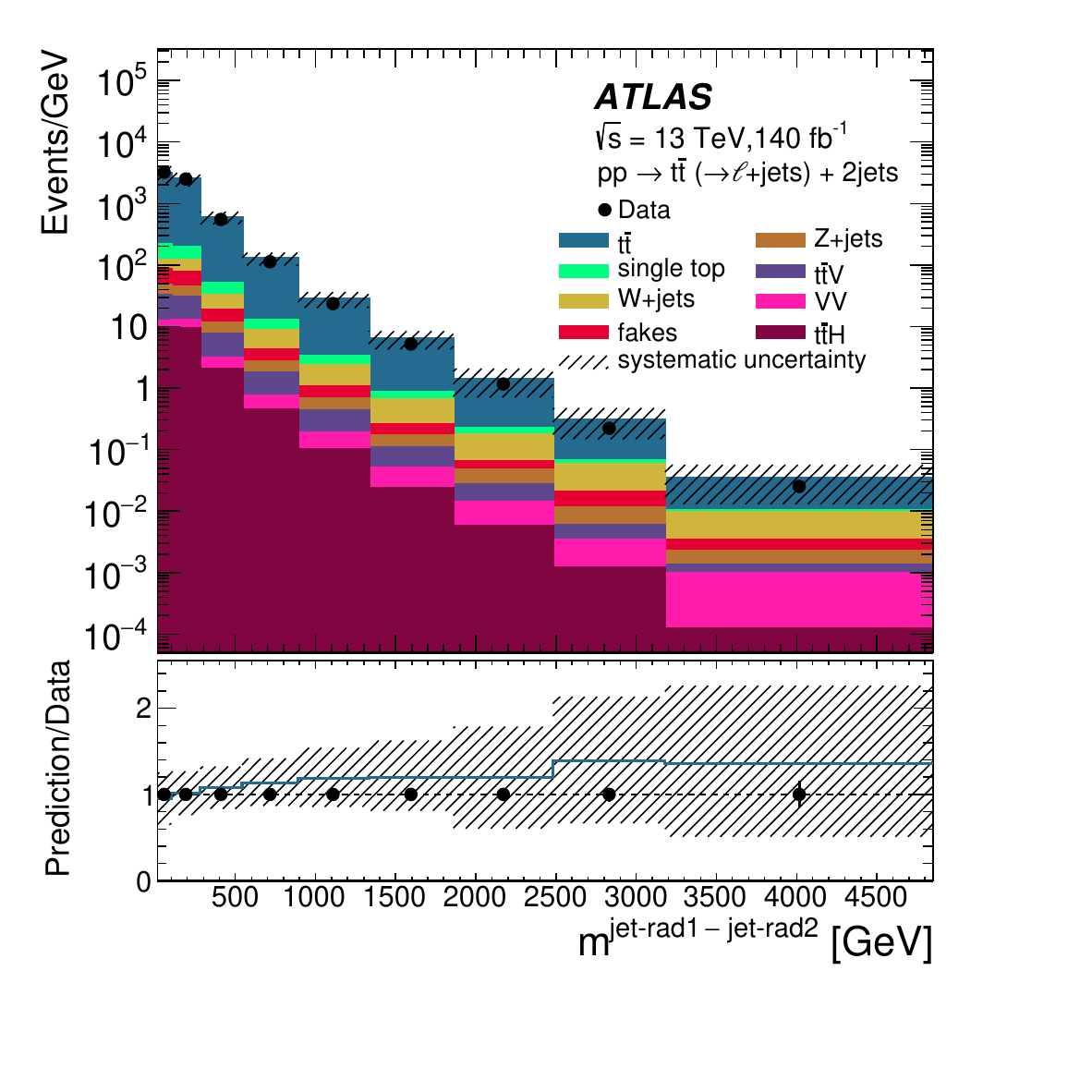}}
\put (3.5,7.2){\small (a)}
\put (11.5,7.2){\small (b)}
\put (3.5,0.2){\small (c)}
\put (11.5,0.2){\small (d)}
\end{picture}
\caption{
The measured (a) $|\deltaphitlf|$, (b) $|\deltaphithf|$,
(c) $|\deltaphiaf|$ and (d) $\mef$
distributions divided by the bin width (dots) in the $\tt+$2jets channel.
The error bars represent the statistical uncertainty in the data; for many
points, the uncertainty is smaller than the marker size and, thus,
not visible.
The hatched bands represent the systematic uncertainties (see
Section~\ref{syst}).
The prediction of the signal and various
backgrounds (histograms) are
also included. The lower panels of each figure show the ratio of the
total prediction and the data distributions.
}
\label{fig13}
\end{figure}

The following observables are measured in the $\tt+$2jets channel:
\begin{itemize}
\item The kinematics of the second gluon emission is characterised by
measuring the transverse momentum of jet-rad2 ($\Ptjetf$) and the
absolute value of its rapidity ($|\Rapjetf|$).
\item The topology of the second gluon emission is characterised by
measuring $|\deltayef|$, defined as the absolute value of the
difference in rapidity between jet-rad1 and jet-rad2, and
$|\deltaphief|$, defined as the absolute value of the difference in
azimuth between jet-rad1 and jet-rad2. Other angular correlations
provide additional insight into the topology of the $\tt+$2jets events:
$|\deltaphitlf|$, defined as the absolute value of the difference in
azimuth between the leptonically decaying top quark and jet-rad2, $|\deltaphithf|$,
defined as the absolute value of the difference in azimuth between
the hadronically decaying top quark and jet-rad2, and $|\deltaphiaf|$, defined as the
absolute value of the difference in azimuth between jet-W1 and
jet-rad2.
\item The dynamics of the $\tt+$2jets events is characterised by
measuring the invariant mass of the jet-rad1 and jet-rad2 system
($\mef$).
\end{itemize}
Figures~\ref{fig12} and \ref{fig13} show the
data, backgrounds and signal MC distributions at detector level for
these observables. There is an adequate description of
the measured observables by the SM predictions.
 
\FloatBarrier
 
\newpage
 
\FloatBarrier
 
\newpage
 
\FloatBarrier
 
\newpage
 
\FloatBarrier
 
\newpage
 
\FloatBarrier
\section{Cross-section measurement}
\label{xcross}
Differential cross-sections are measured at particle level as
functions of the observables defined in Section~\ref{observables} for
the three channels separately. The data distributions, after background
subtraction (see Section~\ref{backcon}), are unfolded to the particle
level as explained below.
 
\subsection{Unfolding method}
The Bayesian unfolding~\cite{nim:a362:487} method as implemented in
RooUnfold~\cite{1105.1160} is used to obtain the measured differential
cross-sections. In this method, repeated applications of the Bayes'
theorem are used to invert the unfolding matrix. The regularisation
parameter is the number of iterations ($N_{\mathrm{iter}}$), therefore
regularisation is achieved by stopping the iterative procedure at a
given value of $N_{\mathrm{iter}}$. The results are found to be fairly
insensitive to $N_{\mathrm{iter}}$ for most of the phase space
measured; two iterations, i.e. $N_{\mathrm{iter}}=2$, are
used. Uncertainties related to this choice of number of iterations
are considered (see Sections~\ref{nonclo} and \ref{syst}).
 
The differential cross-section as a function of the observable $X$ in bin
$i$ is computed using the equation
 
\begin{equation}
\frac{\textrm{d}\sigma}{\textrm{d}X_i} \equiv \frac{1}{\mathcal{L}
\cdot \Delta X_i} \cdot  N^{\mathrm{unf}}_{i},
\nonumber
\end{equation}
where the index $i$ iterates over bins of observable $X$, $\Delta X$
is the bin width, $\mathcal{L}$ is the integrated luminosity and
$N^{\mathrm{unf}}$ represents the unfolded data value in each bin
(overflow and underflow events are not considered when evaluating
$N^{\mathrm{unf}}$). It is calculated as
 
\begin{equation}
N^{\mathrm{unf}}_i \equiv \frac{1}{\varepsilon^i} \cdot \sum_j
f_\textrm{match}^j \cdot \mathcal{M}_{ij}^{-1}  \cdot
f_\textrm{acc}^j \cdot \left(N^j_{\textrm{data}}- N_\textrm{bkg}^j\right)\hbox{,}
\nonumber
\end{equation}
where $N_{\textrm{data}}$ ($N_\textrm{bkg}$) is the number of data
(background) events in each bin of observable $X$. The acceptance
fraction $f_{\mathrm{acc}}$ is defined as
 
\begin{equation}
f_{\mathrm{acc}}^j=\frac{N_{\mathrm{particle}\,\wedge\mathrm{detector}}^j}{N_{\mathrm{detector}}^j},
\nonumber
\end{equation}
where
$N_{\mathrm{particle}\,\wedge\mathrm{detector}}$ is the number
of detector-level signal MC events that satisfy the particle-level
selection in each bin; this factor corrects for events that are
generated outside the fiducial phase-space but satisfy the
detector-level selection. The unfolding matrix $\mathcal{M}$
is constructed from the correlation matrix, which maps the
generated particle-level signal MC events to the detector-level events
in their overlap region. Thus, the probability for particle-level
events to be reconstructed in the same bin is represented by the
diagonal elements, and the off-diagonal elements describe the fraction
of particle-level events that migrate into other bins. To suppress
combinatorial background in $\mathcal{M}$, angular matching
requirements are applied to the reconstructed objects.
For all events, the reconstructed top quarks are required to satisfy
$\Delta R($top-x$_{\mathrm{part}},$ top-x$_{\mathrm{det}})<0.8$,
where top-x represents the leptonically or hadronically decaying top
quark. In the $\tt$ inclusive channel, matching of the jets is
performed by requiring $\Delta R($jet-y$_{\mathrm{part}},$
jet-y$_{\mathrm{det}}) < 0.3$, where jet-y represents jet-W1,
jet-W2. In the $\tt$+1jet and $\tt+2$jets channels, this requirement
is applied to jet-rad1, jet-rad2 or both, where necessary. In
addition, the reconstructed objects (top quarks or jets) that are not
matched to a generated object as well as reconstruction inefficiencies
(generated objects that are not matched to a reconstructed object) are
taken into account in the Bayesian method. In addition, the inclusion
of a matching scheme ensures the stability of the iterative process,
so that the results are insensitive to $N_{\mathrm{iter}}$, except for
statistical fluctuations, as mentioned above. The effect of the
matching procedure in the unfolding is accounted for by the matching
fraction $f_{\mathrm{match}}$, which is defined as
 
\begin{equation}
f_{\mathrm{match}}^j=\frac{N_{\mathrm{particle}\,\wedge\mathrm{detector}\,\wedge\mathrm{match}}^j}{N_{\mathrm{particle}\,\wedge\mathrm{detector}}^j},
\nonumber
\end{equation}
 
where
$N_{\mathrm{particle}\,\wedge\mathrm{detector}\,\wedge\mathrm{match}}$
is the number of signal MC detector-level events that satisfy the particle-level
selection and matching requirements and
$N_{\mathrm{particle}\,\wedge\mathrm{detector}}$ is the number of
particle-level events that satisfy the detector-level selection
without the matching requirements. Finally, the efficiency $\varepsilon$ is defined as
 
\begin{equation}
\varepsilon^i = \frac{N_{\mathrm{particle}\,\wedge\,\mathrm{detector}}^i}{N_{\mathrm{particle}}^i},
\nonumber
\end{equation}
where $N_{\mathrm{particle}}$ is the number of signal MC events at
particle level in each bin; this factor corrects for events that
satisfy the particle-level selection but are not selected at the
detector level.
 
\begin{figure}[h]
\setlength{\unitlength}{1.0cm}
\begin{picture} (18.0,20.5)
\put (0.0,13.8){\includegraphics[width=7cm]{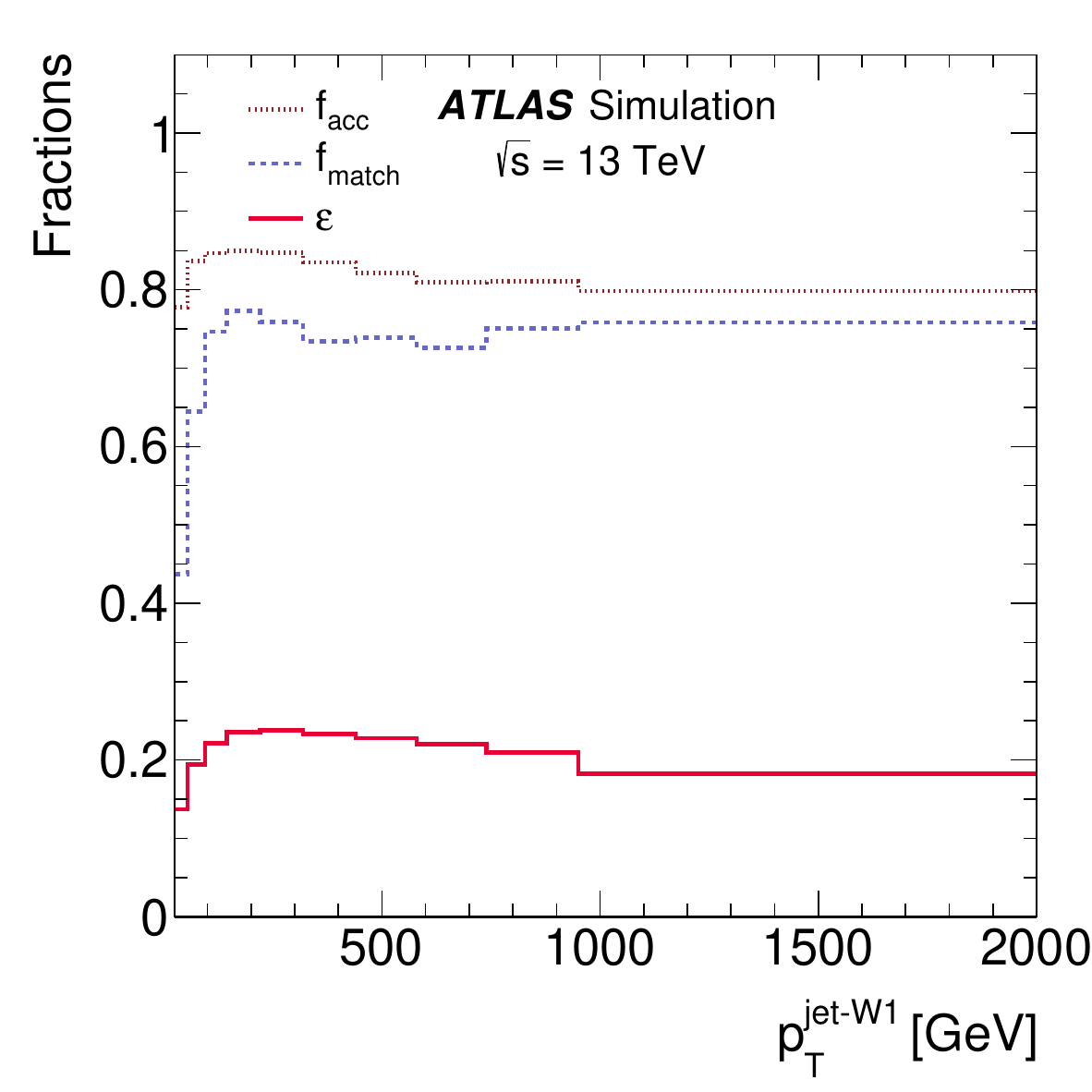}}
\put (8.0,13.8){\includegraphics[width=7cm]{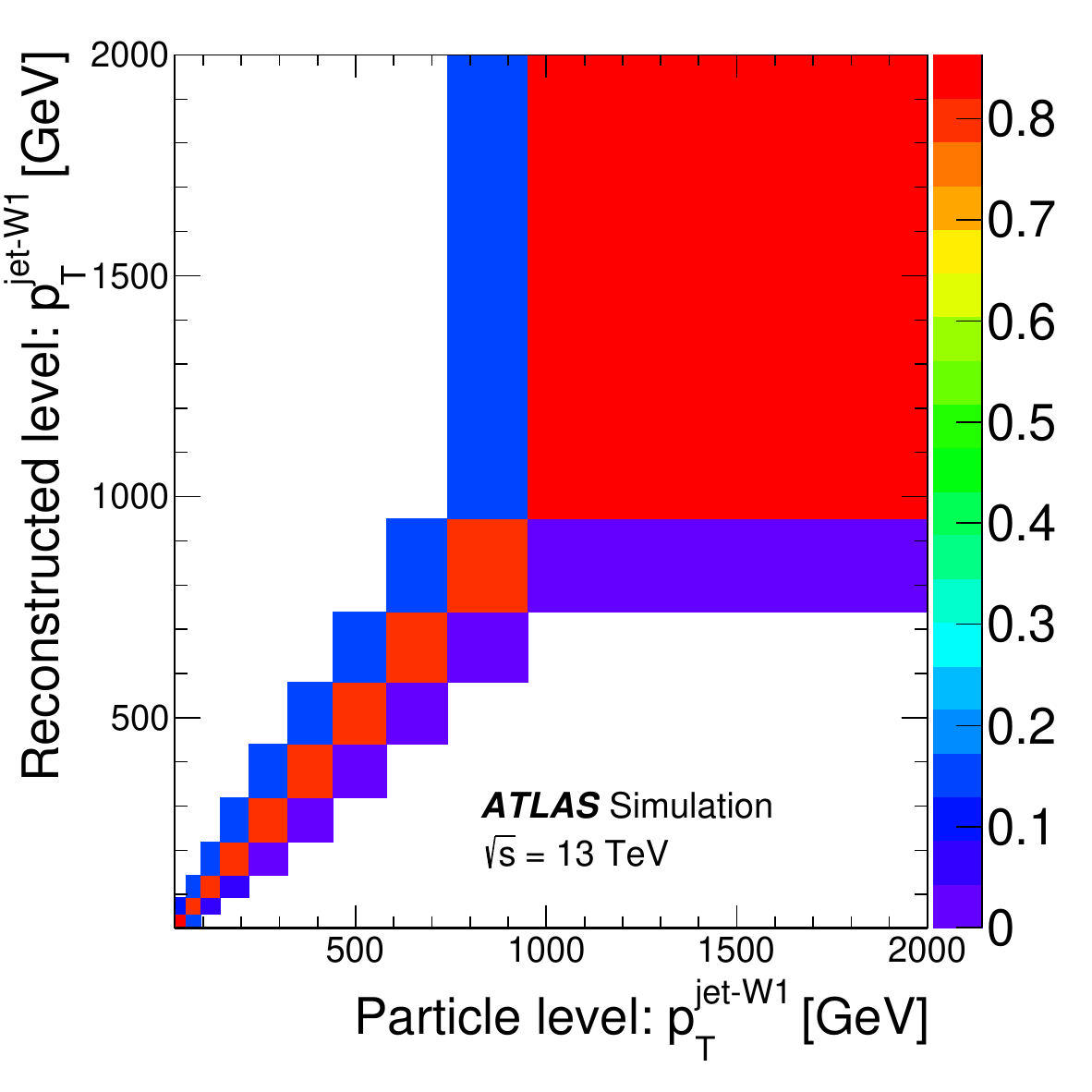}}
\put (0.0,6.9){\includegraphics[width=7cm]{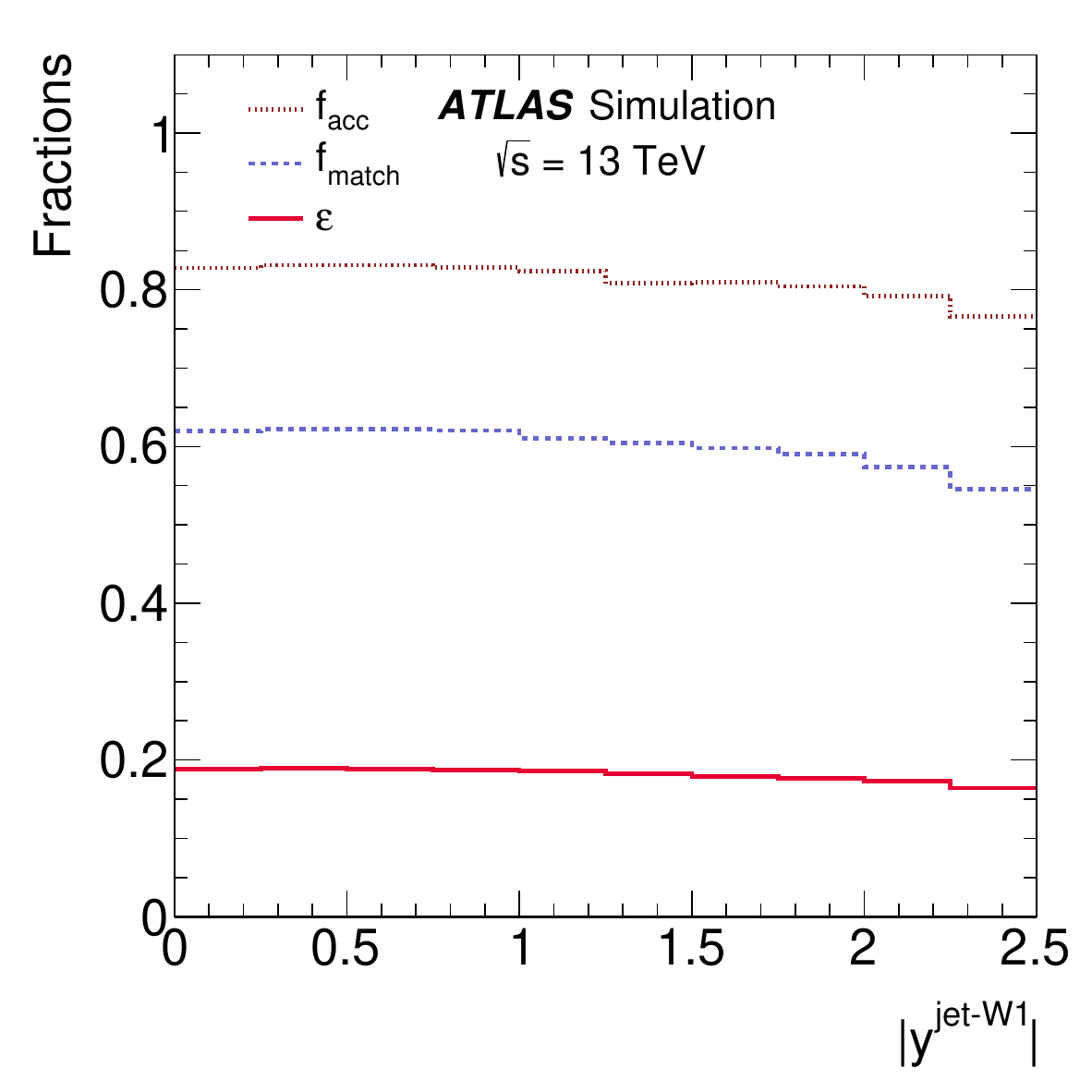}}
\put (8.0,6.9){\includegraphics[width=7cm]{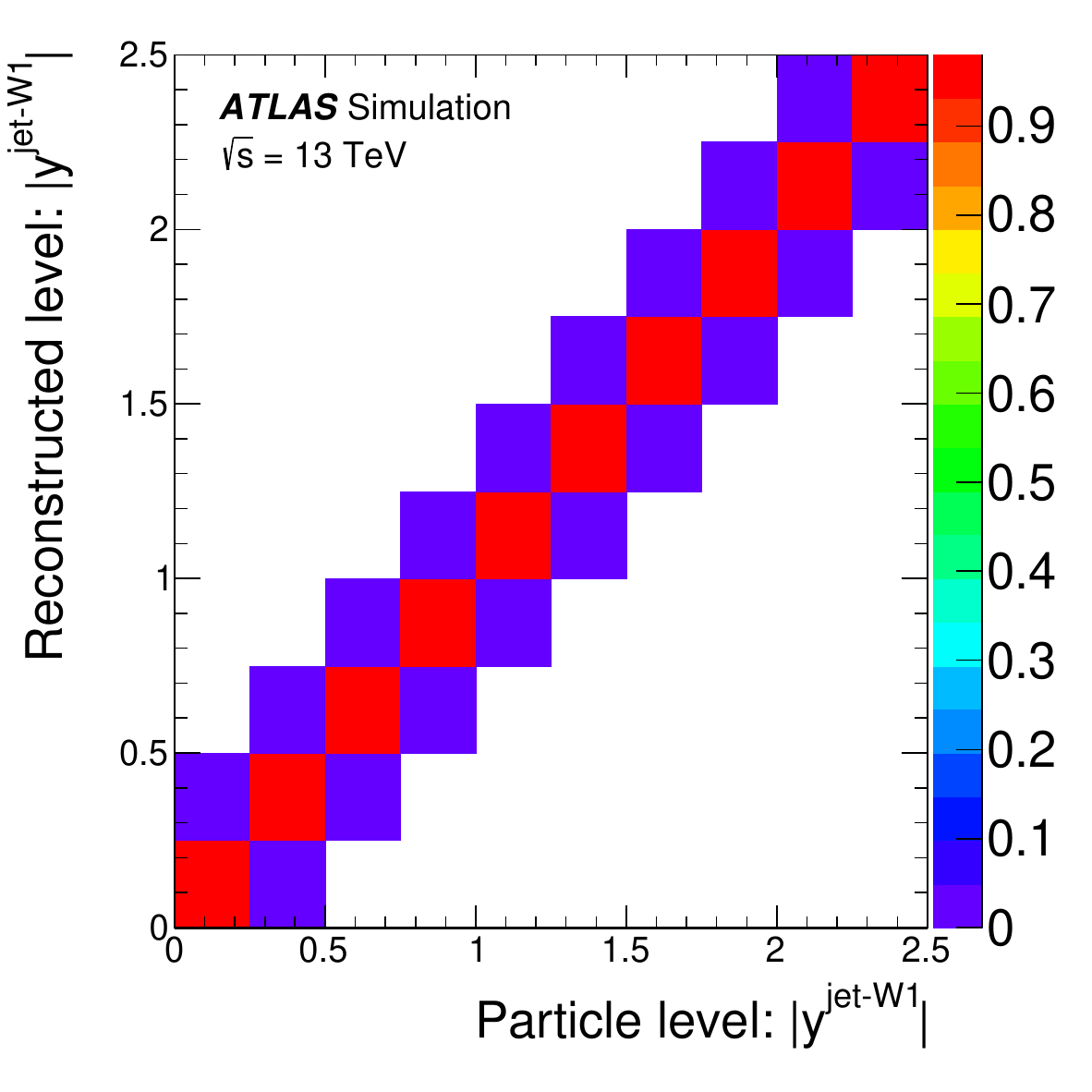}}
\put (0.0,0.0){\includegraphics[width=7cm]{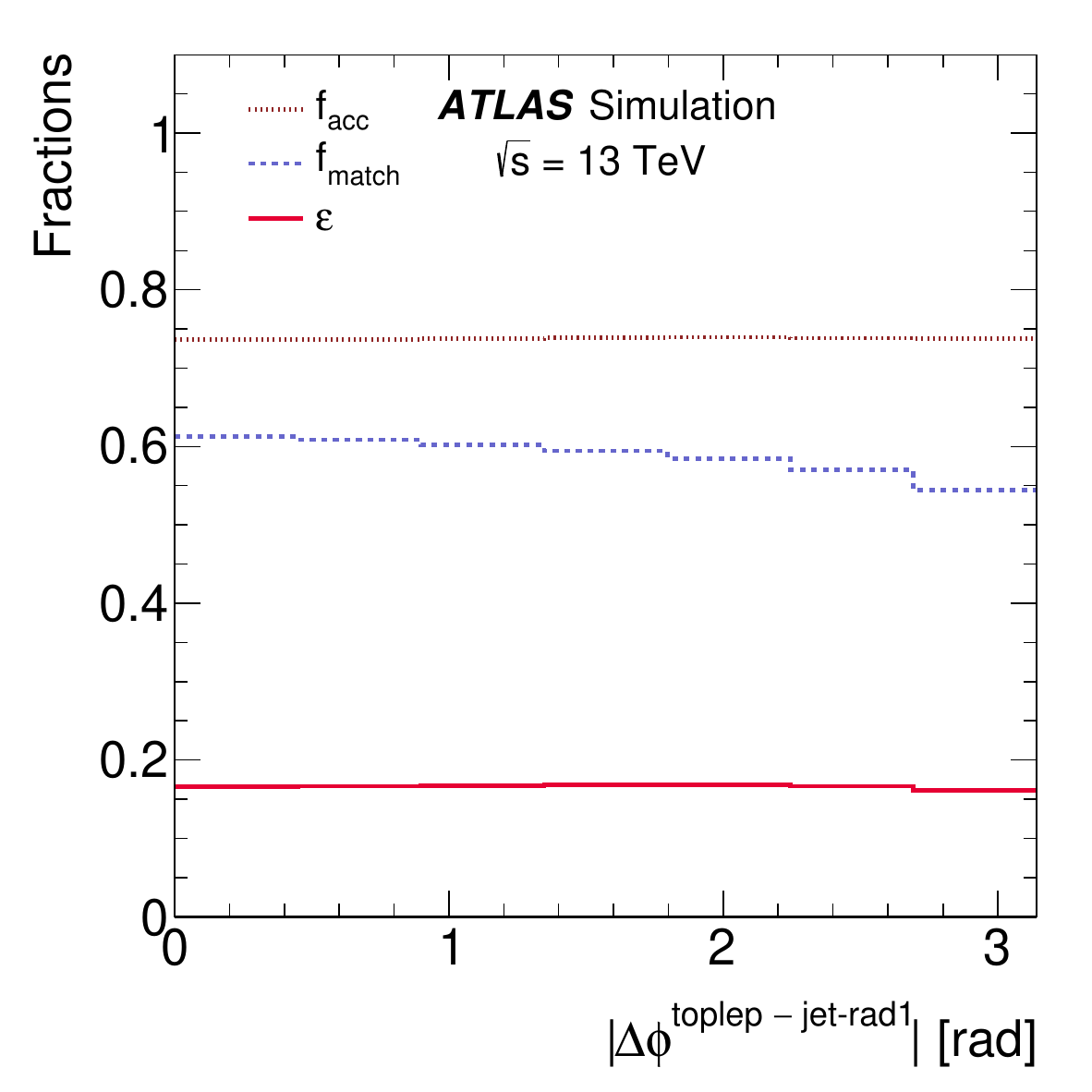}}
\put (8.0,0.0){\includegraphics[width=7cm]{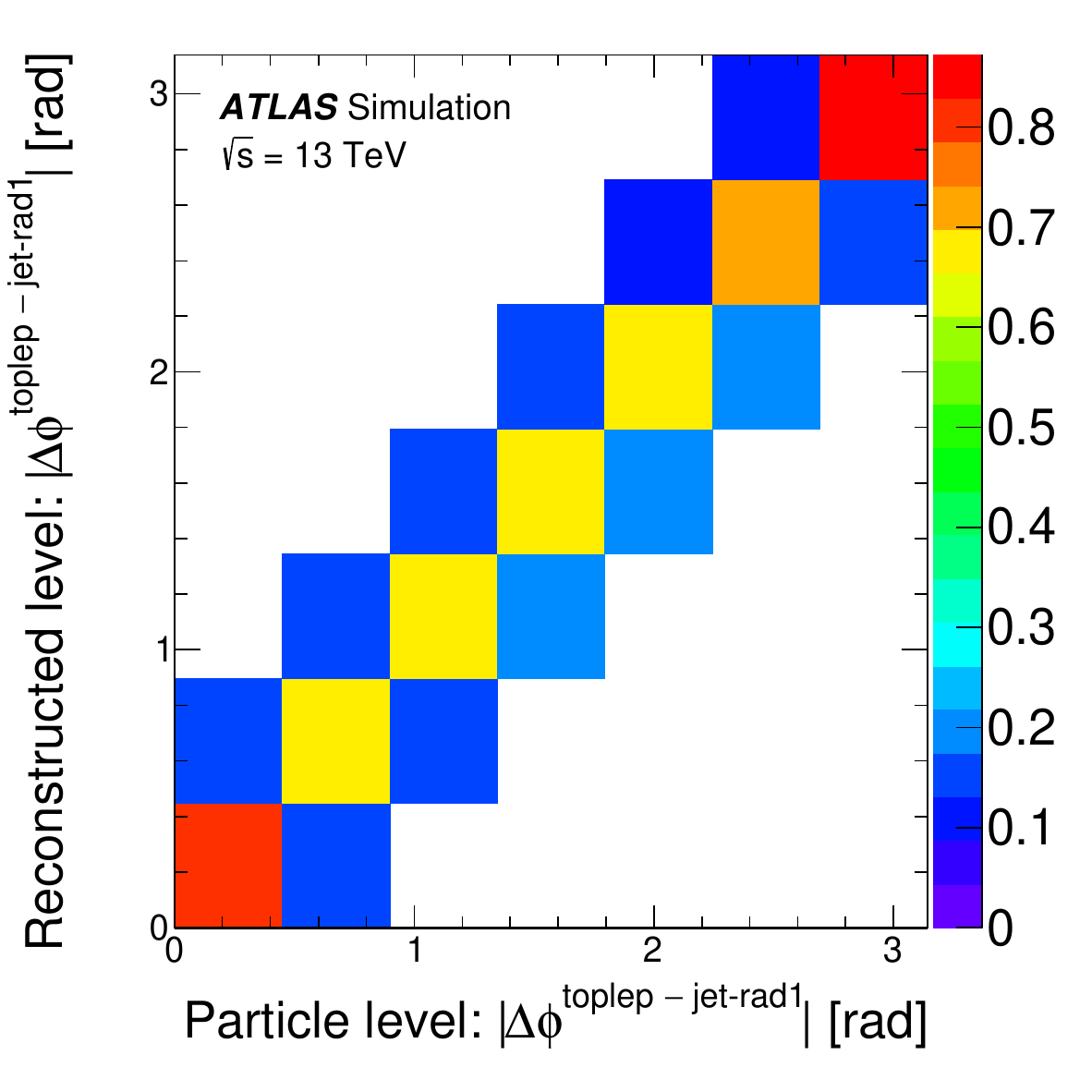}}
\put (3.5,13.8){\small (a)}
\put (11.5,13.8){\small (b)}
\put (3.5,6.9){\small (c)}
\put (11.5,6.9){\small (d)}
\put (3.5,0.0){\small (e)}
\put (11.5,0.0){\small (f)}
\end{picture}
\caption{
The efficiency $\varepsilon$ (solid lines), matching fraction
$f_{\mathrm{match}}$ (dashed lines) and acceptance fraction
$f_{\mathrm{acc}}$ (dotted lines) for
(a) $\Ptjeta$,
(c) $|\Rapjeta|$ and
(e) $|\deltaphitle|$.
The unfolding matrix $\mathcal{M}_{ij}$ for
(b) $\Ptjeta$,
(d) $|\Rapjeta|$ and
(f) $|\deltaphitle|$;
the distributions are normalised such that the sum of the entries in
each column is equal to unity. The signal samples of \pwp\ are used.
}
\label{fig_cor}
\end{figure}

As examples, Figure~\ref{fig_cor} shows $\varepsilon$,
$f_{\mathrm{match}}$ and $f_{\mathrm{acc}}$ for $\Ptjeta$,
$|\Rapjeta|$ and $|\deltaphitle|$ and the unfolding matrices for the
same observables. The signal samples of \pwp\ are used in this figure.
 
The normalised differential cross-sections are measured by dividing
the differential cross-sections by the total cross-section in the
region of phase space considered for each channel.
 
\FloatBarrier
 
\newpage
 
\subsection{Closure and stress tests}
\label{nonclo}
The unfolding procedure is checked by performing several `closure' and stress
tests, which validate the consistency of the procedure.
 
The first closure test is a cross-check of the reconstruction-generation level
matching procedure: the cross-sections are unfolded with (as in the nominal
cross-sections) and without matching in the transfer matrices. The
ratio of these two cross-sections is calculated; the ratios are
typically consistent with unity within $1\%$. Thus, it is verified that the
matching procedure does not affect the unfolding.
 
A second closure test is performed to check that the unfolding procedure
recovers the generated result in a statistically independent
sample. For this test, a MC signal sample is divided into two
subsamples A and B; sample A is used as pseudo-data and unfolded
to particle level using sample B (pseudo-data cross-section). The
pseudo-data cross-sections are compared with the generated results of
sample A, B or A+B; the differences observed in these
comparisons are typically less than $1\%$. Therefore, the generated result
is recovered in a statistically independent sample.
 
A first stress test is performed by weighting the nominal MC signal sample
so that it matches the data distributions after background
subtraction at reconstruction level. This weighted MC sample is used
to unfold the data (weighted cross-sections); the ratios of the
weighted and nominal cross-sections are typically consistent with
unity within $1\%$. The differences observed in this test are
considered as a non-closure systematic uncertainty (see
Section~\ref{syst}).
 
A second stress test is performed by using the weighted
MC sample as pseudo-data and unfolding it with the nominal MC sample
(weighted pseudo-data cross-section); the ratio of the weighted
pseudo-data cross-section and the generated result gives back, as
expected, the input fitting function for each observable.
Therefore, it is concluded that differences between data and MC at
reconstruction level do not affect the unfolded measured cross-sections.
 
A third stress test is performed to ascertain the level of
recoverability of the signal for the possible presence of new physics
in the data. The existence of a $\tt$ resonance is hypothesised using
a Gaussian distribution introduced at particle level and propagated to
the reconstruction level using a weighting procedure (stressed pseudo-data).
To determine whether the observables studied are sensitive to this hypothetical
resonance and whether the unfolding procedure is able to recover the
stressed shape and normalisation, the stressed pseudo-data is unfolded
using the nominal MC sample and compared with the stressed generated
result. The ratio of these cross-sections shows that the shape and
normalisation of the effects due to the resonance are recovered for all observables.
 
\section{Systematic uncertainties}
\label{syst}
The sources of systematic uncertainties that affect the measurements
of the differential cross-sections include the $b$-tagging
calibration and efficiency, the detector energy scale and resolution, the
reconstruction efficiency, the background and signal modelling,
the statistics of the MC samples and the uncertainty
arising from the non-closure.
Each source of uncertainty is separated into independent contributions
and evaluated by varying the background and the signal
MC samples in a correlated way, where necessary. The varied
background samples are used to subtract the background and the varied
signal samples are used to unfold the background-subtracted data
distributions, as explained in Section~\ref{xcross}. Systematic
uncertainties are evaluated relative to the nominal measured
cross-sections separately for the upward and downward variations (or
in the case of a single variation by symmetrising the single
deviation) for each bin of each observable.
For the normalised cross-sections, each contribution to the systematic
uncertainty is assumed to be fully correlated between the differential
cross-section in a given bin and the integrated cross-section.
Each source of systematic uncertainty is discussed separately in
the following.
 
\textbf{$b$-tagging calibration and efficiency.}
The uncertainties in the $b$-tagging calibration and efficiency are
determined separately for $b$-jets, $c$-jets and
light-flavour-jets~\cite{epj:c79:970,epj:c82:95,epc:c83:728}
using a component breakdown with 19 independent contributions. They
depend on the $\pt$ for $b$- and $c$-jets, and on the $\pt$ and $\eta$
for light-flavour jets, and they account for differences between data
and simulation. The dominant contribution arises from the $b$-quark
eigenvectors and amounts to typically $3\%$ for the absolute
cross-sections and decreases to typically below $0.5\%$ for the
normalised cross-sections.
 
\textbf{Detector energy scale and resolution.}
The jet energy scale (JES) is calibrated using a combination of test
beam data, simulation and in situ
techniques~\cite{epj:c81:689}. Its uncertainty is decomposed into a
set of $30$ uncorrelated components, with contributions from pile-up,
jet flavour composition and response, single-particle response, and effects of
jets not contained within the calorimeter. The uncertainty in the jet
energy resolution (JER) is represented by eight components accounting
for jet-$\pt$ and $\eta$-dependent differences between simulation and
data~\cite{epj:c81:689}.
The uncertainty in $\etmiss$ due to a possible miscalibration of its
soft-track component is derived from data and simulation comparisons of
the $\pt$ balance between the hard and the soft \met\
components~\cite{2402.05858}. To account for the difference in
pile-up distributions between the simulation and data, the pile-up
profile in the simulation is corrected to match the one in the data. The
uncertainty associated with the correction factor is applied.
The uncertainty in the electron~\cite{jinst:19:P02009} and
muon~\cite{epj:c83:686} energy scale and resolution
is also included. The resulting 49 up and down independent
contributions are added in quadrature and the dominant contribution
comes from the effect of the pile-up on the JES. This uncertainty is
typically in the range of $3\%$ to $10\%$, depending on the
observable, for the absolute cross-sections and decreases to the range
of $0.3\%$ to $9\%$ for the normalised cross-sections.
 
\textbf{Reconstruction efficiency.}
Reconstruction, identification, isolation and trigger performance for
electrons and muons differ between data and simulations. Scale
factors are applied to simulated events to correct for these differences.
These scale factors are assessed using $Z\rightarrow \ell^+\ell^-$
events in simulation and data with methods similar to those described
in Refs.~\cite{jhep:05:162,epj:c81:578}. The associated systematic
uncertainties are propagated to the measured cross-sections. The JVT
requirement and pile-up systematic uncertainties are also included.
The resulting 16 up and down independent contributions are added in
quadrature and the dominant contribution comes from the JVT contribution.
This uncertainty is in the range of $0.5\%$ to $2\%$ for both
the absolute and normalised cross-sections.
 
\textbf{Background modelling.}
The uncertainties in the modelling of the background are estimated by
changing the normalisation of the MC samples according to the
uncertainty in the predicted cross-section used for the normalisation
of these samples (see Table~\ref{tab:MC}). For the \wjets, \zjets\ and
fakes background, an uncertainty of $50\%$ is included for the
normalisation. This value for the uncertainty in the normalisation
covers the differences observed between data and
simulations in control regions  with exactly three jets and at most
one $b$-tagged jet or  at least four jets and at most one $b$-tagged
jet; these control regions are away from the signal region and
enriched in these background processes.
In addition, for the single-top-quark background, since the $tW$
contribution is the largest among the three single-top-quark
production channels, the uncertainty arising from the method used to
handle the interference between $tW$ and $\tt$ production is determined
by comparing the $tW$ simulated sample that uses the DS method with
the nominal one based on the DR technique.
The resulting uncertainties are symmetrised and added in quadrature.
This uncertainty is typically in the range of $1.5\%$ to $15\%$,
depending on the observable, for the absolute cross-sections and
decreases to the range of $0.1\%$ to $10\%$ for the normalised
cross-sections.
 
\textbf{Signal modelling.}
The uncertainties that affect the signal modelling and the method of
estimation are discussed below. The effects of the PS and
hadronisation models are estimated by unfolding the data after
background subtraction using the \pwh\ sample and calculating the
difference relative to the nominal cross-section (typically
$<2\%$). Since the single-top-quark processes are the dominant background,
\pwh\ samples are also used to model these processes when estimating
this uncertainty.
The impact of the matching procedure between the ME calculation and
the PS model is evaluated using the \amcnloh\ sample to
unfold the data after background subtraction and calculating the
difference relative to the cross-section unfolded using the \pwh\
samples, both for signal and single-top-quark background (typically
$\approx 1\%$).
The uncertainties affecting the description of the hard gluon
radiation are evaluated by unfolding the data after background
subtraction using the \pwp\ $\hdamp$ sample and calculating the
difference relative to the nominal cross-section (typically $<0.5\%$).
The effect of the assumed value of $\mt$ ($172.5$~\GeV) is estimated
using samples of \pwp\ generated with $\mt=169$~\GeV\ or
$\mt=176$~\GeV to unfold the data after background subtraction and
calculating the difference relative to the nominal cross-section;
the value of the uncertainty obtained in this way is then scaled by
$1/3.5$, to account for the current uncertainty in $\mt$ ($1$~\GeV)
and amounts to typically $<0.3\%$. The first three contributions listed above
are symmetrised. The dominant contribution arises from the
PS and hadronisation models.
The resulting uncertainty is typically in the range of $0.5\%$ to $10\%$,
depending on the observable, for the absolute cross-sections and
decreases to the range of $0.2\%$ to $8\%$ for the normalised
cross-sections.
The theoretical uncertainties
that affect the unfolding procedure and the method of estimate are
discussed below. These uncertainties affect the $\tt$ signal and the
single-top-quark background samples and are taken as correlated between the
two MC samples. The uncertainty due to missing higher-order QCD
corrections in the ME computation is estimated by independently
varying $\muR$ and $\muF$ by factors of $2$ and $0.5$ relative to
the central value (typically $<1\%$). The uncertainties in the amount
of initial- and final-state radiation from the PS are
assessed by varying the corresponding parameter of the
A14 PS tune (Var3c)~\cite{ATL-PHYS-PUB-2014-021} (typically
$<0.5\%$) and by varying the scale $\muR^{\mathrm{FSR}}$
by factors of $2$ and $0.5$ (typically $\approx 2\%$). The uncertainties
due to the PDFs are estimated following the PDF4LHC
prescription~\cite{jp:g43:023001} (typically $< 0.5\%$).
The dominant contribution to the systematic uncertainty arising from
the theoretical uncertainties in the unfolding
comes from the variation of the $\muR^{\mathrm{FSR}}$ scale.
The resulting uncertainty is typically in the range of $1\%$ to $5\%$,
depending on the observable, for the absolute cross-sections and
decreases to the range of $0.5\%$ to $4\%$ for the normalised
cross-sections.
 
\textbf{Non-closure.} As explained in Section~\ref{nonclo}, the
differences observed in the stress test are considered as a
non-closure systematic uncertainty. This uncertainty is typically
below $0.5\%$. For the $\mef$ observable, the uncertainty increases up
to $25\%$ in the last two bins due to the limited statistics in these
bins and affects the weighting procedure.
 
\subsection{Total systematic uncertainty}
\begin{figure}[h]
\setlength{\unitlength}{1.0cm}
\begin{picture} (18.0,20.5)
\put (0.0,13.8){\includegraphics[width=7cm]{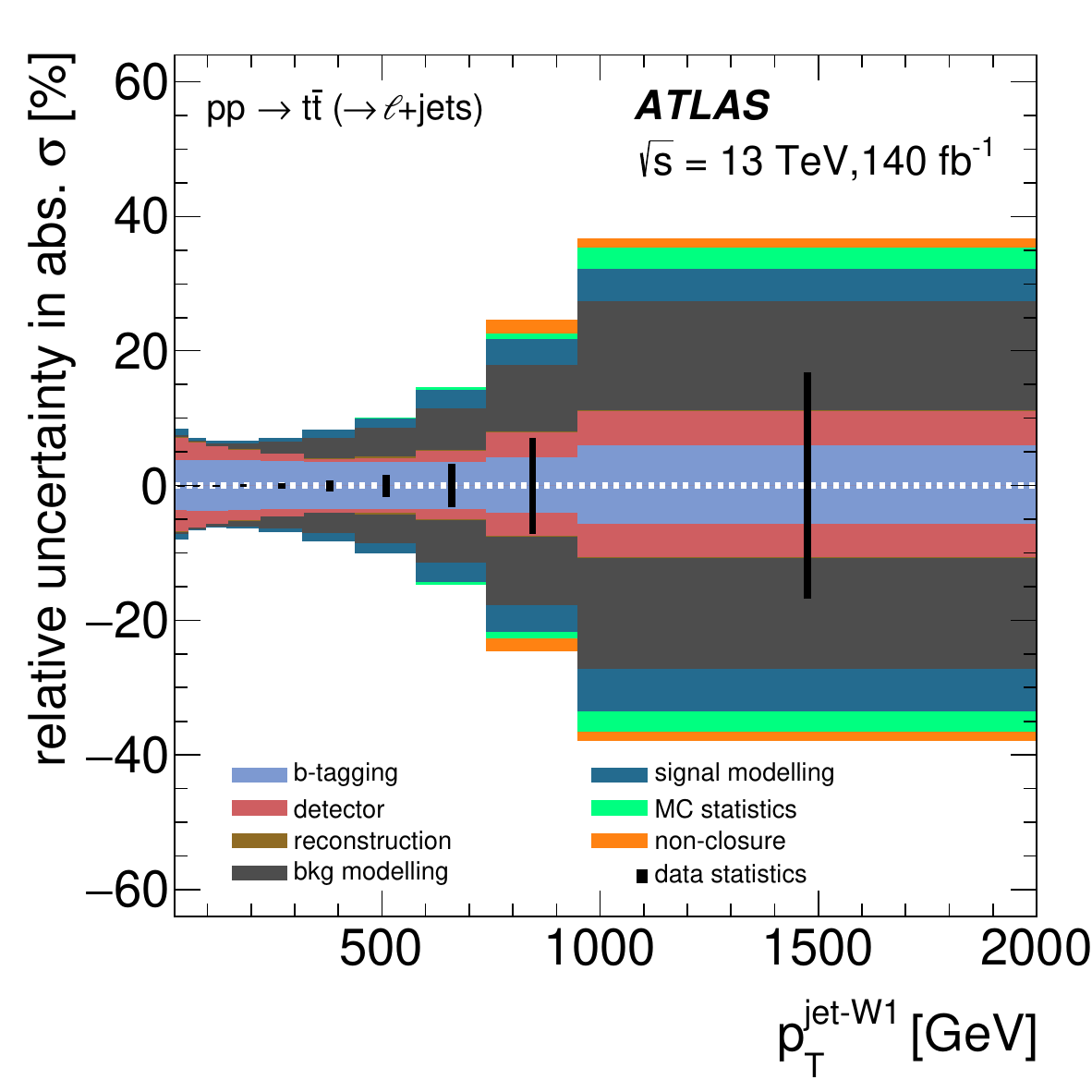}}
\put (8.0,13.8){\includegraphics[width=7cm]{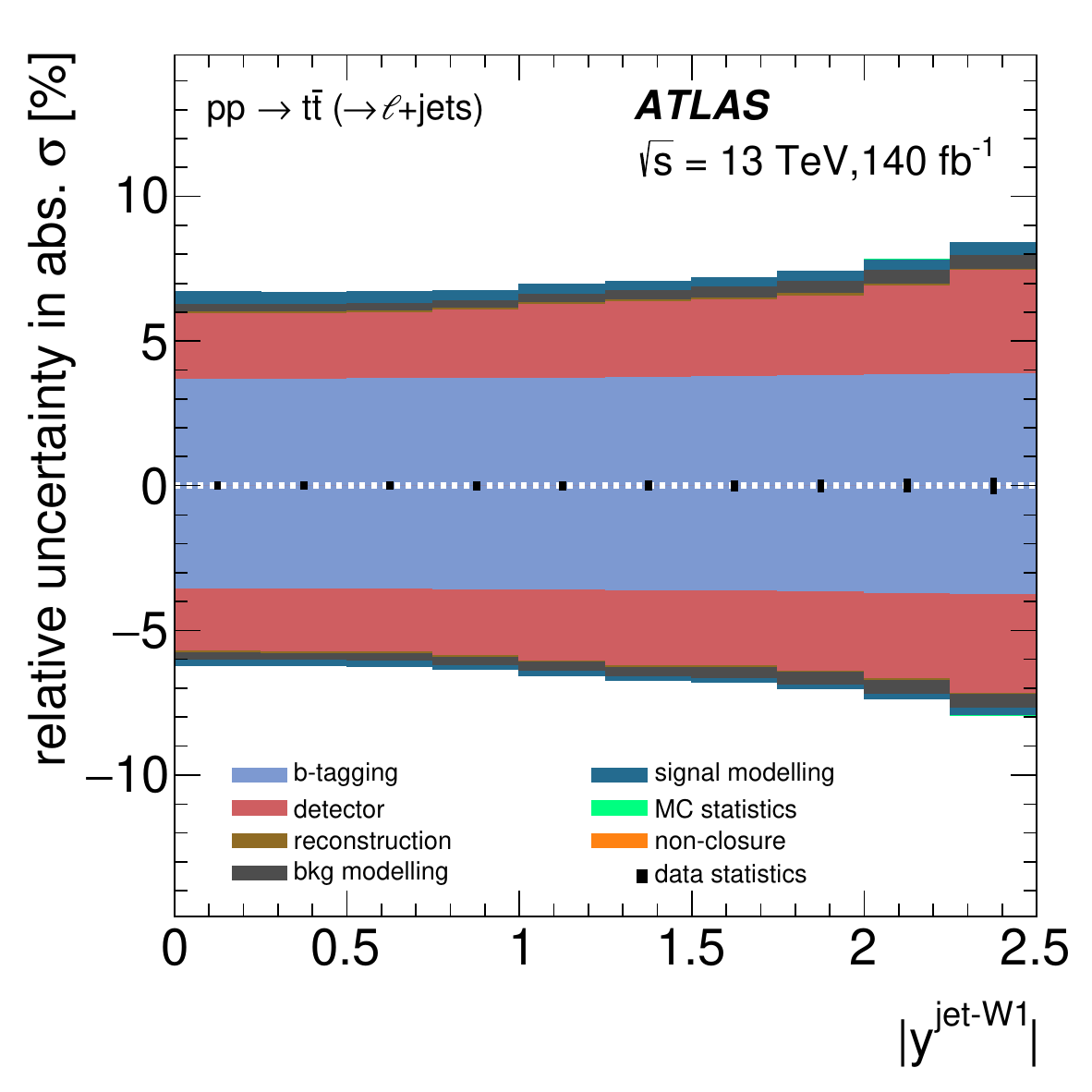}}
\put (0.0,6.9){\includegraphics[width=7cm]{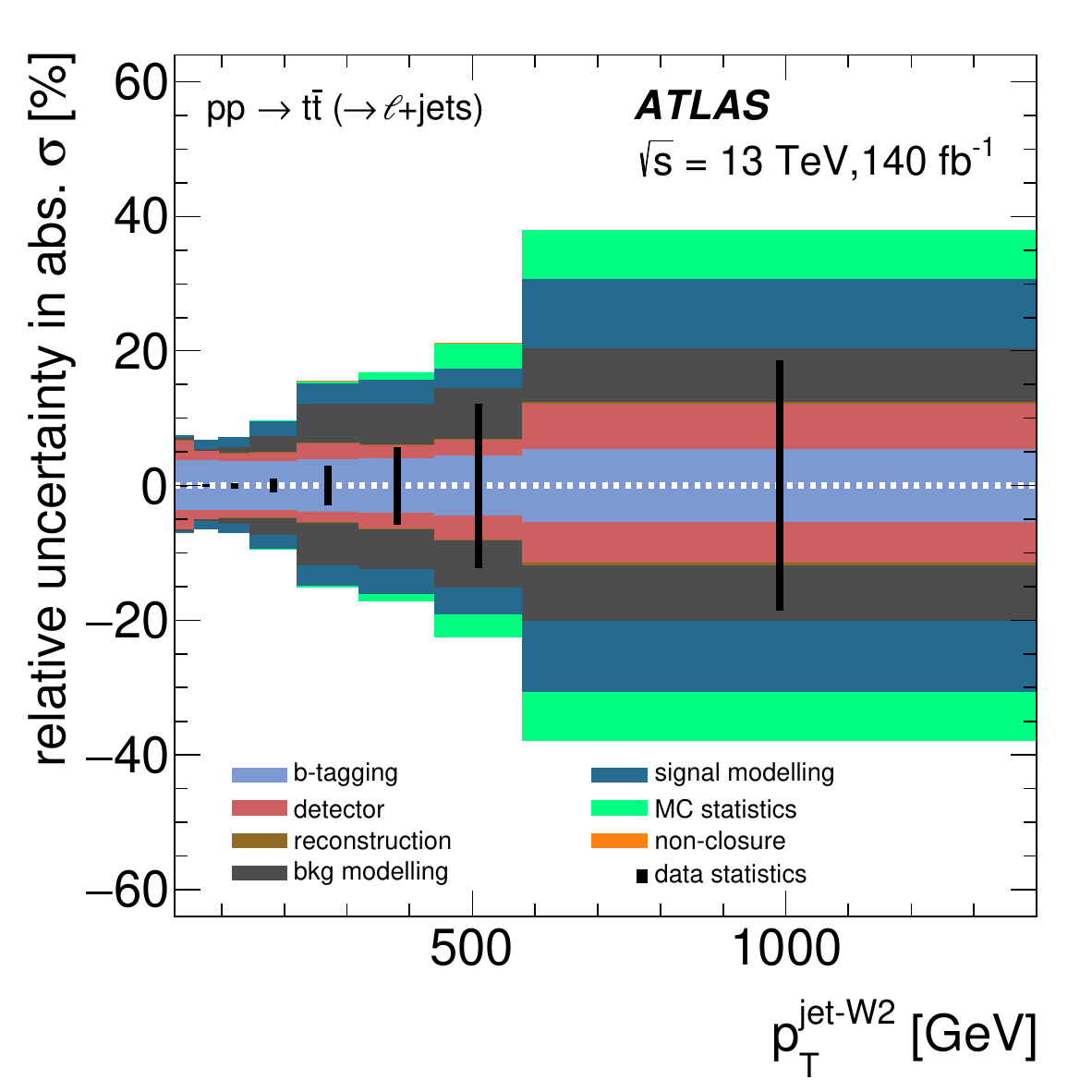}}
\put (8.0,6.9){\includegraphics[width=7cm]{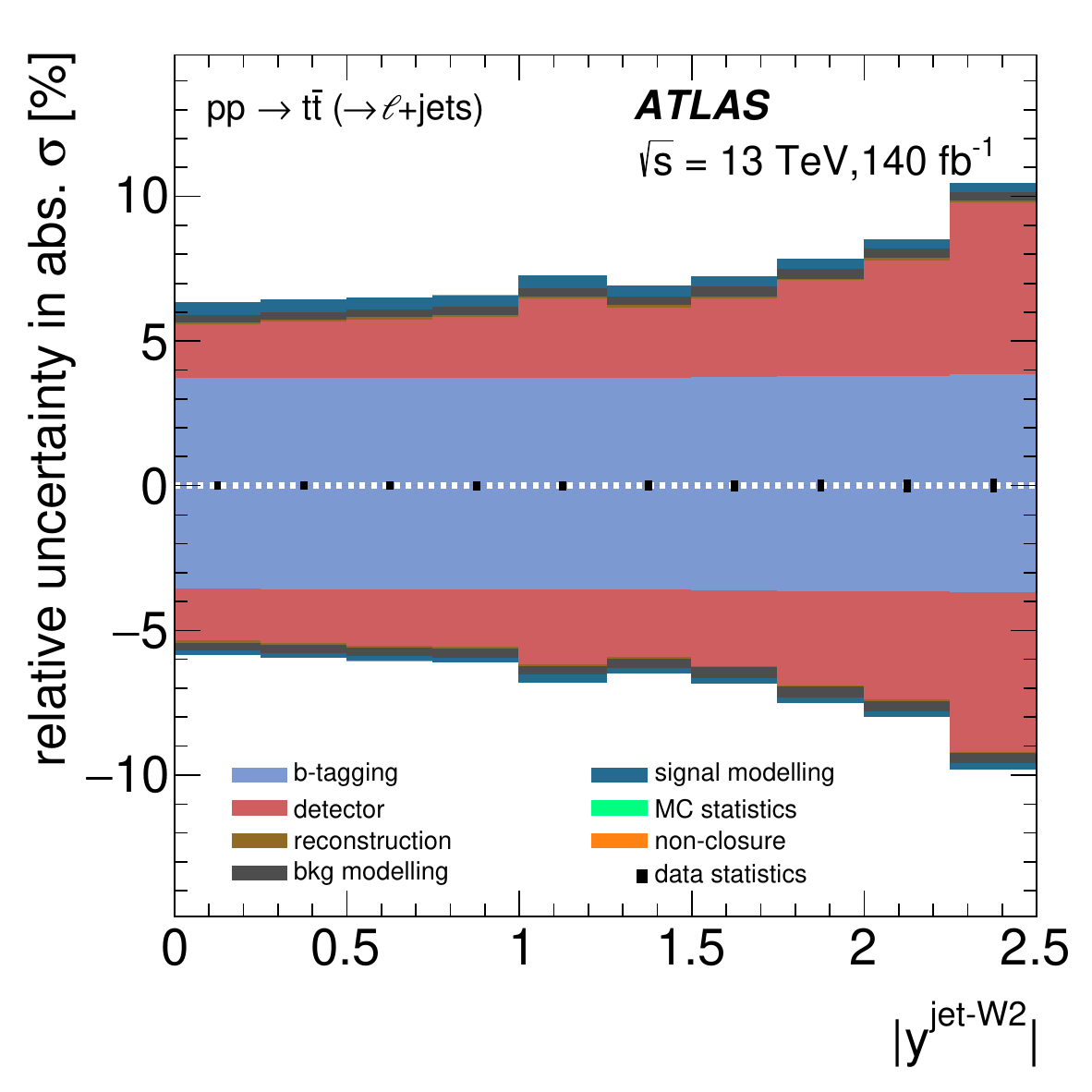}}
\put (0.0,0.0){\includegraphics[width=7cm]{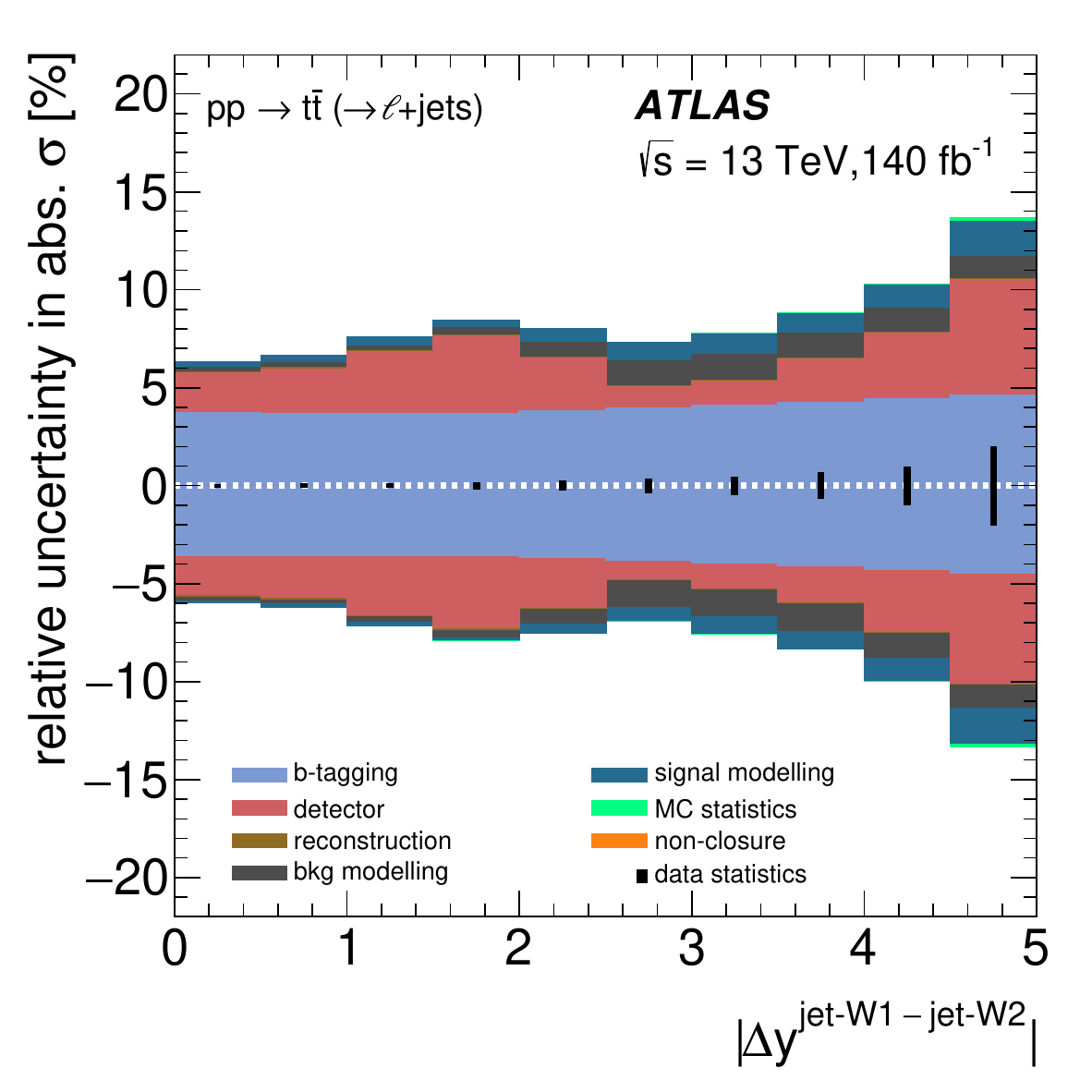}}
\put (8.0,0.0){\includegraphics[width=7cm]{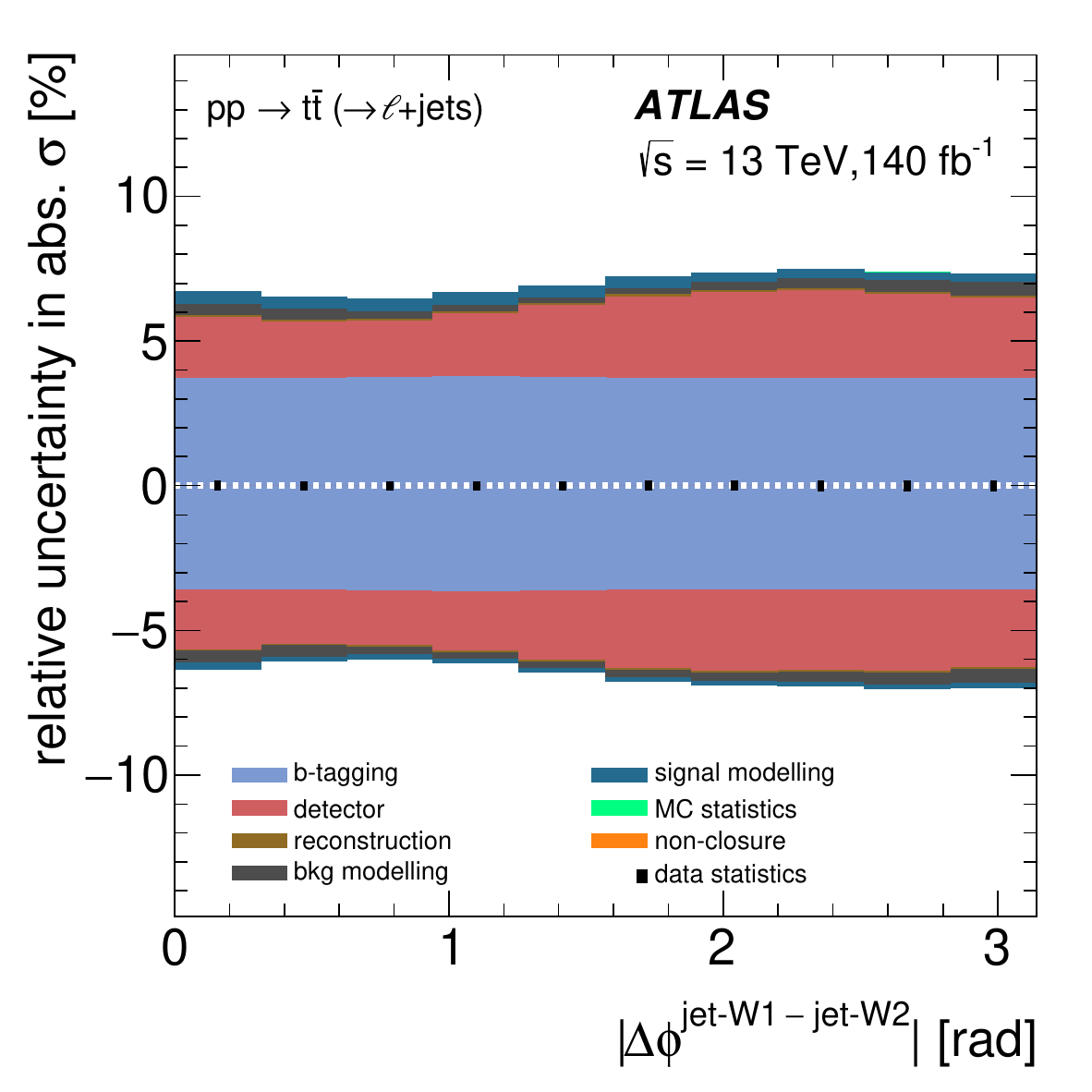}}
\put (3.0,14.0){\small (a)}
\put (11.0,14.0){\small (b)}
\put (3.0,7.1){\small (c)}
\put (11.0,7.1){\small (d)}
\put (3.0,0.2){\small (e)}
\put (11.0,0.2){\small (f)}
\end{picture}
\caption{
Total relative systematic uncertainties in the
measured cross-sections in percentages for
(a) $\Ptjeta$, (b) $|\Rapjeta|$,
(c) $\Ptjetb$, (d) $|\Rapjetb|$,
(e) $|\deltayab|$ and (f) $|\deltaphiab|$
in the $\tt$ inclusive channel.
The individual sources of systematic uncertainty (plotted in stacked
form) and the statistical uncertainty in the data are also included.
}
\label{fig74}
\end{figure}

The total systematic uncertainty is computed by adding in quadrature
the sources of uncertainty listed in the previous section.
Figures~\ref{fig74} to \ref{fig77} (\ref{fig_appm37} to
\ref{fig_appm40}) show the resulting total systematic uncertainty,
together with the contribution from each source, the statistical
uncertainty in the MC samples and the statistical uncertainty in the
data for the absolute (normalised) cross-sections.
The statistical uncertainties in data and MC
samples are evaluated using `pseudo-data experiments' (bootstrap technique)
with 1000 statistically independent replicas. In the figures, the systematic
contributions are stacked, except for the statistical uncertainty in
the data. In addition, there is an uncertainty from the combined 2015--2018
integrated luminosity, which amounts to
$0.83\%$~\cite{epj:c83:982} (not included in the figures).
 
In the case of the angular observables, for the $\tt$ inclusive
channel, the total relative systematic uncertainty
is $\approx 7\%$ and dominated by the $b$-tagging calibration uncertainty. In
the case of the $\tt+$1jet channel, the total relative
systematic uncertainty amounts to $\approx 10\%$ and is dominated by
the detector energy scale and resolution. The $\tt+$2jets channel
has a total relative systematic uncertainty of $\approx 13\%$ and is
also dominated by the detector energy scale and resolution.
For the absolute differential cross sections in the $\tt$ inclusive
channel (see Figure~\ref{fig74}), the total relative uncertainty as a
function of $|\deltaphiab|$ is approximately constant, whereas the
total relative uncertainties as functions of $|\Rapjeta|$ and $|\Rapjetb|$
increase as the rapidity increases, though the uncertainty as a
function of $|\Rapjetb|$ is larger at large values than for
$|\Rapjeta|$. This is driven by the detector energy scale and
resolution contribution and leads to a minimum in the uncertainty as a
function of $|\deltayab|$ in the region $2.5<|\deltayab|<3.5$. For the
$\tt+$1jet and $\tt+$2jets channels, the total relative uncertainties
are approximately constant as functions of the measured angular
observables. For the normalised cross-sections, the total relative
systematic uncertainties are $\approx 1\%$, $\approx 1.5\%$ and
$\approx 2\%$ for the $\tt$ inclusive, $\tt+$1jet and $\tt+$2jets
channels, respectively.
 
In the case of the transverse momentum and invariant-mass observables,
the values for the uncertainties quoted above are only valid at low $\pt$
or invariant mass. The total relative systematic uncertainty in the
absolute (normalised) cross-sections increases up to
$\approx 38\%$ ($\approx 37\%$) for $\Ptjeta$,
$\approx 38\%$ ($\approx 37\%$) for $\Ptjetb$,
$\approx 57\%$ ($\approx 55\%$) for $\Ptjete$,
$\approx 77\%$ ($\approx 74\%$) for $\mtte$,
$\approx 50\%$ ($\approx 52\%$) for $\Ptjetf$
and
$\approx 48\%$ ($\approx 46\%$) for $\mef$
in the last bins of these distributions. In these regions of phase
space, the background modelling and statistical uncertainties dominate.
 
The statistical uncertainty in the data is below $0.3\%$ for the
angular observables in the $\tt$ inclusive channel;
in the tails of the $\Ptjeta$ and $\Ptjetb$ cross-sections, it
increases up to $17\%$ and $19\%$, respectively. For the $\tt+$1jet
channel, the statistical uncertainty in the data is
$\approx 0.3\%$ for the angular observables and increases up to $36\%$
and $23\%$ in the last bins of the $\Ptjete$ and $\mtte$
cross-sections, respectively. In the $\tt+$2jets channel, the
statistical uncertainty in the data is $\approx 0.4\%$ for the angular observables
and increases up to $19\%$ and $28\%$ in the last bins of the $\Ptjetf$
and $\mef$ cross-sections, respectively.
 
\FloatBarrier
 
\newpage
 
\FloatBarrier
 
\newpage
 
\begin{figure}[h]
\setlength{\unitlength}{1.0cm}
\begin{picture} (18.0,20.5)
\put (0.0,13.8){\includegraphics[width=7cm]{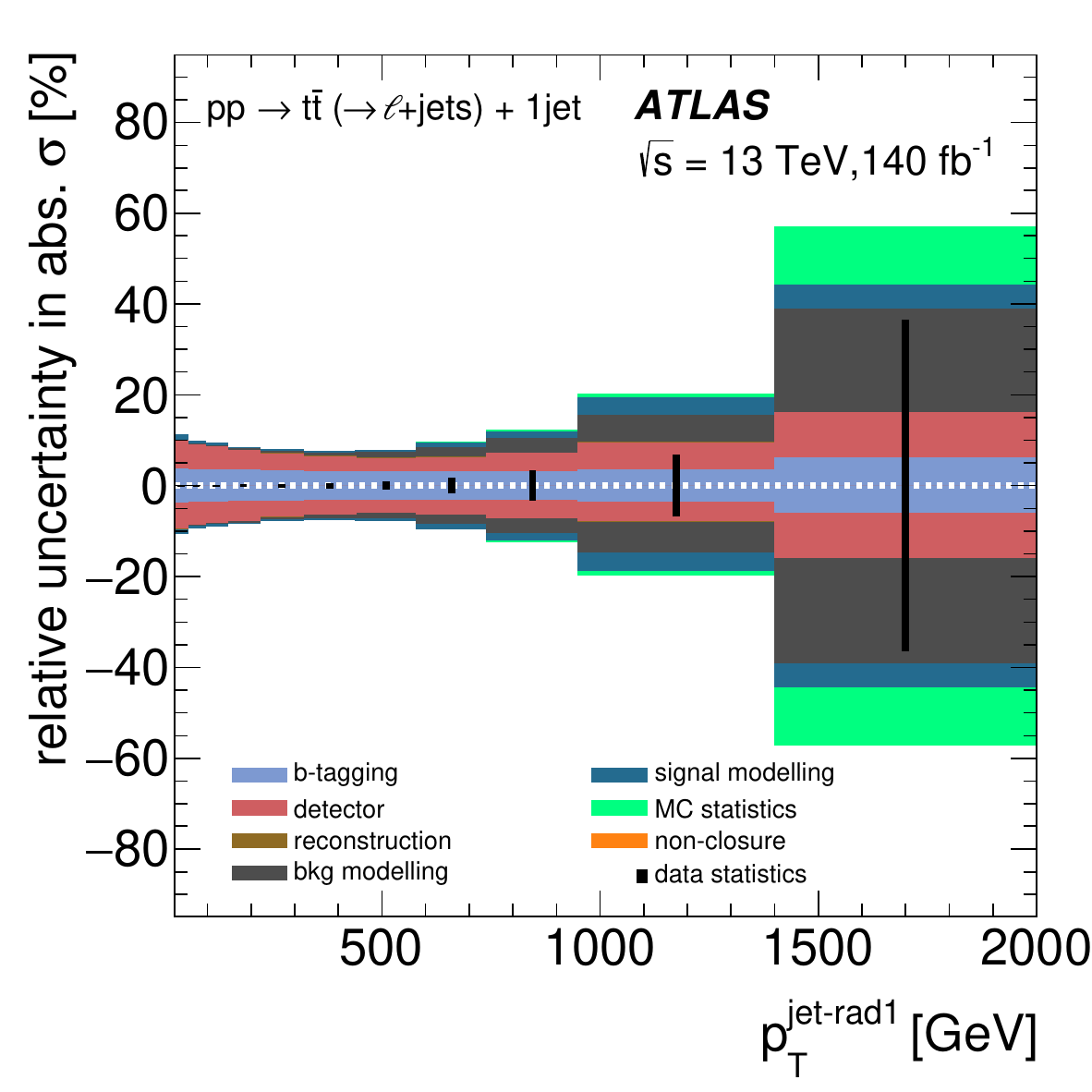}}
\put (8.0,13.8){\includegraphics[width=7cm]{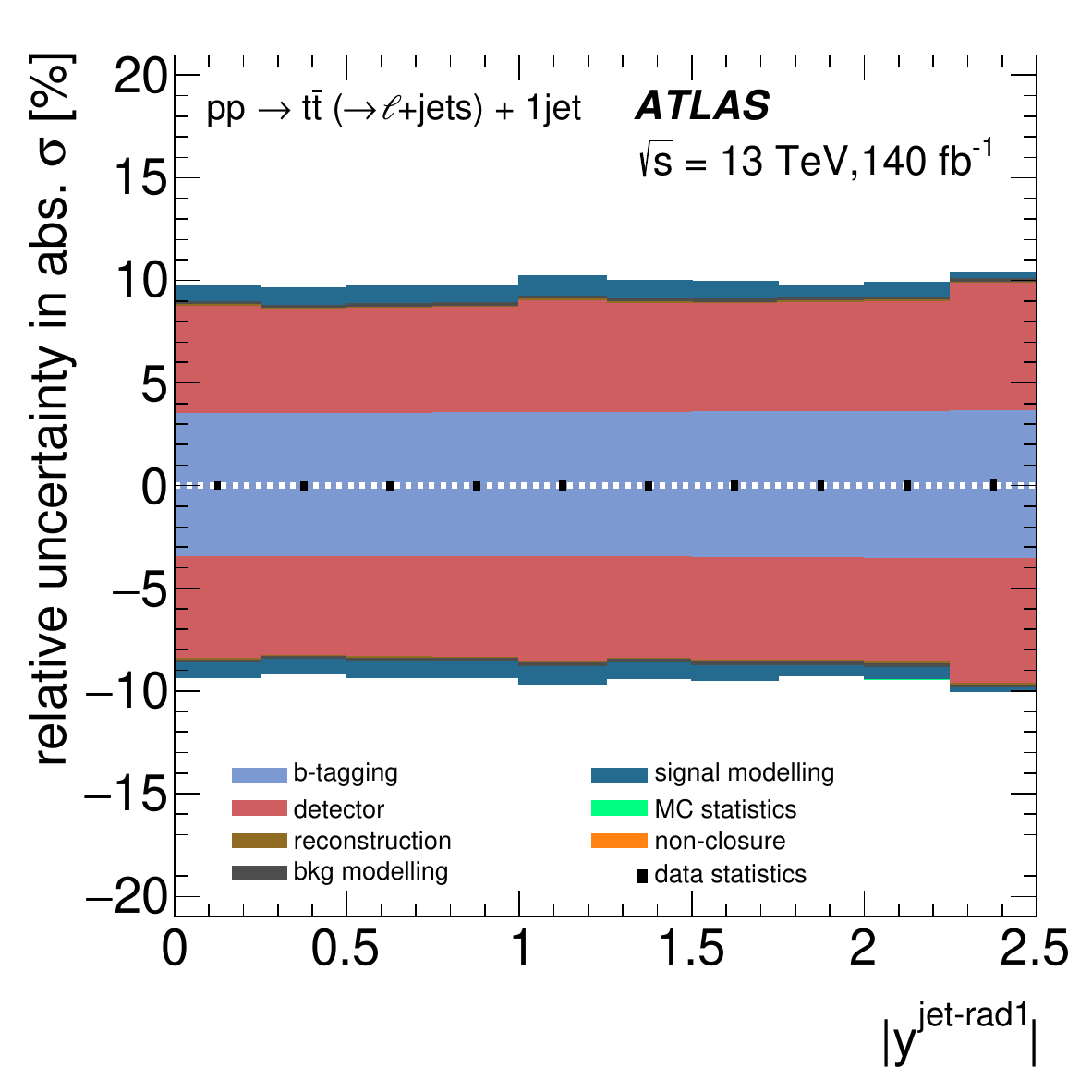}}
\put (0.0,6.9){\includegraphics[width=7cm]{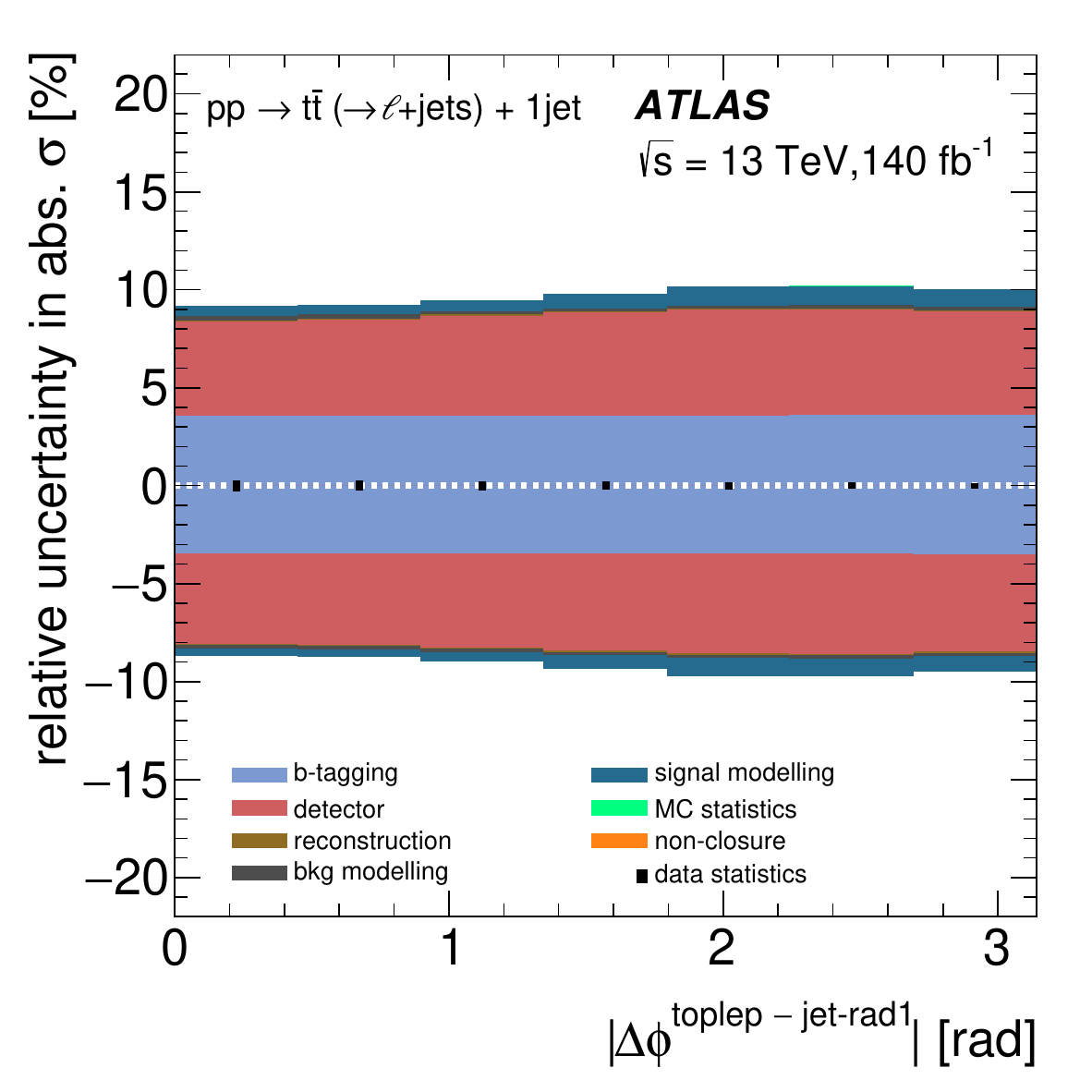}}
\put (8.0,6.9){\includegraphics[width=7cm]{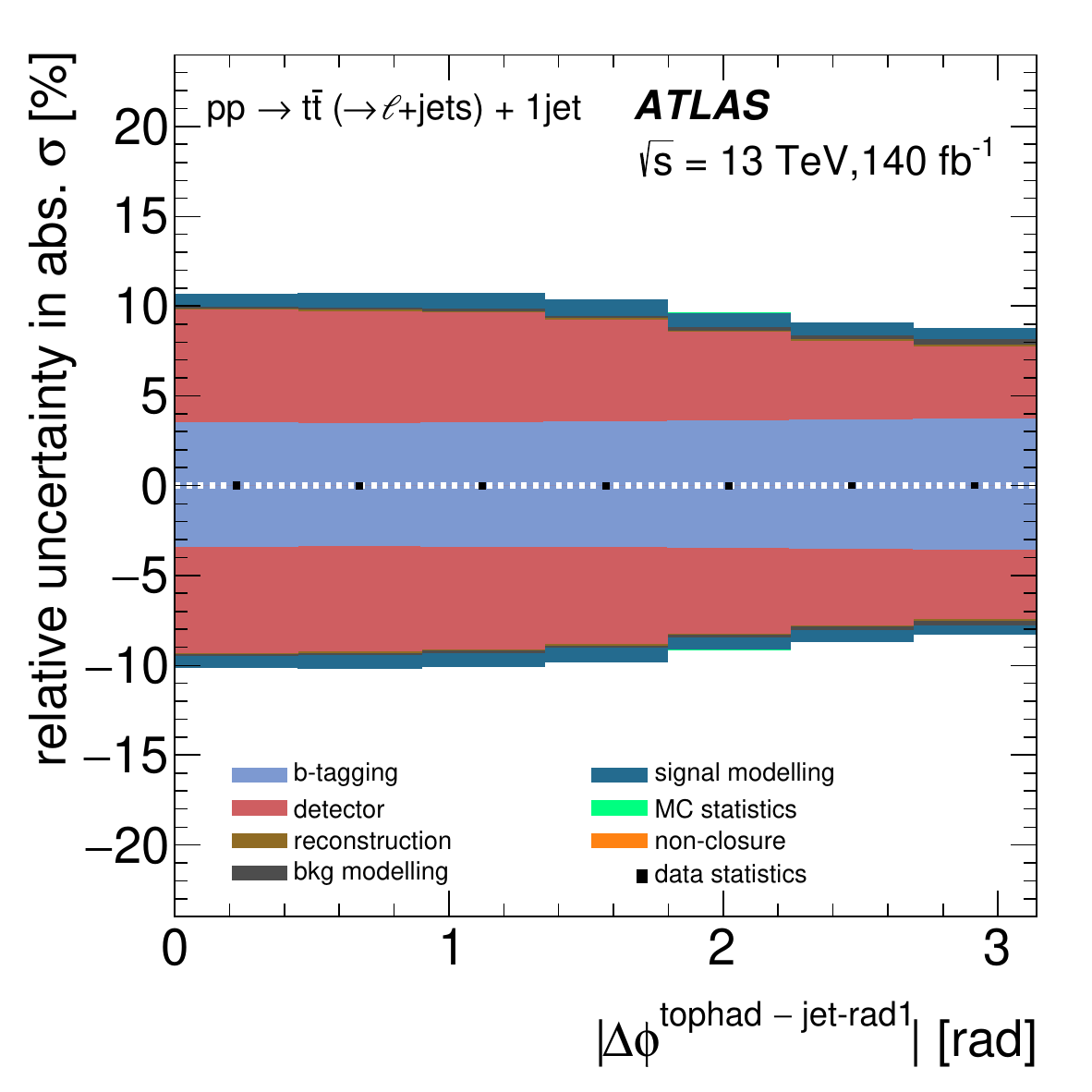}}
\put (0.0,0.0){\includegraphics[width=7cm]{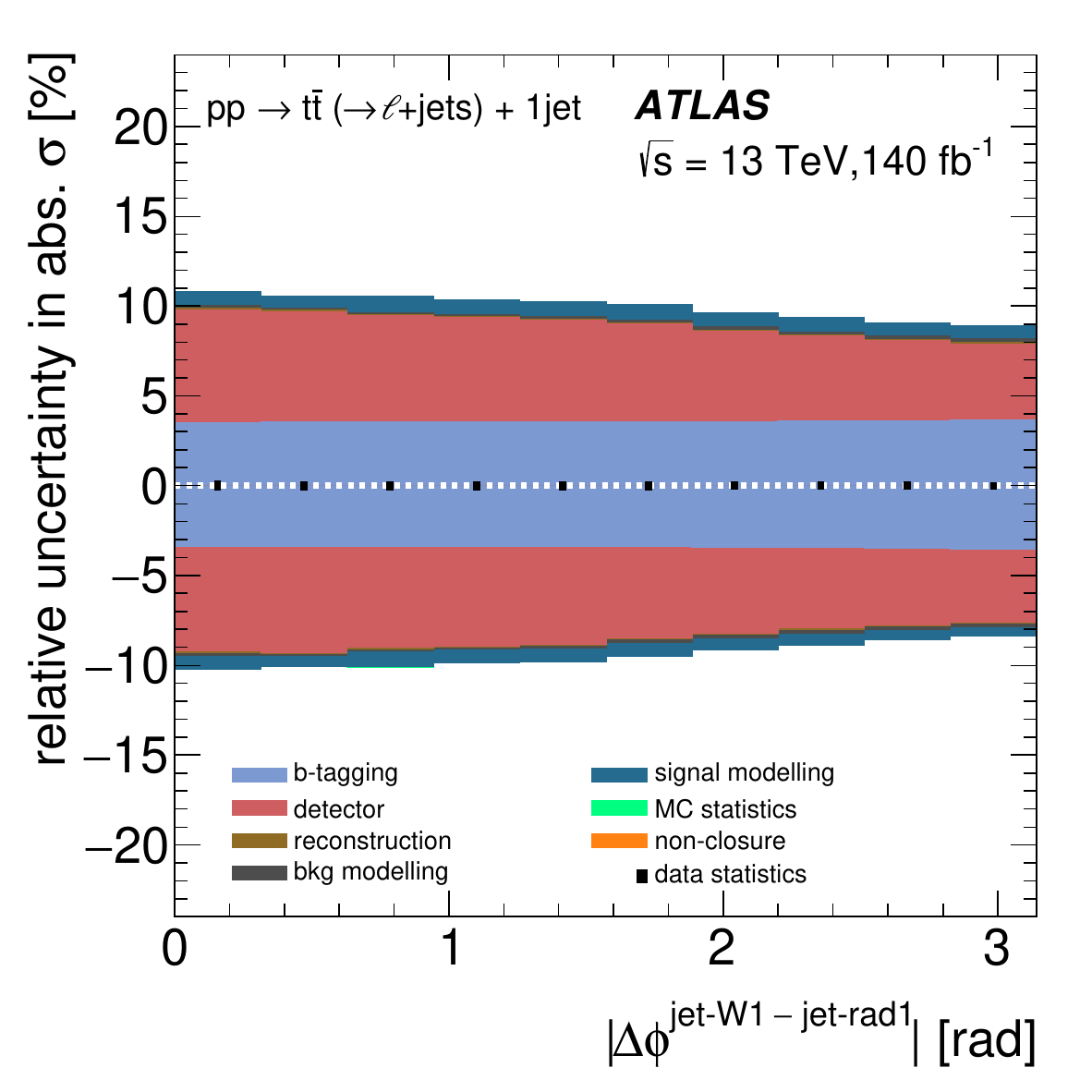}}
\put (8.0,0.0){\includegraphics[width=7cm]{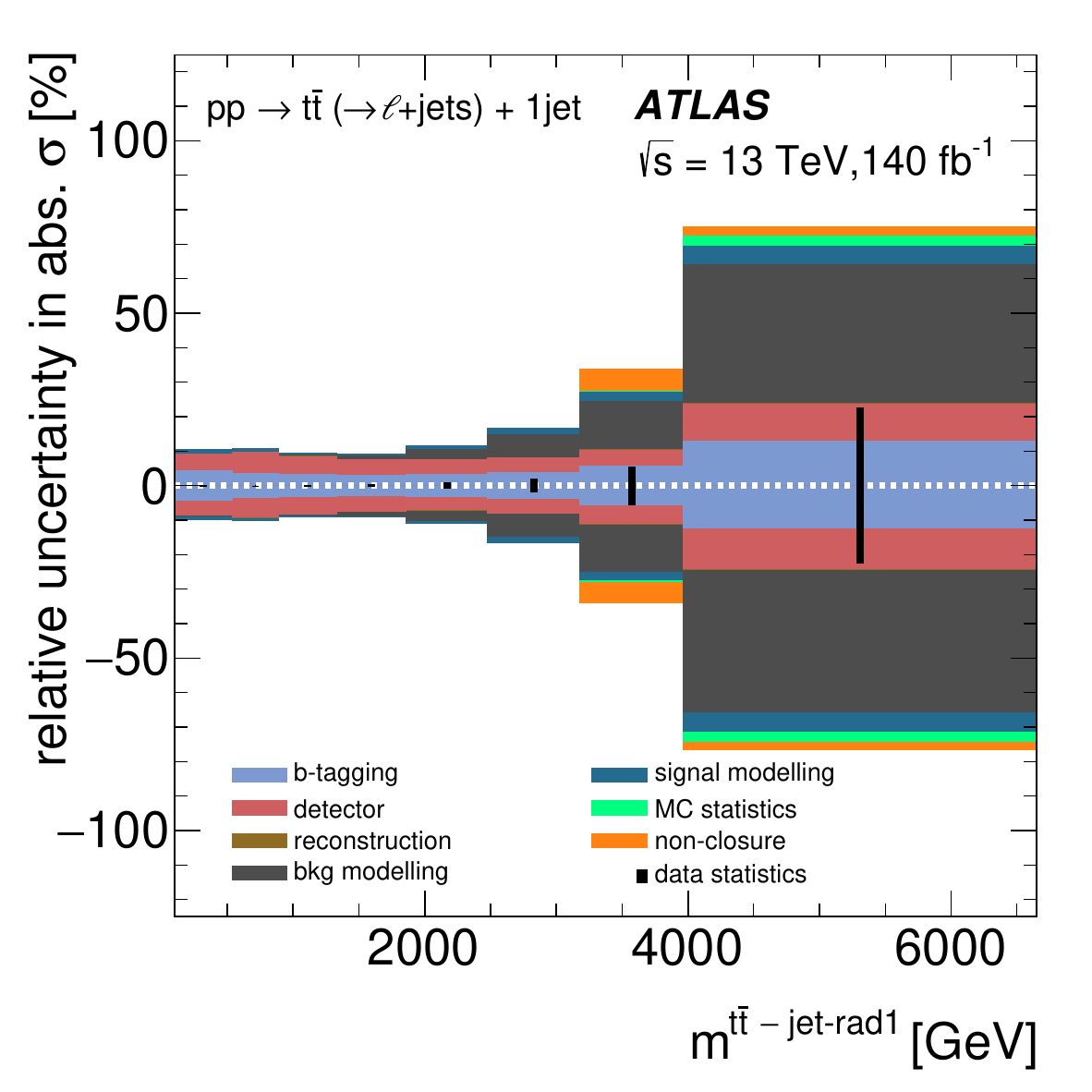}}
\put (3.0,14.0){\small (a)}
\put (11.0,14.0){\small (b)}
\put (3.0,7.1){\small (c)}
\put (11.0,7.1){\small (d)}
\put (3.0,0.2){\small (e)}
\put (11.0,0.2){\small (f)}
\end{picture}
\caption{
Total relative systematic uncertainties in the
measured cross-sections in percentages for
(a) $\Ptjete$, (b) $|\Rapjete|$,
(c) $|\deltaphitle|$, (d) $|\deltaphithe|$,
(e) $|\deltaphiae|$ and (f) $\mtte$
in the $\tt+$1jet channel.
The individual sources of systematic uncertainty (plotted in stacked
form) and the statistical uncertainty in the data are also included.
}
\end{figure}
 
\FloatBarrier
 
\newpage
 
\begin{figure}[h]
\setlength{\unitlength}{1.0cm}
\begin{picture} (18.0,17.5)
\put (0.0,6.9){\includegraphics[width=7cm]{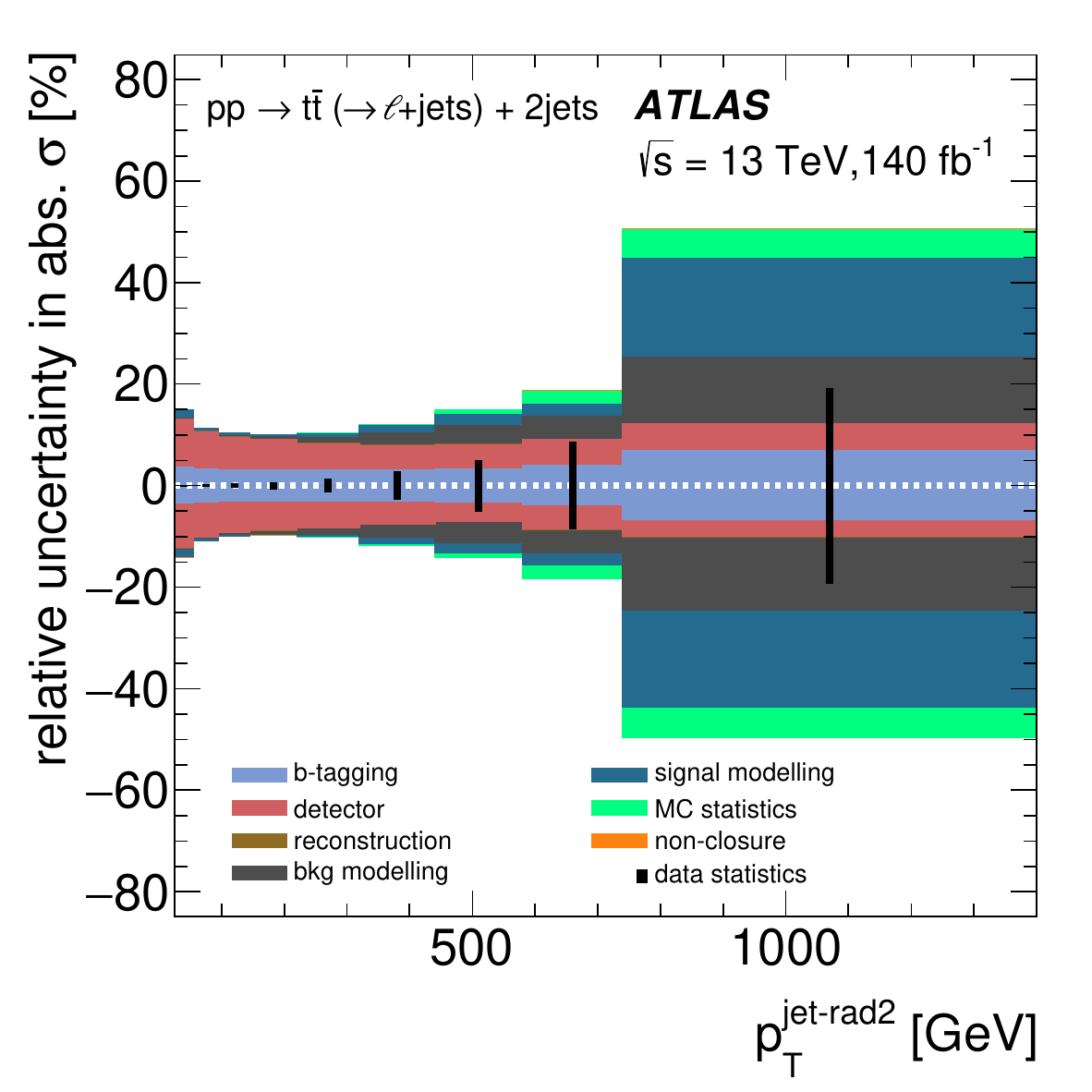}}
\put (8.0,6.9){\includegraphics[width=7cm]{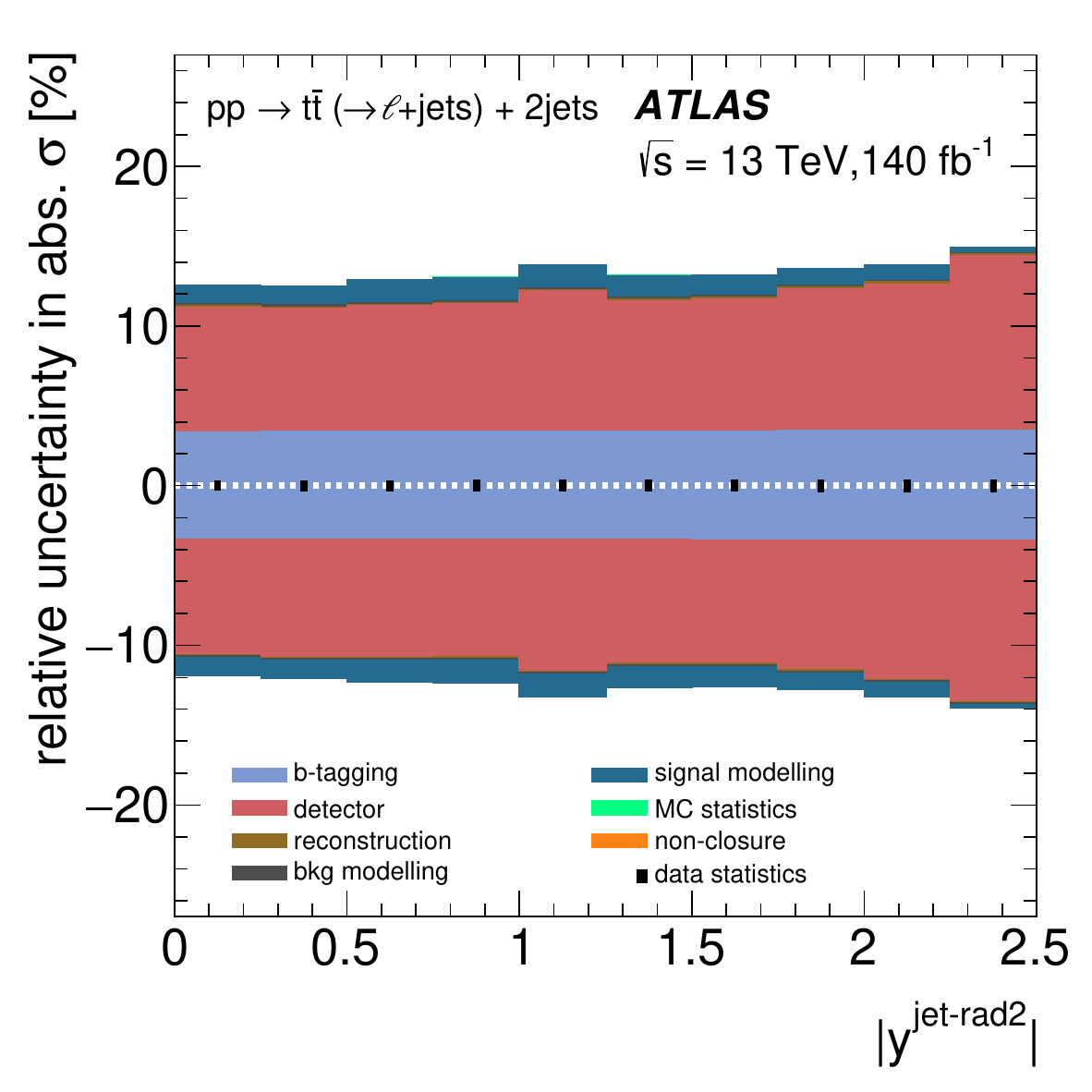}}
\put (0.0,0.0){\includegraphics[width=7cm]{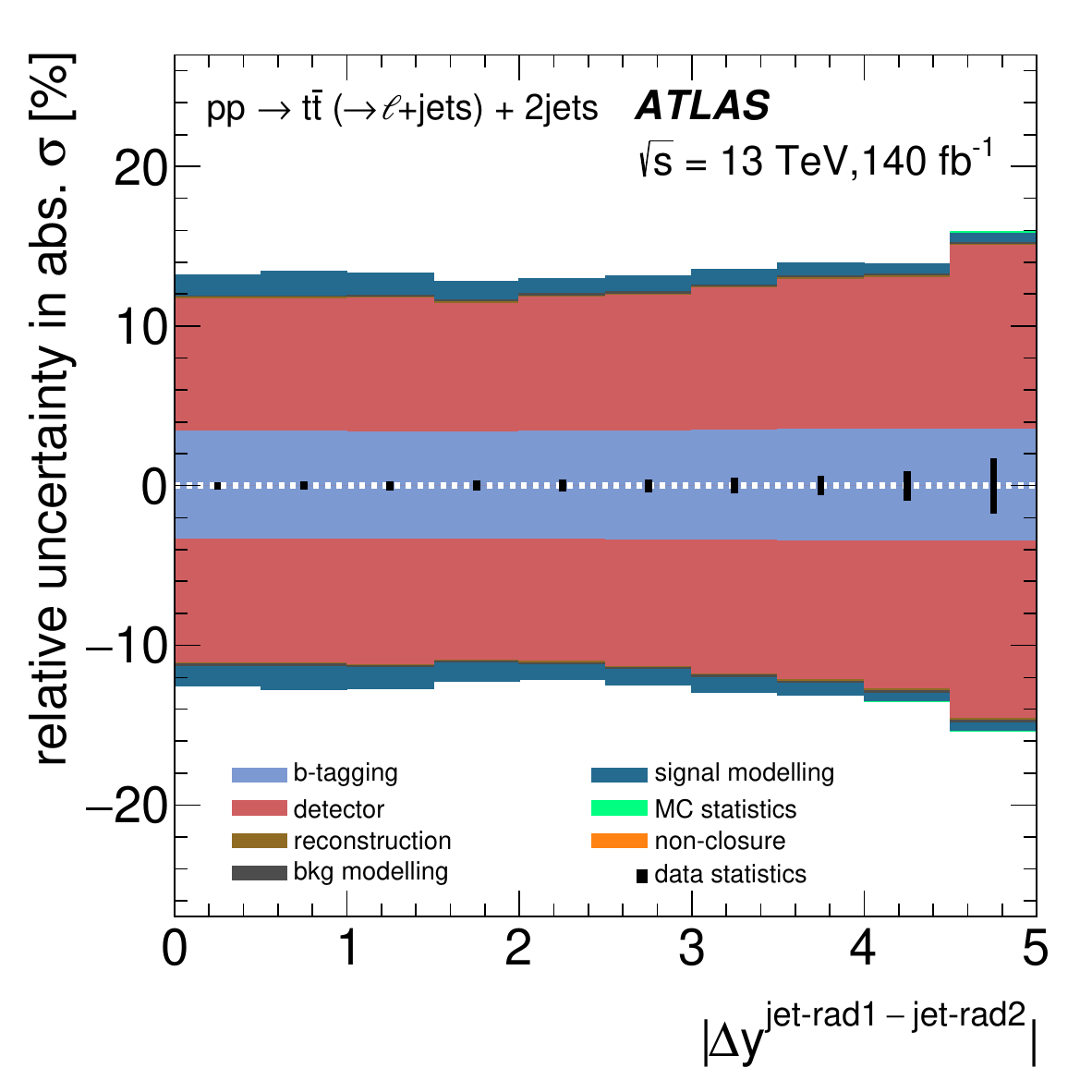}}
\put (8.0,0.0){\includegraphics[width=7cm]{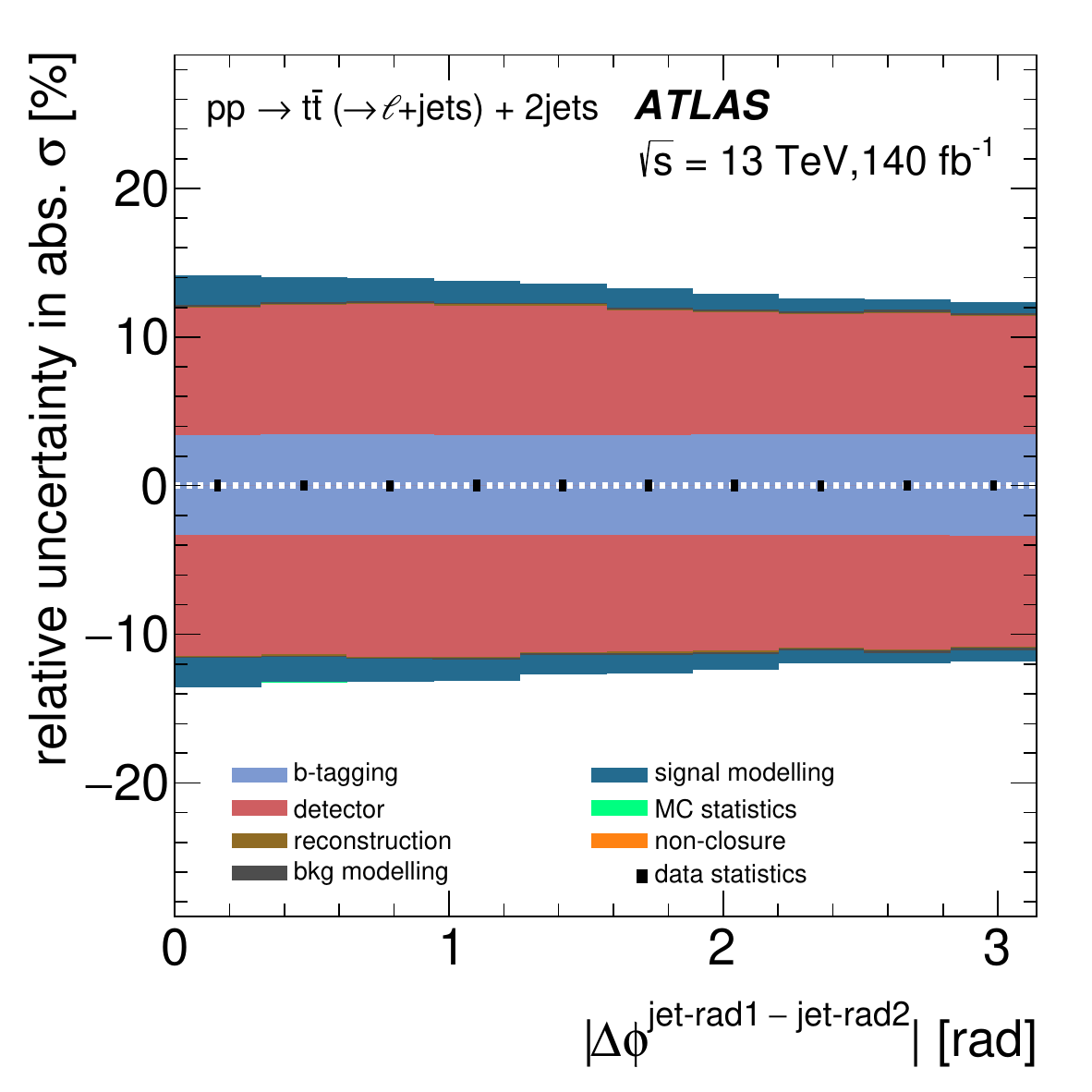}}
\put (3.0,7.1){\small (a)}
\put (11.0,7.1){\small (b)}
\put (3.0,0.2){\small (c)}
\put (11.0,0.2){\small (d)}
\end{picture}
\caption{
Total relative systematic uncertainties in the
measured cross-sections in percentages for
(a) $\Ptjetf$, (b) $|\Rapjetf|$,
(c) $|\deltayef|$ and (d) $|\deltaphief|$
in the $\tt+$2jets channel.
The individual sources of systematic uncertainty (plotted in stacked
form) and the statistical uncertainty in the data are also included.
}
\end{figure}
 
\FloatBarrier
 
\newpage
 
\begin{figure}[h]
\setlength{\unitlength}{1.0cm}
\begin{picture} (18.0,17.5)
\put (0.0,6.9){\includegraphics[width=7cm]{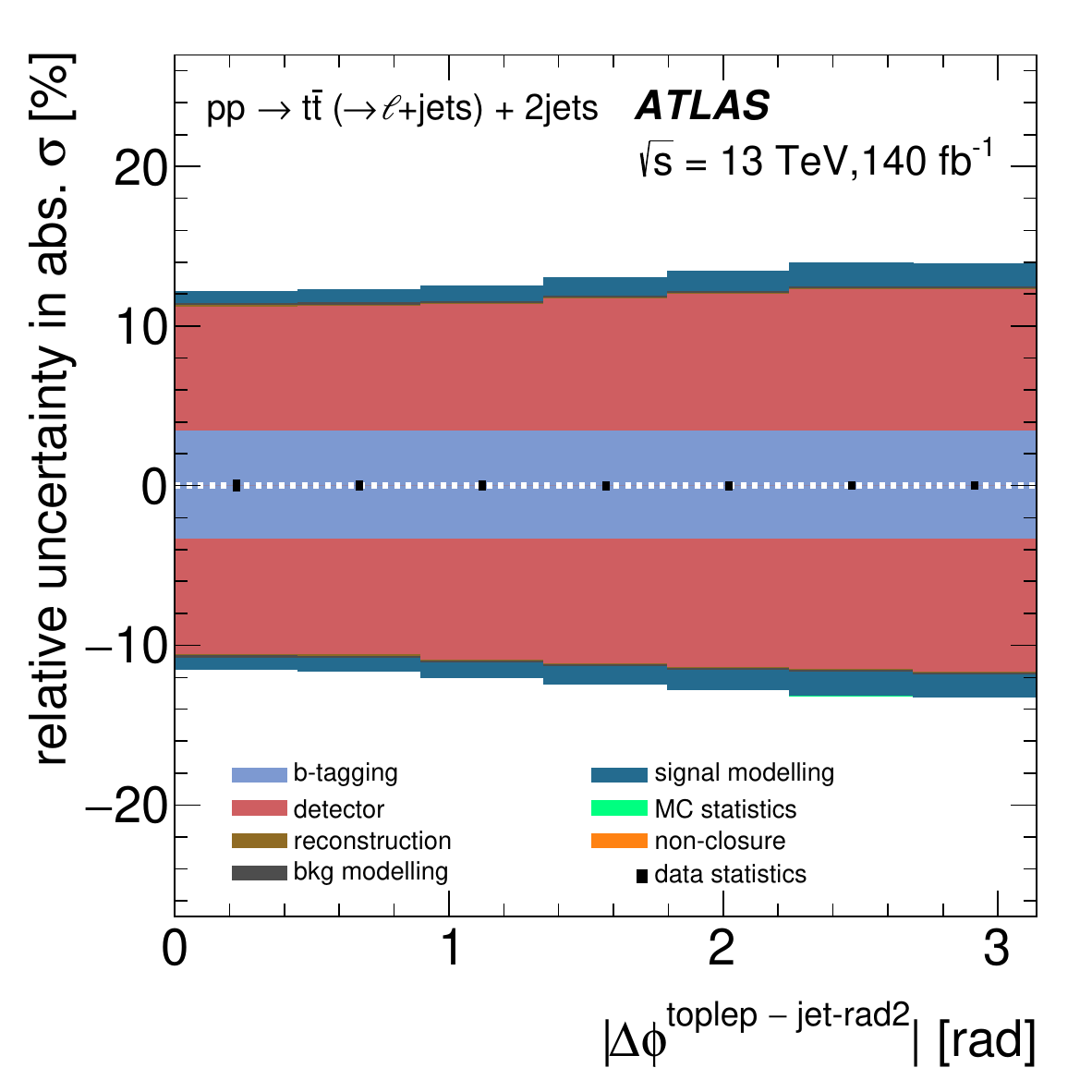}}
\put (8.0,6.9){\includegraphics[width=7cm]{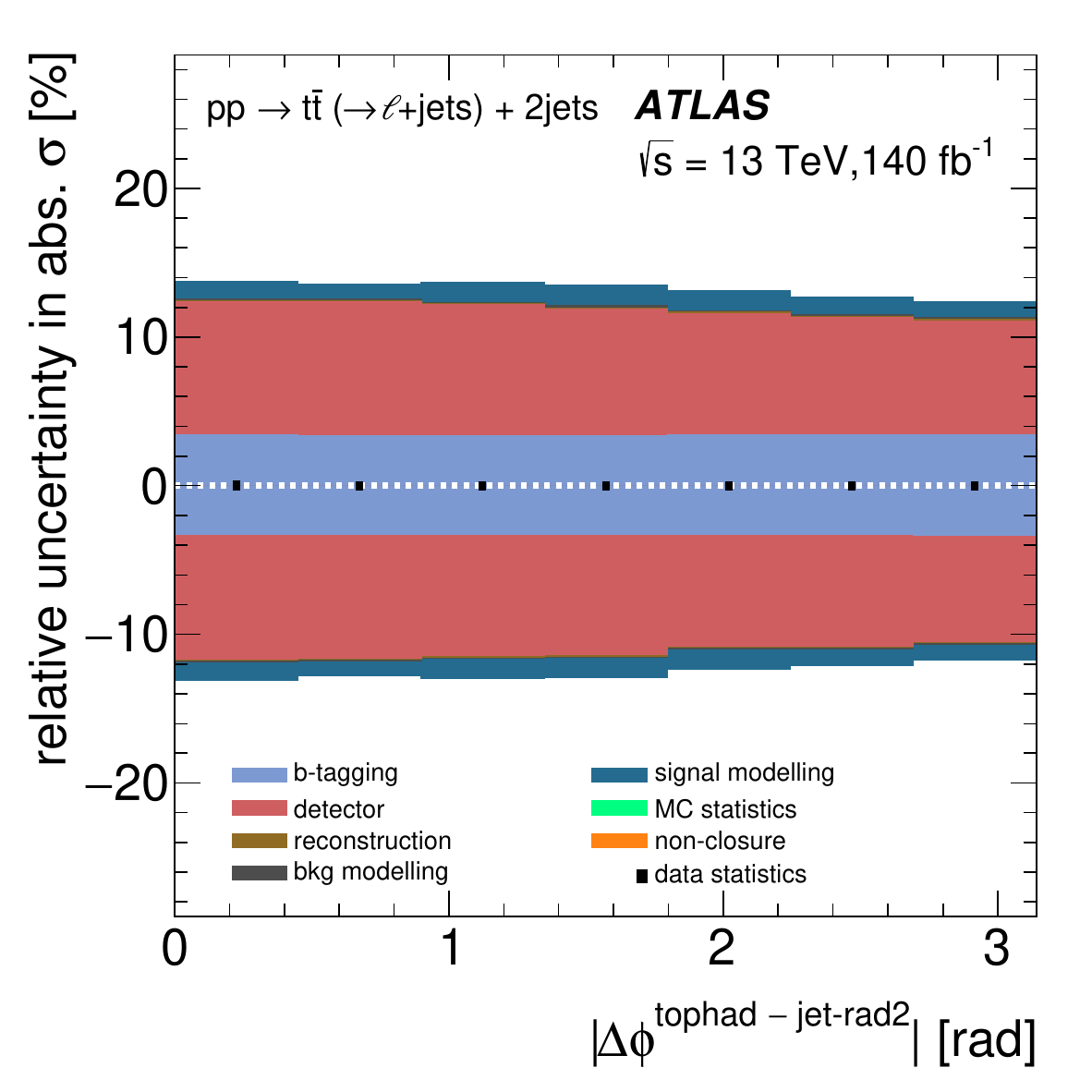}}
\put (0.0,0.0){\includegraphics[width=7cm]{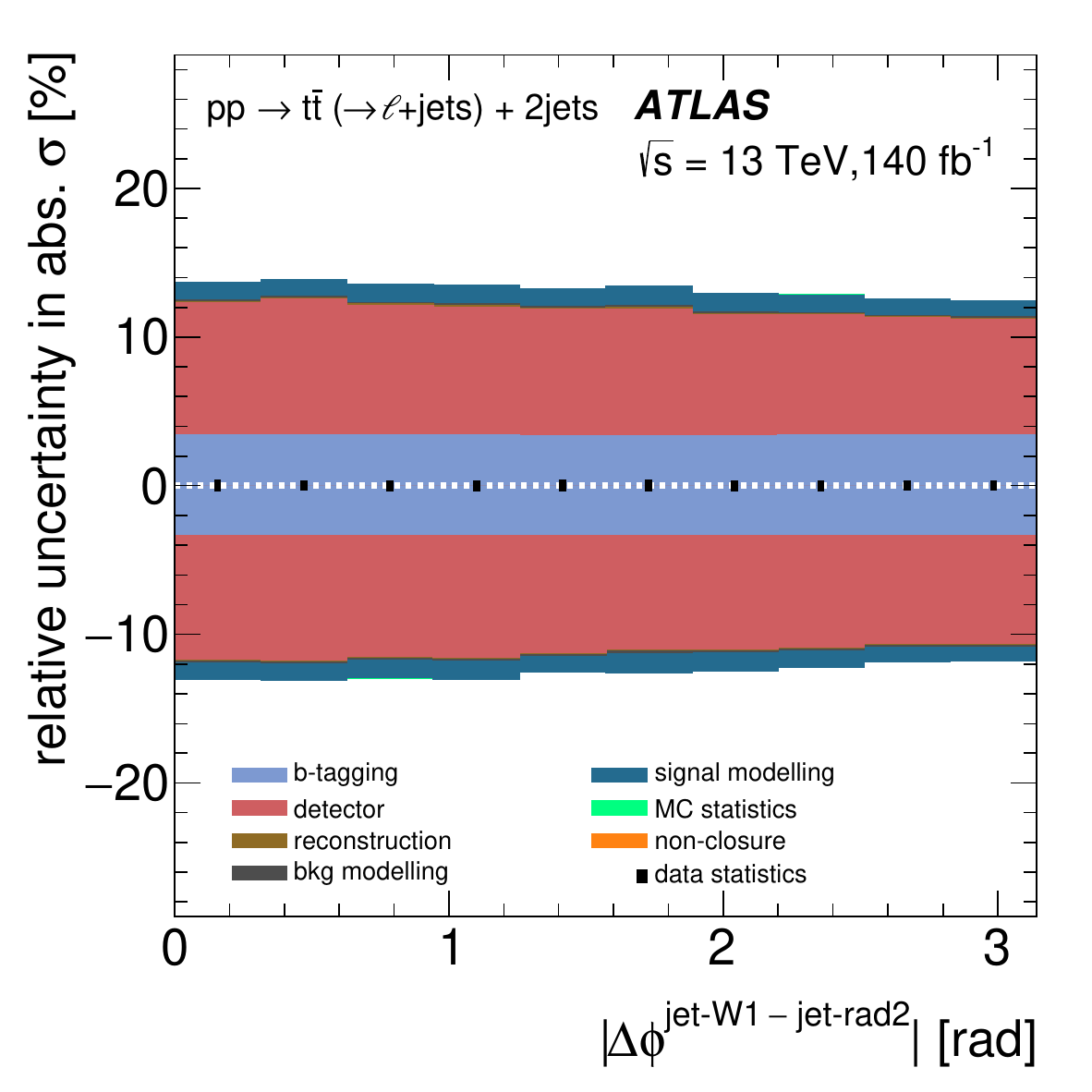}}
\put (8.0,0.0){\includegraphics[width=7cm]{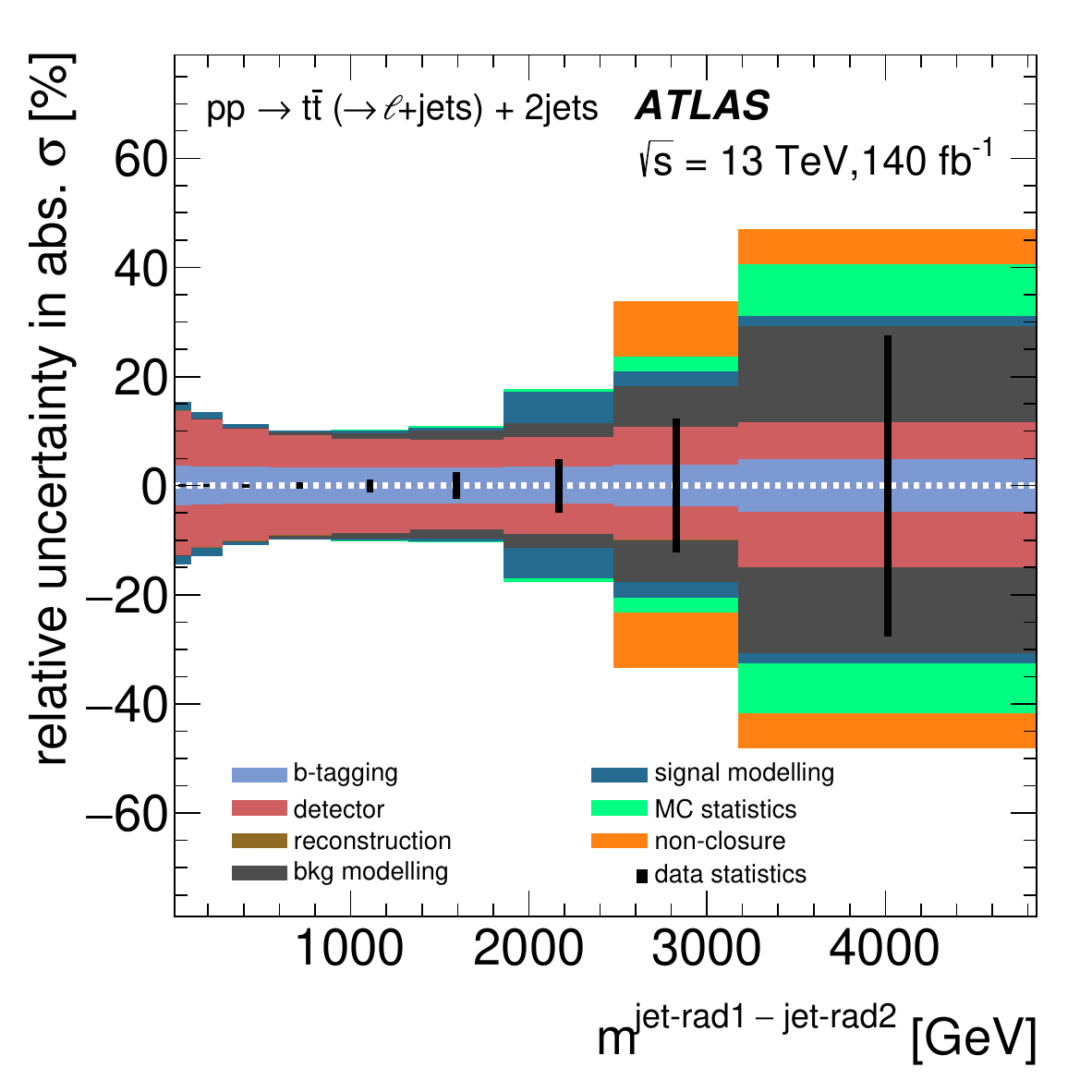}}
\put (3.0,7.1){\small (a)}
\put (11.0,7.1){\small (b)}
\put (3.0,0.2){\small (c)}
\put (11.0,0.2){\small (d)}
\end{picture}
\caption{
Total relative systematic uncertainties in the
measured cross-sections in percentages for
(a) $|\deltaphitlf|$, (b) $|\deltaphithf|$,
(c) $|\deltaphiaf|$ and (d) $\mef$
in the $\tt+$2jets channel.
The individual sources of systematic uncertainty (plotted in stacked
form) and the statistical uncertainty in the data are also included.
}
\label{fig77}
\end{figure}
 
\FloatBarrier
 
\newpage
 
\begin{figure}[h]
\setlength{\unitlength}{1.0cm}
\begin{picture} (18.0,20.5)
\put (0.0,13.8){\includegraphics[width=7cm]{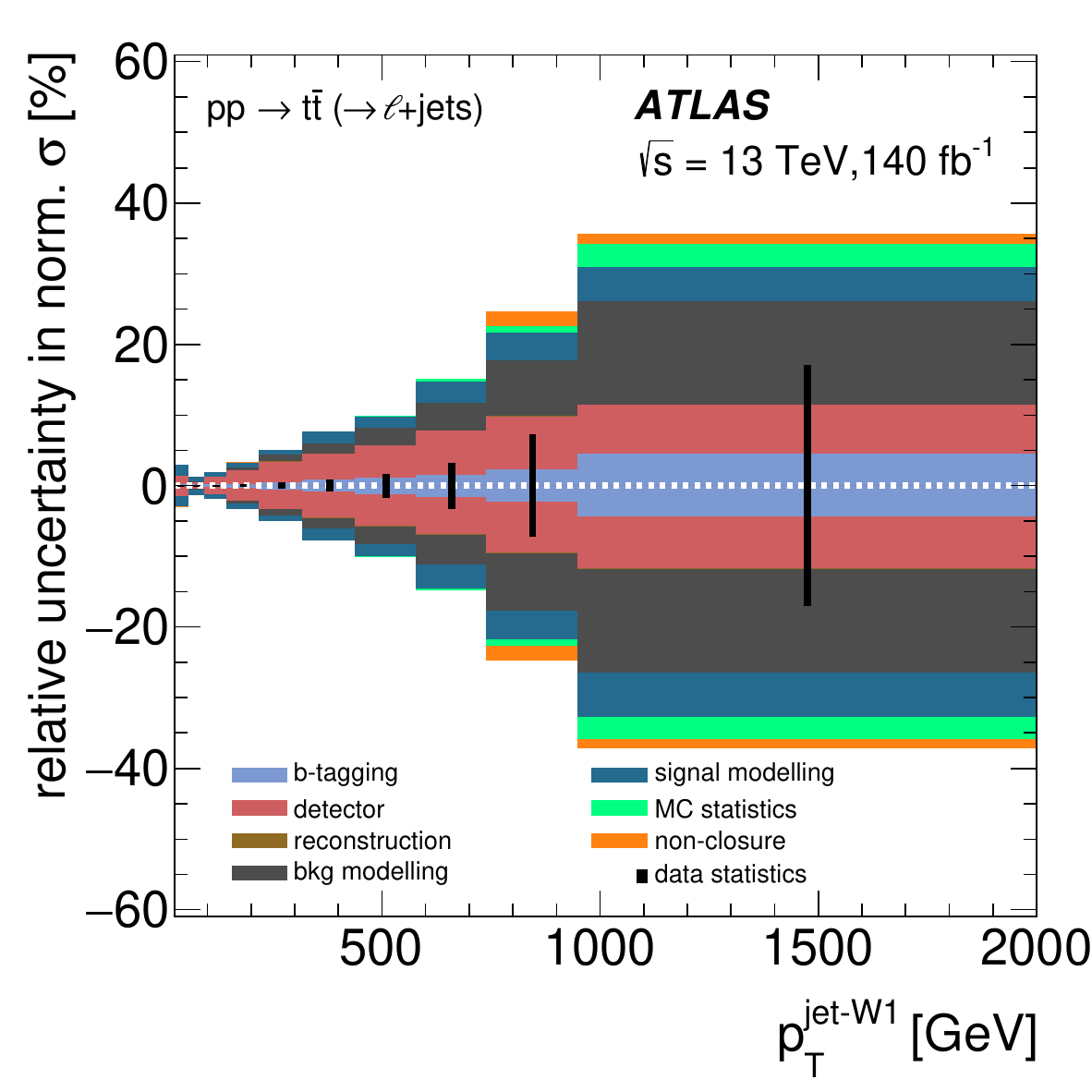}}
\put (8.0,13.8){\includegraphics[width=7cm]{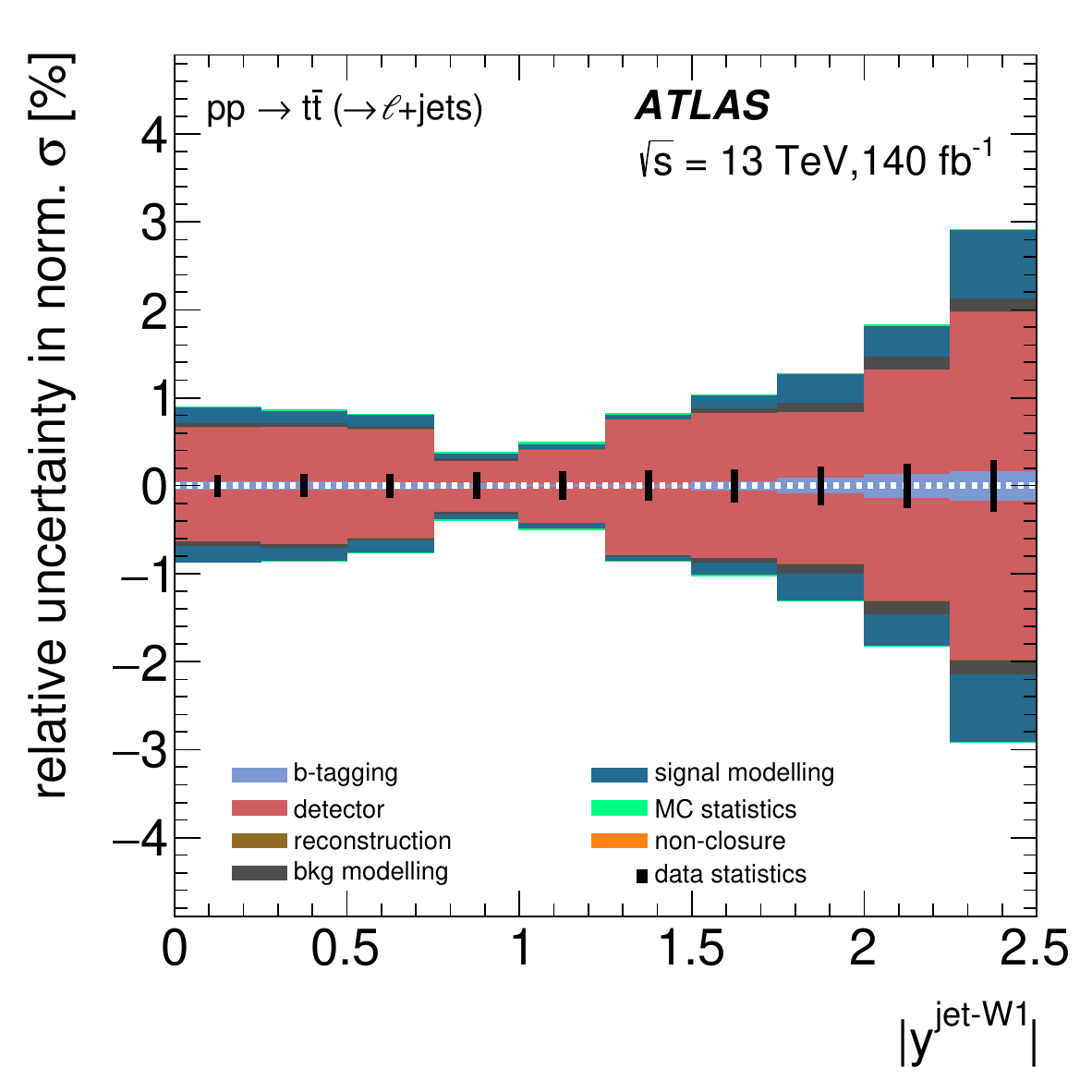}}
\put (0.0,6.9){\includegraphics[width=7cm]{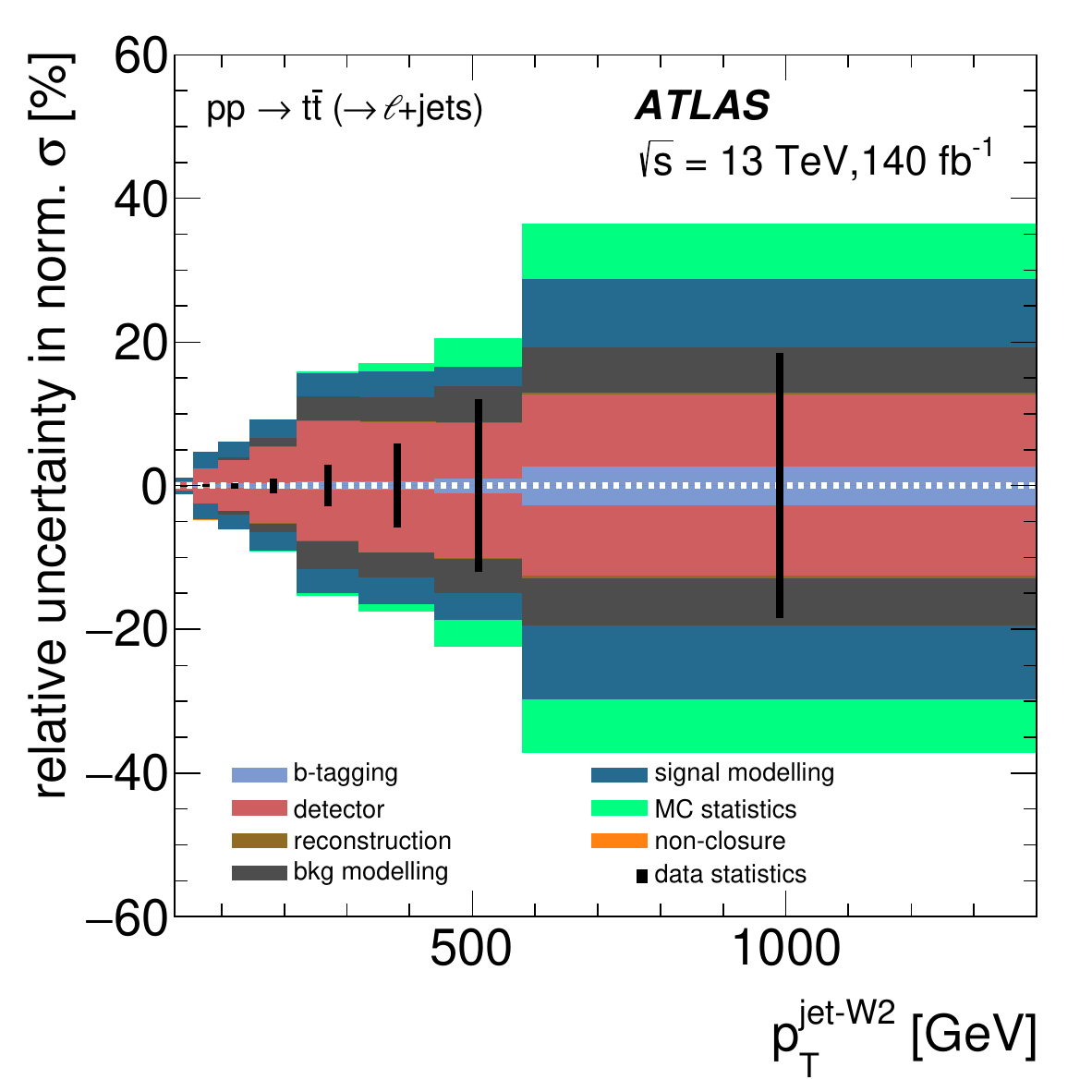}}
\put (8.0,6.9){\includegraphics[width=7cm]{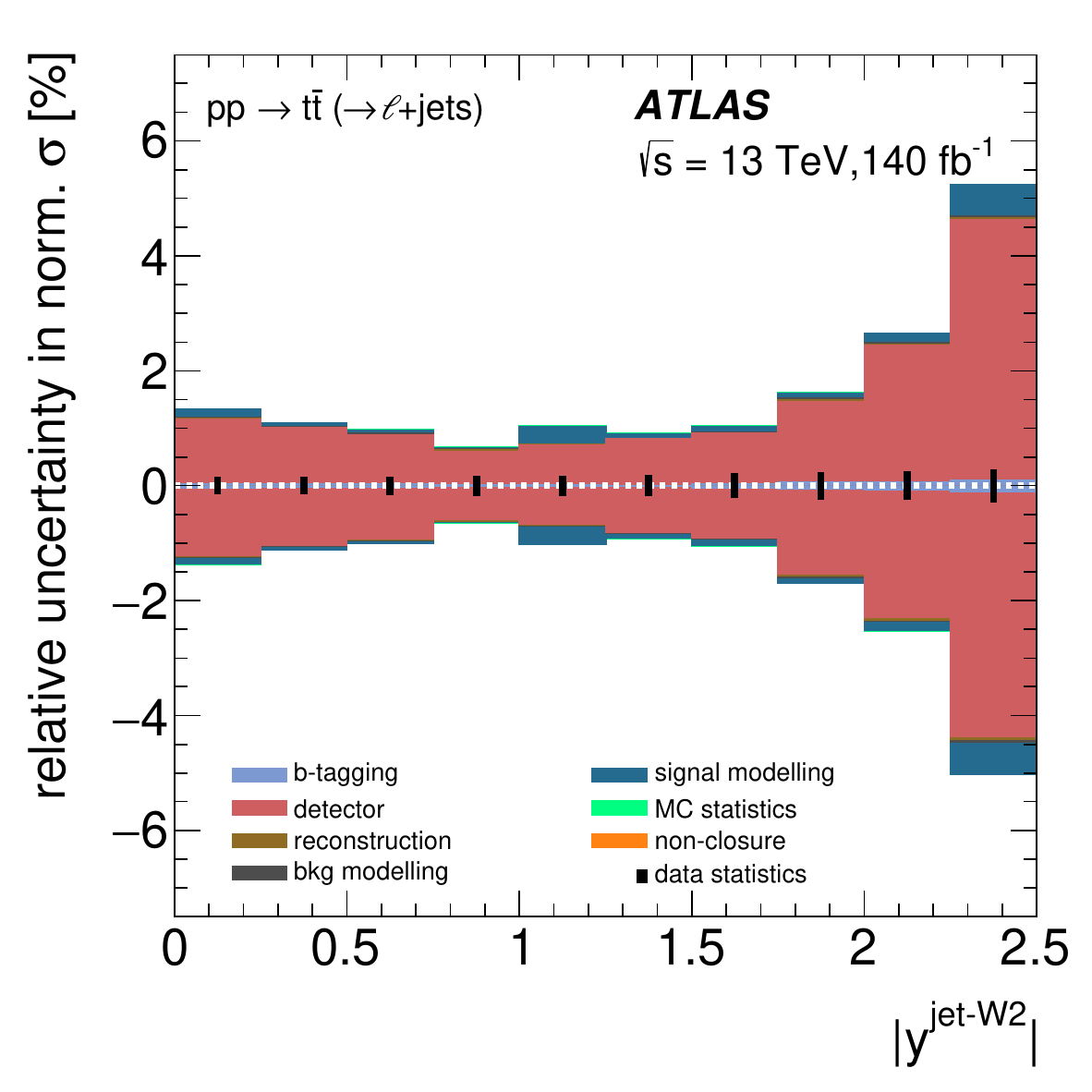}}
\put (0.0,0.0){\includegraphics[width=7cm]{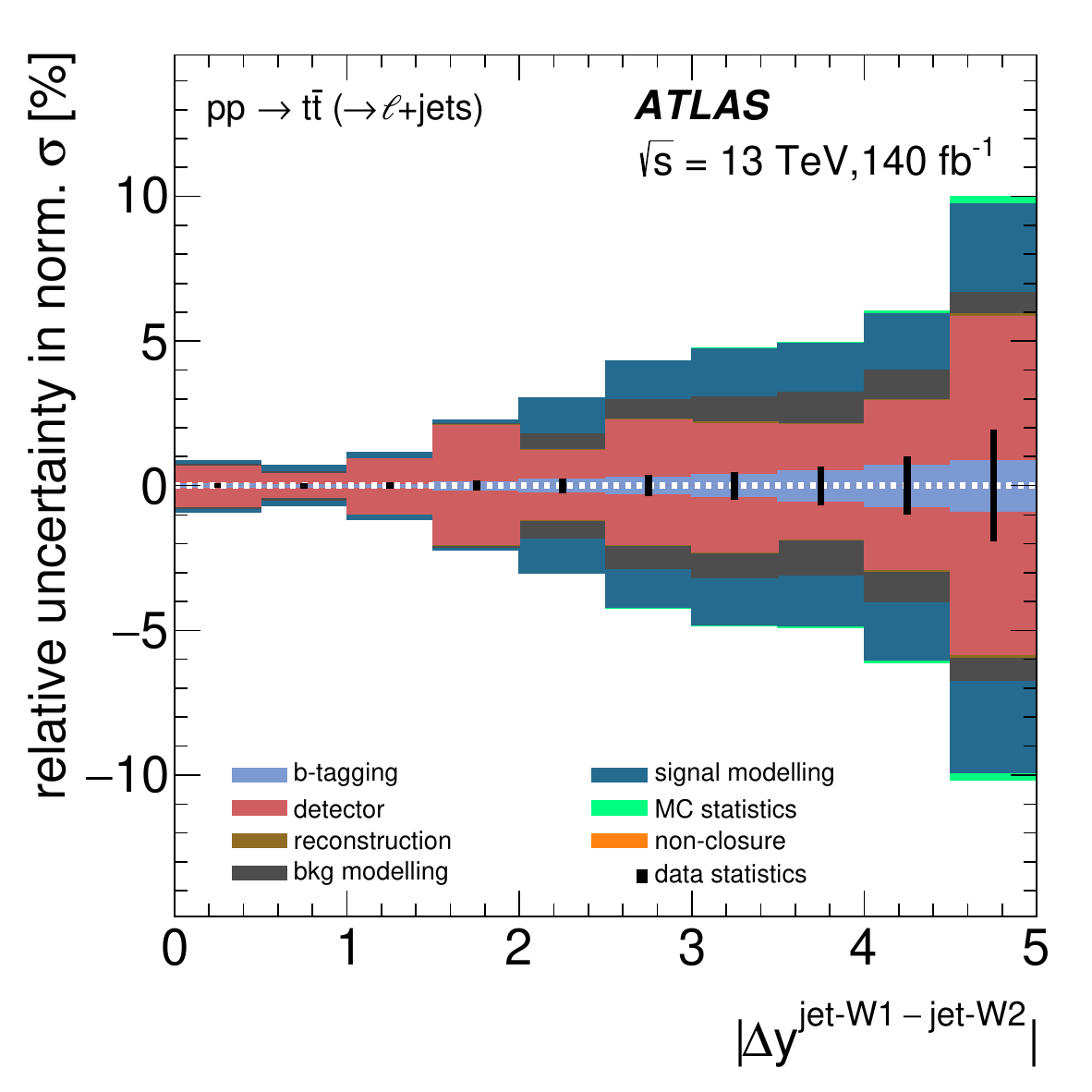}}
\put (8.0,0.0){\includegraphics[width=7cm]{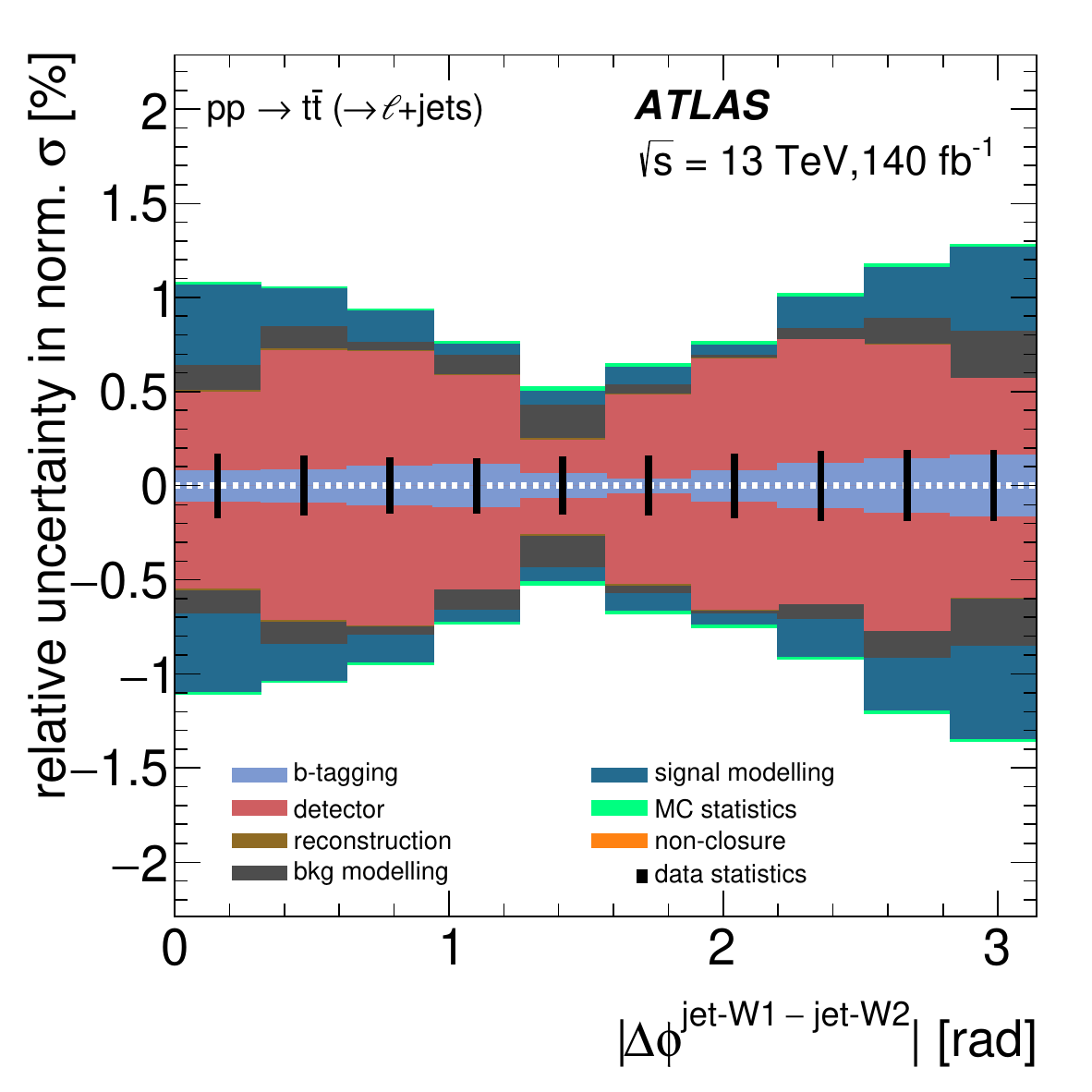}}
\put (3.0,14.0){\small (a)}
\put (11.0,14.0){\small (b)}
\put (3.0,7.1){\small (c)}
\put (11.0,7.1){\small (d)}
\put (3.0,0.2){\small (e)}
\put (11.0,0.2){\small (f)}
\end{picture}
\caption{
Total relative systematic uncertainties in the
measured normalised cross-sections in percentages for
(a) $\Ptjeta$, (b) $|\Rapjeta|$,
(c) $\Ptjetb$, (d) $|\Rapjetb|$,
(e) $|\deltayab|$ and (f) $|\deltaphiab|$
in the $\tt$ inclusive channel.
The individual sources of systematic uncertainty (plotted in stacked
form) and the statistical uncertainty in the data are also included.
}
\label{fig_appm37}
\end{figure}
 
\FloatBarrier
 
\newpage
 
\begin{figure}[h]
\setlength{\unitlength}{1.0cm}
\begin{picture} (18.0,20.5)
\put (0.0,13.8){\includegraphics[width=7cm]{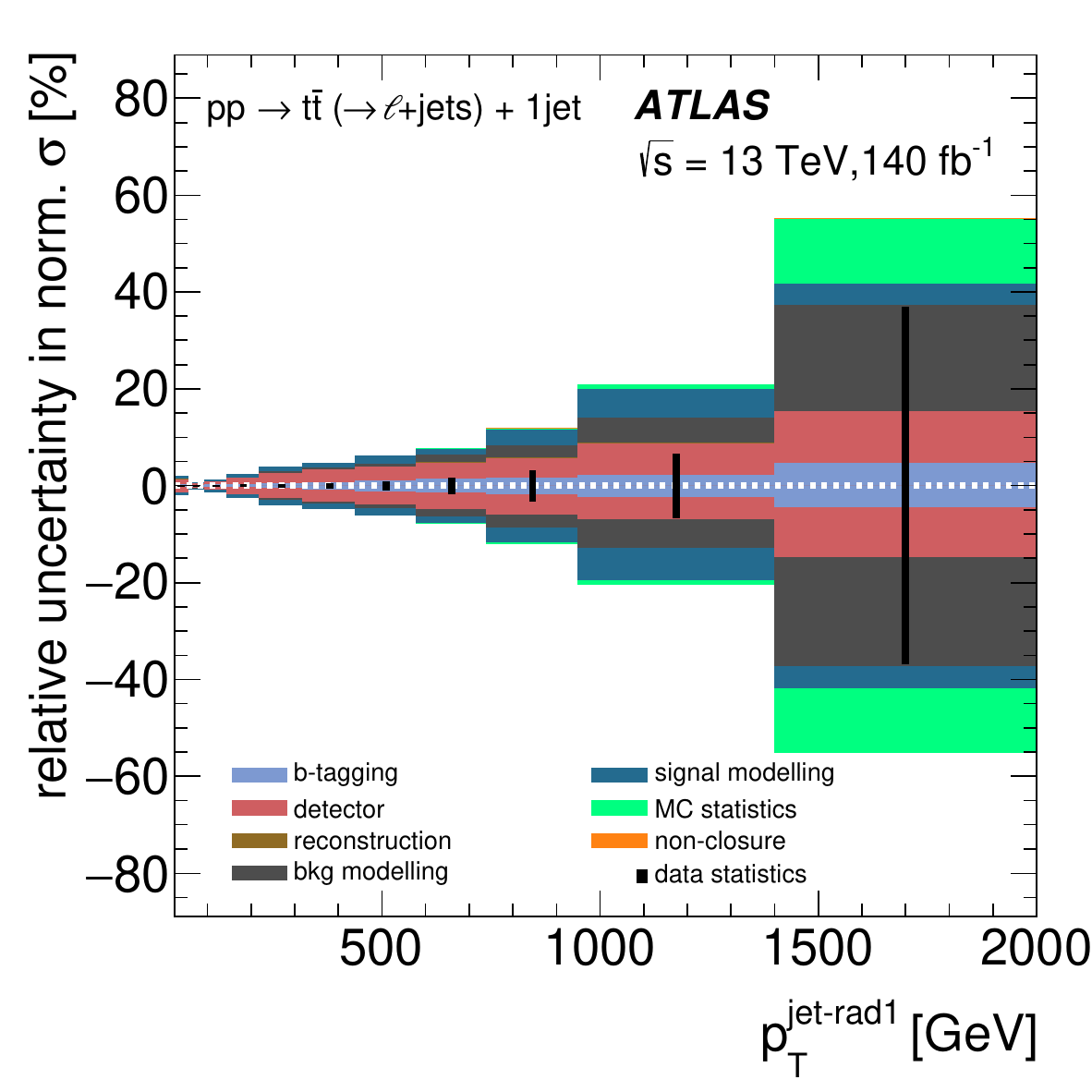}}
\put (8.0,13.8){\includegraphics[width=7cm]{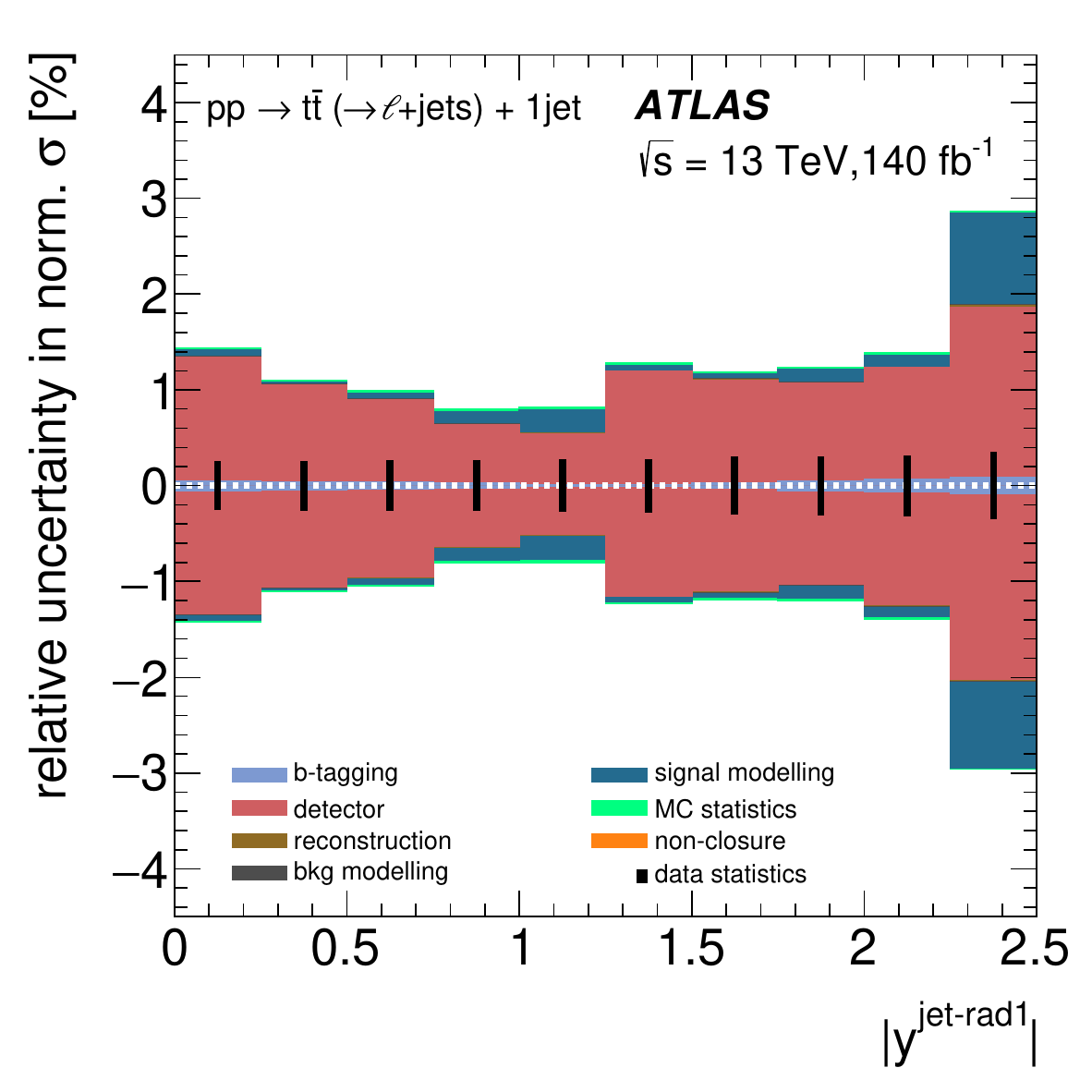}}
\put (0.0,6.9){\includegraphics[width=7cm]{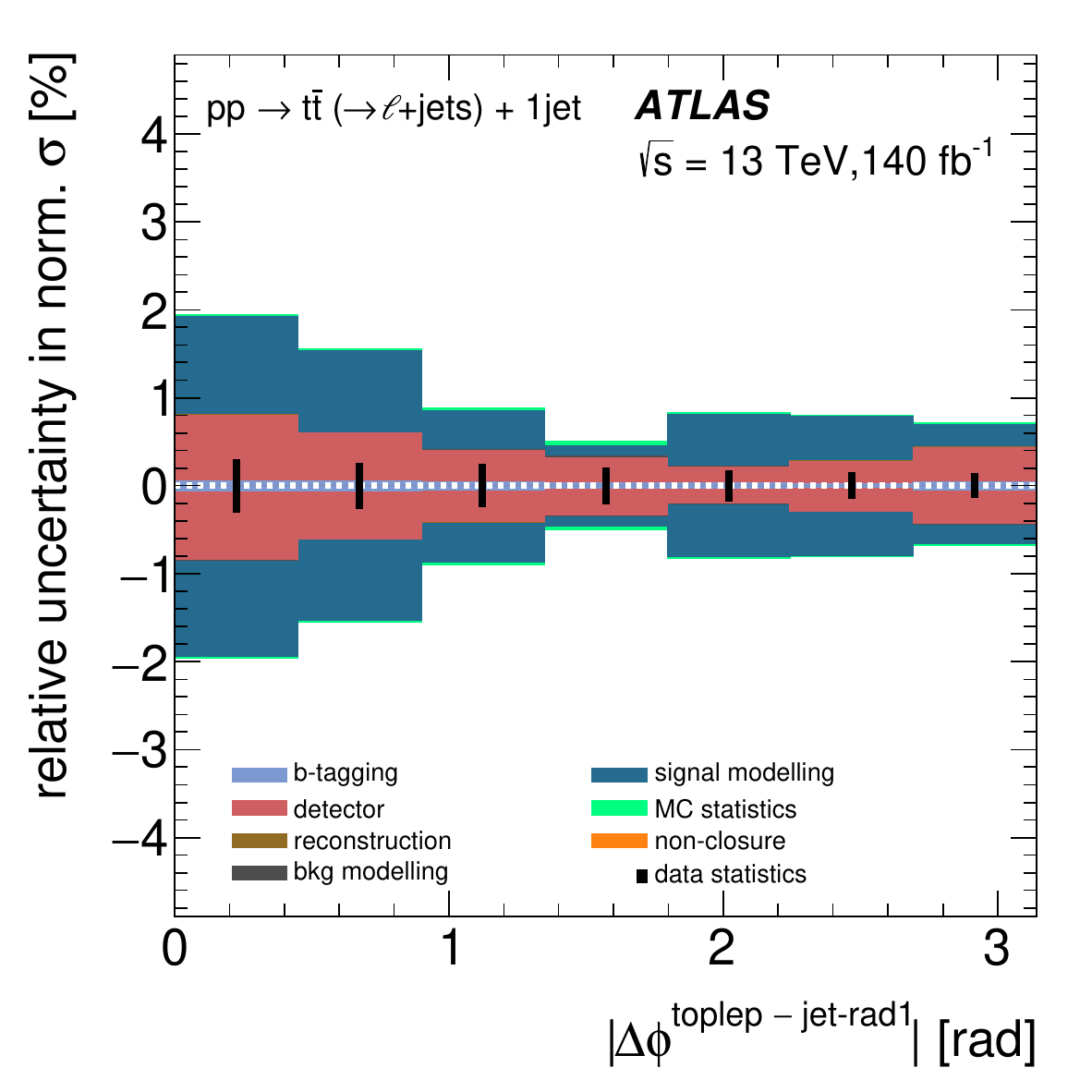}}
\put (8.0,6.9){\includegraphics[width=7cm]{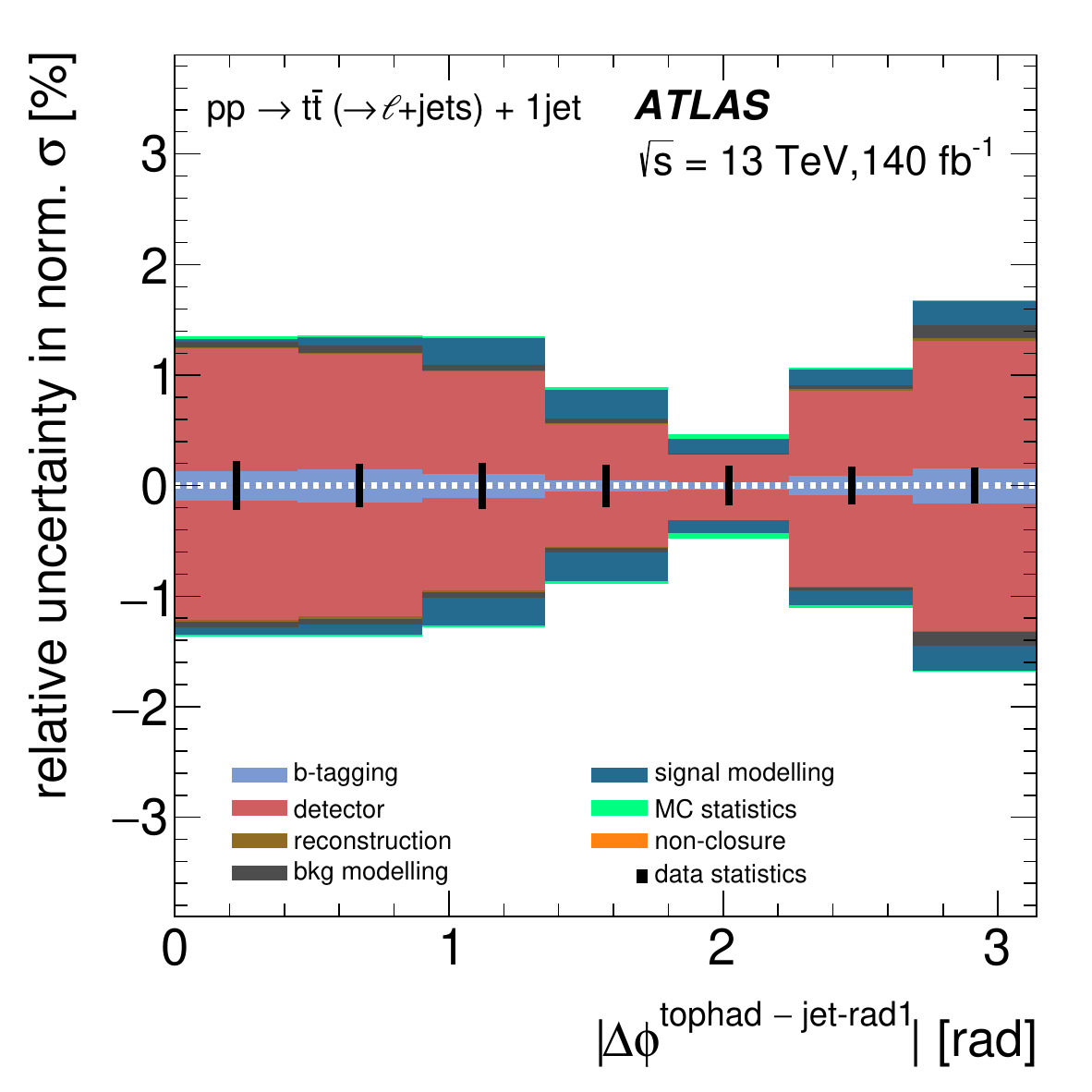}}
\put (0.0,0.0){\includegraphics[width=7cm]{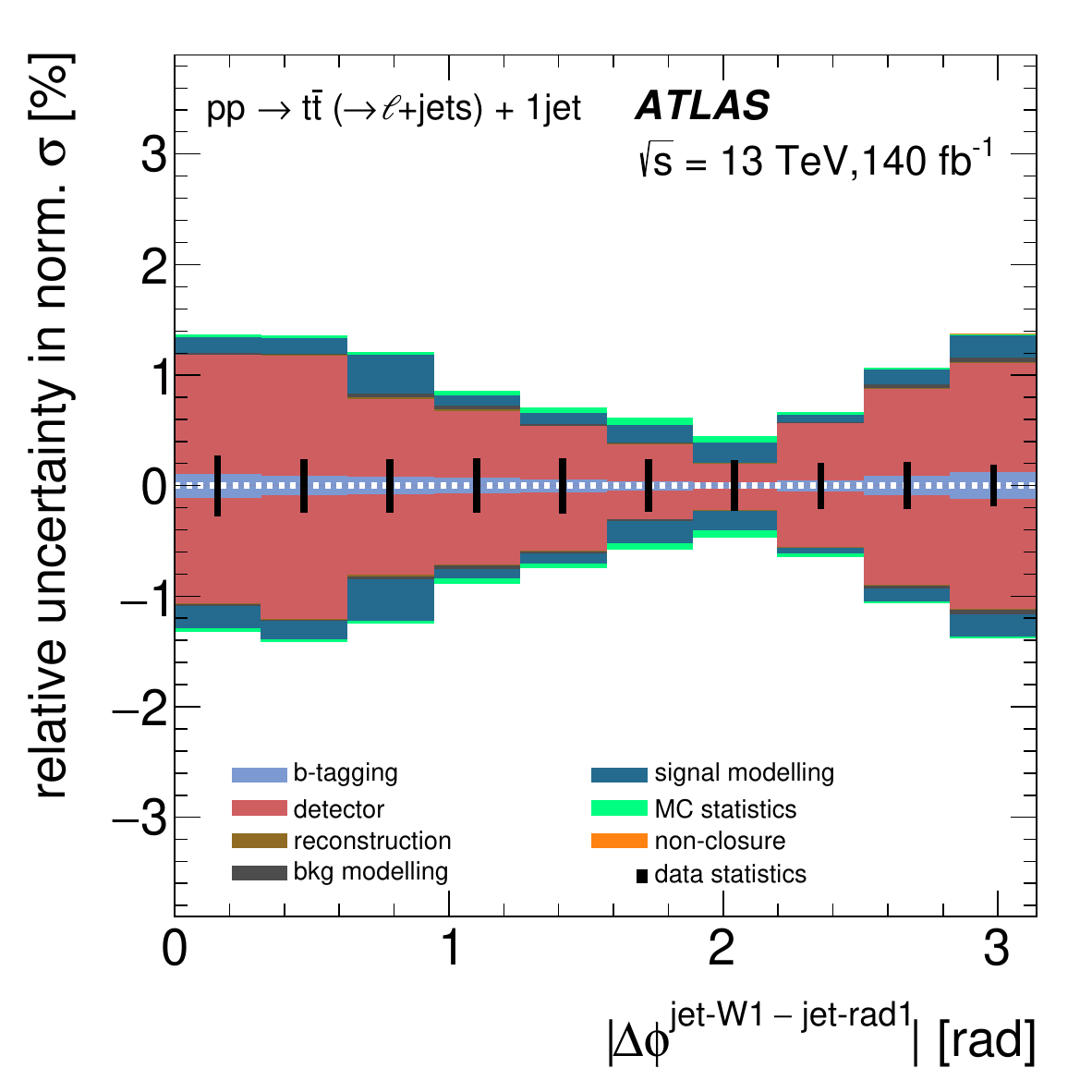}}
\put (8.0,0.0){\includegraphics[width=7cm]{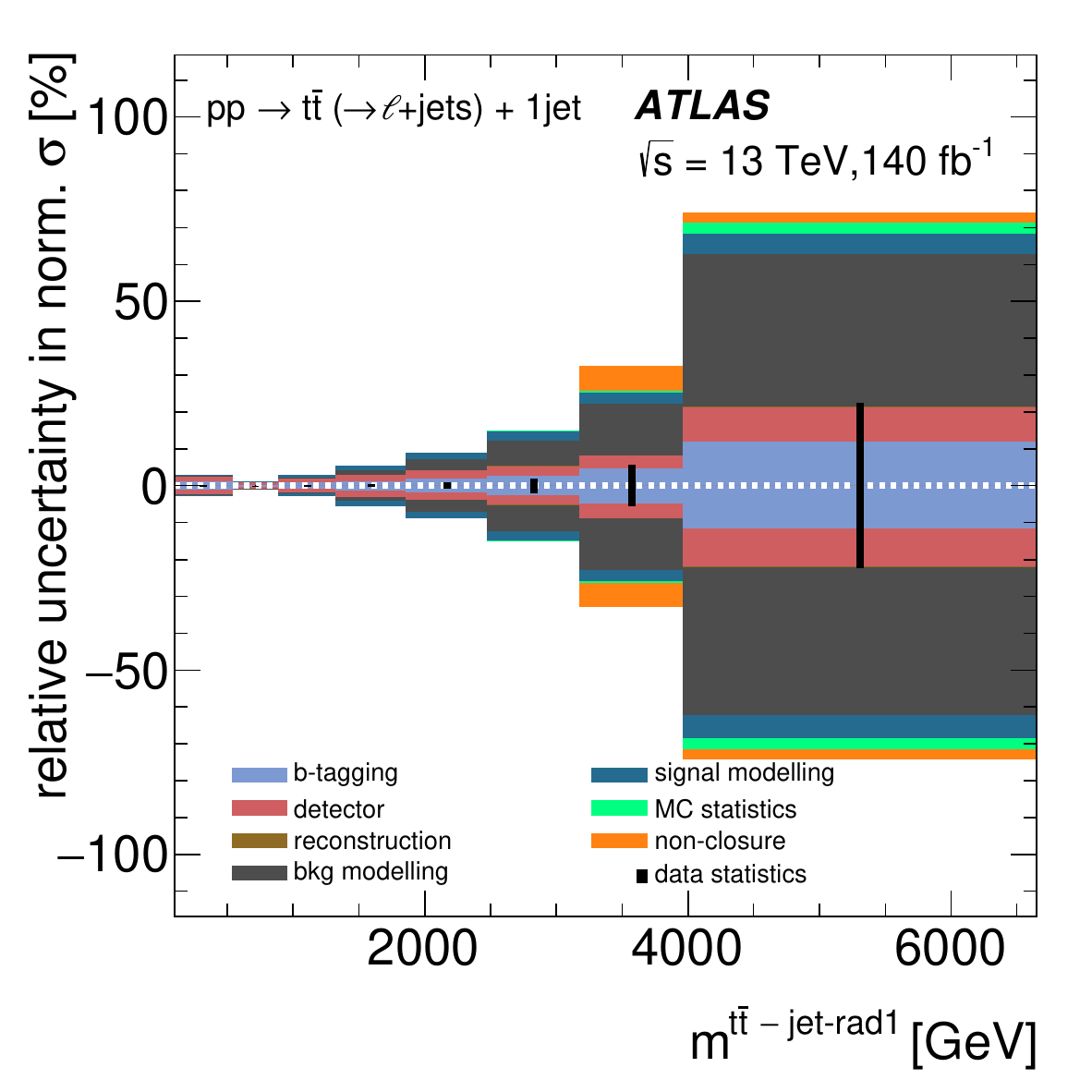}}
\put (3.0,14.0){\small (a)}
\put (11.0,14.0){\small (b)}
\put (3.0,7.1){\small (c)}
\put (11.0,7.1){\small (d)}
\put (3.0,0.2){\small (e)}
\put (11.0,0.2){\small (f)}
\end{picture}
\caption{
Total relative systematic uncertainties in the
measured normalised cross-sections in percentages for
(a) $\Ptjete$, (b) $|\Rapjete|$,
(c) $|\deltaphitle|$, (d) $|\deltaphithe|$,
(e) $|\deltaphiae|$ and (f) $\mtte$
in the $\tt+$1jet channel.
The individual sources of systematic uncertainty (plotted in stacked
form) and the statistical uncertainty in the data are also included.
}
\end{figure}
 
\FloatBarrier
 
\newpage
 
\begin{figure}[h]
\setlength{\unitlength}{1.0cm}
\begin{picture} (18.0,17.5)
\put (0.0,6.9){\includegraphics[width=7cm]{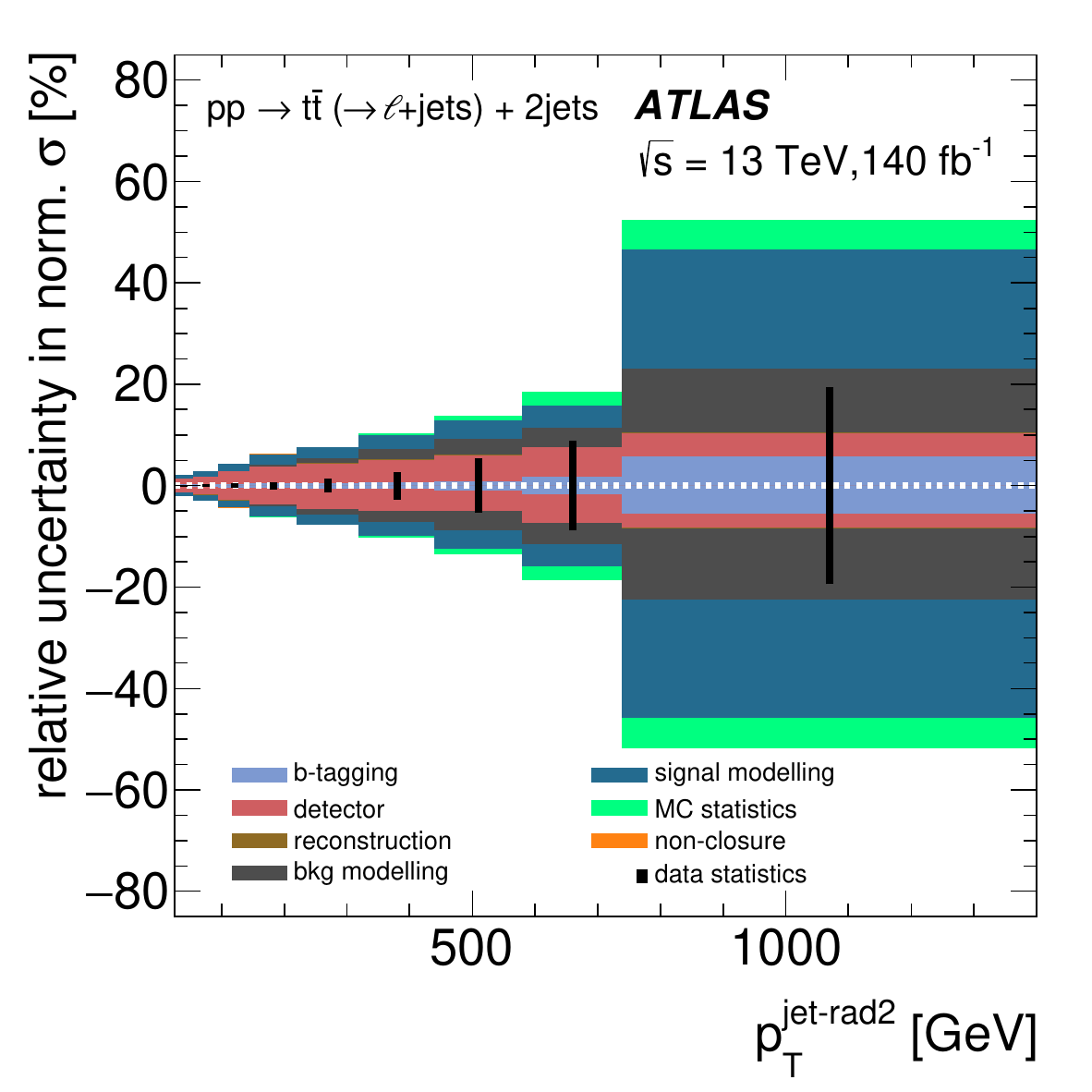}}
\put (8.0,6.9){\includegraphics[width=7cm]{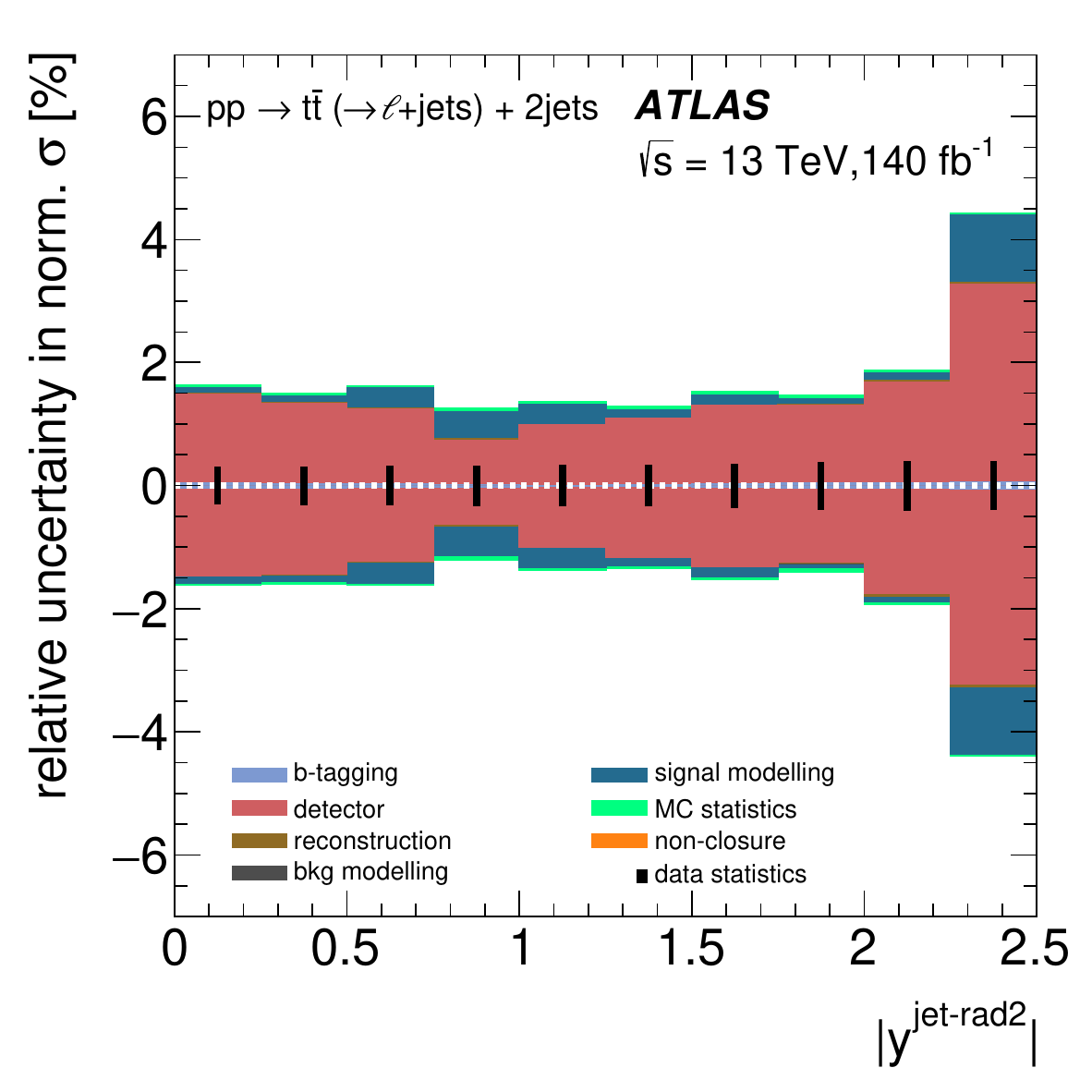}}
\put (0.0,0.0){\includegraphics[width=7cm]{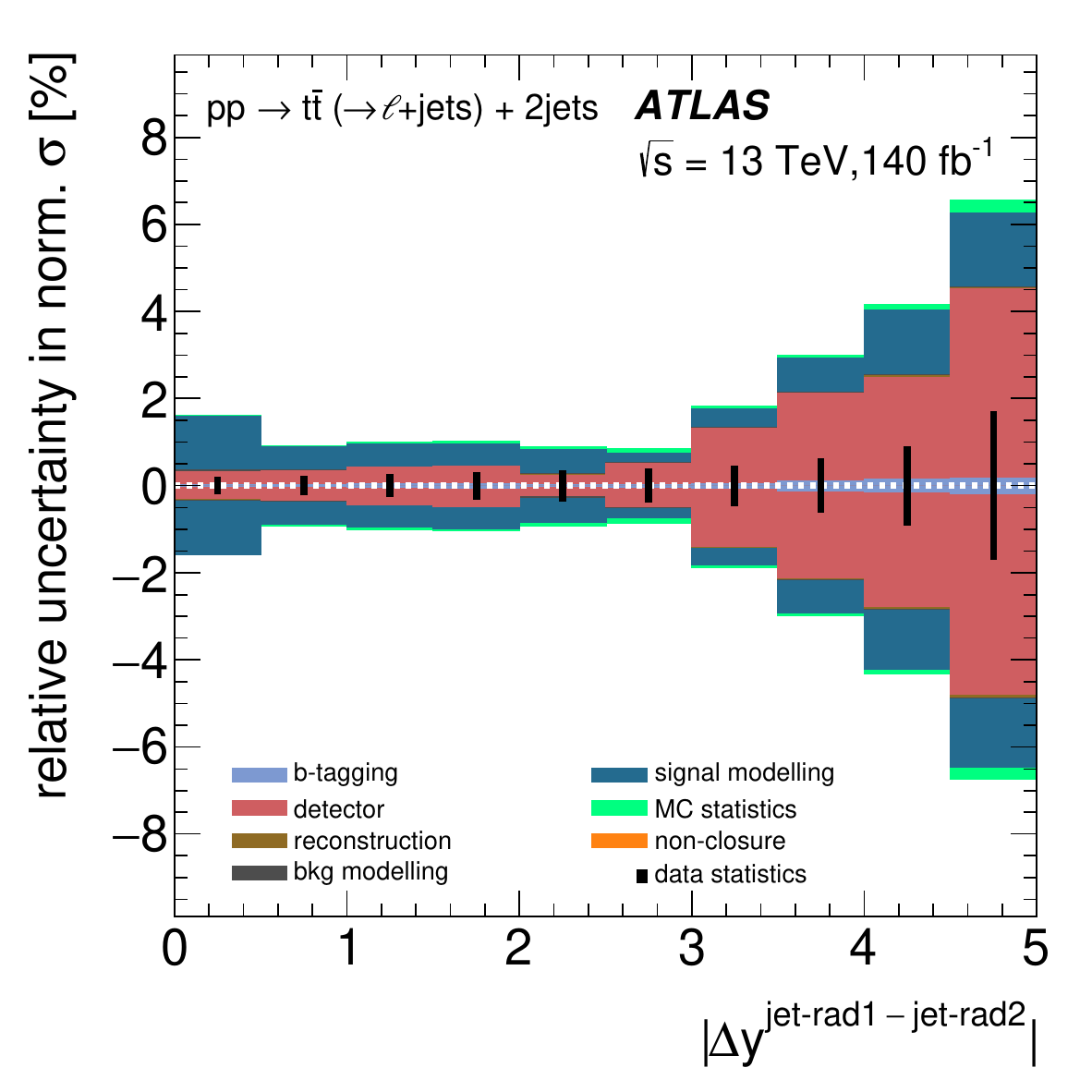}}
\put (8.0,0.0){\includegraphics[width=7cm]{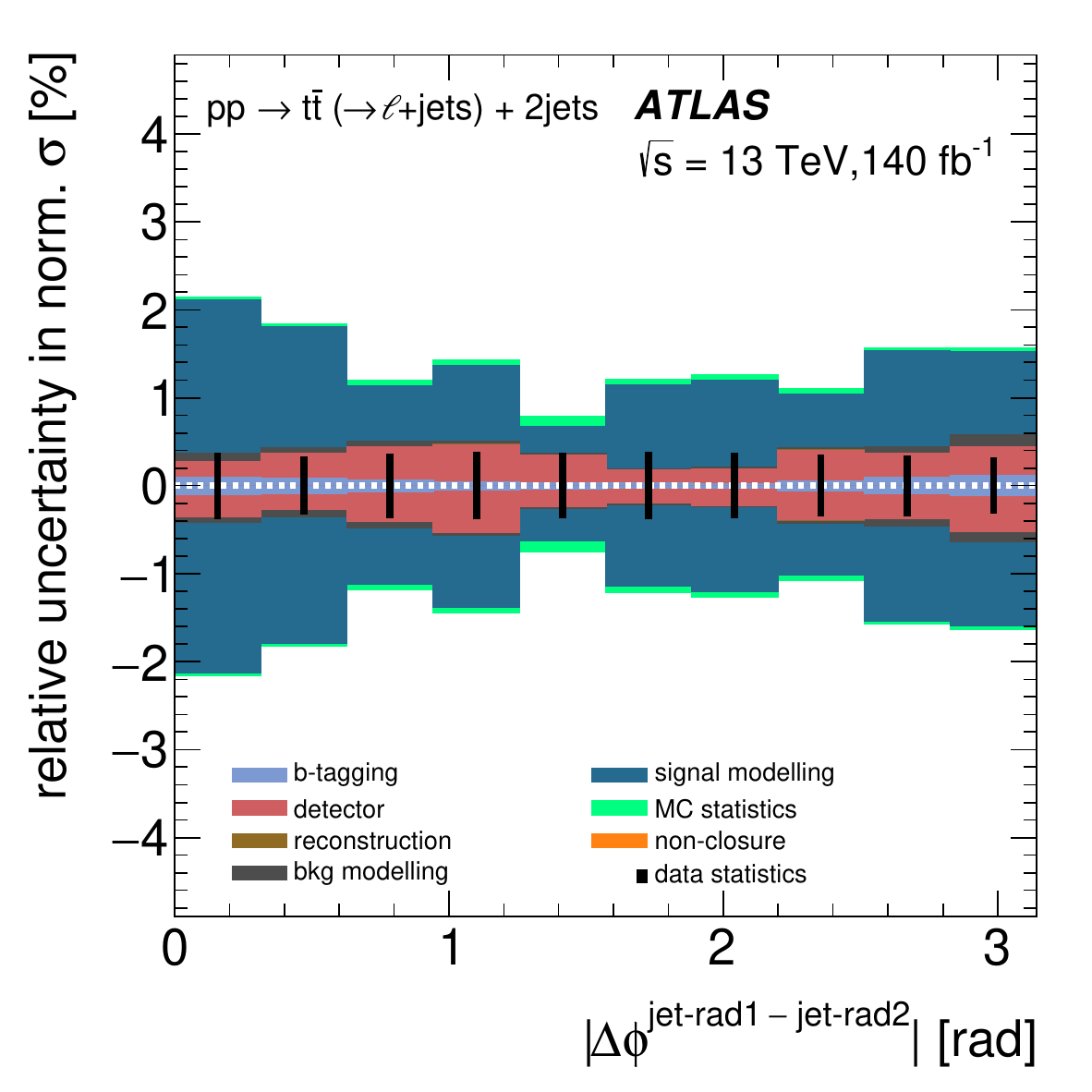}}
\put (3.0,7.1){\small (a)}
\put (11.0,7.1){\small (b)}
\put (3.0,0.2){\small (c)}
\put (11.0,0.2){\small (d)}
\end{picture}
\caption{
Total relative systematic uncertainties in the
measured normalised cross-sections in percentages for
(a) $\Ptjetf$, (b) $|\Rapjetf|$,
(c) $|\deltayef|$ and (d) $|\deltaphief|$
in the $\tt+$2jets channel.
The individual sources of systematic uncertainty (plotted in stacked
form) and the statistical uncertainty in the data are also included.
}
\end{figure}
 
\FloatBarrier
 
\newpage
 
\begin{figure}[h]
\setlength{\unitlength}{1.0cm}
\begin{picture} (18.0,17.5)
\put (0.0,6.9){\includegraphics[width=7cm]{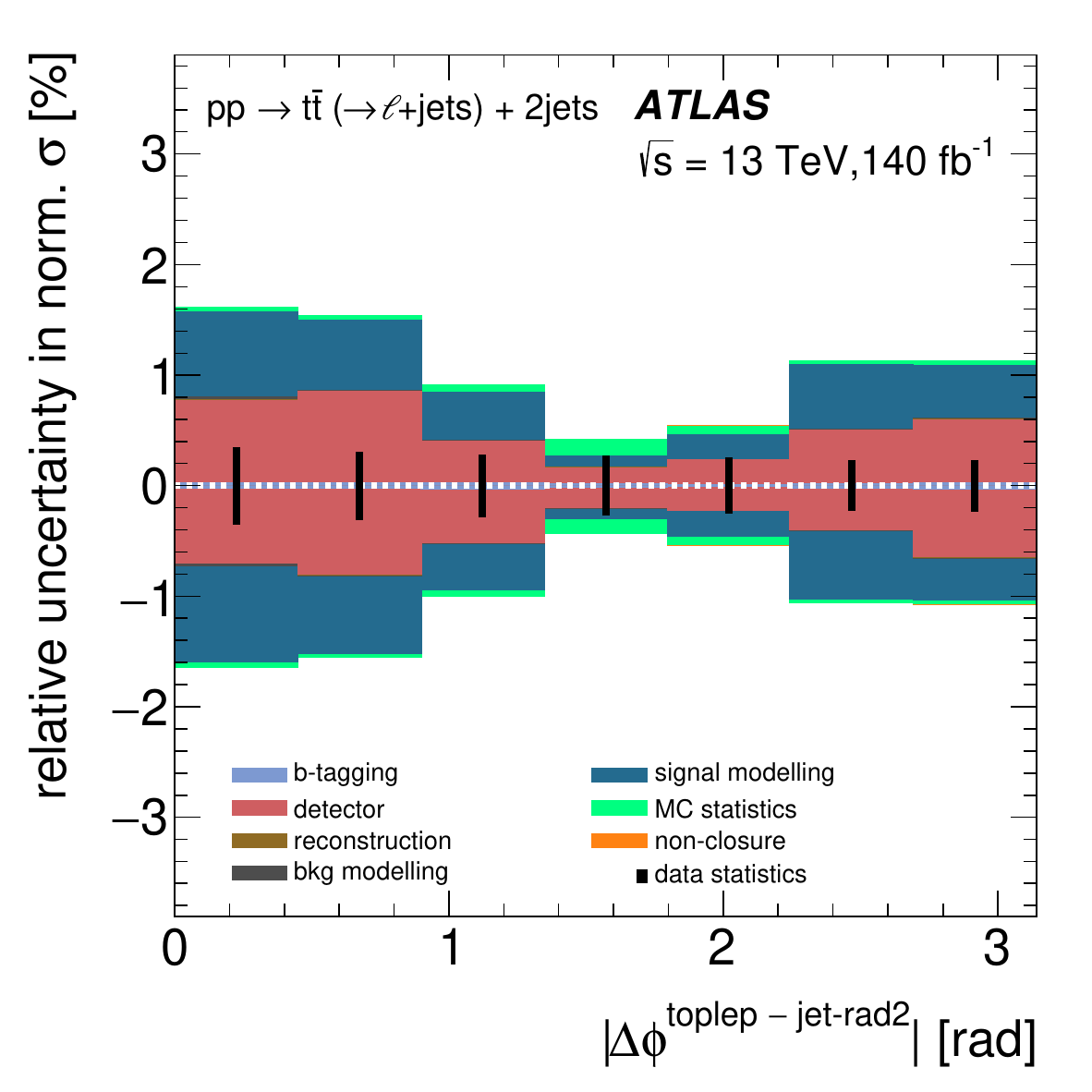}}
\put (8.0,6.9){\includegraphics[width=7cm]{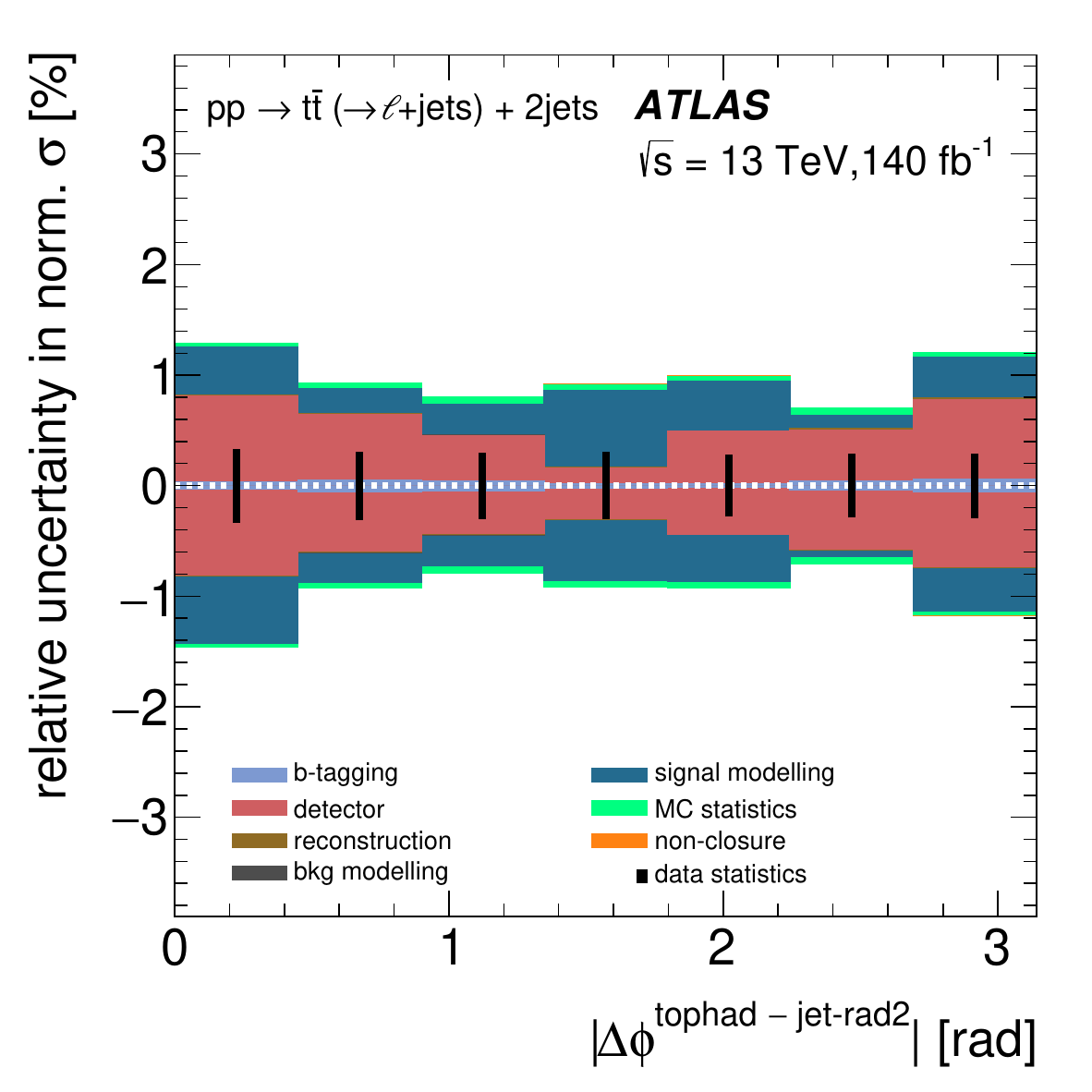}}
\put (0.0,0.0){\includegraphics[width=7cm]{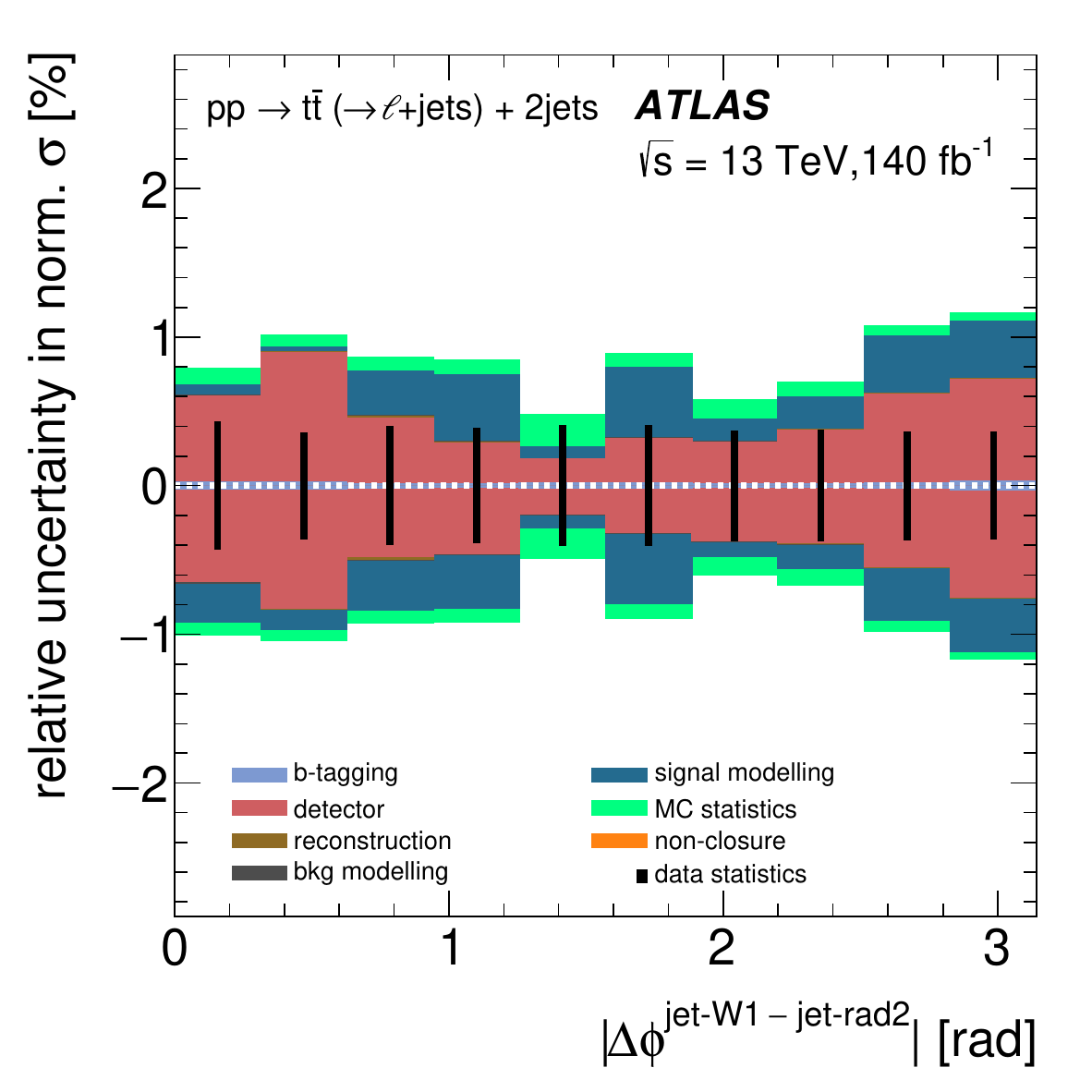}}
\put (8.0,0.0){\includegraphics[width=7cm]{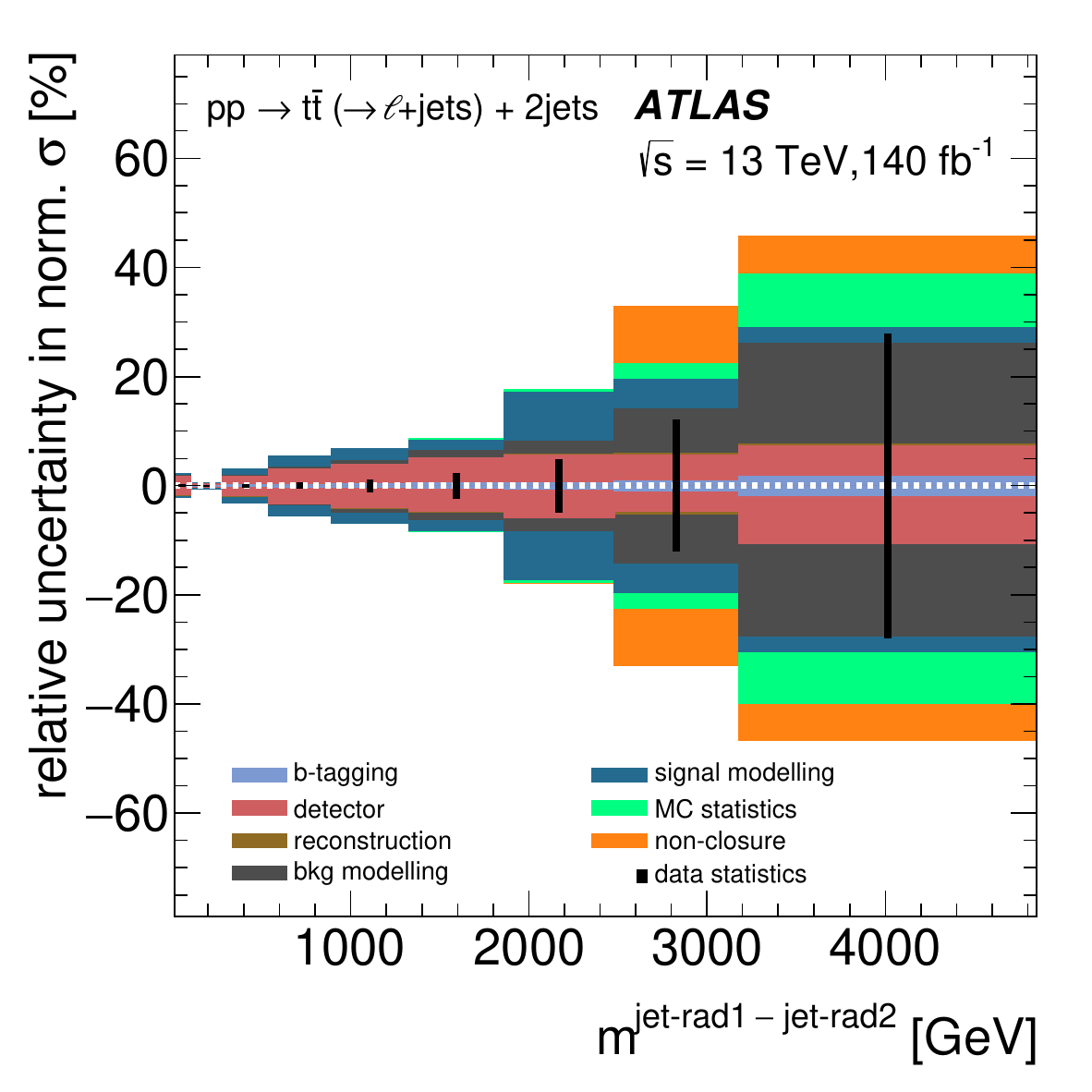}}
\put (3.0,7.1){\small (a)}
\put (11.0,7.1){\small (b)}
\put (3.0,0.2){\small (c)}
\put (11.0,0.2){\small (d)}
\end{picture}
\caption{
Total relative systematic uncertainties in the
measured normalised cross-sections in percentages for
(a) $|\deltaphitlf|$, (b) $|\deltaphithf|$,
(c) $|\deltaphiaf|$ and (d) $\mef$
in the $\tt+$2jets channel.
The individual sources of systematic uncertainty (plotted in stacked
form) and the statistical uncertainty in the data are also included.
}
\label{fig_appm40}
\end{figure}
 
\FloatBarrier
 
\section{Results}
\label{res}
 
The measured absolute and normalised differential cross-sections for
$\tt$ production at particle level in the $\tt$ inclusive, the
$\tt+$1jet and the $\tt+$2jets channels are presented in
Sections~\ref{xsecinc}, \ref{xsecone} and \ref{xsectwo}, respectively.
Next-to-leading-order QCD predictions from \pwp, \pwh, \amcnloh\ and
\sher~2.2.12  and NNLO QCD predictions from \pwpm\ are compared with
the measurements.
The level of agreement between the measured differential cross-section
as a function of each observable and the different theoretical
predictions is quantified by the $\chi^2$ and $p$-values (see
Section~\ref{pval}).
 
\subsection{Differential cross-sections for the $\tt$ inclusive channel}
\label{xsecinc}
\begin{figure}[h]
\setlength{\unitlength}{1.0cm}
\begin{picture} (18.0,20.0)
\put (0.0,13.0){\includegraphics[width=8cm]{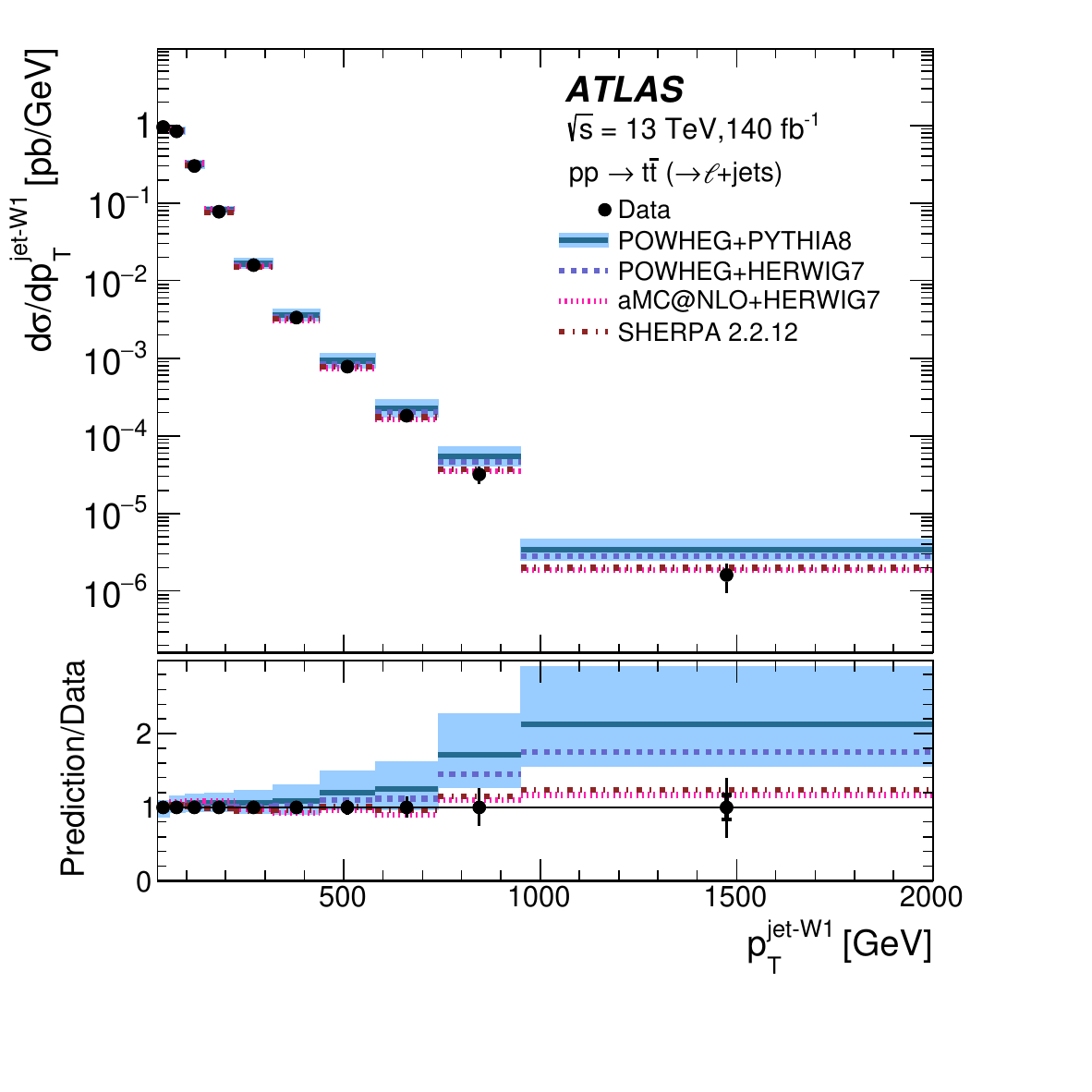}}
\put (8.0,13.0){\includegraphics[width=8cm]{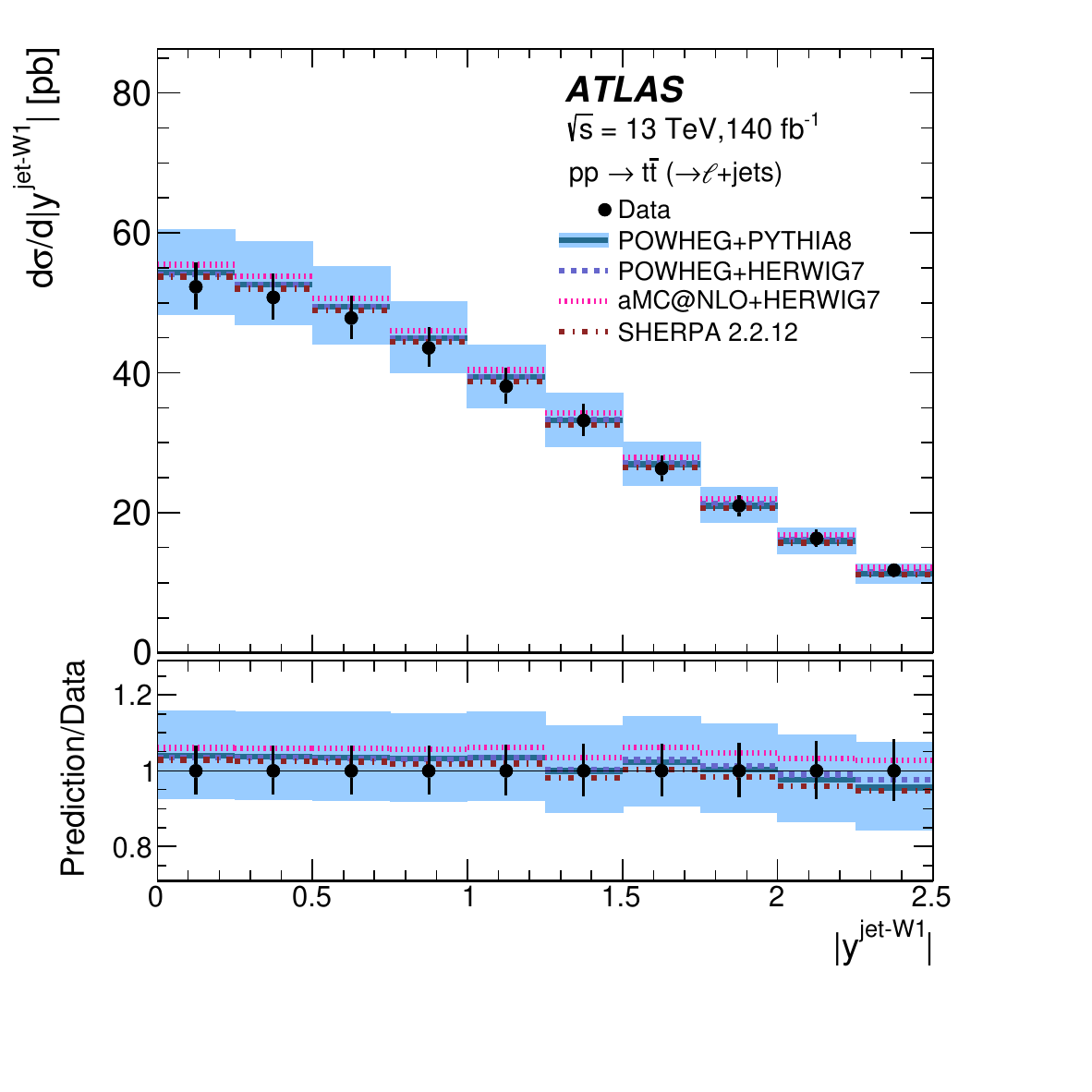}}
\put (0.0,6.0){\includegraphics[width=8cm]{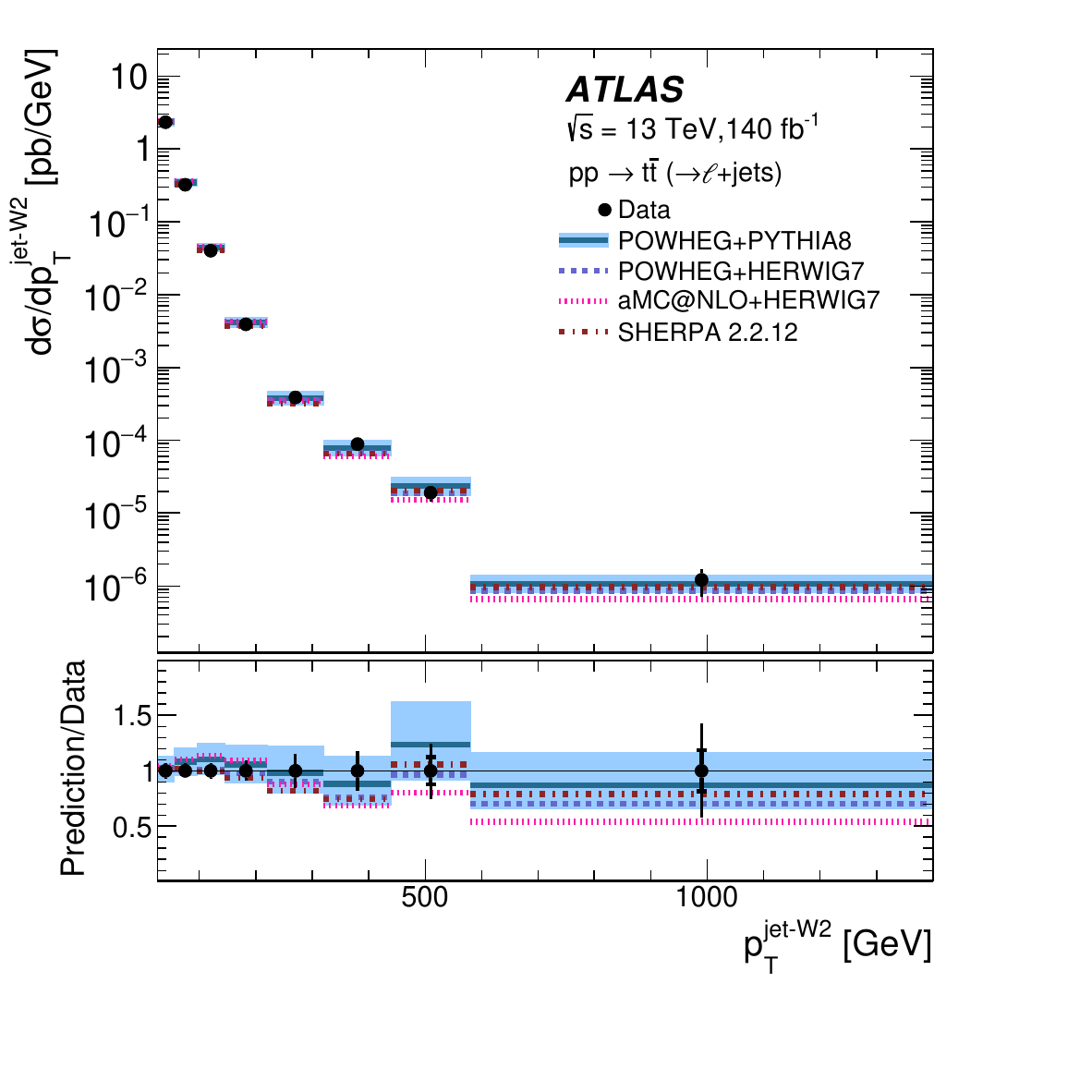}}
\put (8.0,6.0){\includegraphics[width=8cm]{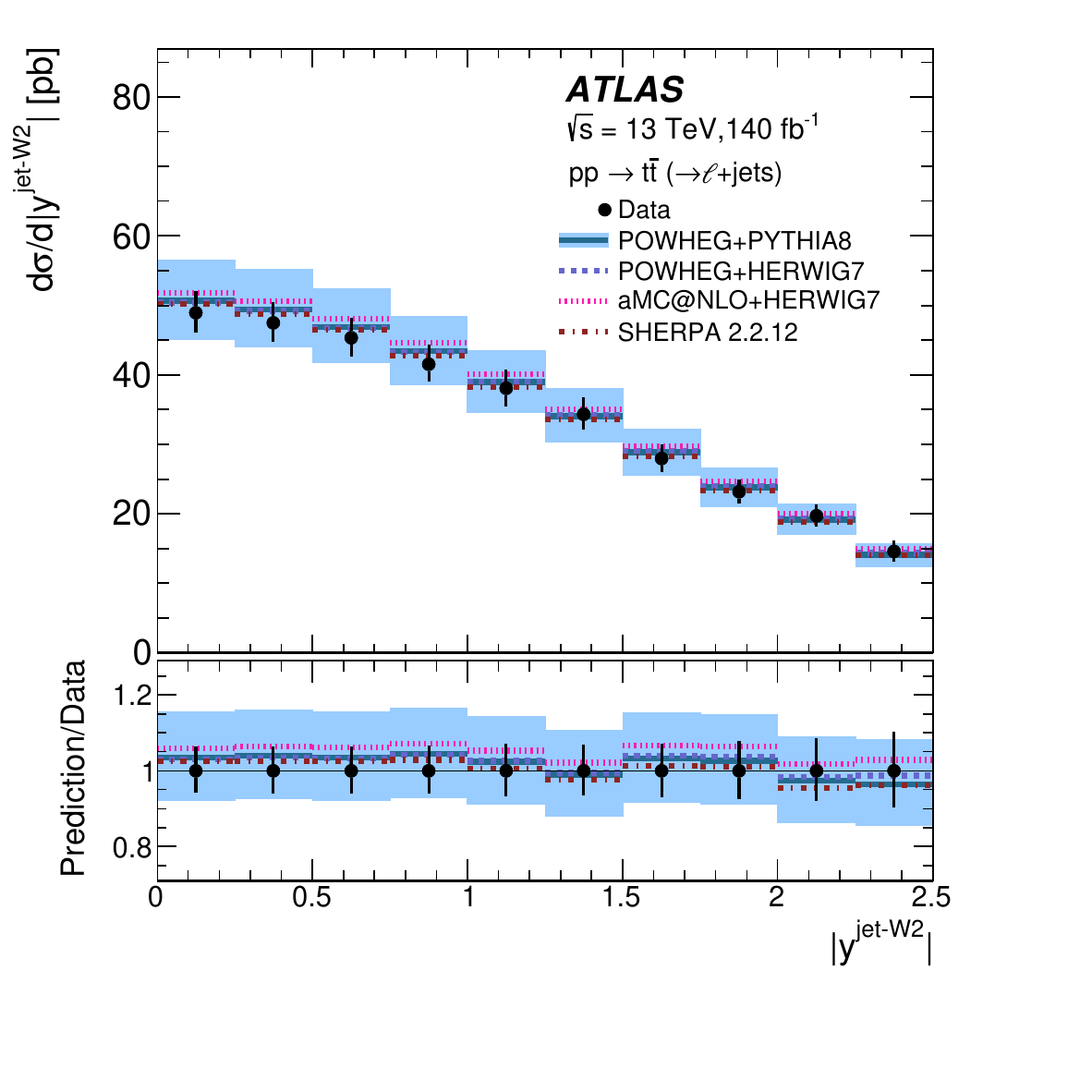}}
\put (0.0,-1.0){\includegraphics[width=8cm]{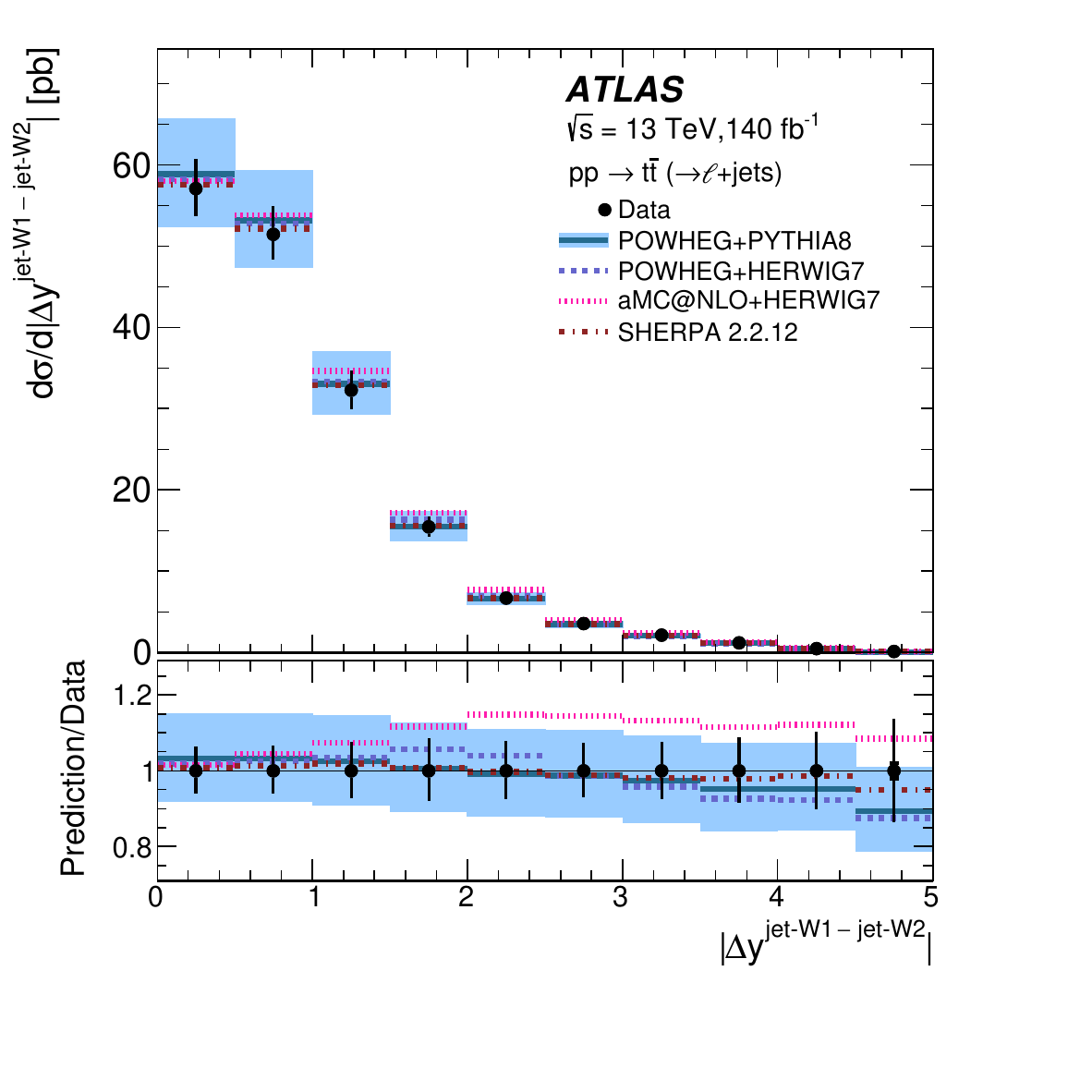}}
\put (8.0,-1.0){\includegraphics[width=8cm]{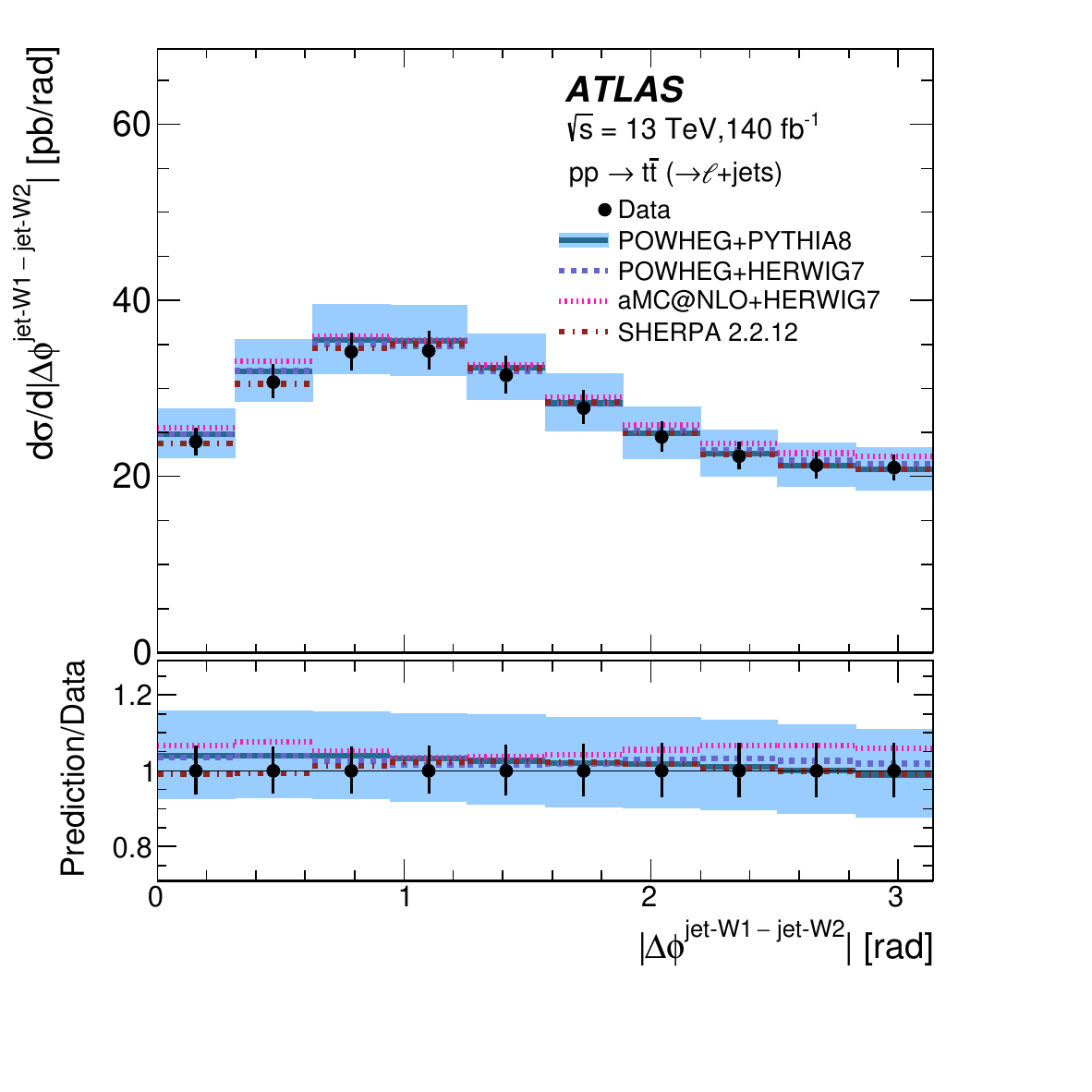}}
\put (3.5,13.9){\small (a)}
\put (11.5,13.9){\small (b)}
\put (3.5,6.9){\small (c)}
\put (11.5,6.9){\small (d)}
\put (3.5,-0.1){\small (e)}
\put (11.5,-0.1){\small (f)}
\end{picture}
\caption{
The measured differential cross-sections for $\tt$ production in
the \ljets\ decay mode as functions of
(a) $\Ptjeta$, (b) $|\Rapjeta|$,
(c) $\Ptjetb$, (d) $|\Rapjetb|$,
(e) $|\deltayab|$ and (f) $|\deltaphiab|$ (dots)
in the $\tt$ inclusive channel.
The NLO QCD predictions from \pwp\ (solid lines), \pwh\ (dashed lines),
\amcnloh\ (dotted lines) and \sher~2.2.12 (dot-dashed lines)
normalised to ${\cal O}$(NNLO+NNLL) are also shown.
The inner (outer) error bars represent the statistical uncertainties
(the statistical and systematic uncertainties added in
quadrature). For most of the points, the inner error bars are smaller than
the marker size and, thus, not visible.
The shaded bands represent
the theoretical uncertainty in the \pwp\ prediction. The lower panels of
the figures show the ratios of the predicted and the measured cross-sections.
}
\label{figres1}
\end{figure}

\begin{figure}[h]
\setlength{\unitlength}{1.0cm}
\begin{picture} (18.0,20.0)
\put (0.0,13.0){\includegraphics[width=8cm]{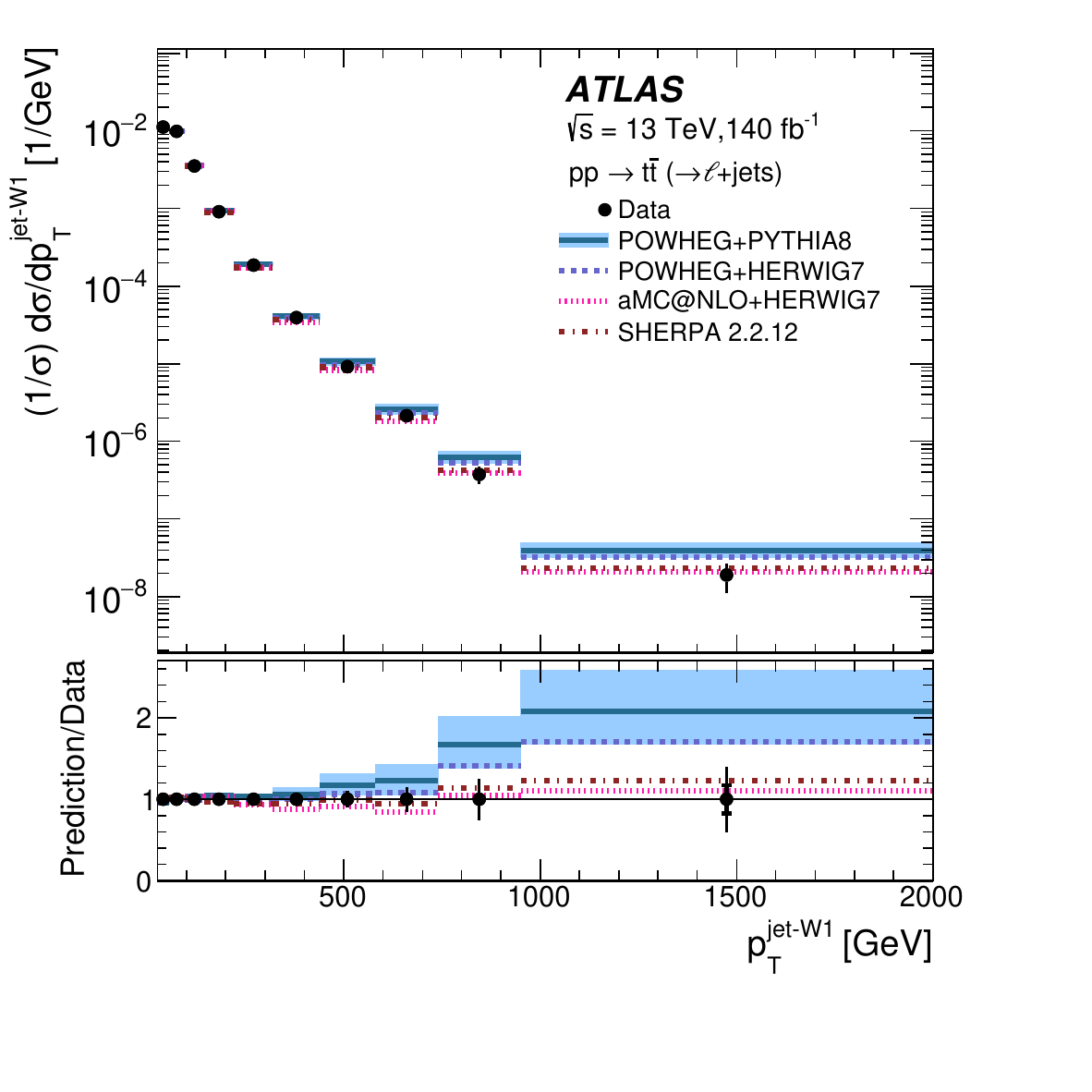}}
\put (8.0,13.0){\includegraphics[width=8cm]{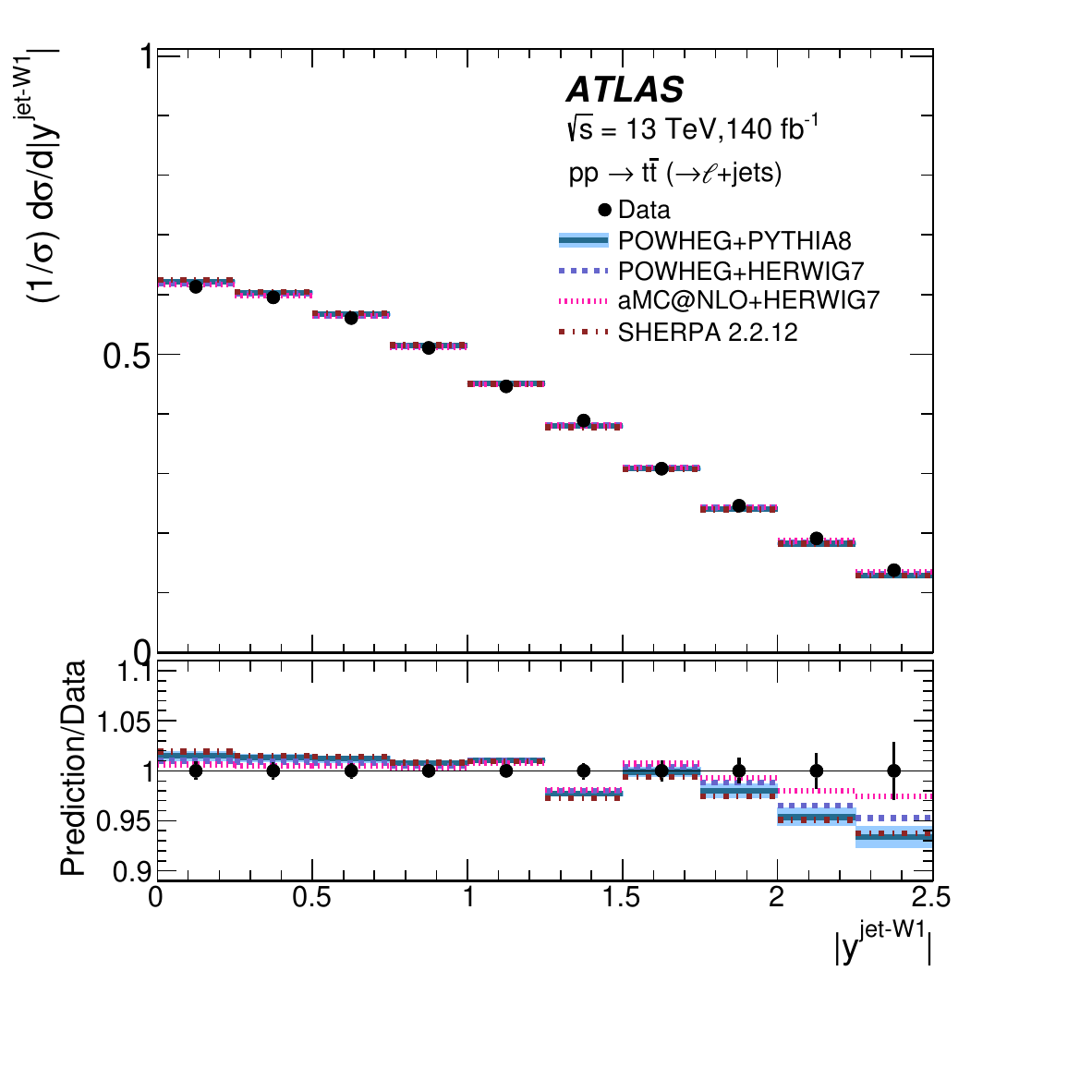}}
\put (0.0,6.0){\includegraphics[width=8cm]{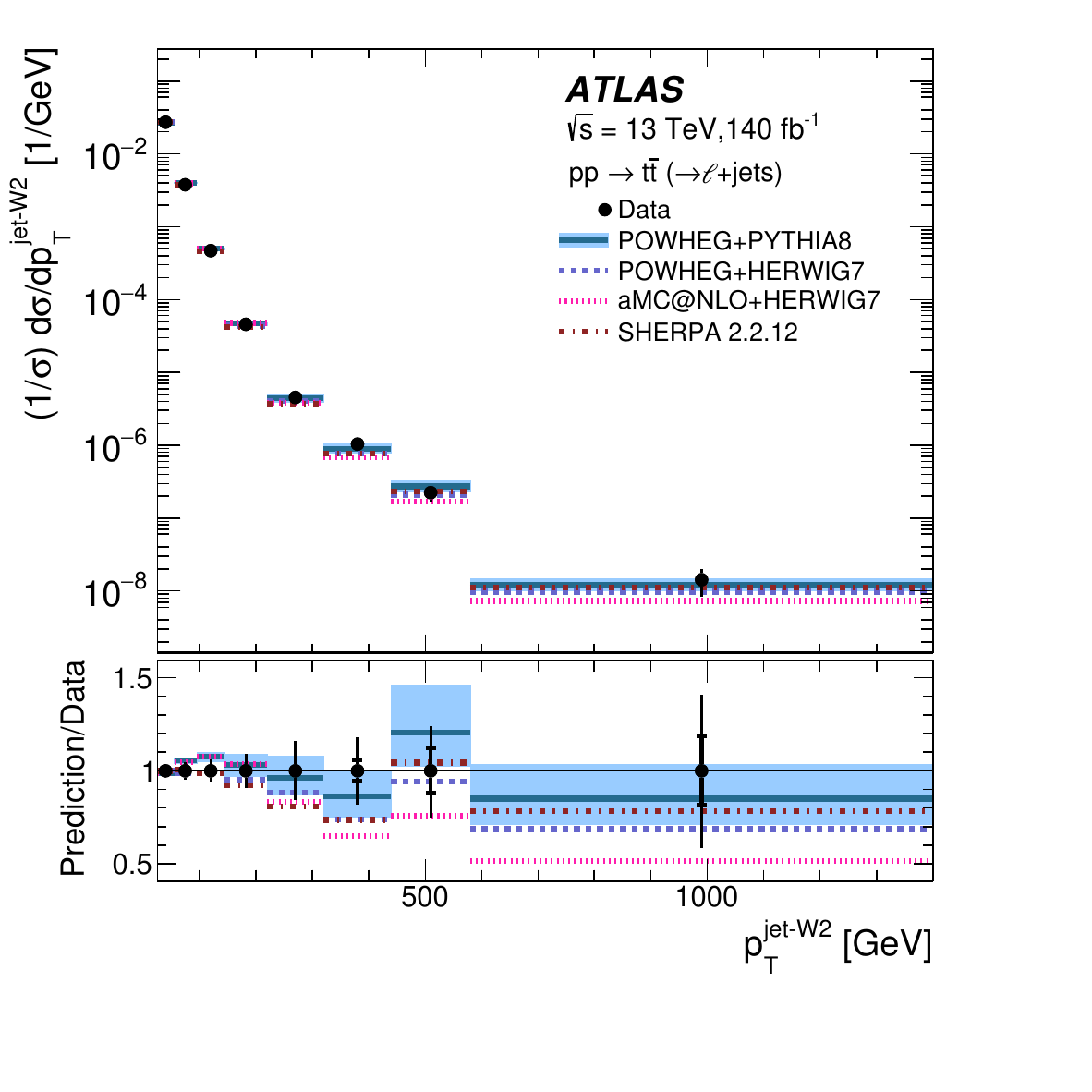}}
\put (8.0,6.0){\includegraphics[width=8cm]{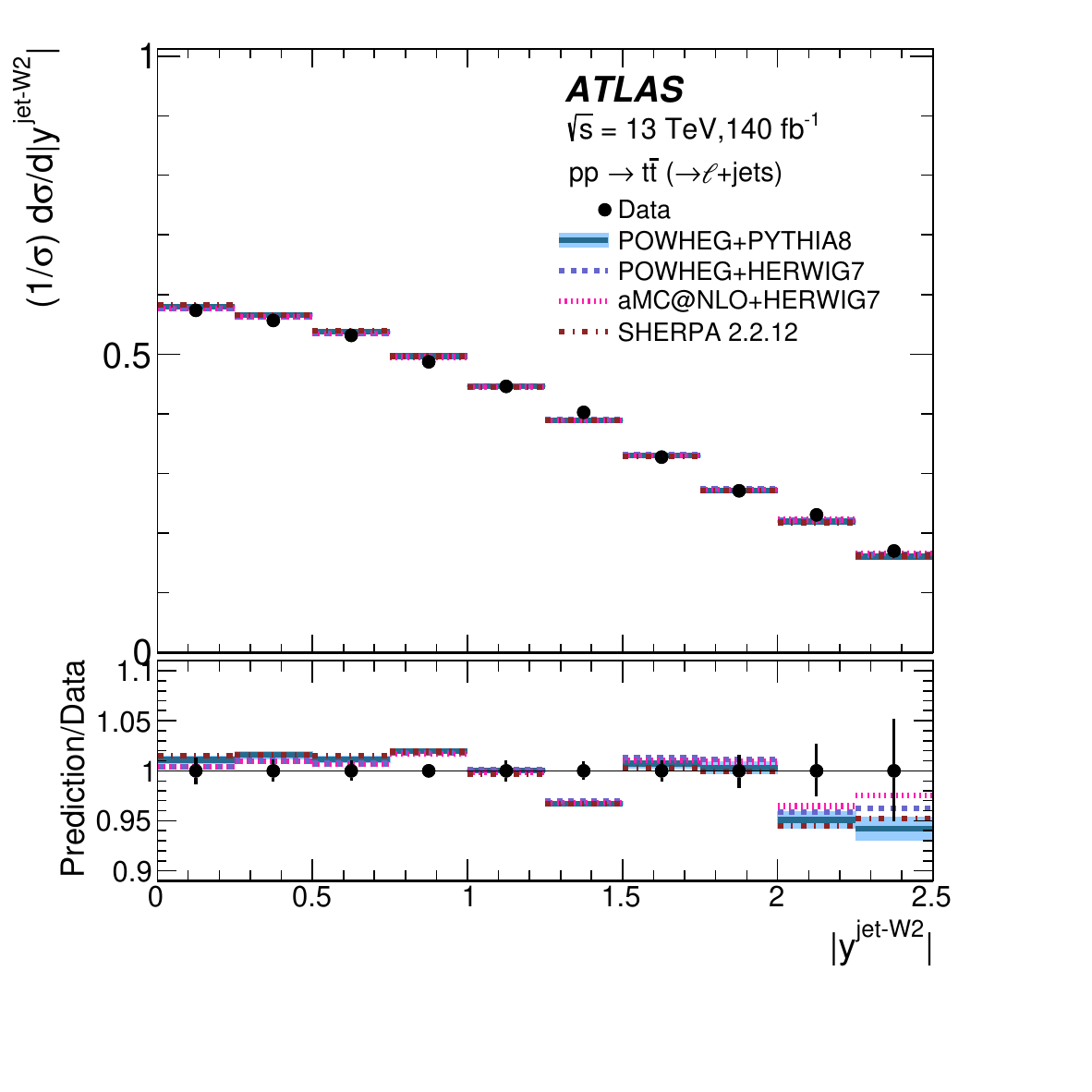}}
\put (0.0,-1.0){\includegraphics[width=8cm]{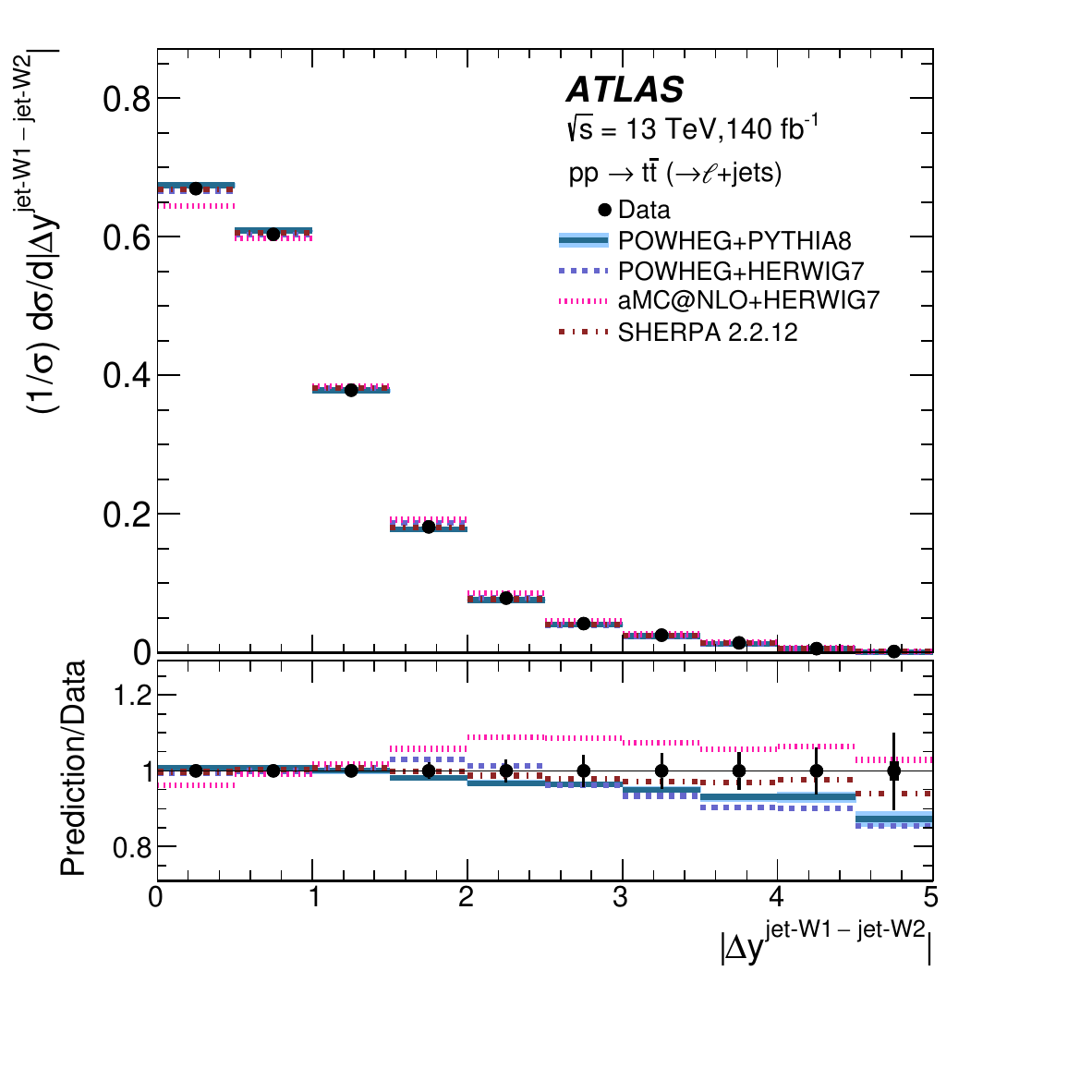}}
\put (8.0,-1.0){\includegraphics[width=8cm]{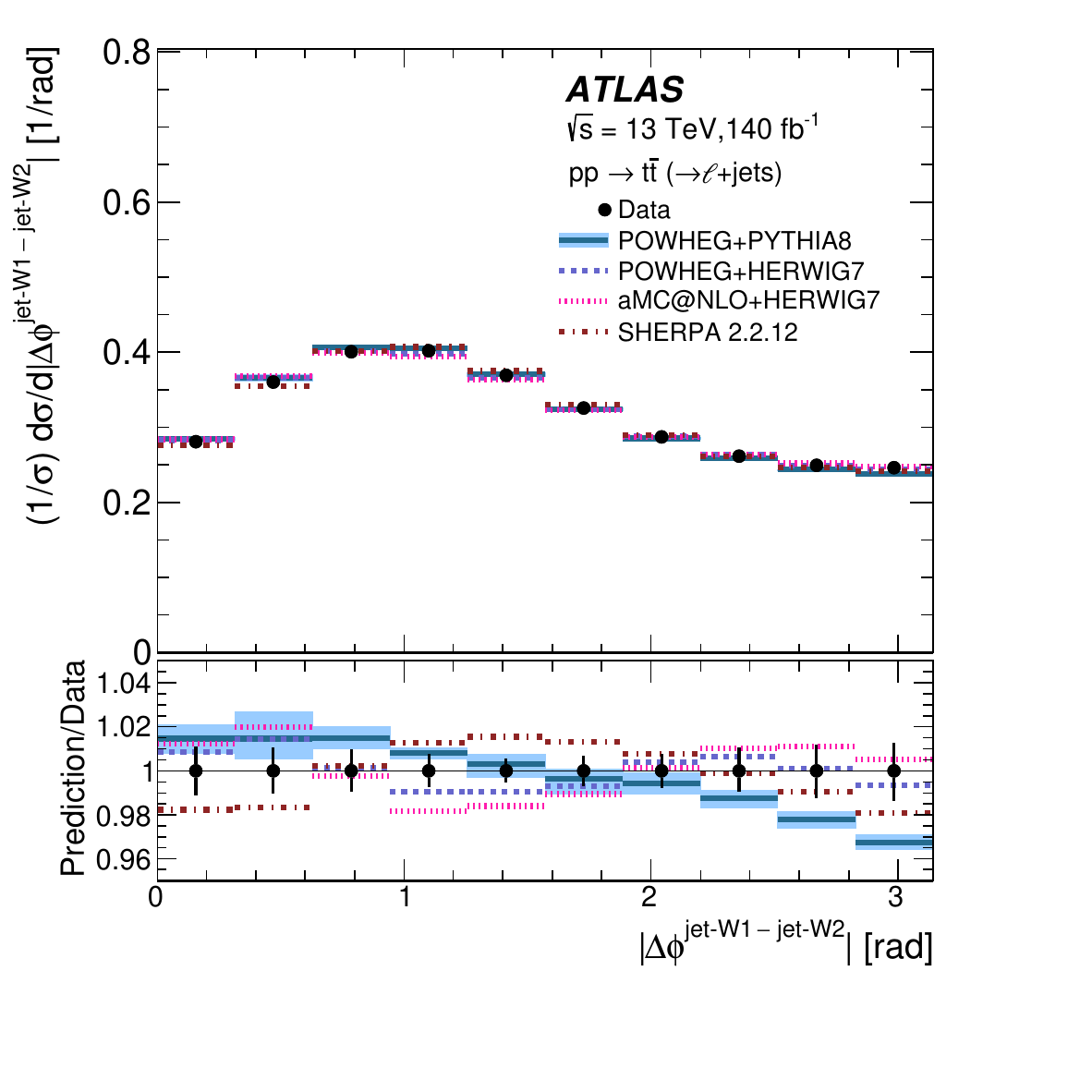}}
\put (3.5,13.9){\small (a)}
\put (11.5,13.9){\small (b)}
\put (3.5,6.9){\small (c)}
\put (11.5,6.9){\small (d)}
\put (3.5,-0.1){\small (e)}
\put (11.5,-0.1){\small (f)}
\end{picture}
\caption{
The measured normalised differential cross-sections for $\tt$ production in
the \ljets\ channel as functions of
(a) $\Ptjeta$, (b) $|\Rapjeta|$,
(c) $\Ptjetb$, (d) $|\Rapjetb|$,
(e) $|\deltayab|$ and (f) $|\deltaphiab|$ (dots)
in the $\tt$ inclusive channel.
The NLO QCD predictions from \pwp\ (solid lines), \pwh\ (dashed lines),
\amcnloh\ (dotted lines) and \sher~2.2.12 (dot-dashed lines)
are also shown.
The inner (outer) error bars represent the statistical uncertainties
(the statistical and systematic uncertainties added in
quadrature). For most of the points, the inner error bars are smaller than
the marker size and, thus, not visible.
The shaded bands represent
the theoretical uncertainty in the \pwp\ prediction. The lower panels of
the figures show the ratios of the predicted and the measured cross-sections.
}
\label{fig_appm1}
\end{figure}

Figures~\ref{figres1} and \ref{fig_appm1} show the absolute and
normalised differential $\tt$ cross-sections in the \ljets\
decay mode for the $\tt$ inclusive channel as functions of $\Ptjeta$,
$|\Rapjeta|$, $\Ptjetb$, $|\Rapjetb|$, $|\deltayab|$ and
$|\deltaphiab|$. Values of $\Ptjeta$ and $\Ptjetb$ up to $\sim
1.5$~\TeV\ and $\sim 1.1$~\TeV, respectively, are measured. The
measured cross-section as a function of $\Ptjeta$ ($\Ptjetb$)
decreases by more than five (six) orders of magnitude within the
measured range. The measured cross-section as function of $\Ptjeta$
has a harder spectrum than that of $\Ptjetb$, which decreases more
rapidly with increasing $\Ptjetb$. The measured cross-sections as
functions of $|\Rapjeta|$ and $|\Rapjetb|$ decrease as the rapidity
increases and have very similar shape and normalisation. The measured
cross-section as a function of $|\deltayab|$ ($|\deltaphiab|$)
exhibits a peak around $|\deltayab|=0$ ($|\deltaphiab|=1$~rad).
 
The predictions of \pwp, \amcnloh, \pwh\ and \sher~2.2.12
are compared to the measurements in Figure~\ref{figres1}. As explained
in Section~\ref{sec:signalMC}, these predictions are calculated at NLO,
however, the normalisation is performed at ${\cal O}$(NNLO+NNLL). The
$|\Rapjeta|$, $|\Rapjetb|$, $|\deltayab|$ and $|\deltaphiab|$ measured
cross-sections are well described in shape and normalisation by all
the predictions, except for the prediction from \amcnloh\ which has a
tendency to be above the data for $|\deltayab|>2$. The measured
cross-section as a function of $\Ptjeta$ is described by the
predictions of \pwp\
and \pwh\ only for $\Ptjeta<500$~\GeV, whereas \amcnloh\ and
\sher~2.2.12 provide a good description of the data in the whole
measured range. The measured cross-section as a function of $\Ptjetb$
is well described by the NLO predictions; the differences between the
measurement and the prediction from \amcnloh\ are within the
theoretical and experimental uncertainties. From the comparison of the
normalised differential predictions to the measurements (see
Figure~\ref{fig_appm1}), the same conclusions are drawn.
 
\begin{figure}[h]
\setlength{\unitlength}{1.0cm}
\begin{picture} (18.0,20.0)
\put (0.0,13.0){\includegraphics[width=8cm]{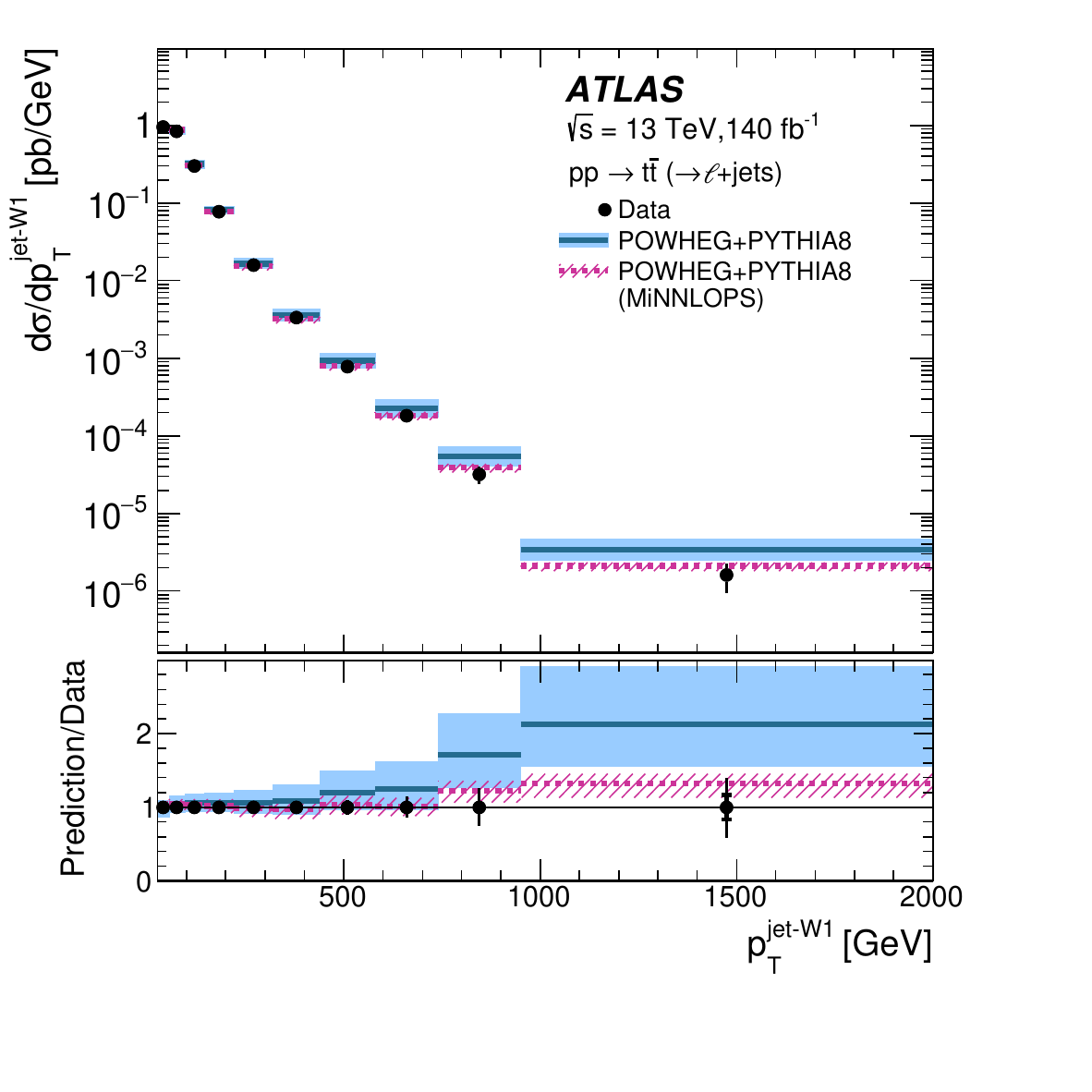}}
\put (8.0,13.0){\includegraphics[width=8cm]{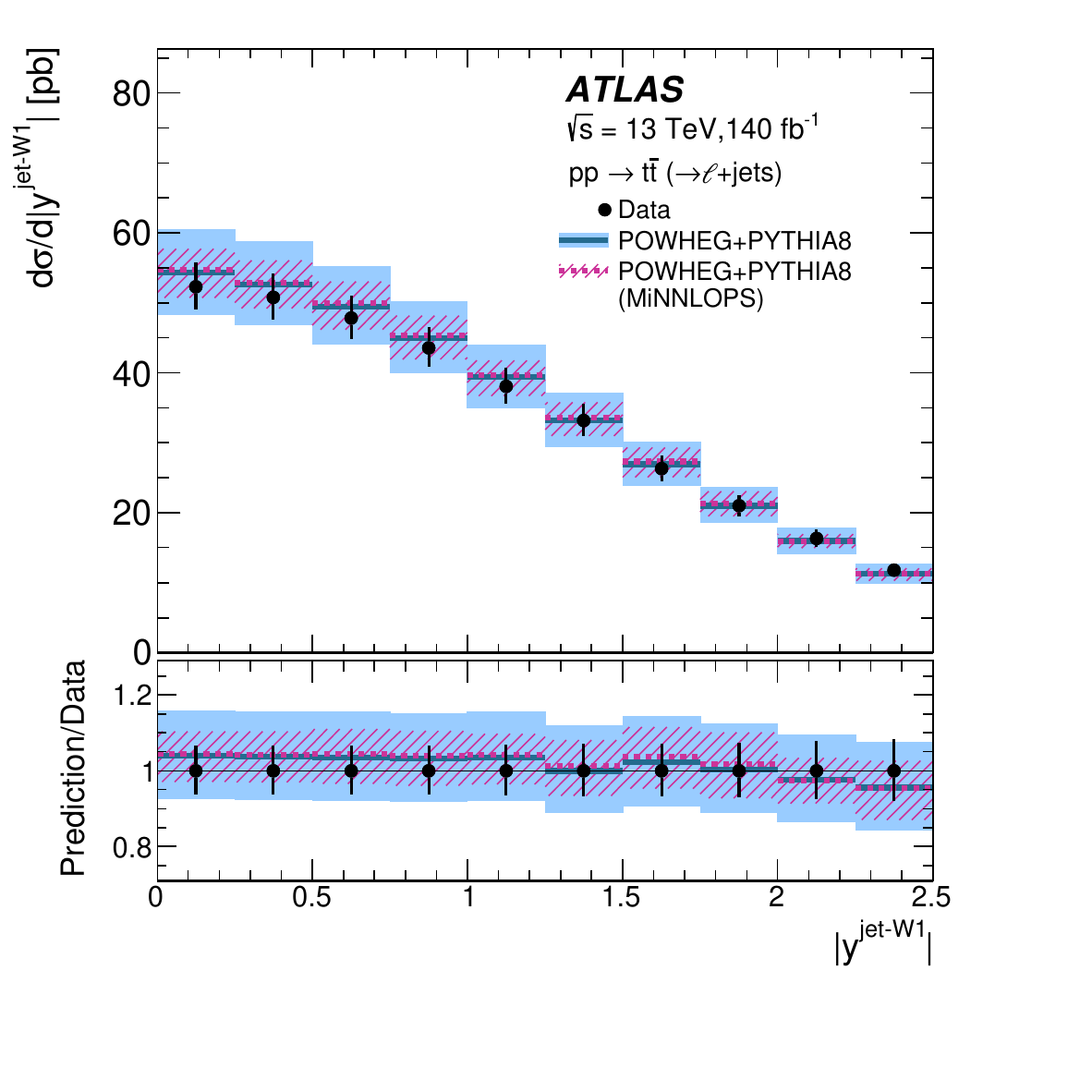}}
\put (0.0,6.0){\includegraphics[width=8cm]{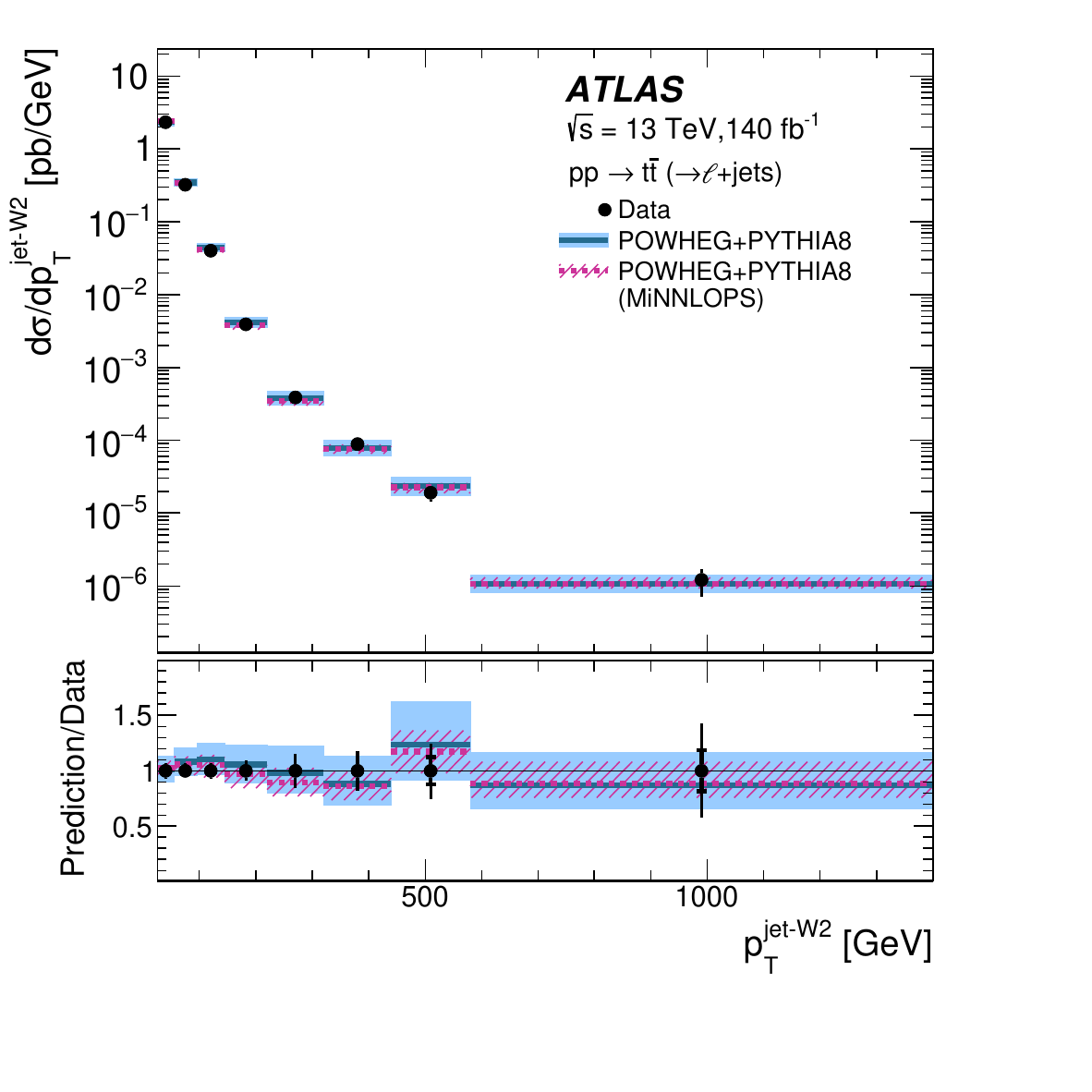}}
\put (8.0,6.0){\includegraphics[width=8cm]{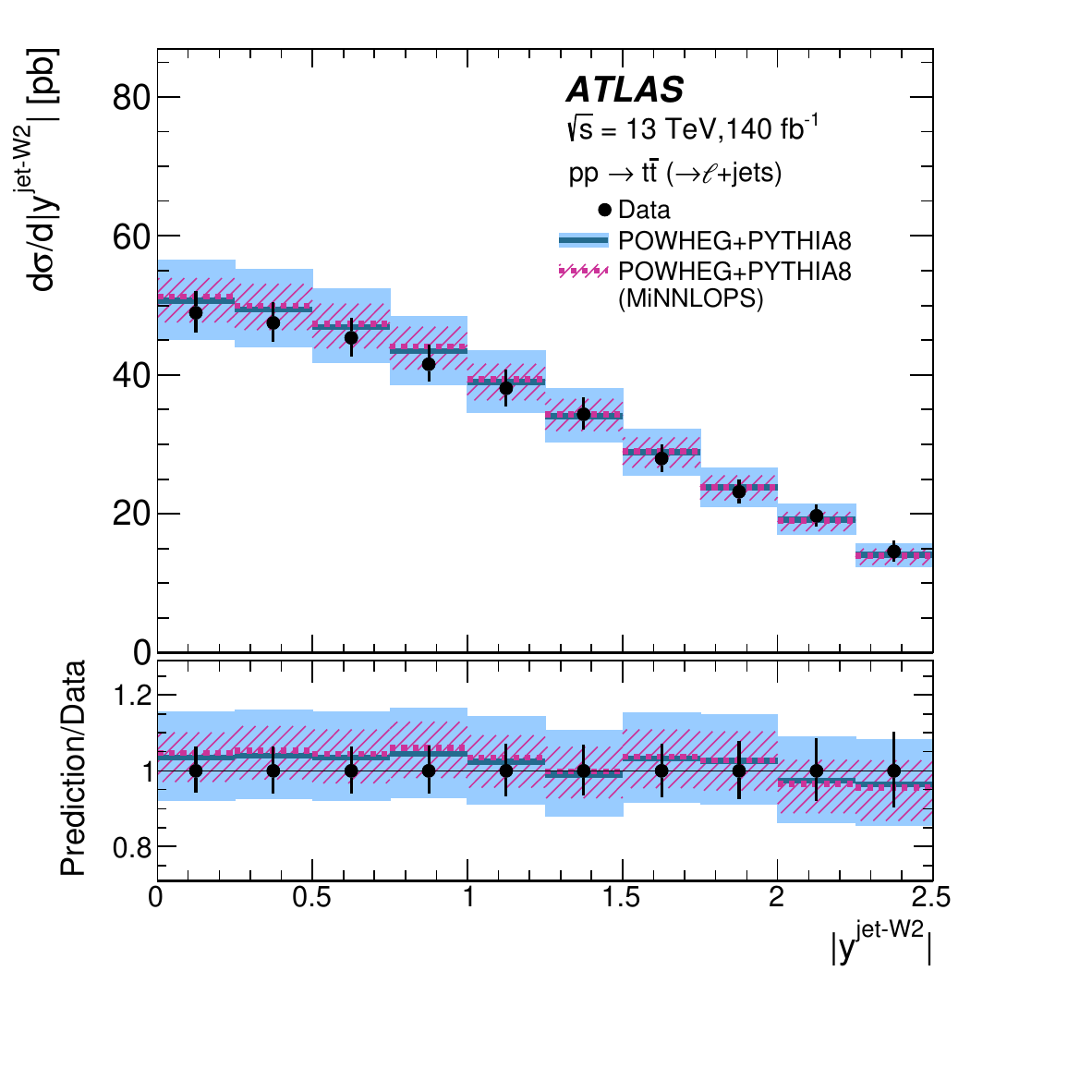}}
\put (0.0,-1.0){\includegraphics[width=8cm]{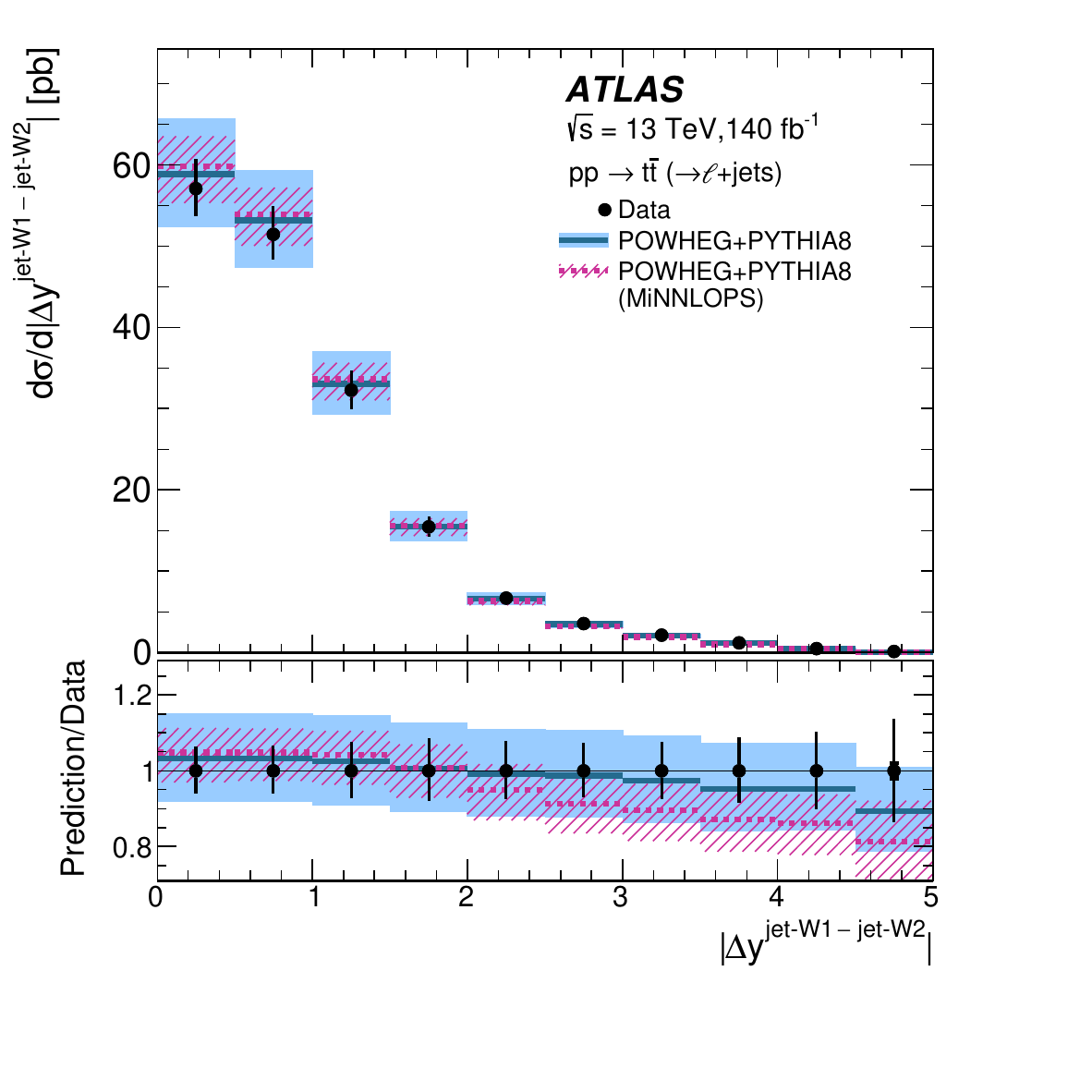}}
\put (8.0,-1.0){\includegraphics[width=8cm]{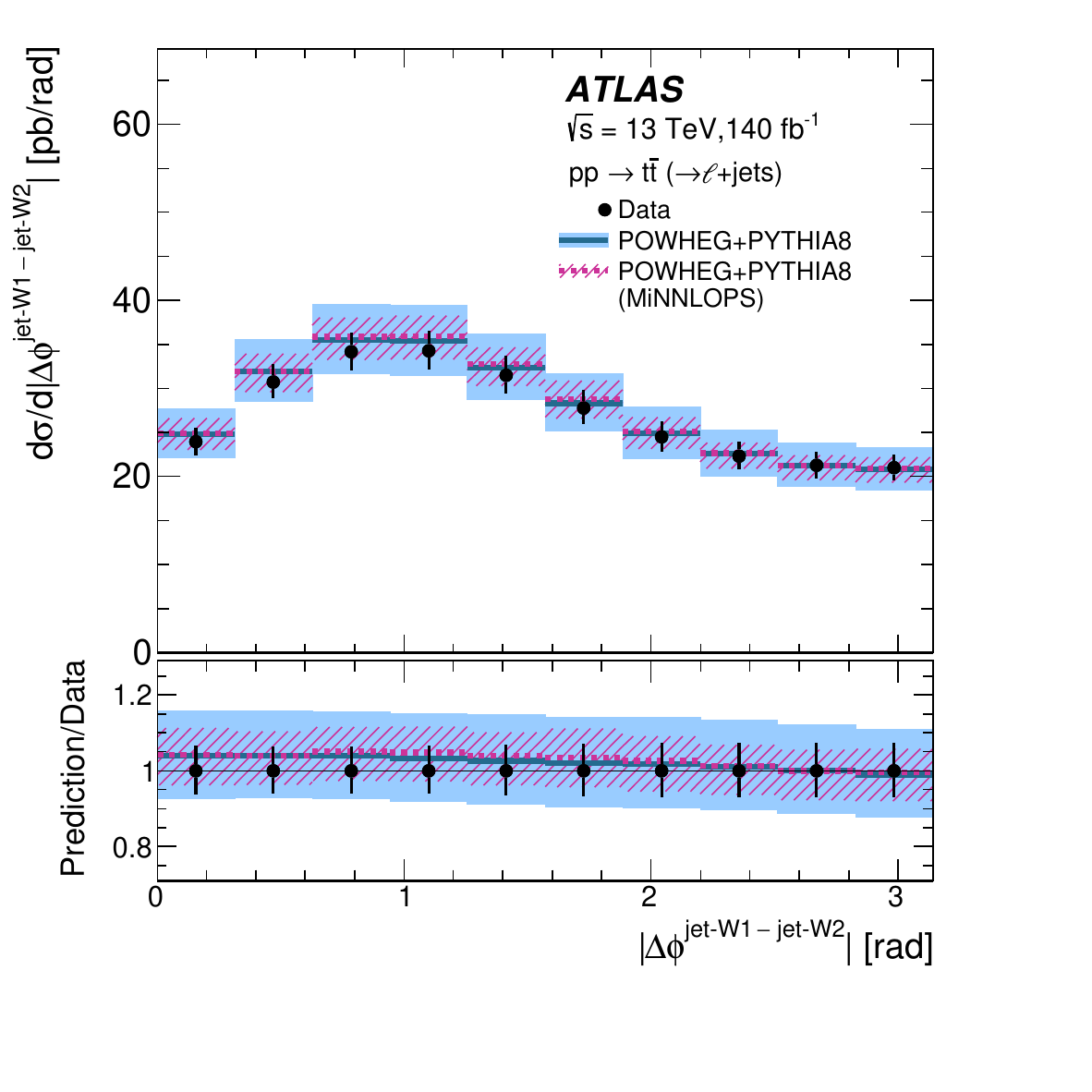}}
\put (3.5,13.9){\small (a)}
\put (11.5,13.9){\small (b)}
\put (3.5,6.9){\small (c)}
\put (11.5,6.9){\small (d)}
\put (3.5,-0.1){\small (e)}
\put (11.5,-0.1){\small (f)}
\end{picture}
\caption{
The measured differential cross-sections for $\tt$ production in
the \ljets\ decay mode as functions of
(a) $\Ptjeta$, (b) $|\Rapjeta|$,
(c) $\Ptjetb$, (d) $|\Rapjetb|$,
(e) $|\deltayab|$ and (f) $|\deltaphiab|$ (dots)
in the $\tt$ inclusive channel.
The NLO QCD predictions from \pwp\ (solid lines) and the NNLO QCD
predictions from \pwpm\ (dashed lines) are also shown. Both
predictions are normalised to the ${\cal O}$(NNLO+NNLL) total cross-section.
The inner (outer) error bars represent the statistical uncertainties
(the statistical and systematic uncertainties added in
quadrature). For most of the points, the inner error bars are smaller than
the marker size and, thus, not visible.
The shaded (hatched) bands represent
the theoretical uncertainty in the \pwp\ (\pwpm) prediction. The lower panels of
the figures show the ratios of the predicted and the measured cross-sections.
}
\label{figres5}
\end{figure}

\begin{figure}[h]
\setlength{\unitlength}{1.0cm}
\begin{picture} (18.0,20.0)
\put (0.0,13.0){\includegraphics[width=8cm]{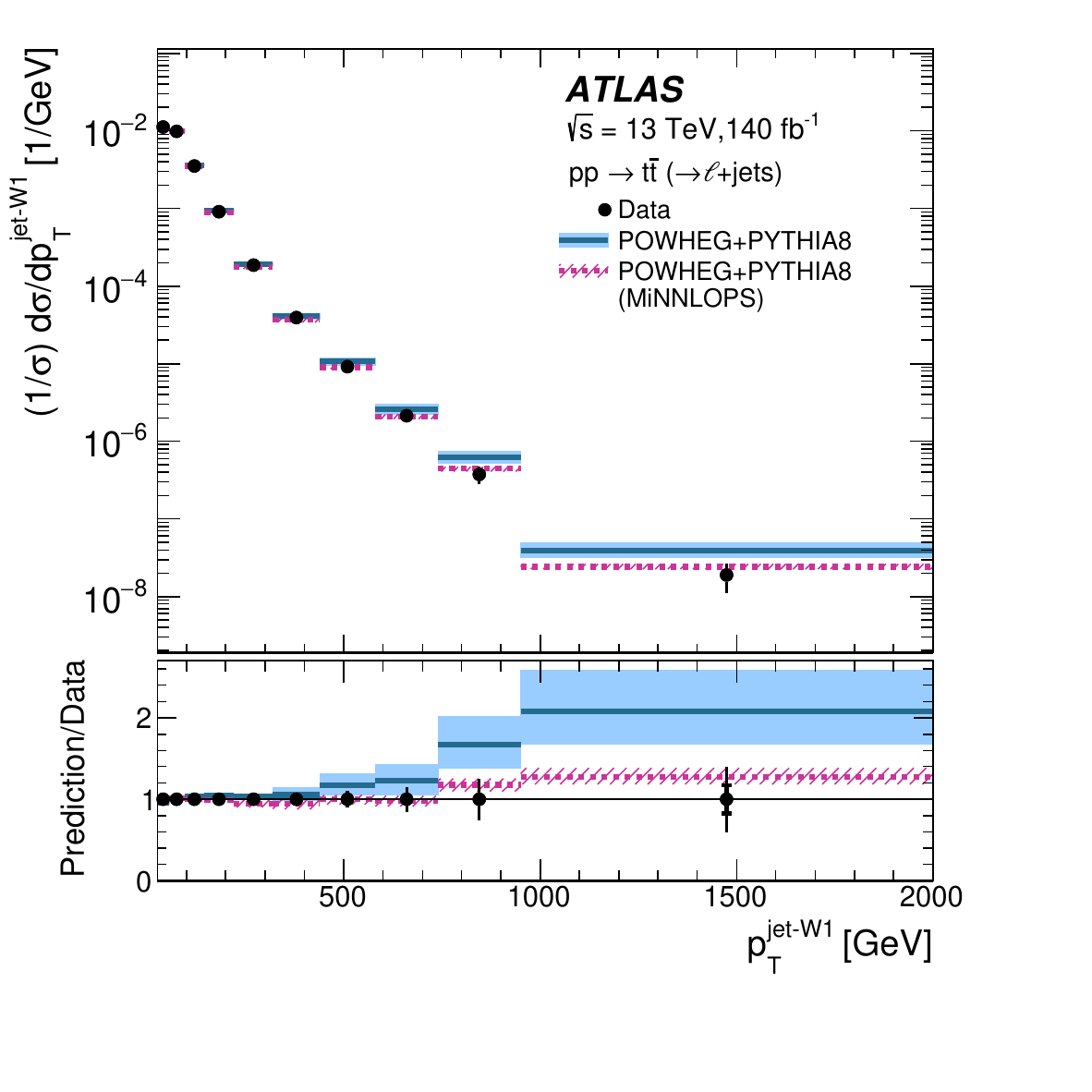}}
\put (8.0,13.0){\includegraphics[width=8cm]{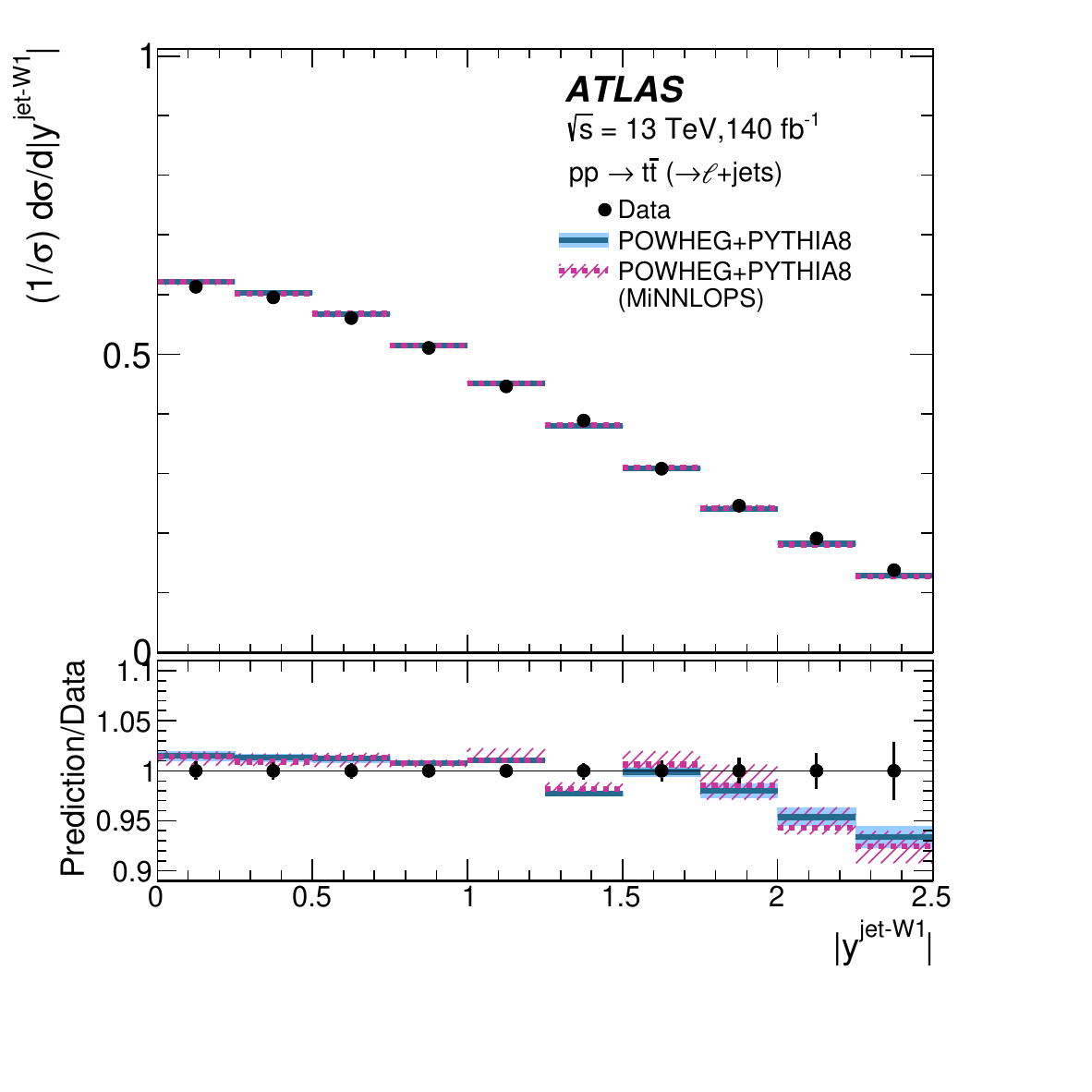}}
\put (0.0,6.0){\includegraphics[width=8cm]{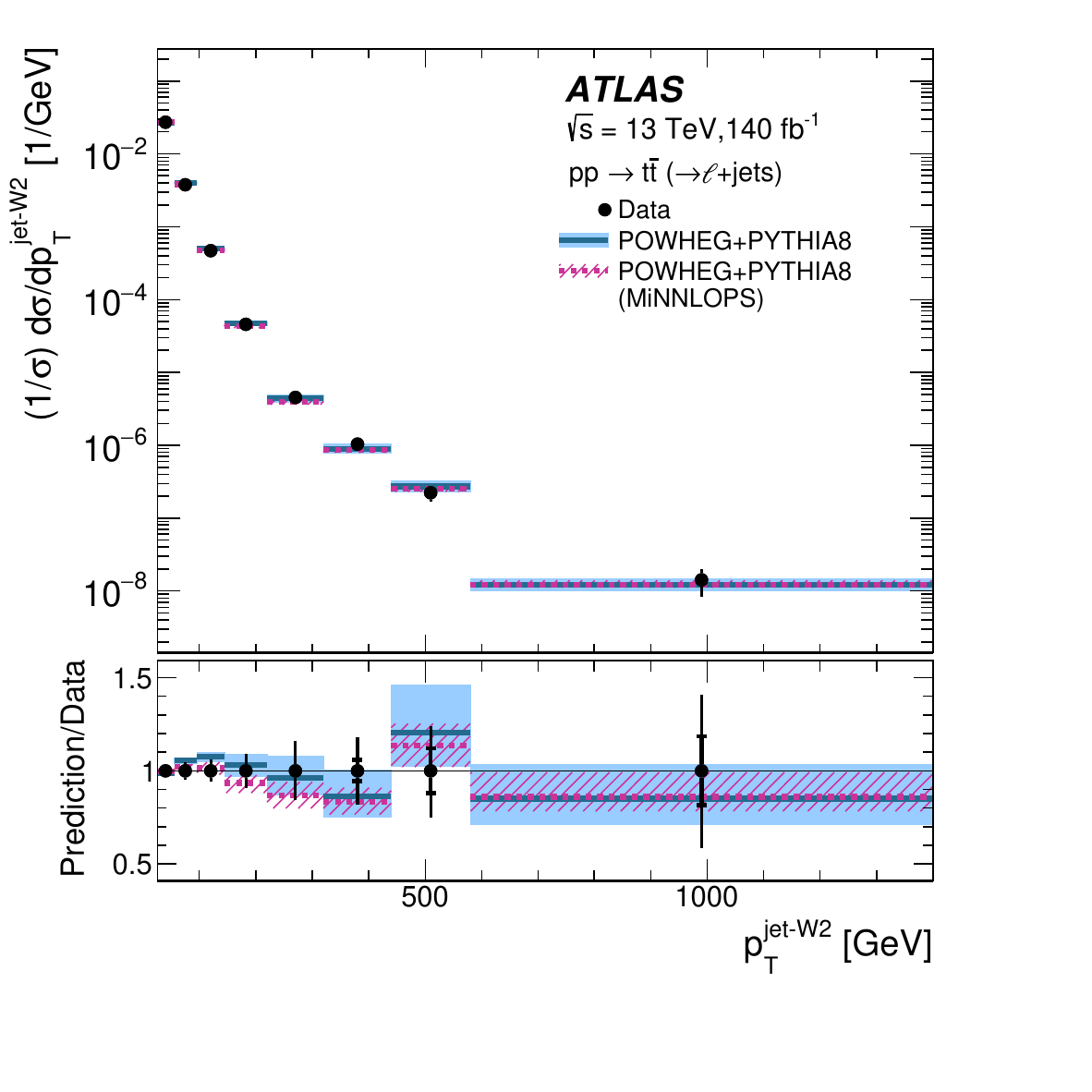}}
\put (8.0,6.0){\includegraphics[width=8cm]{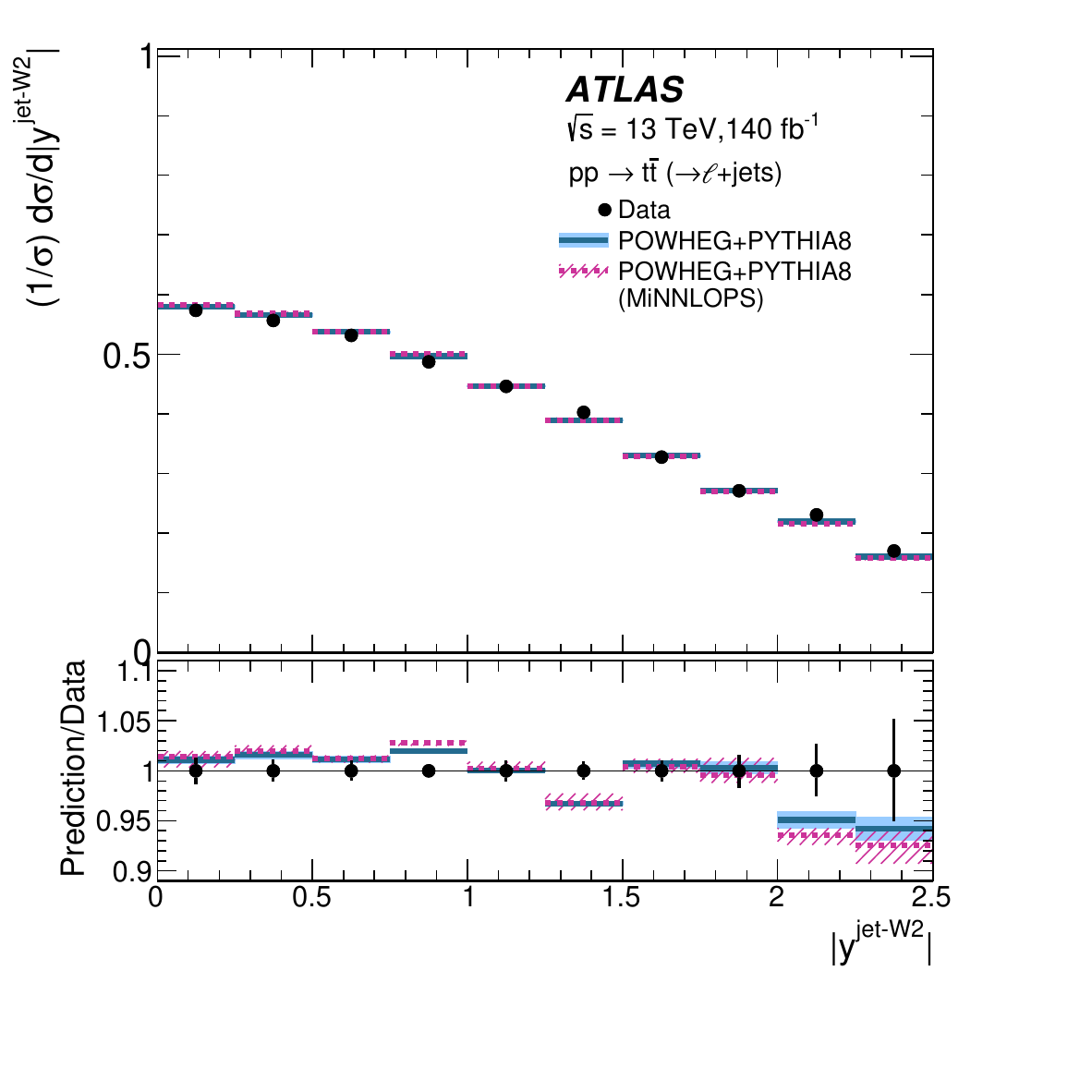}}
\put (0.0,-1.0){\includegraphics[width=8cm]{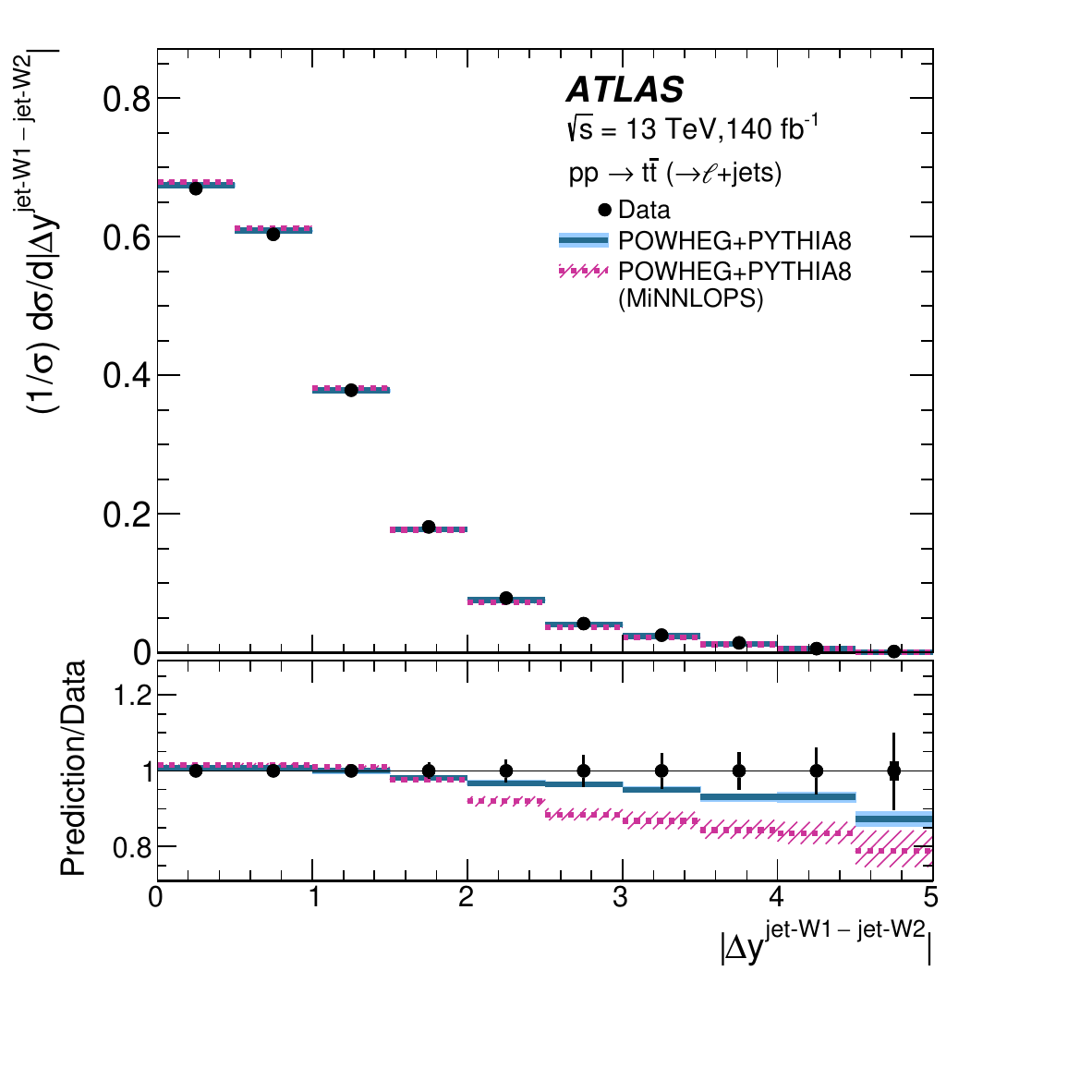}}
\put (8.0,-1.0){\includegraphics[width=8cm]{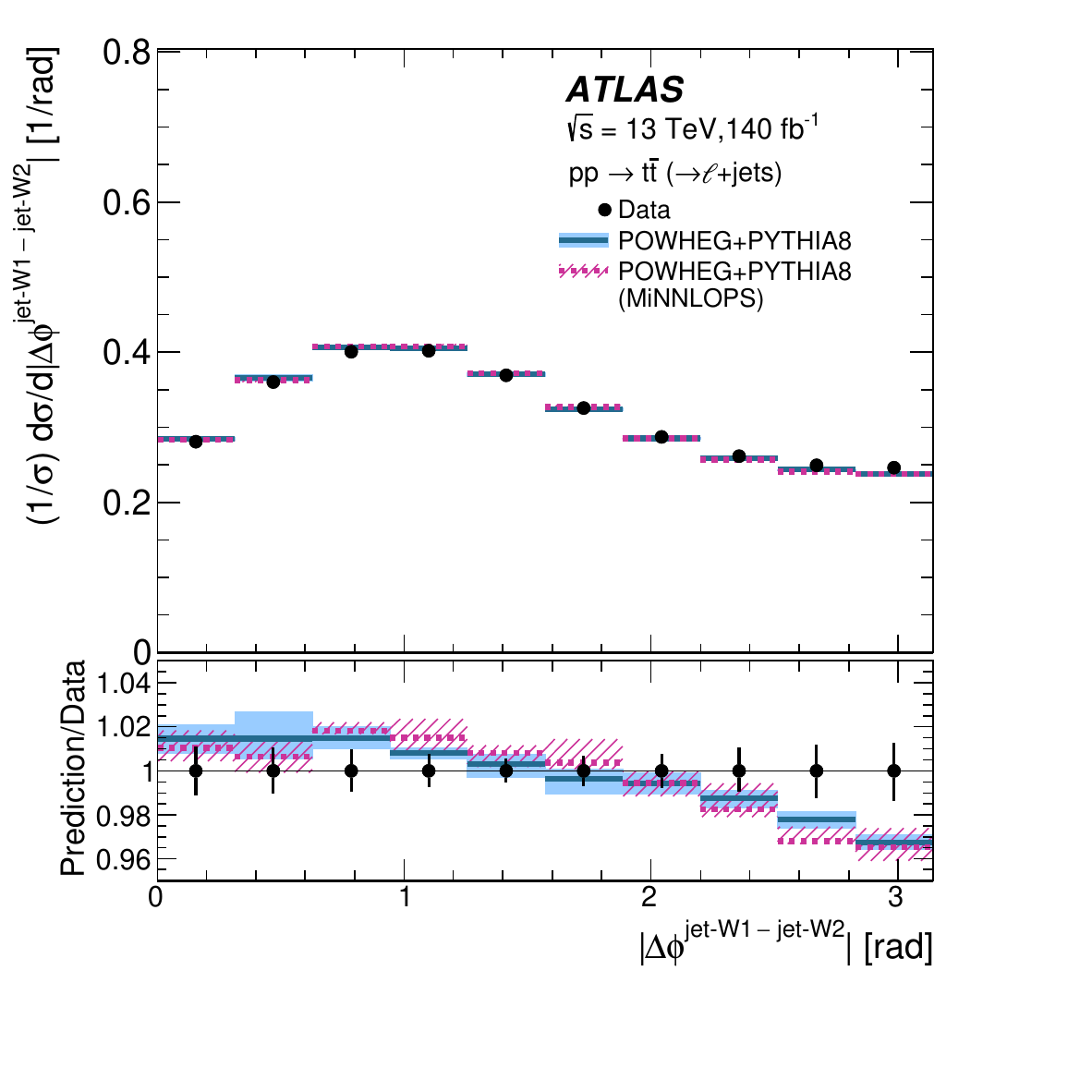}}
\put (3.5,13.9){\small (a)}
\put (11.5,13.9){\small (b)}
\put (3.5,6.9){\small (c)}
\put (11.5,6.9){\small (d)}
\put (3.5,-0.1){\small (e)}
\put (11.5,-0.1){\small (f)}
\end{picture}
\caption{
The measured normalised  differential cross-sections for $\tt$ production in
the \ljets\ decay mode as functions of
(a) $\Ptjeta$, (b) $|\Rapjeta|$,
(c) $\Ptjetb$, (d) $|\Rapjetb|$,
(e) $|\deltayab|$ and (f) $|\deltaphiab|$ (dots)
in the $\tt$ inclusive channel.
The NLO QCD predictions from \pwp\ (solid lines) and the NNLO QCD
predictions from \pwpm\ (dashed lines) are also shown.
The inner (outer) error bars represent the statistical uncertainties
(the statistical and systematic uncertainties added in
quadrature). For most of the points, the inner error bars are smaller than
the marker size and, thus, not visible.
The shaded (hatched) bands represent
the theoretical uncertainty in the \pwp\ (\pwpm) prediction. The lower panels of
the figures show the ratios of the predicted and the measured
cross-sections.
}
\label{fig_appm5}
\end{figure}

The NNLO predictions of \pwpm\ are compared with the measured absolute
and normalised differential cross-sections in Figures~\ref{figres5} and
\ref{fig_appm5}, respectively. For comparison, the predictions of
\pwp\ are also included in these figures. For all observables, the
theoretical uncertainty in the NNLO calculations is smaller than for
the NLO predictions. The angular observables are well described by the
NNLO predictions within the experimental and theoretical
uncertainties, though there is a tendency for the NNLO predictions to
be below the data for $|\deltayab|>3$, which is clearly observed in
the comparison of the normalised cross-section (Figure~\ref{fig_appm5}(e)).
For the NNLO prediction, a significant improvement in the description
of the shape and normalisation of the measured differential
cross-section as a function of $\Ptjeta$ is observed.
 
\FloatBarrier
 
\newpage
 
\FloatBarrier
 
\newpage
 
\FloatBarrier
 
\newpage
 
\FloatBarrier
 
\newpage
 
\FloatBarrier
 
\subsection{Differential cross-sections for the $\tt+$1jet channel}
\label{xsecone}
\begin{figure}[h]
\setlength{\unitlength}{1.0cm}
\begin{picture} (18.0,20.0)
\put (0.0,13.0){\includegraphics[width=8cm]{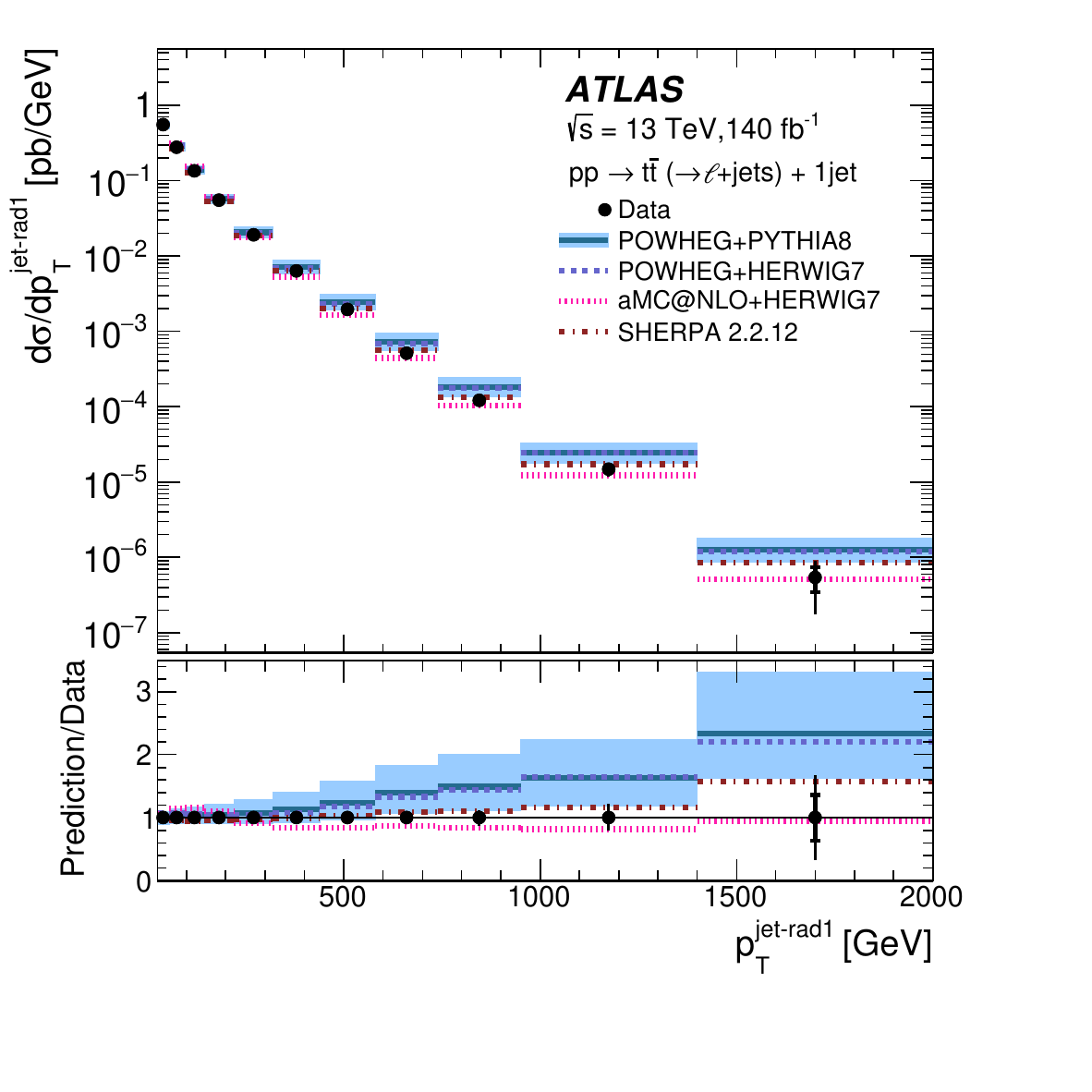}}
\put (8.0,13.0){\includegraphics[width=8cm]{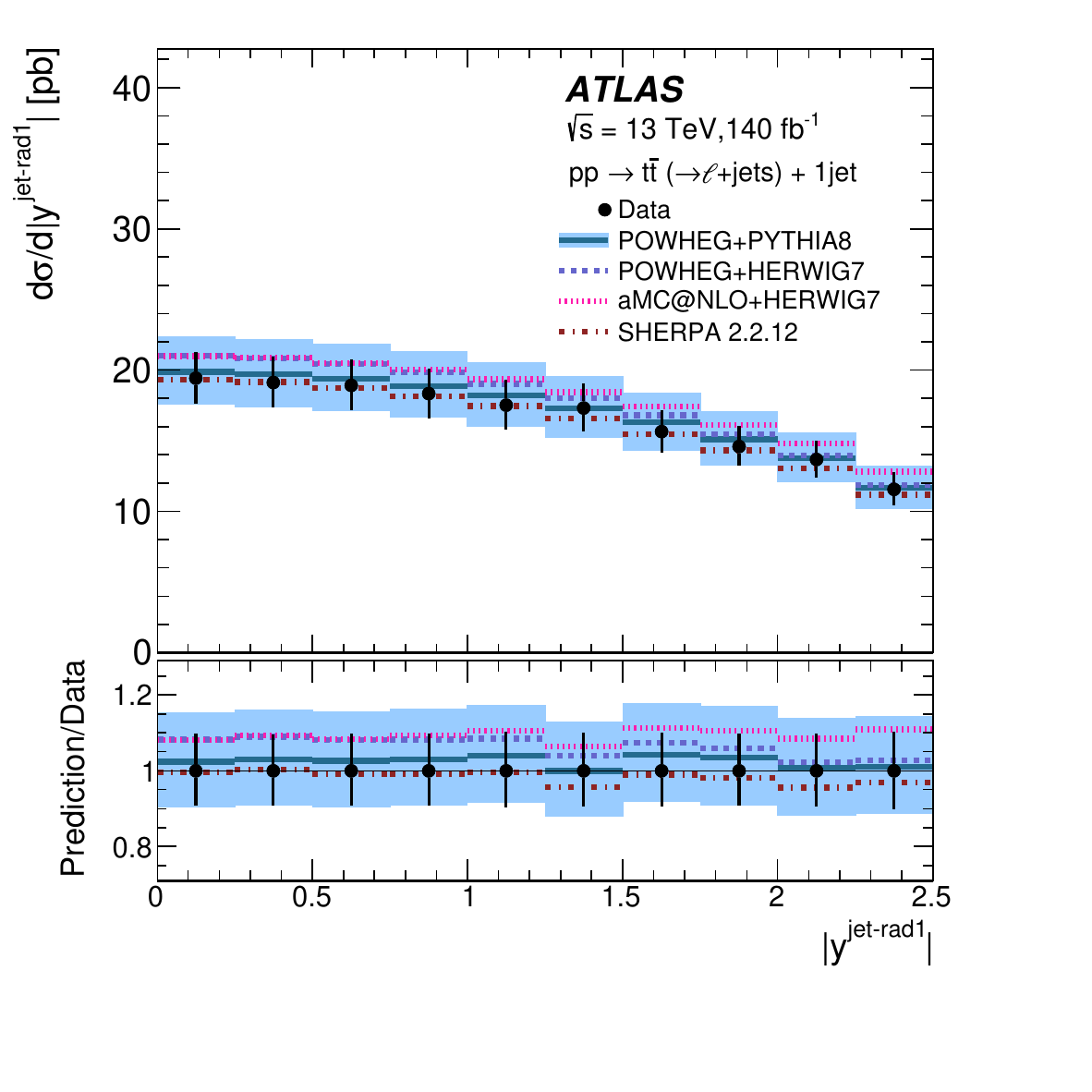}}
\put (0.0,6.0){\includegraphics[width=8cm]{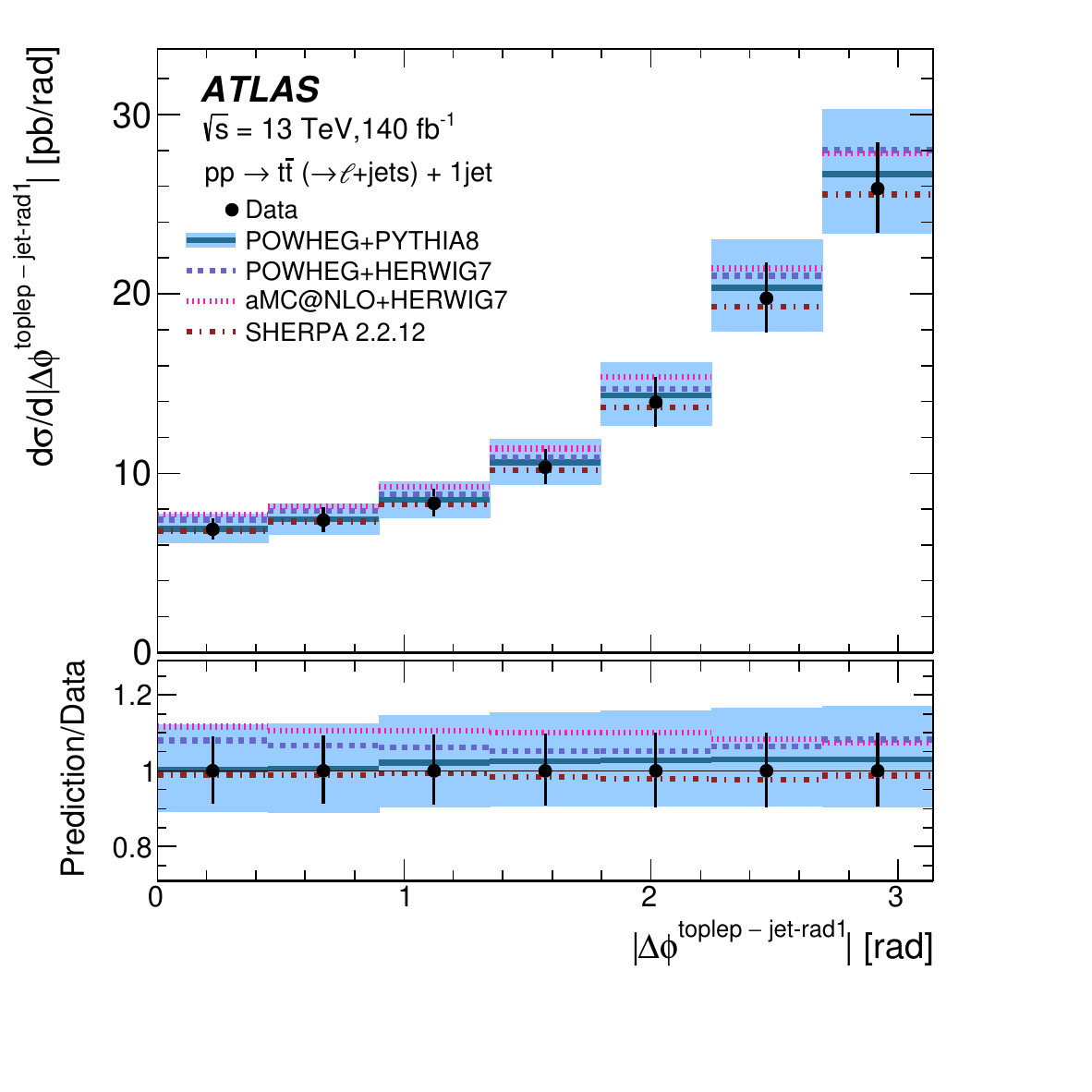}}
\put (8.0,6.0){\includegraphics[width=8cm]{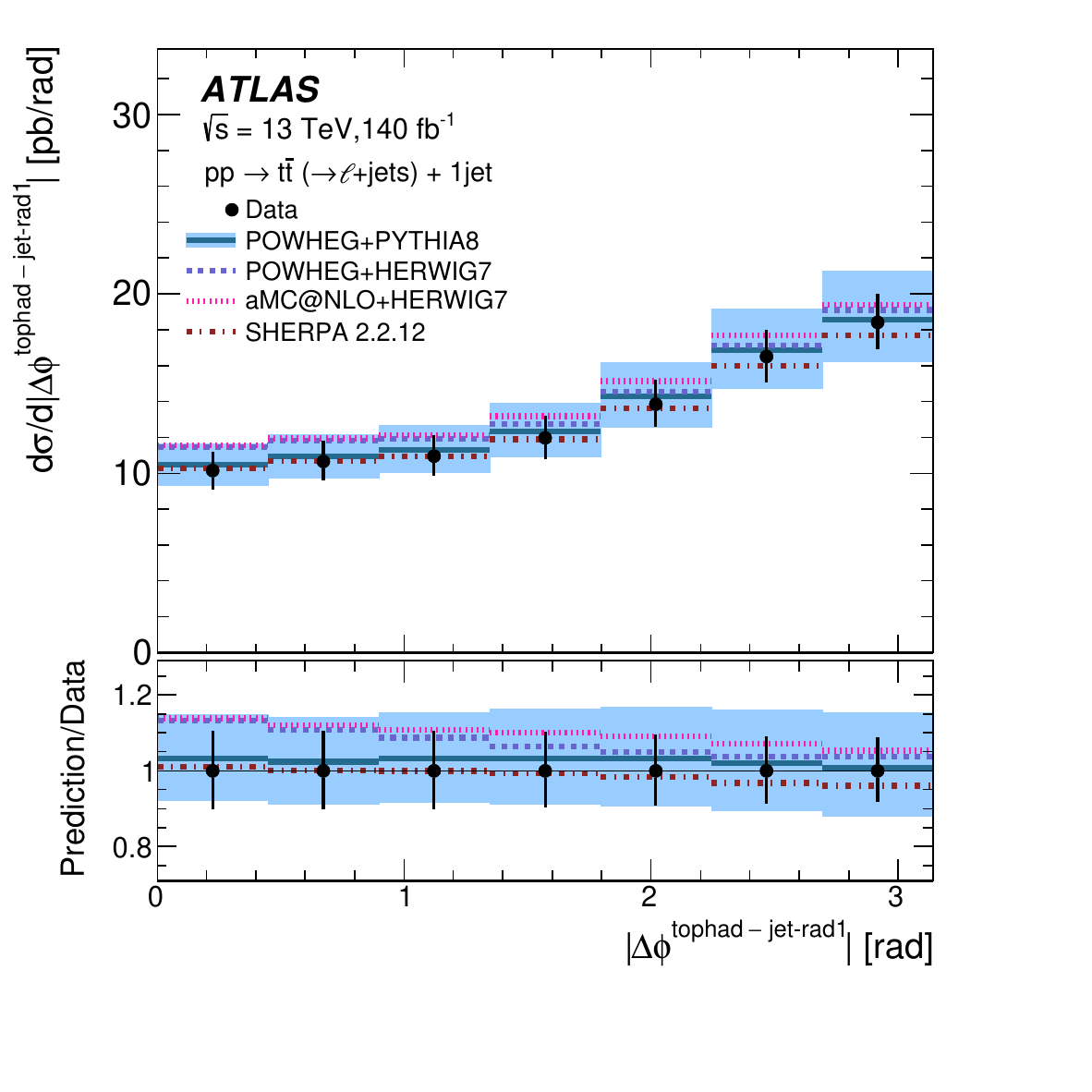}}
\put (0.0,-1.0){\includegraphics[width=8cm]{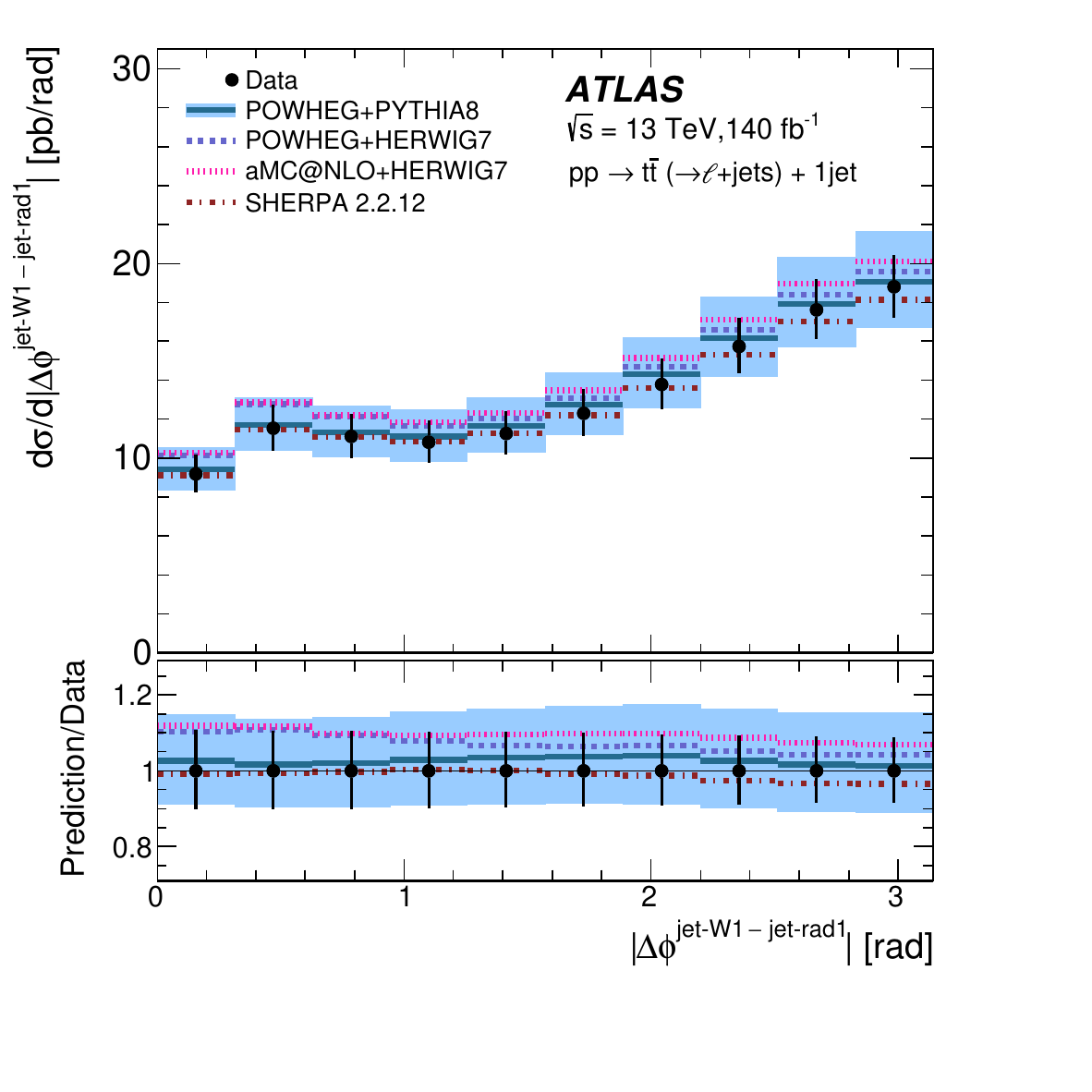}}
\put (8.0,-1.0){\includegraphics[width=8cm]{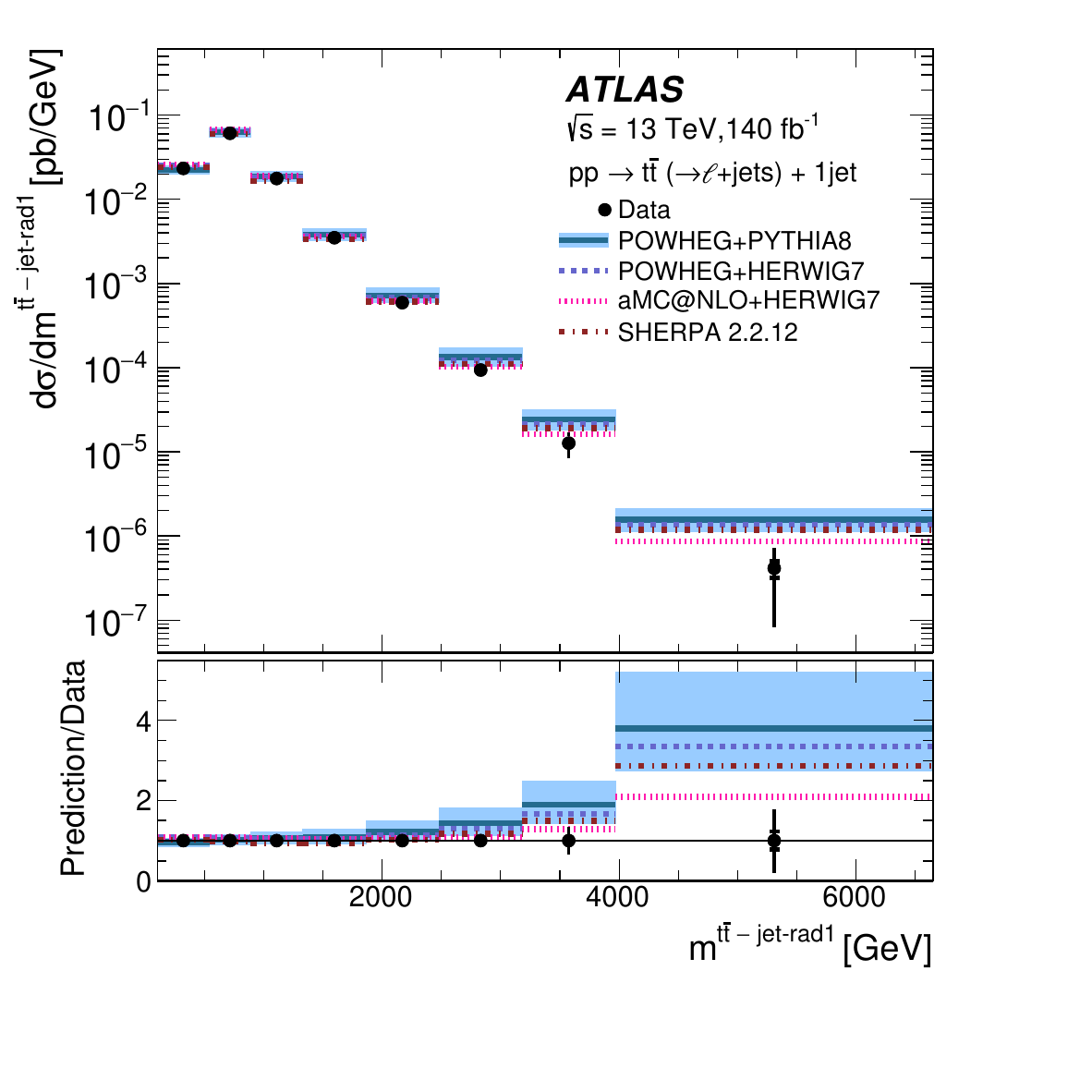}}
\put (3.5,13.9){\small (a)}
\put (11.5,13.9){\small (b)}
\put (3.5,6.9){\small (c)}
\put (11.5,6.9){\small (d)}
\put (3.5,-0.1){\small (e)}
\put (11.5,-0.1){\small (f)}
\end{picture}
\caption{
The measured differential cross-sections for $\tt$ production in
the \ljets\ decay mode as functions of
(a) $\Ptjete$, (b) $|\Rapjete|$,
(c) $|\deltaphitle|$, (d) $|\deltaphithe|$,
(e) $|\deltaphiae|$ and (f) $\mtte$ (dots)
in the $\tt+$1jet channel.
The NLO QCD predictions from \pwp\ (solid lines), \pwh\ (dashed lines),
\amcnloh\ (dotted lines) and \sher~2.2.12 (dot-dashed lines)
normalised to ${\cal O}$(NNLO+NNLL) are also shown.
The inner (outer) error bars represent the statistical uncertainties
(the statistical and systematic uncertainties added in
quadrature). For most of the points, the inner error bars are smaller than
the marker size and, thus, not visible.
The shaded bands represent
the theoretical uncertainty in the \pwp\ prediction. The lower panels of
the figures show the ratios of the predicted and the measured
cross-sections.
}
\label{figres2}
\end{figure}

\begin{figure}[h]
\setlength{\unitlength}{1.0cm}
\begin{picture} (18.0,20.0)
\put (0.0,13.0){\includegraphics[width=8cm]{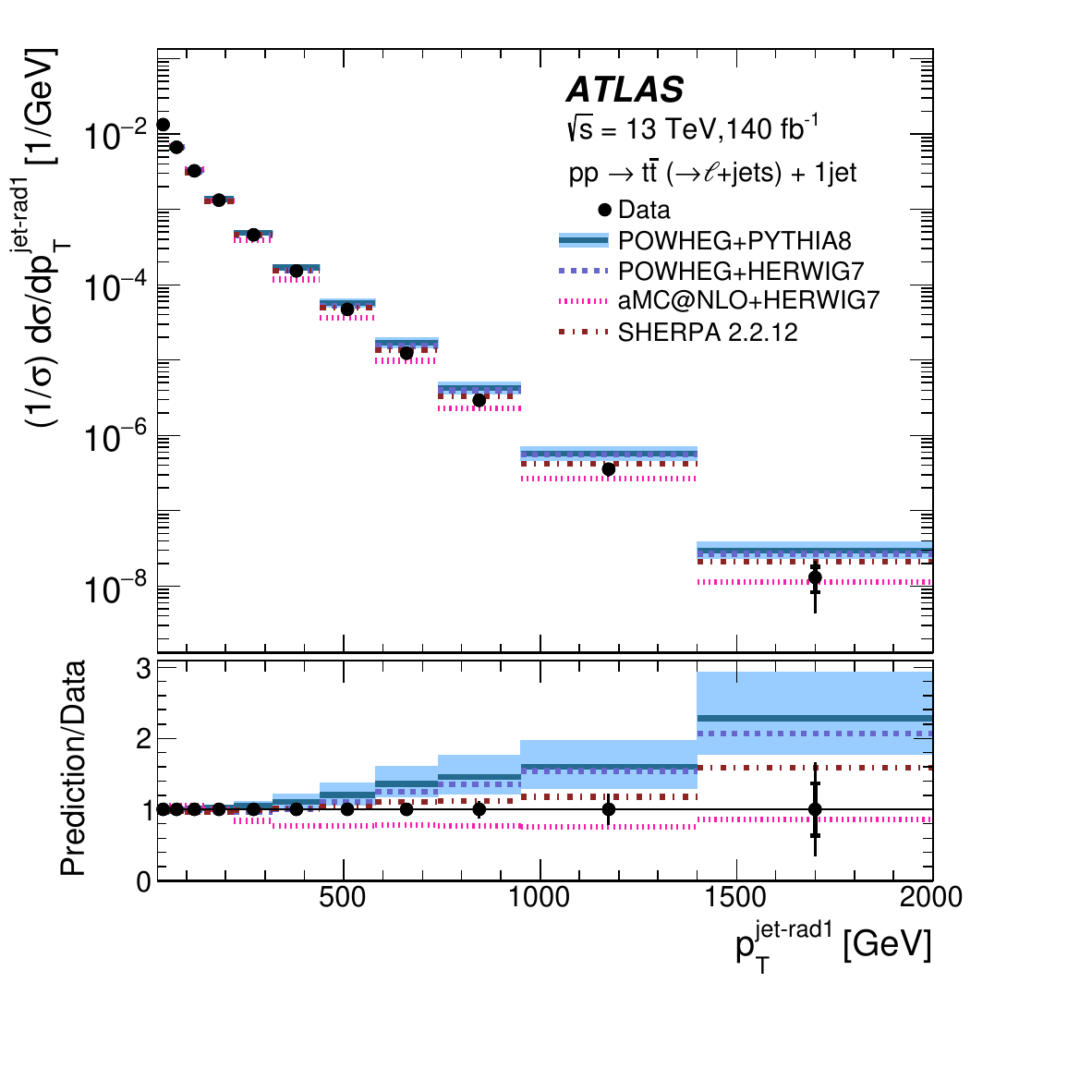}}
\put (8.0,13.0){\includegraphics[width=8cm]{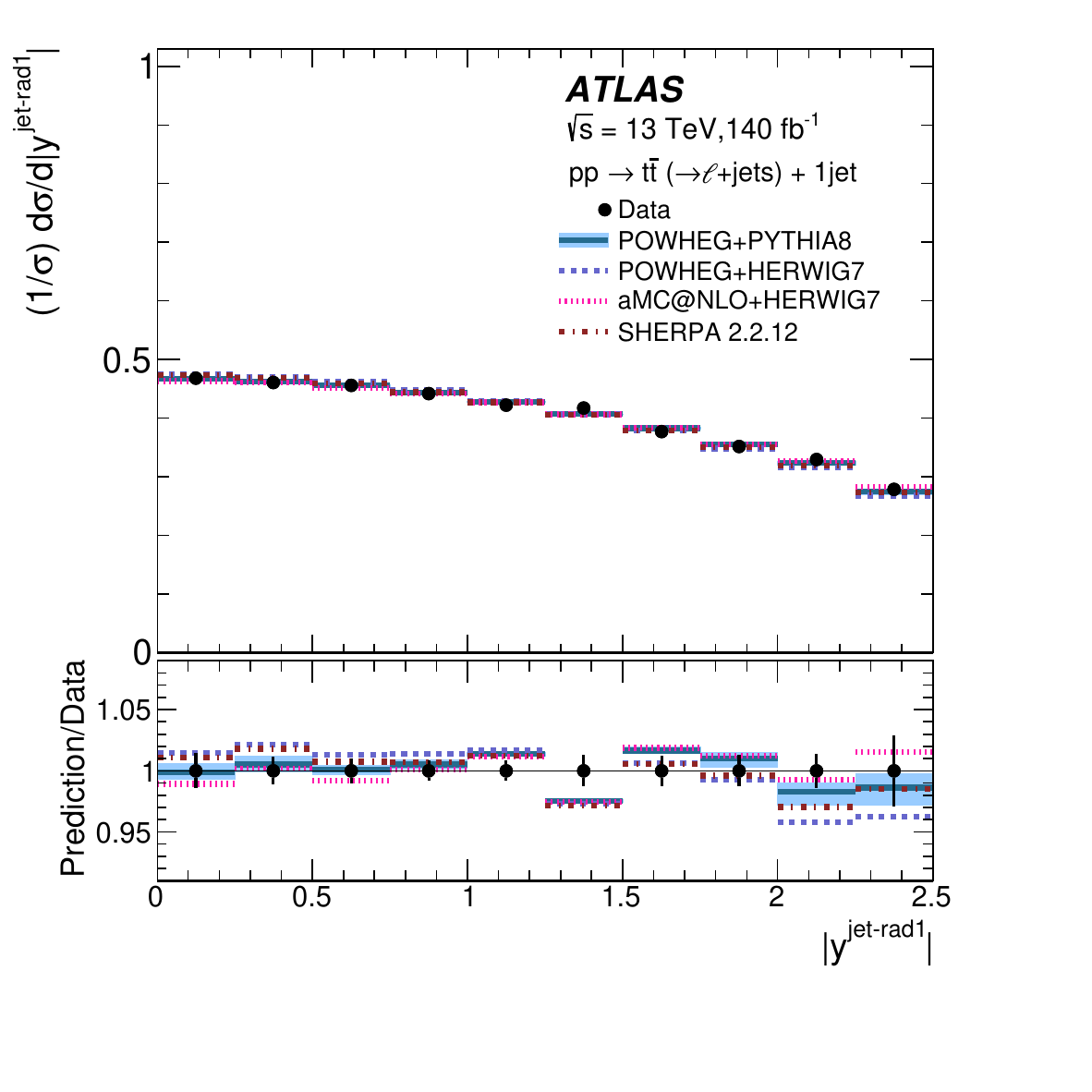}}
\put (0.0,6.0){\includegraphics[width=8cm]{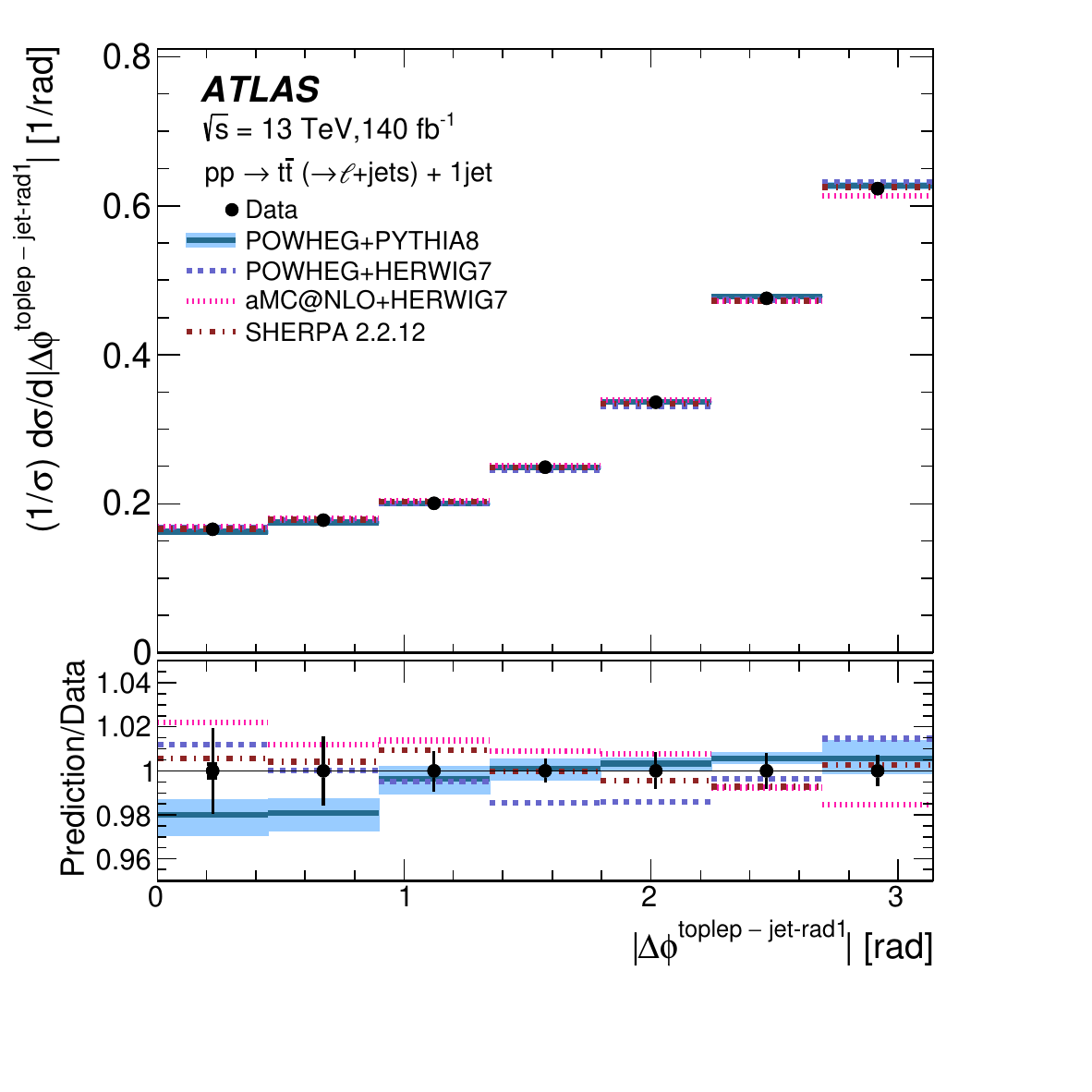}}
\put (8.0,6.0){\includegraphics[width=8cm]{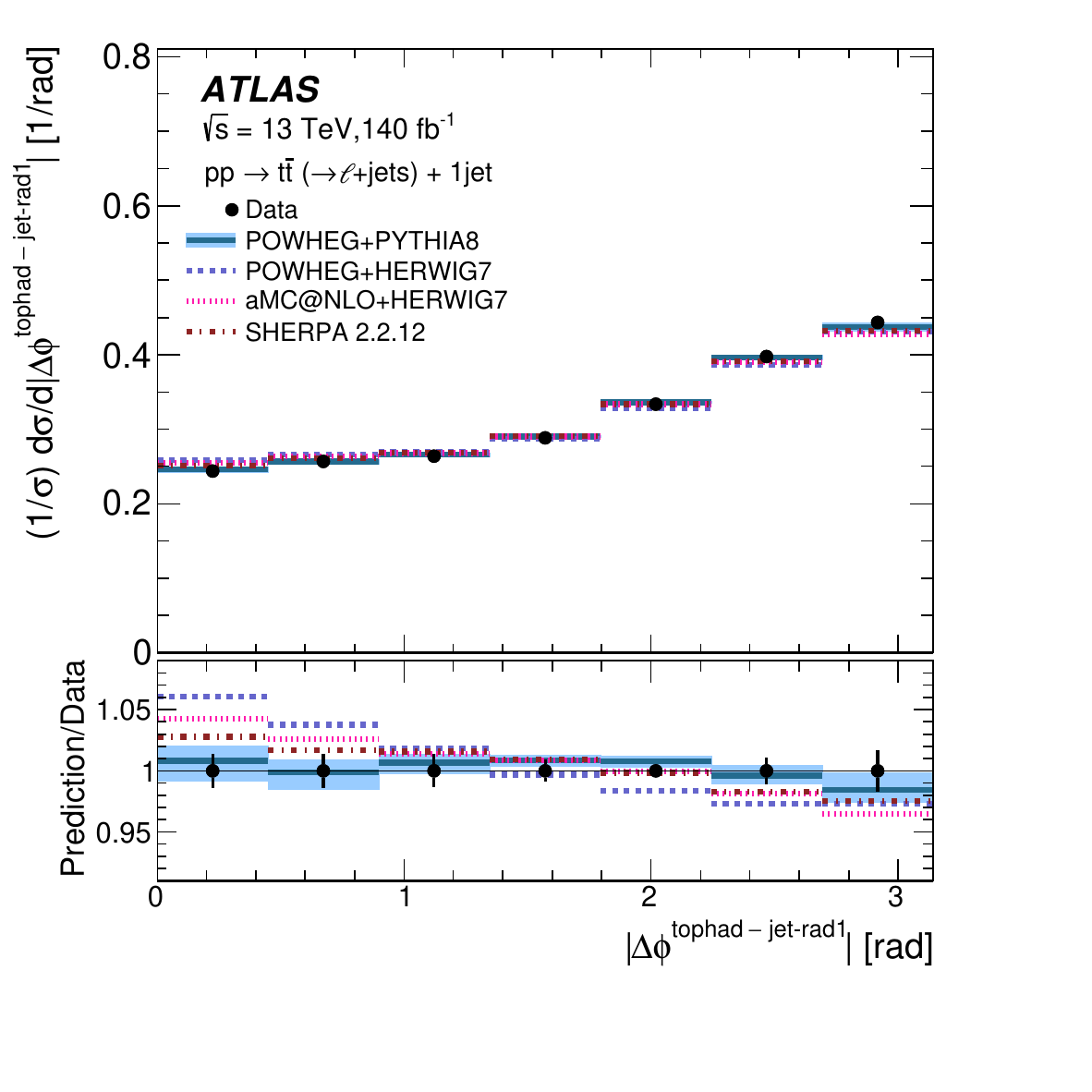}}
\put (0.0,-1.0){\includegraphics[width=8cm]{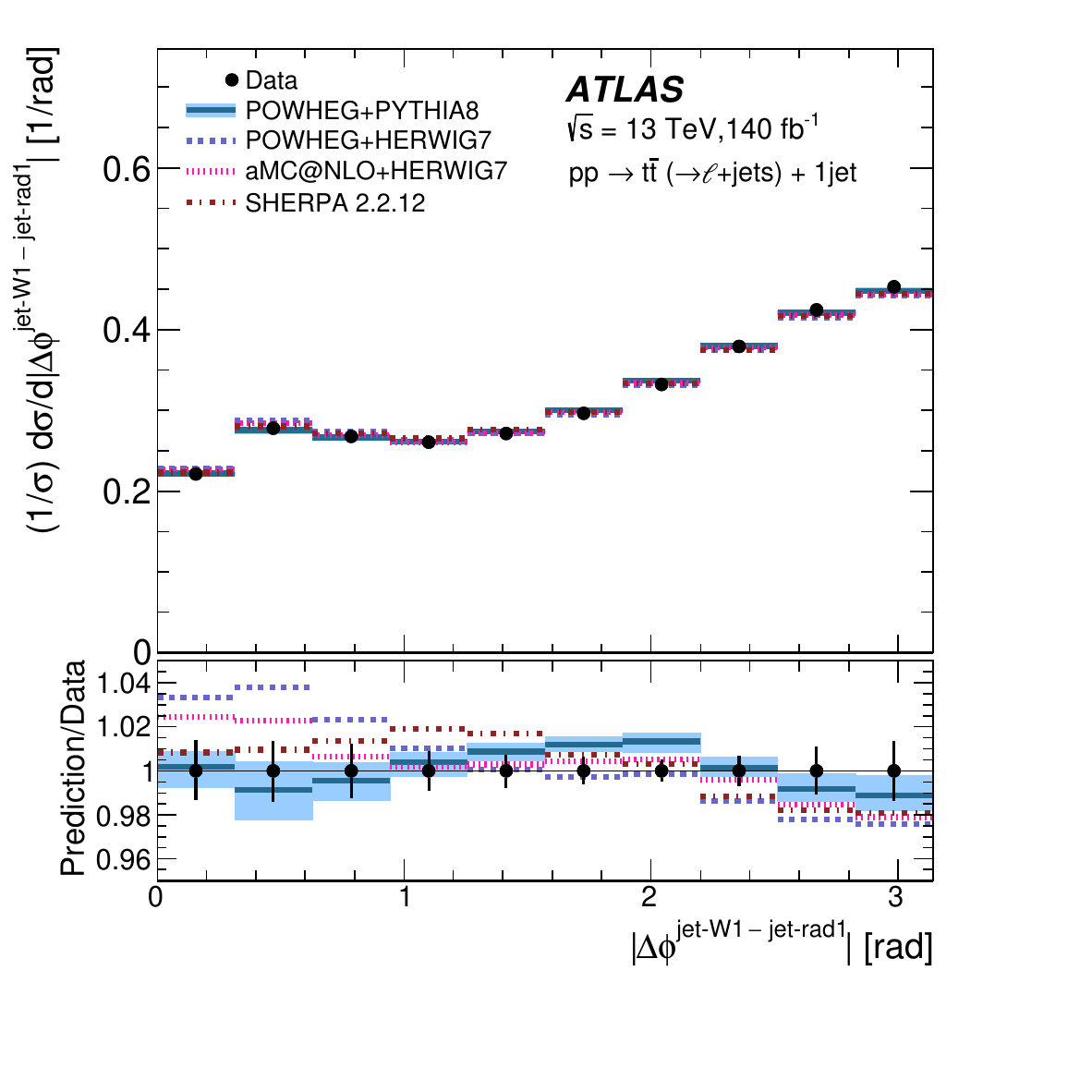}}
\put (8.0,-1.0){\includegraphics[width=8cm]{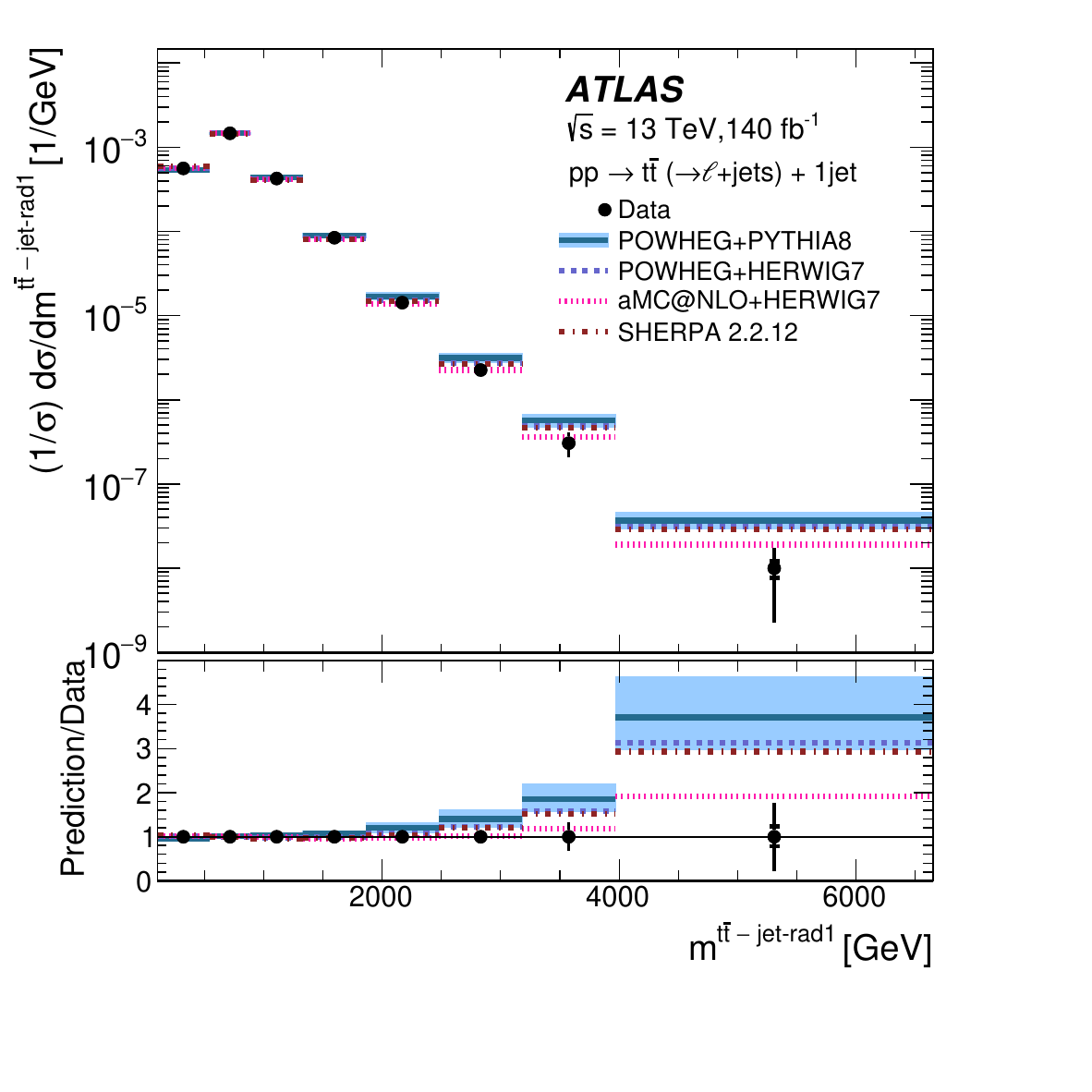}}
\put (3.5,13.9){\small (a)}
\put (11.5,13.9){\small (b)}
\put (3.5,6.9){\small (c)}
\put (11.5,6.9){\small (d)}
\put (3.5,-0.1){\small (e)}
\put (11.5,-0.1){\small (f)}
\end{picture}
\caption{
The measured normalised differential cross-sections for $\tt$ production in
the \ljets\ decay mode as functions of
(a) $\Ptjete$, (b) $|\Rapjete|$,
(c) $|\deltaphitle|$, (d) $|\deltaphithe|$,
(e) $|\deltaphiae|$ and (f) $\mtte$ (dots)
in the $\tt+$1jet channel.
The NLO QCD predictions from \pwp\ (solid lines), \pwh\ (dashed lines),
\amcnloh\ (dotted lines) and \sher~2.2.12 (dot-dashed lines)
are also shown.
The inner (outer) error bars represent the statistical uncertainties
(the statistical and systematic uncertainties added in
quadrature). For most of the points, the inner error bars are smaller than
the marker size and, thus, not visible.
The shaded bands represent
the theoretical uncertainty in the \pwp\ prediction. The lower panels of
the figures show the ratios of the predicted and the measured
cross-sections.
}
\label{fig_appm2}
\end{figure}

Figures~\ref{figres2} and \ref{fig_appm2} show the absolute and
normalised differential $\tt$ cross-sections in the \ljets\
decay mode for the
$\tt+$1jet channel as functions of $\Ptjete$,
$|\Rapjete|$, $|\deltaphitle|$, $|\deltaphithe|$, $|\deltaphiae|$ and
$\mtte$. Values of $\Ptjete$ and $\mtte$ up to $\sim 1.7$~\TeV\ and
$\sim 5.3$~\TeV, respectively, are measured. The measured
cross-section as a function of $\Ptjete$ ($\mtte$) decreases by more
than
six (five) orders of magnitude within the measured range. The measured
cross-section as a function of $\mtte$ decreases for $\mtte<0.5$~\TeV\
due to the kinematic constraints. The shape of the measured
cross-section as a function of $|\Rapjete|$ is different from that of
$|\Rapjeta|$: the value of the cross-section at a rapidity of $2.5$
relative to the value at $0$ is higher for $|\Rapjete|$ than for
$|\Rapjeta|$, which indicates that the first gluon emission tends to be more
isotropic in rapidity than the jets forming the hadronically decaying
top quark. This is investigated further by measuring the cross-sections as
functions of $|\deltaphitle|$ and $|\deltaphithe|$: the measured
cross-sections have very different shapes, which indicates that the
first gluon emission has a higher probability of arising at larger angles from the
leptonically decaying top quark. In addition, the measured cross-section as
a function of $|\deltaphiae|$ shows a peak at $|\deltaphiae|\approx
0.5$~rad and at $\pi$~rad, which indicates that there are more events
with a gluon emission back-to-back with jet-W1, though a fraction
of the events has a gluon emission close to jet-W1.
 
The NLO QCD predictions describe the measurements as functions of
$|\Rapjete|$, $|\deltaphitle|$, \break $|\deltaphithe|$ and
$|\deltaphiae|$ well. As is the case for the measurements in the
$\tt$ inclusive channel, the cross-section as a function of $\Ptjete$
is described by the predictions from \pwp\ and \pwh\ only for
$\Ptjete<500$~\GeV, whereas the predictions from \amcnloh\ and
\sher~2.2.12 provide a good description of the data over the whole
measured range. The cross-section as a function of $\mtte$ is well
described by all predictions for $\mtte<3$~\TeV. The same conclusions
can be drawn from the normalised differential cross-sections in
Figure~\ref{fig_appm2}.
 
\begin{figure}[h]
\setlength{\unitlength}{1.0cm}
\begin{picture} (18.0,20.0)
\put (0.0,13.0){\includegraphics[width=8cm]{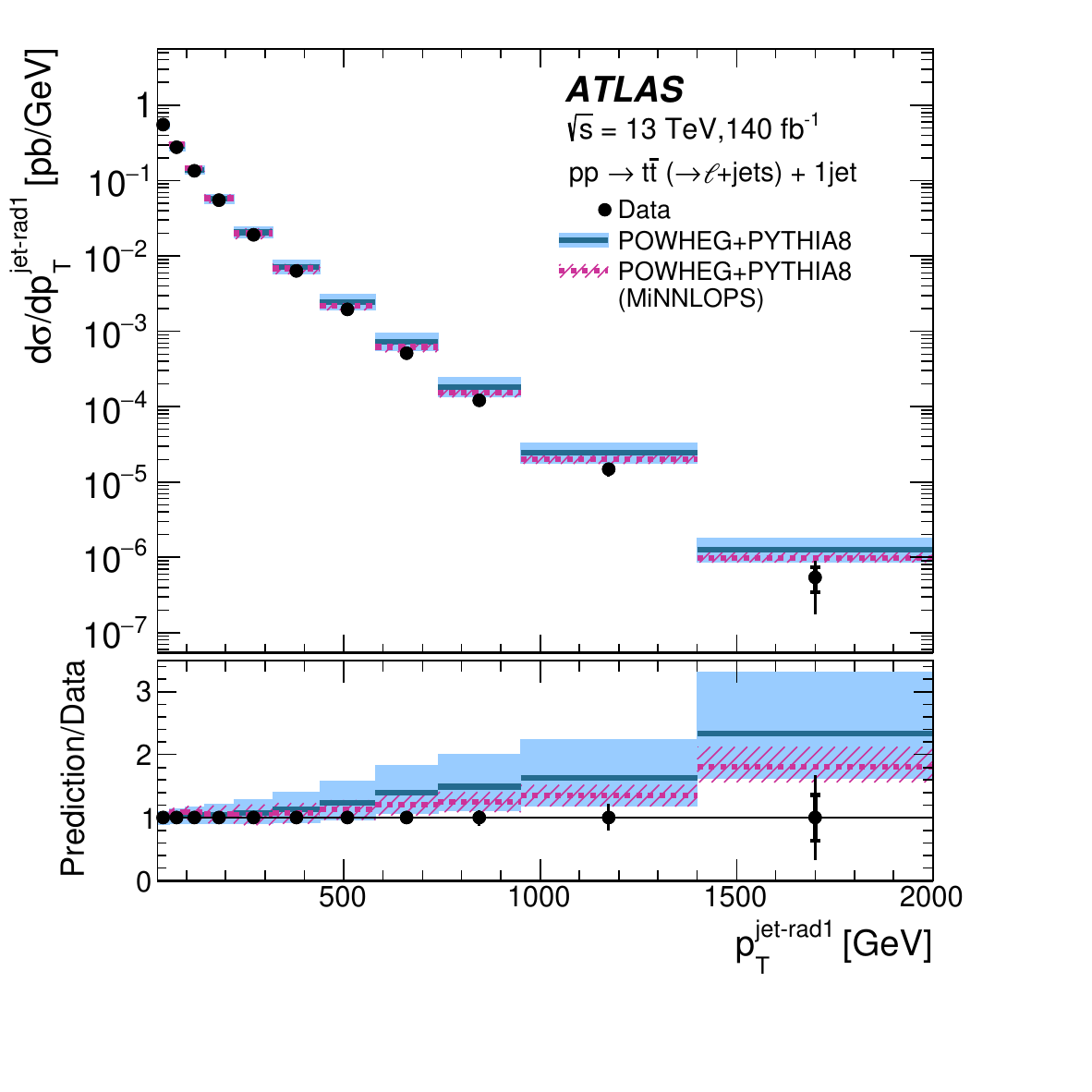}}
\put (8.0,13.0){\includegraphics[width=8cm]{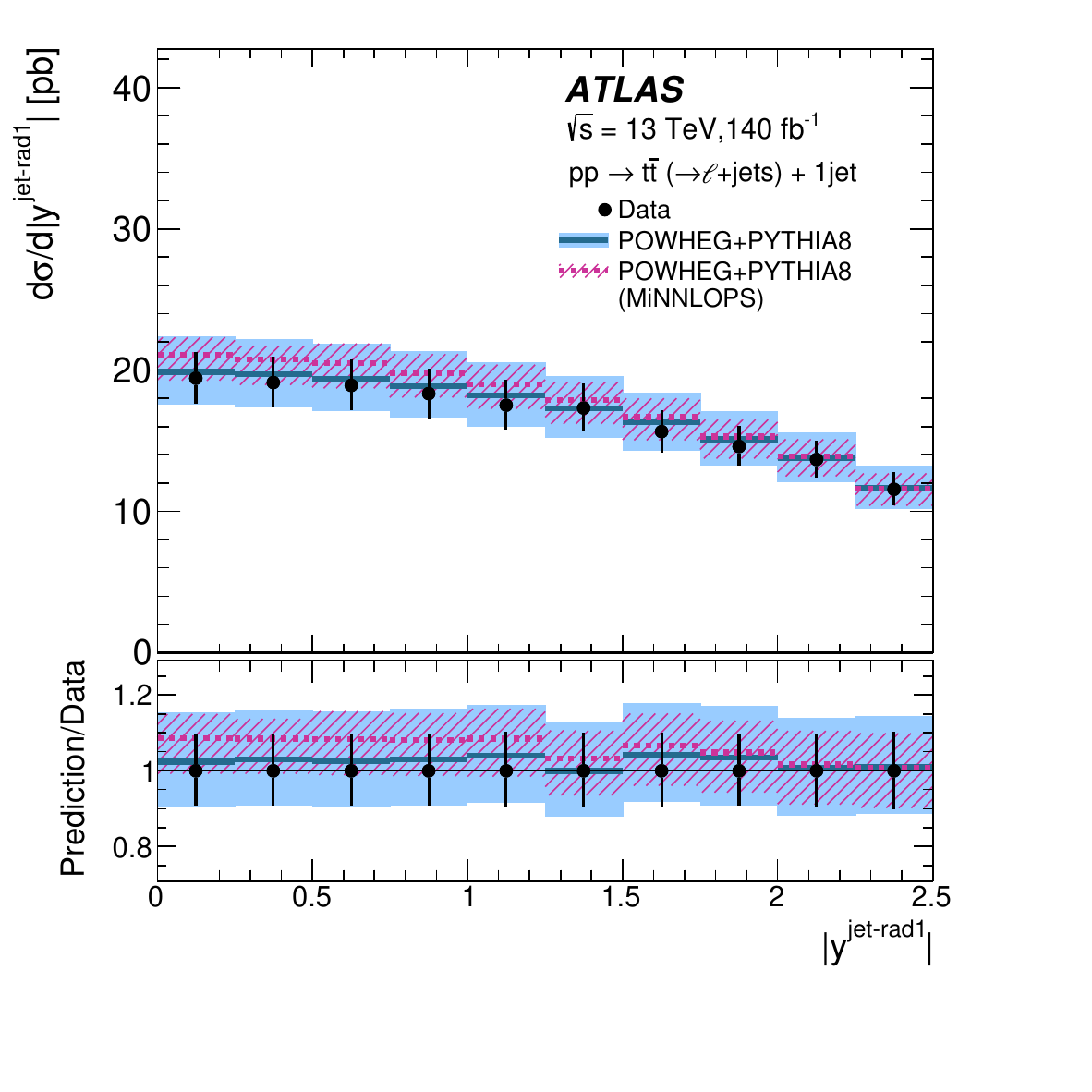}}
\put (0.0,6.0){\includegraphics[width=8cm]{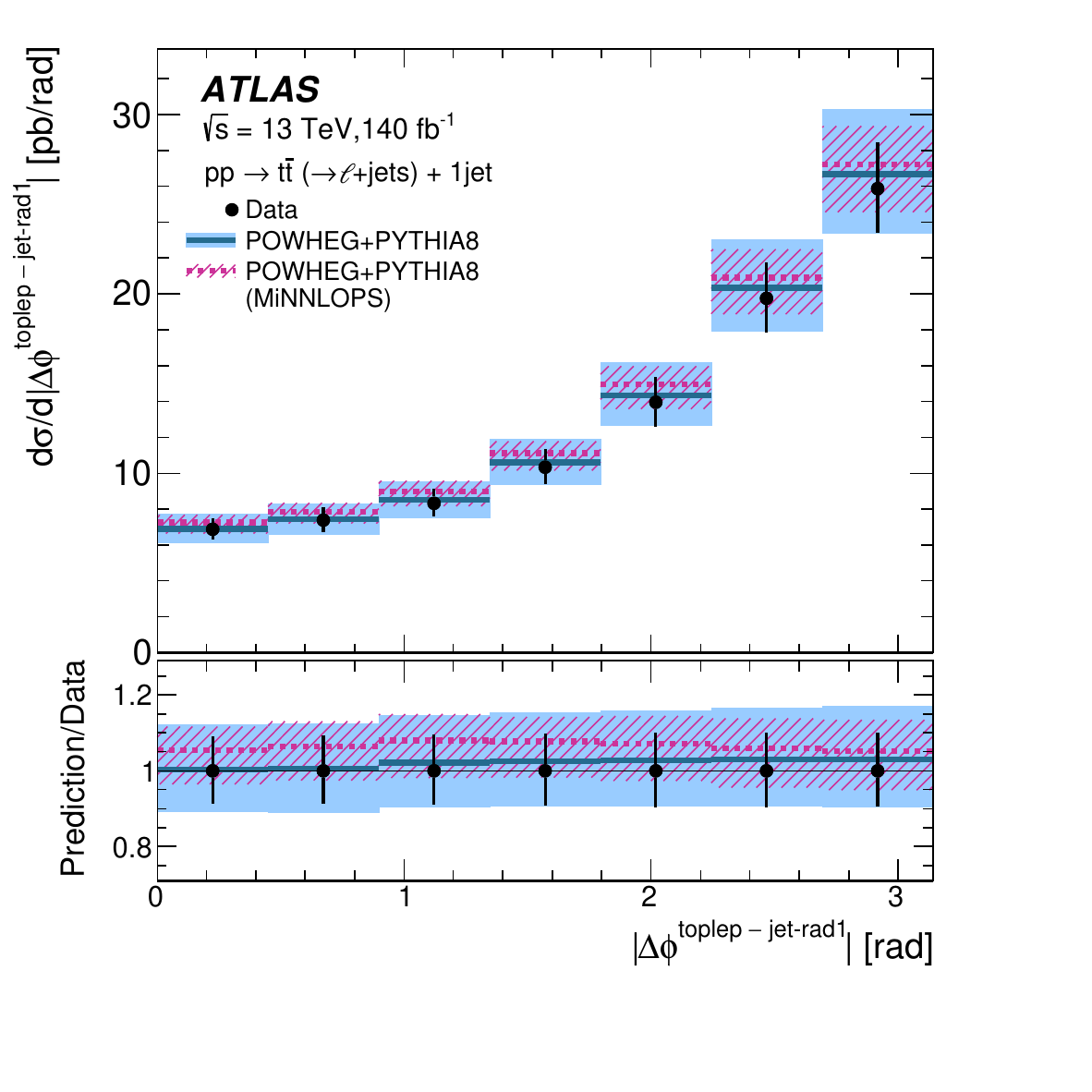}}
\put (8.0,6.0){\includegraphics[width=8cm]{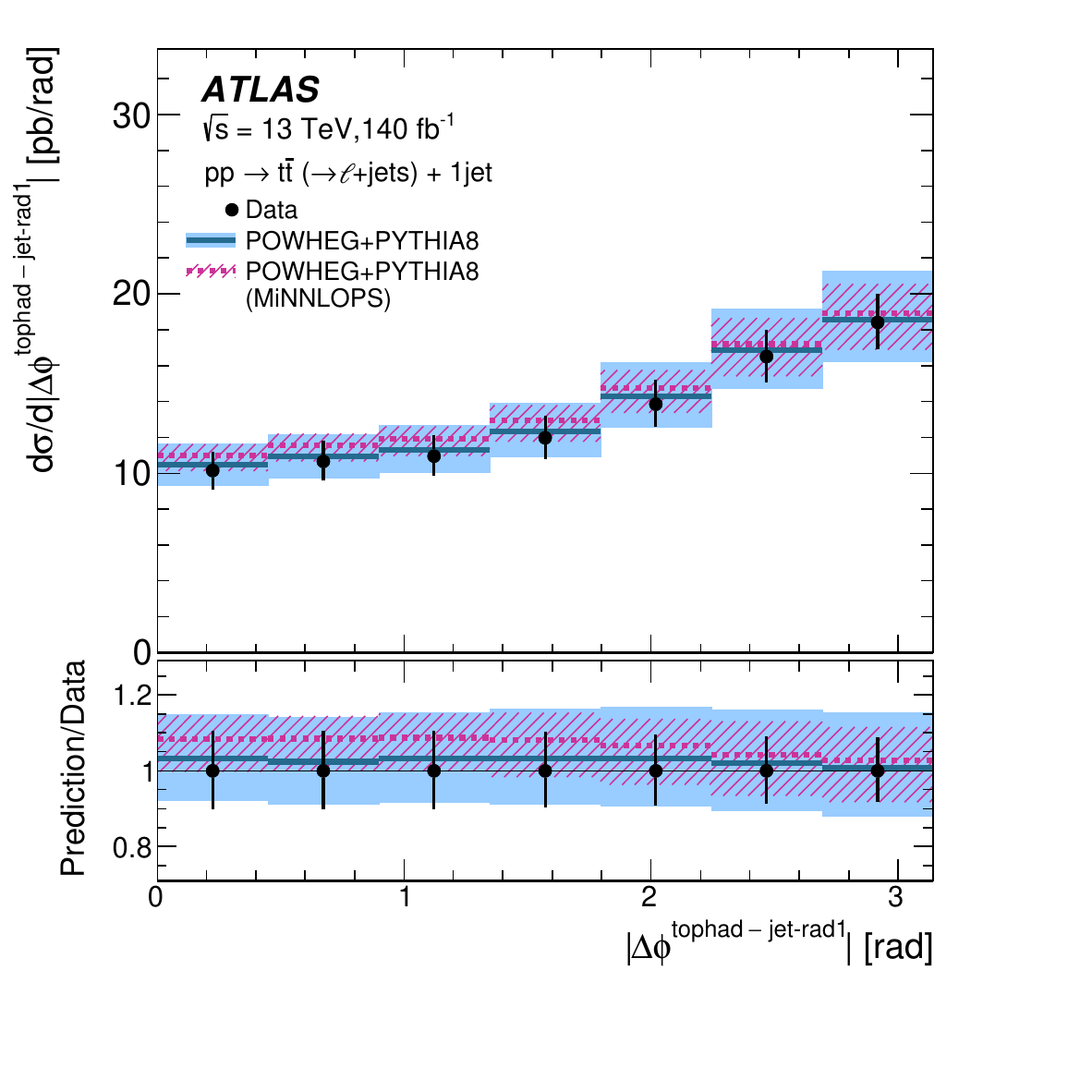}}
\put (0.0,-1.0){\includegraphics[width=8cm]{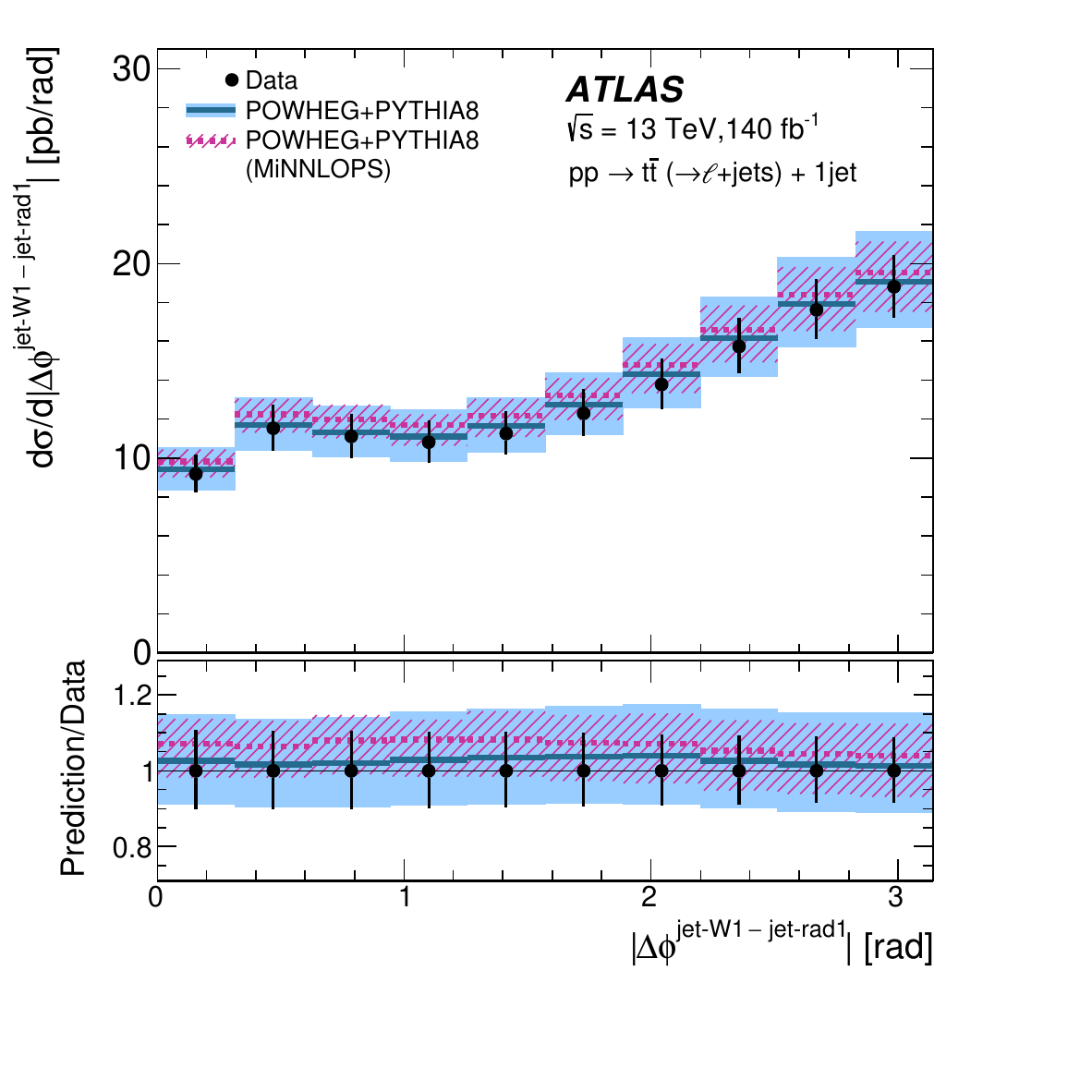}}
\put (8.0,-1.0){\includegraphics[width=8cm]{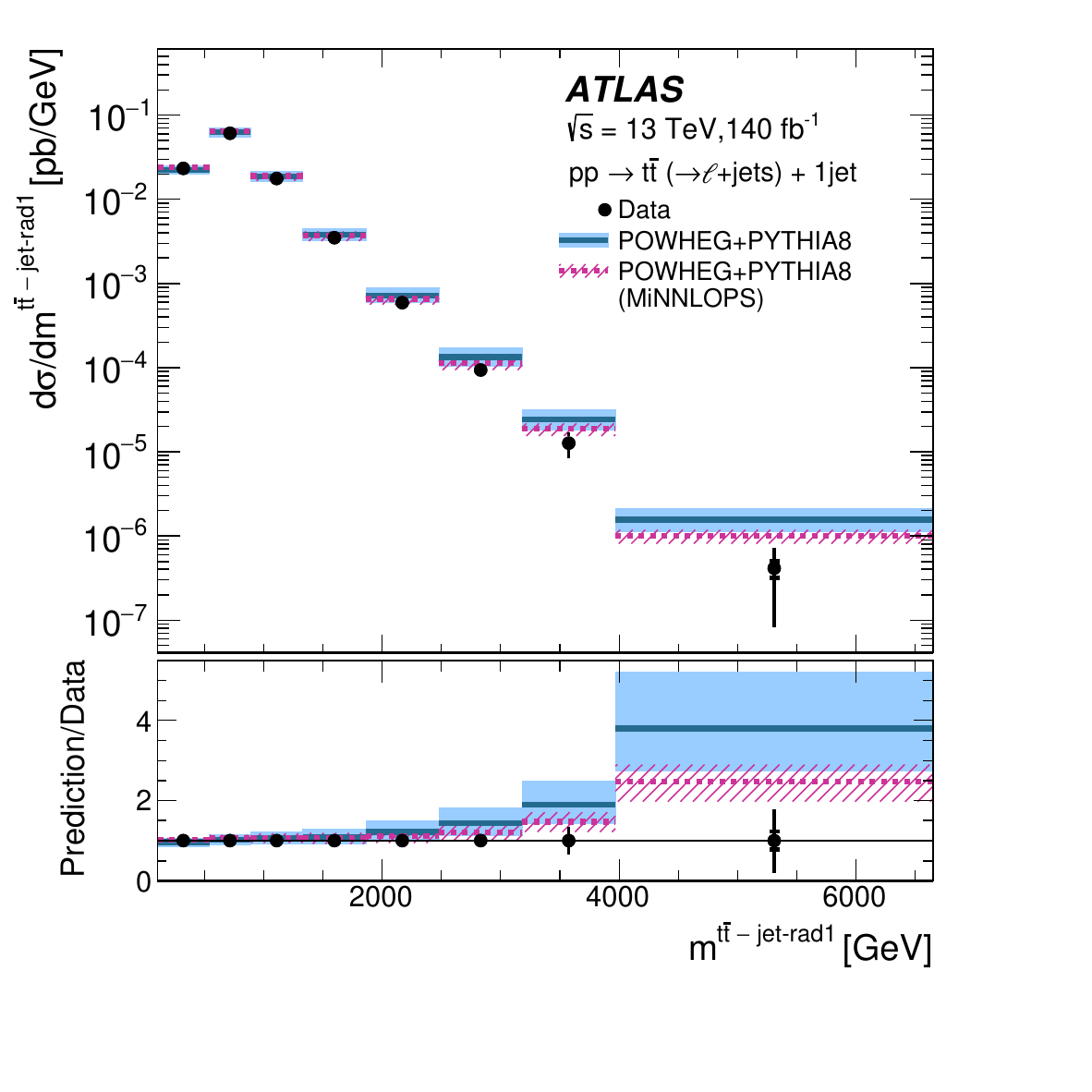}}
\put (3.5,13.9){\small (a)}
\put (11.5,13.9){\small (b)}
\put (3.5,6.9){\small (c)}
\put (11.5,6.9){\small (d)}
\put (3.5,-0.1){\small (e)}
\put (11.5,-0.1){\small (f)}
\end{picture}
\caption{
The measured differential cross-sections for $\tt$ production in
the \ljets\ decay mode as functions of
(a)~$\Ptjete$, (b) $|\Rapjete|$,
(c) $|\deltaphitle|$, (d) $|\deltaphithe|$,
(e) $|\deltaphiae|$ and (f) $\mtte$ (dots)
in the $\tt+$1jet channel.
The NLO QCD predictions from \pwp\ (solid lines) and the NNLO QCD
predictions from \pwpm\ (dashed lines) are also shown. Both
predictions are normalised to the ${\cal O}$(NNLO+NNLL) total cross-section.
The inner (outer) error bars represent the statistical uncertainties
(the statistical and systematic uncertainties added in
quadrature). For most of the points, the inner error bars are smaller than
the marker size and, thus, not visible.
The shaded (hatched) bands represent
the theoretical uncertainty in the \pwp\ (\pwpm) prediction. The lower panels of
the figures show the ratios of the predicted and the measured
cross-sections.
}
\label{figres6}
\end{figure}

\begin{figure}[h]
\setlength{\unitlength}{1.0cm}
\begin{picture} (18.0,20.0)
\put (0.0,13.0){\includegraphics[width=8cm]{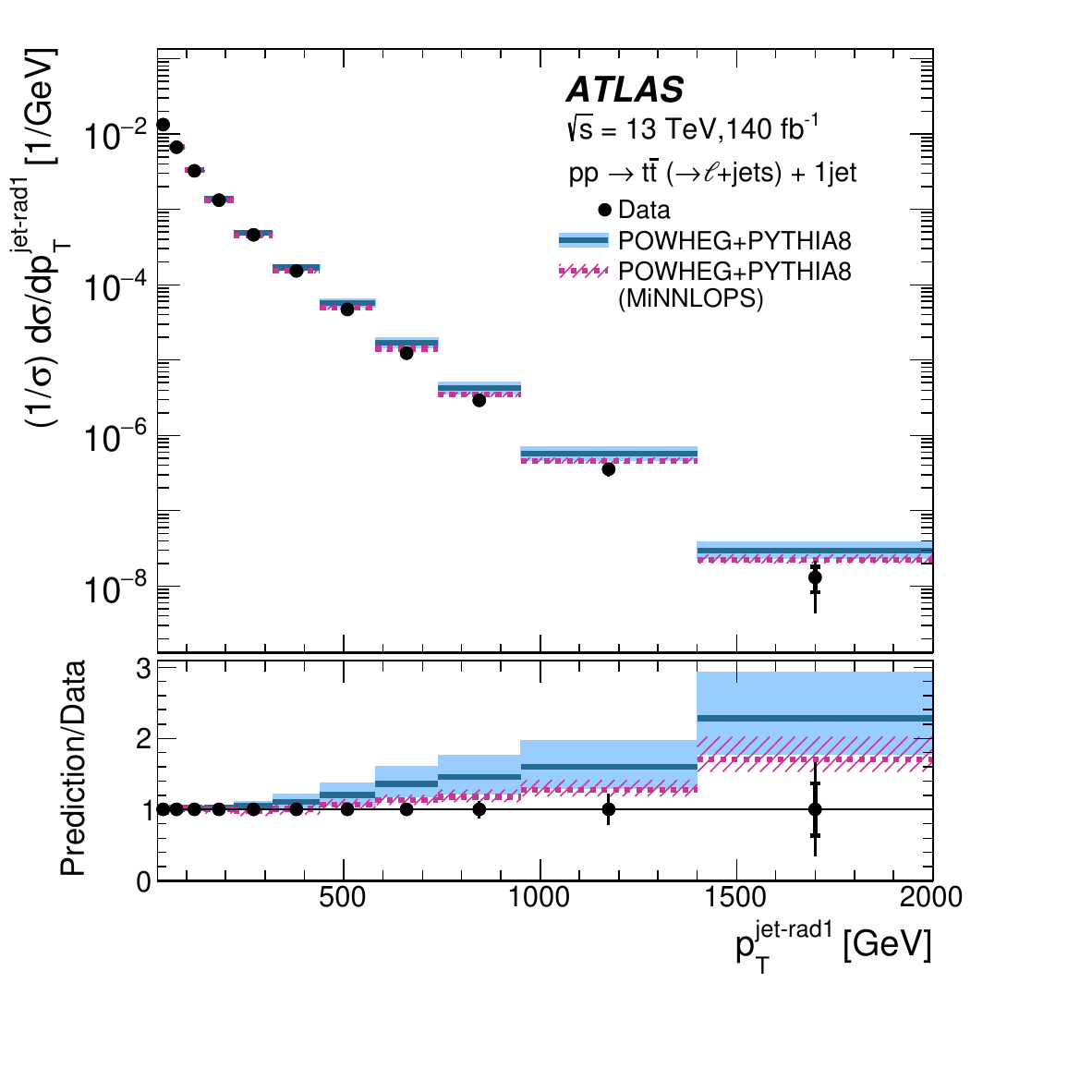}}
\put (8.0,13.0){\includegraphics[width=8cm]{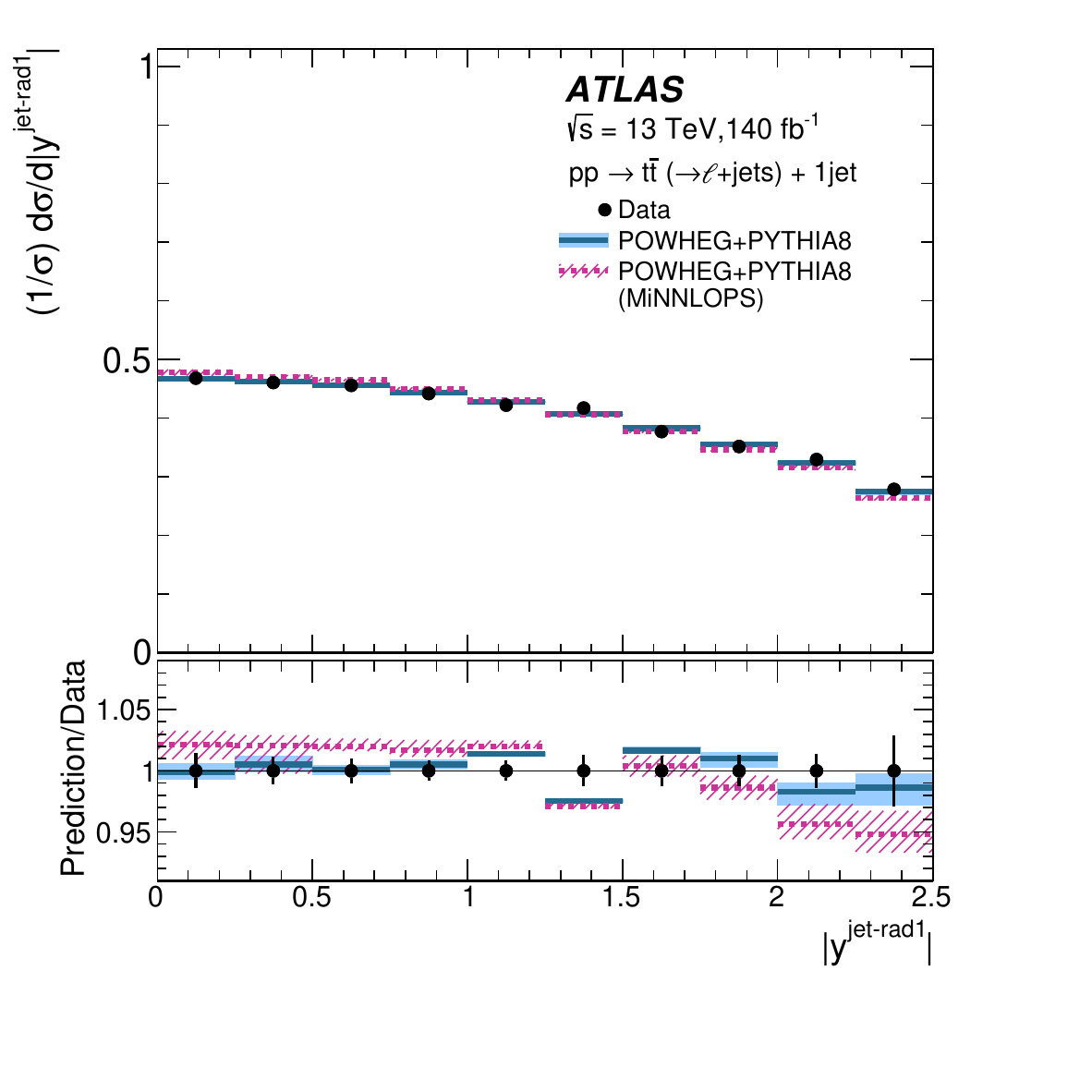}}
\put (0.0,6.0){\includegraphics[width=8cm]{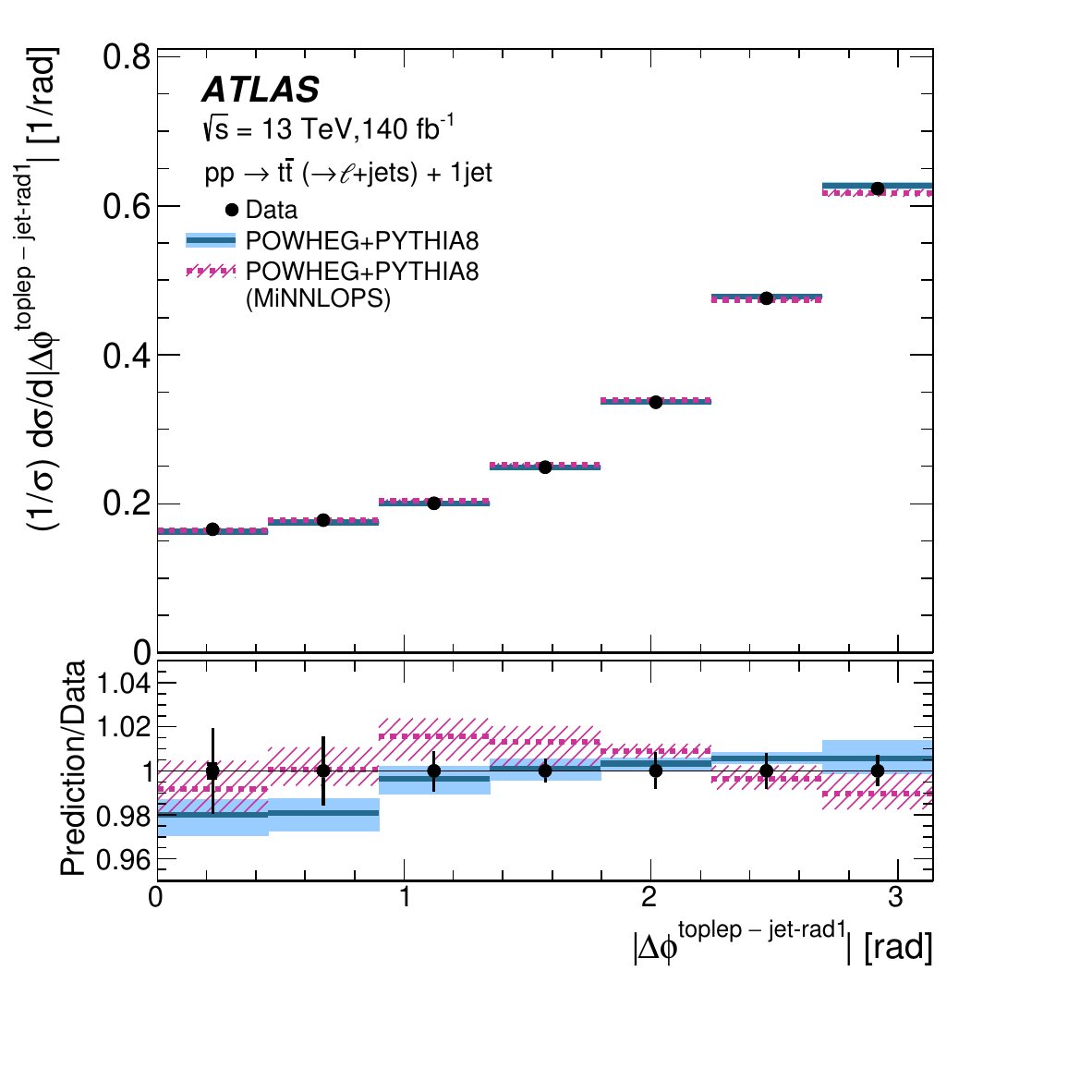}}
\put (8.0,6.0){\includegraphics[width=8cm]{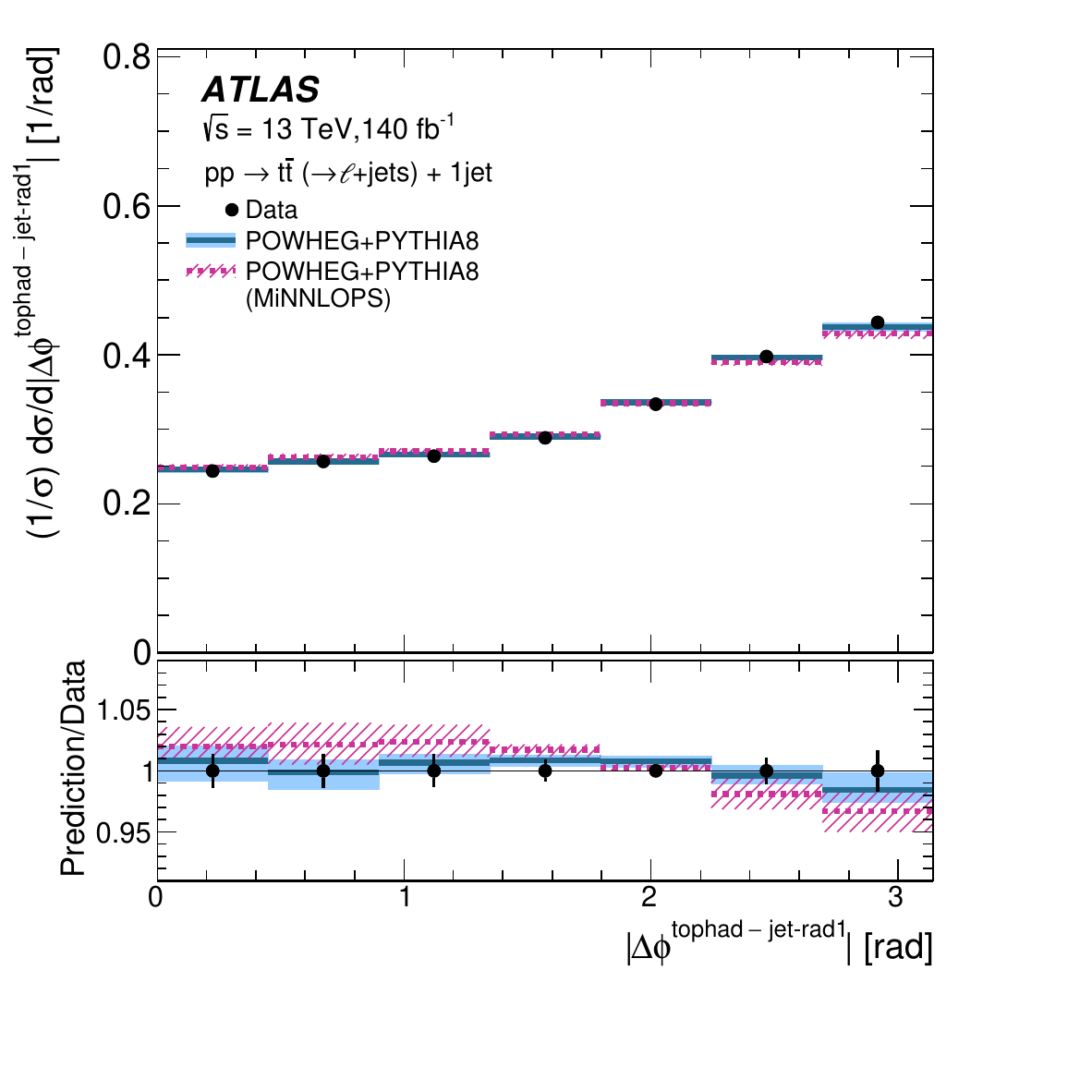}}
\put (0.0,-1.0){\includegraphics[width=8cm]{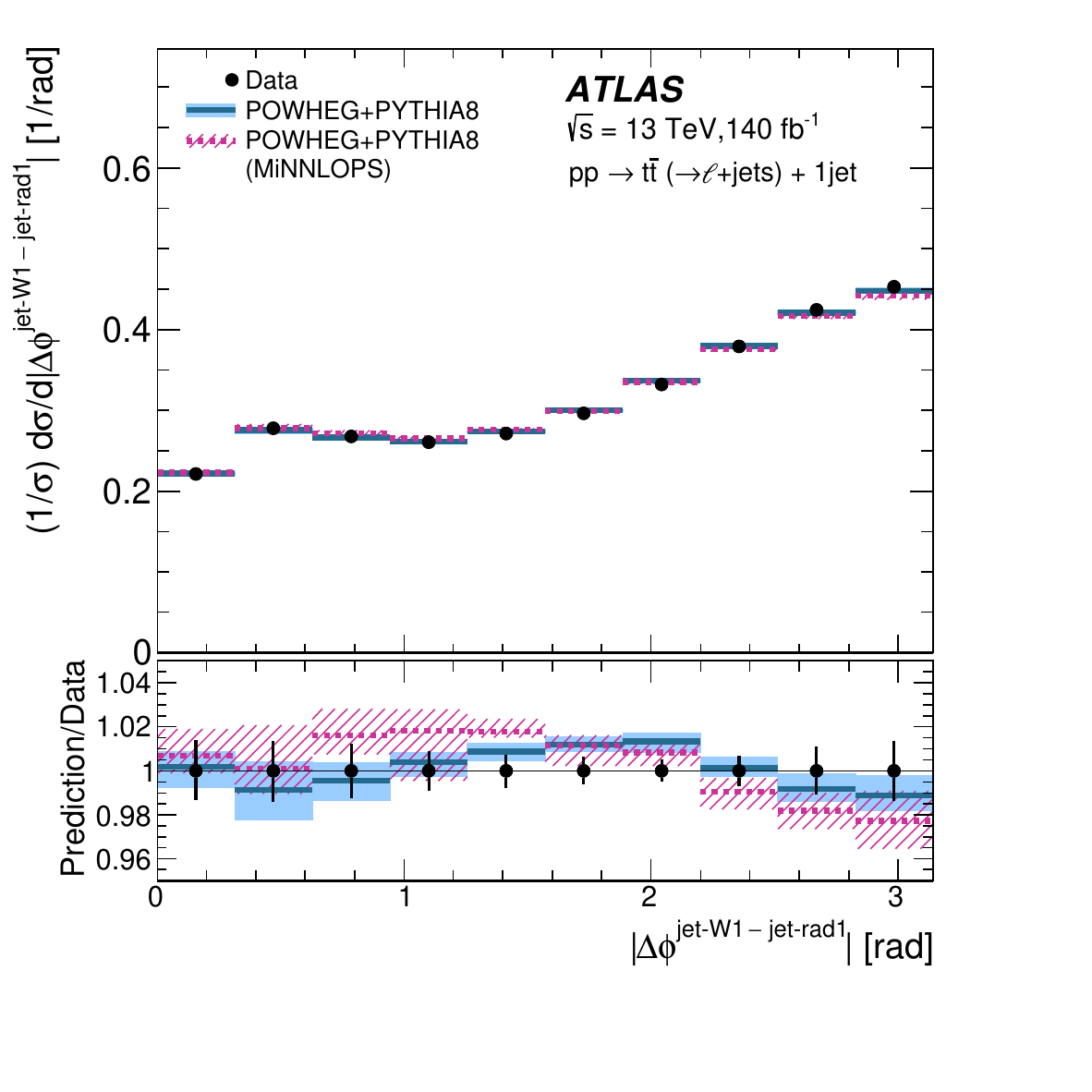}}
\put (8.0,-1.0){\includegraphics[width=8cm]{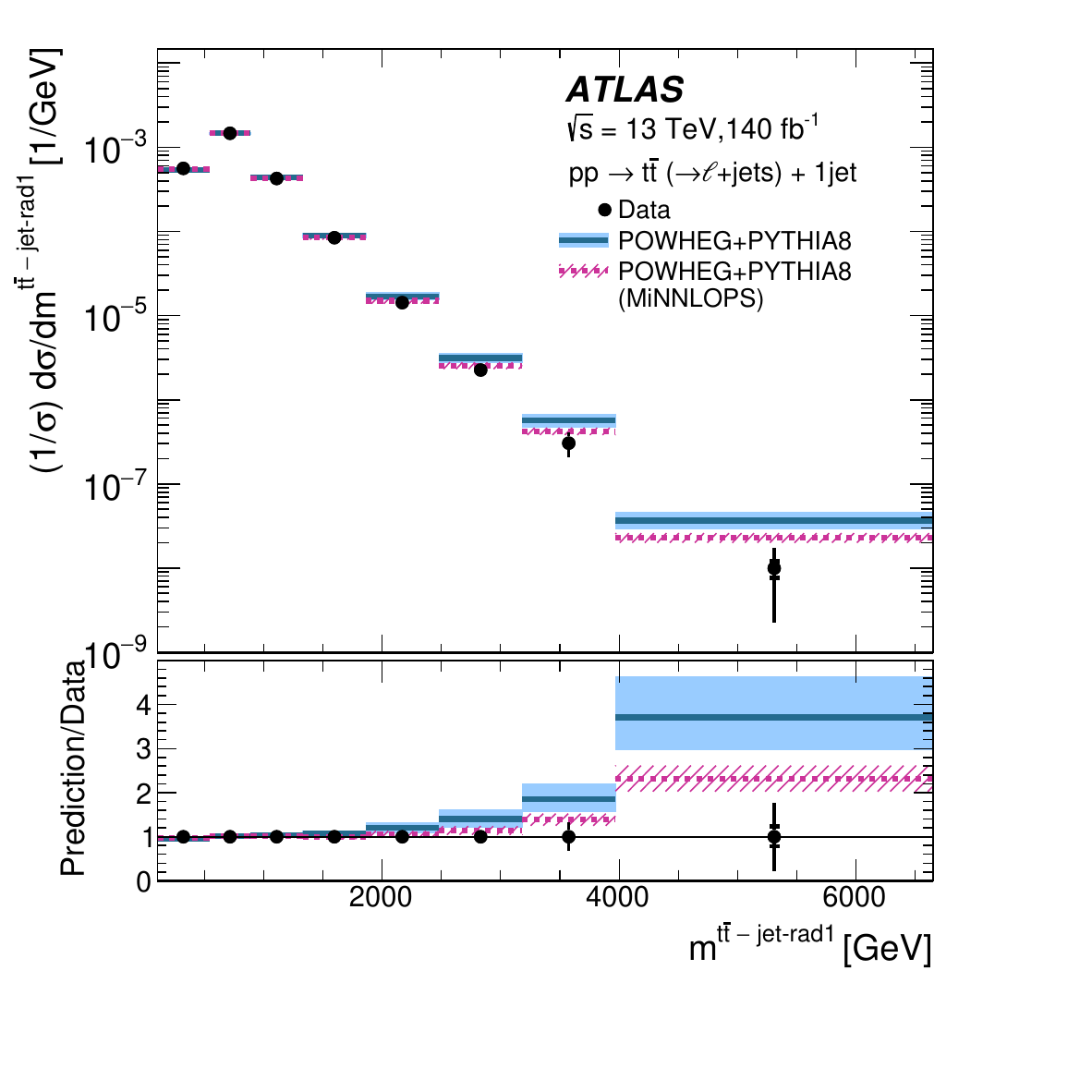}}
\put (3.5,13.9){\small (a)}
\put (11.5,13.9){\small (b)}
\put (3.5,6.9){\small (c)}
\put (11.5,6.9){\small (d)}
\put (3.5,-0.1){\small (e)}
\put (11.5,-0.1){\small (f)}
\end{picture}
\caption{
The measured normalised differential cross-sections for $\tt$ production in
the \ljets\ decay mode as functions of
(a) $\Ptjete$, (b) $|\Rapjete|$,
(c) $|\deltaphitle|$, (d) $|\deltaphithe|$,
(e) $|\deltaphiae|$ and (f) $\mtte$ (dots)
in the $\tt+$1jet channel.
The NLO QCD predictions from \pwp\ (solid lines) and the NNLO QCD
predictions from \pwpm\ (dashed lines) are also shown.
The inner (outer) error bars represent the statistical uncertainties
(the statistical and systematic uncertainties added in
quadrature). For most of the points, the inner error bars are smaller than
the marker size and, thus, not visible.
The shaded (hatched) bands represent
the theoretical uncertainty in the \pwp\ (\pwpm) prediction. The lower panels of
the figures show the ratios of the predicted and the measured
cross-sections.
}
\label{fig_appm6}
\end{figure}

A significant improvement by the NNLO predictions of \pwpm\ is
observed for the description of the measured differential
cross-sections as functions of $\Ptjete$ and $\mtte$ relative to
the predictions of \pwp, as seen in Figures~\ref{figres6} and \ref{fig_appm6}.
 
\FloatBarrier
 
\newpage
 
\FloatBarrier
 
\newpage
 
\FloatBarrier
 
\newpage
 
\FloatBarrier
 
\newpage
 
\FloatBarrier
 
\subsection{Differential cross-sections for the $\tt+$2jets channel}
\label{xsectwo}
\begin{figure}[h]
\setlength{\unitlength}{1.0cm}
\begin{picture} (18.0,17.5)
\put (0.0,6.3){\includegraphics[width=8cm]{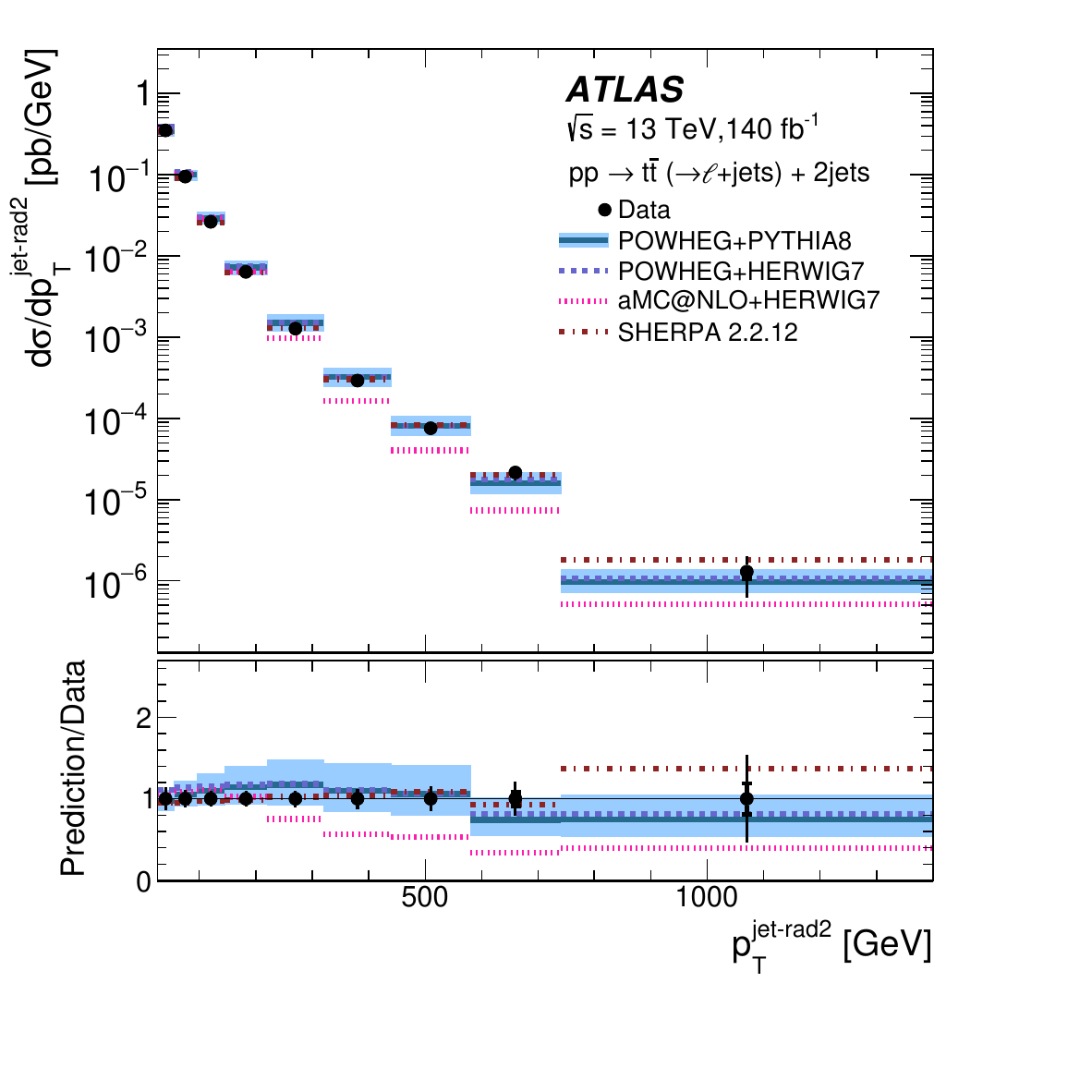}}
\put (8.0,6.3){\includegraphics[width=8cm]{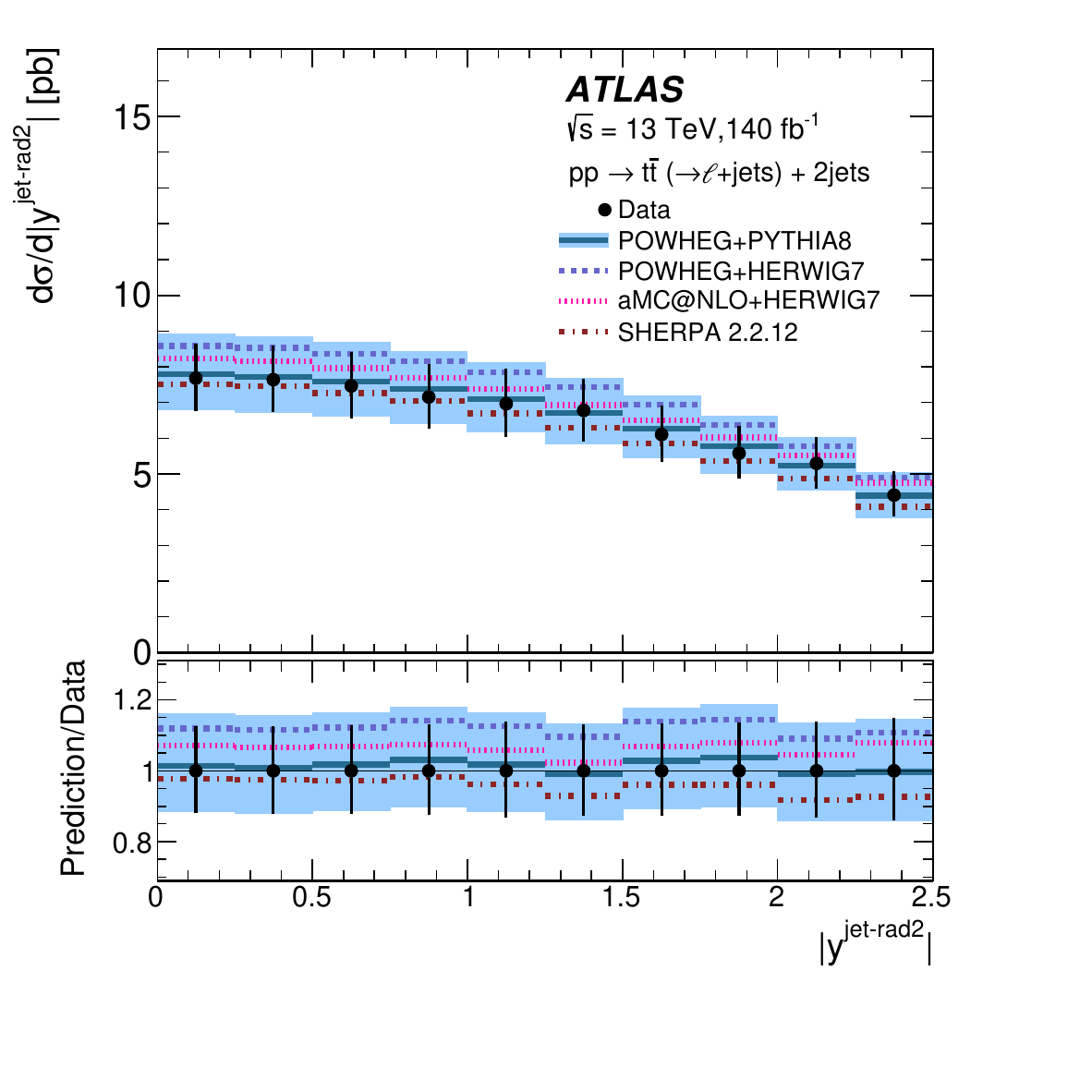}}
\put (0.0,-0.7){\includegraphics[width=8cm]{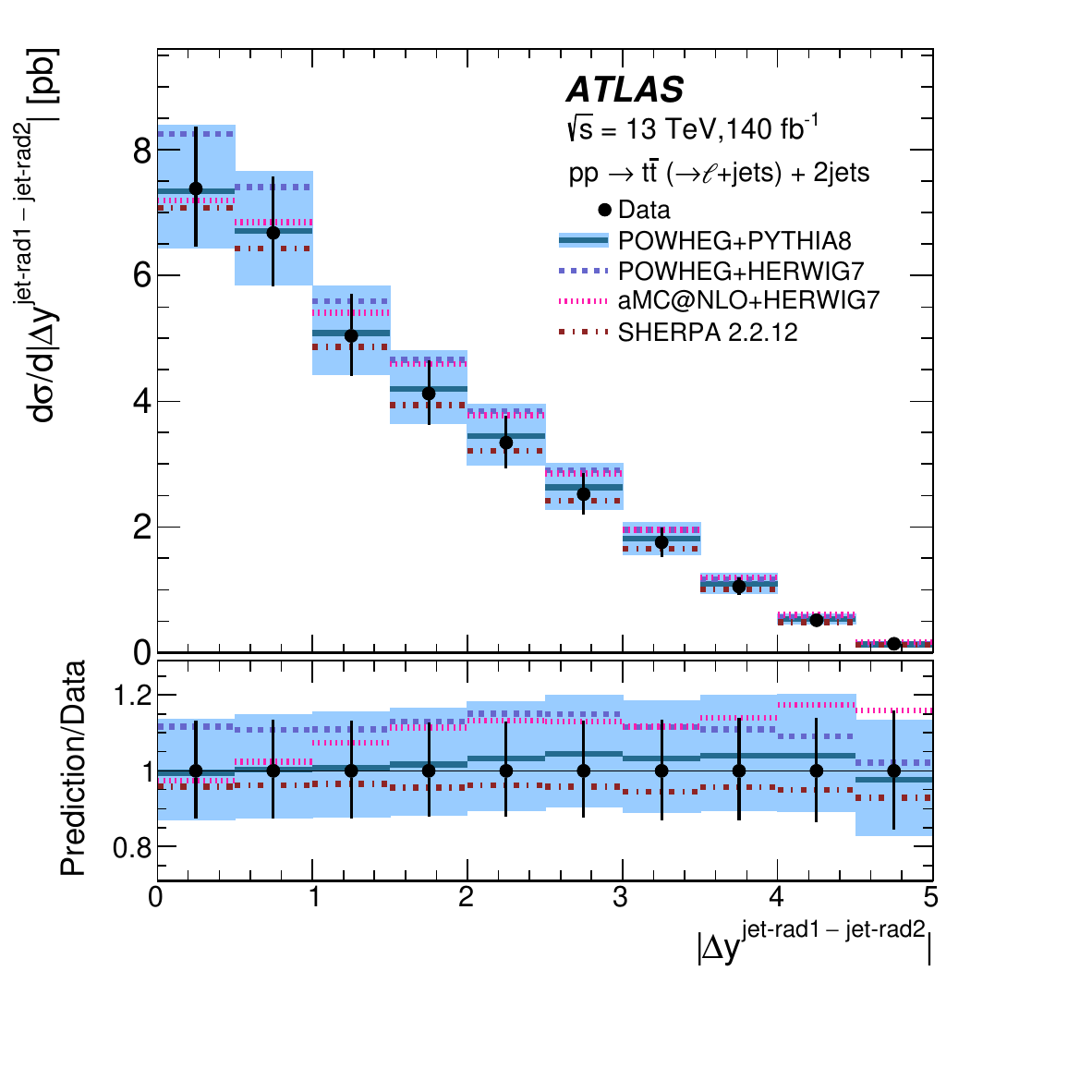}}
\put (8.0,-0.7){\includegraphics[width=8cm]{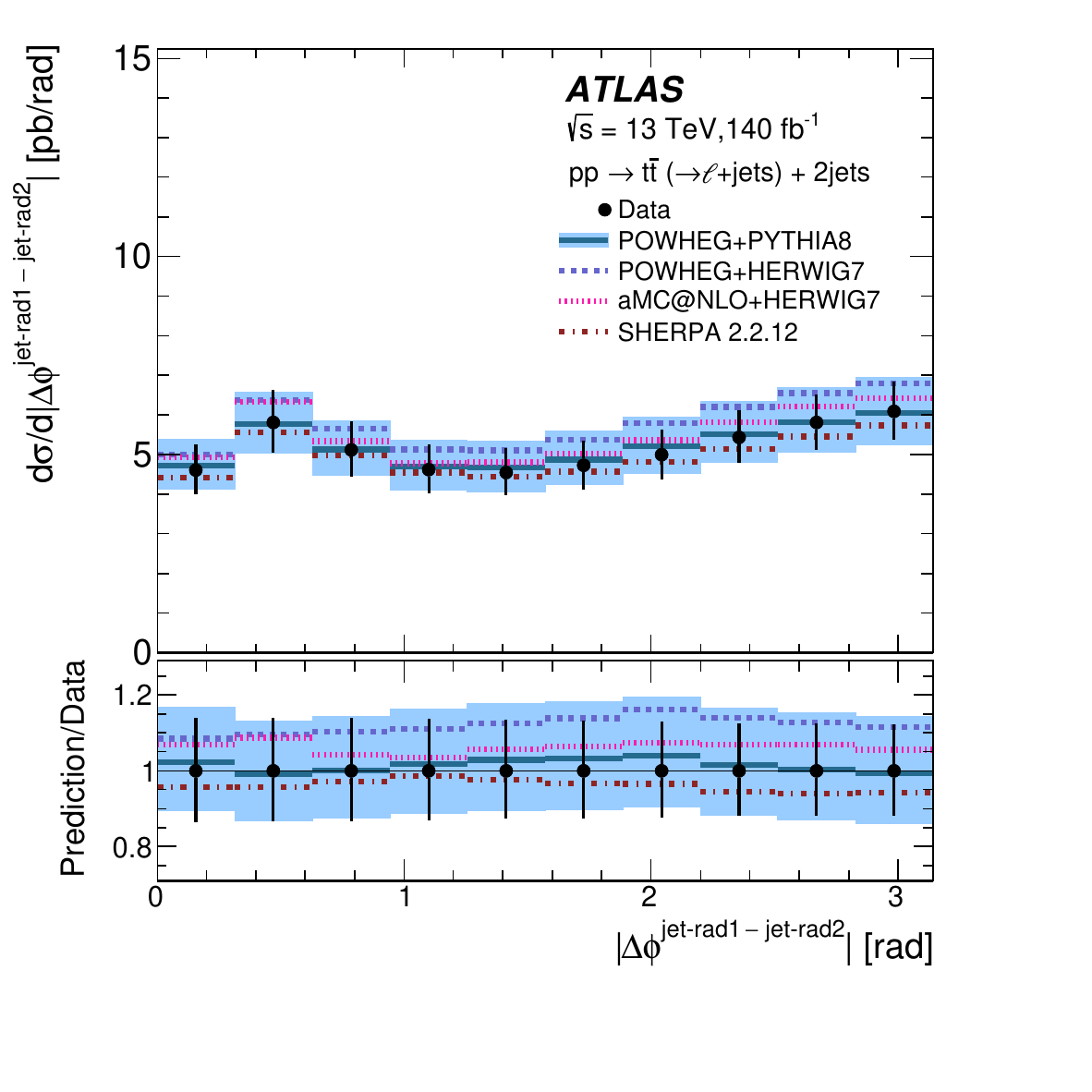}}
\put (3.5,7.2){\small (a)}
\put (11.5,7.2){\small (b)}
\put (3.5,0.2){\small (c)}
\put (11.5,0.2){\small (d)}
\end{picture}
\caption{
The measured differential cross-sections for $\tt$ production in
the \ljets\ decay mode as functions of
(a) $\Ptjetf$, (b) $|\Rapjetf|$,
(c) $|\deltayef|$ and (d) $|\deltaphief|$ (dots)
in the $\tt+$2jets channel.
The NLO QCD predictions from \pwp\ (solid lines), \pwh\ (dashed lines),
\amcnloh\ (dotted lines) and \sher~2.2.12 (dot-dashed lines)
normalised to ${\cal O}$(NNLO+NNLL) are also shown.
The inner (outer) error bars represent the statistical uncertainties
(the statistical and systematic uncertainties added in
quadrature). For most of the points, the inner error bars are smaller than
the marker size and, thus, not visible.
The shaded bands represent
the theoretical uncertainty in the \pwp\ prediction. The lower panels of
the figures show the ratios of the predicted and the measured
cross-sections.
}
\label{figres3}
\end{figure}

\begin{figure}[h]
\setlength{\unitlength}{1.0cm}
\begin{picture} (18.0,17.5)
\put (0.0,6.3){\includegraphics[width=8cm]{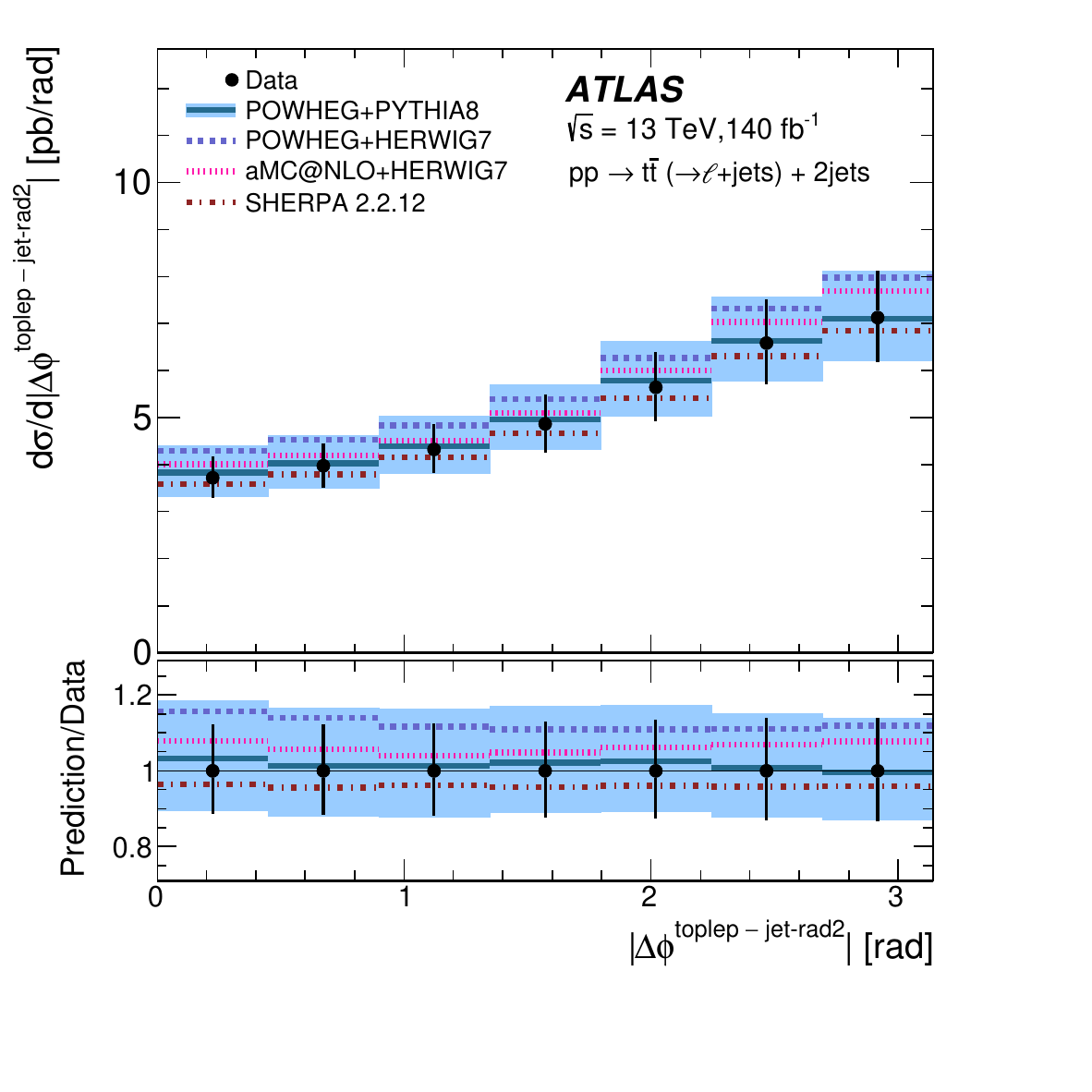}}
\put (8.0,6.3){\includegraphics[width=8cm]{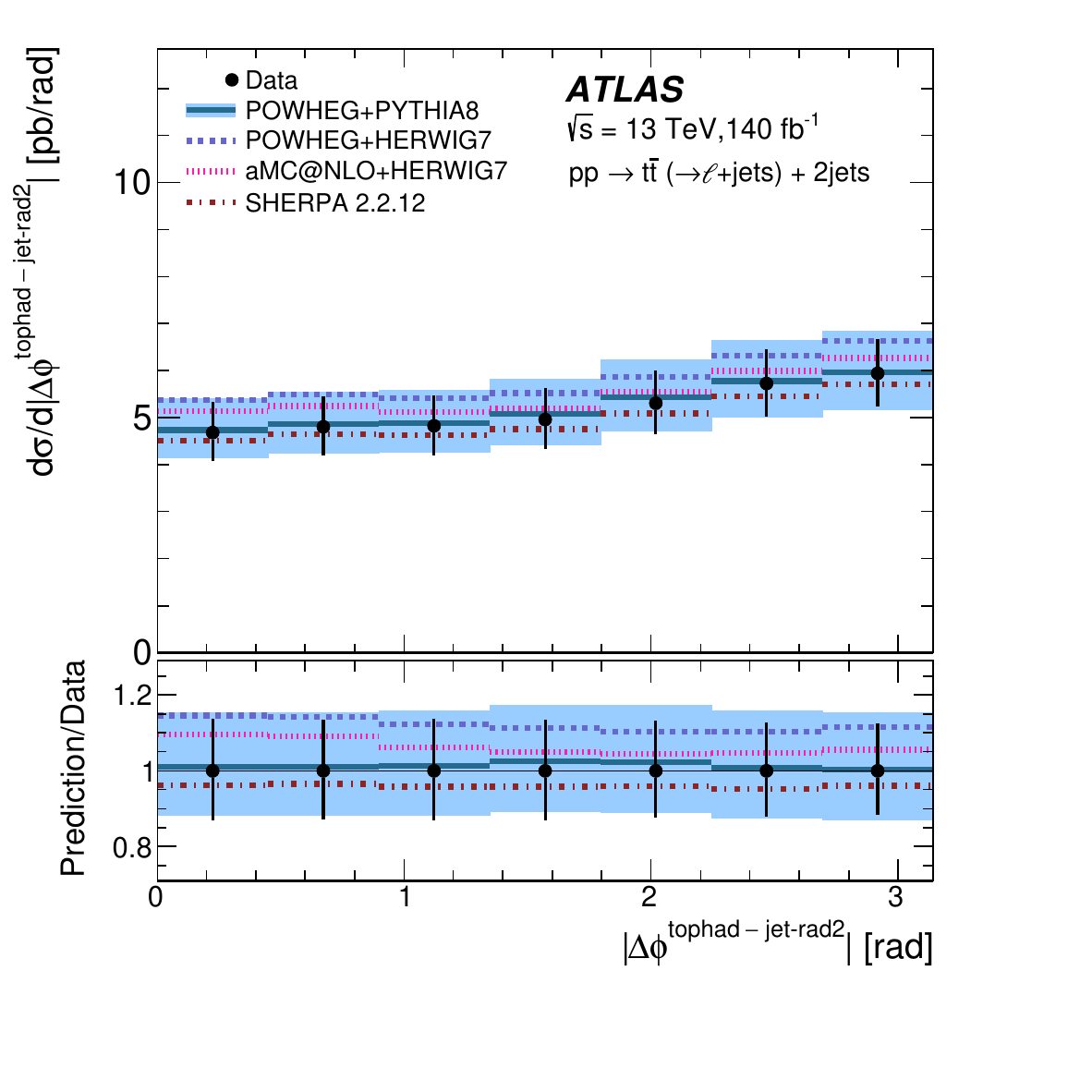}}
\put (0.0,-0.7){\includegraphics[width=8cm]{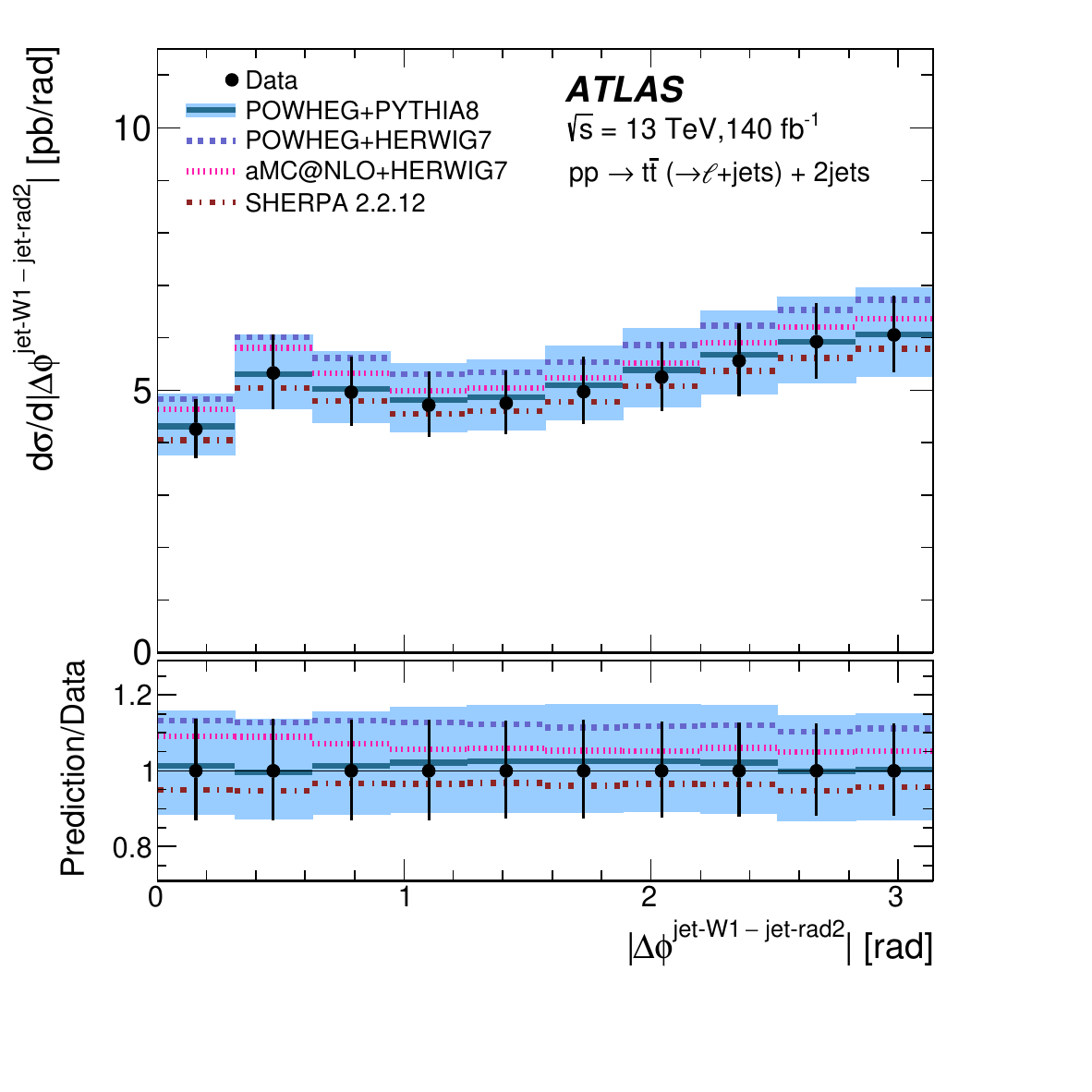}}
\put (8.0,-0.7){\includegraphics[width=8cm]{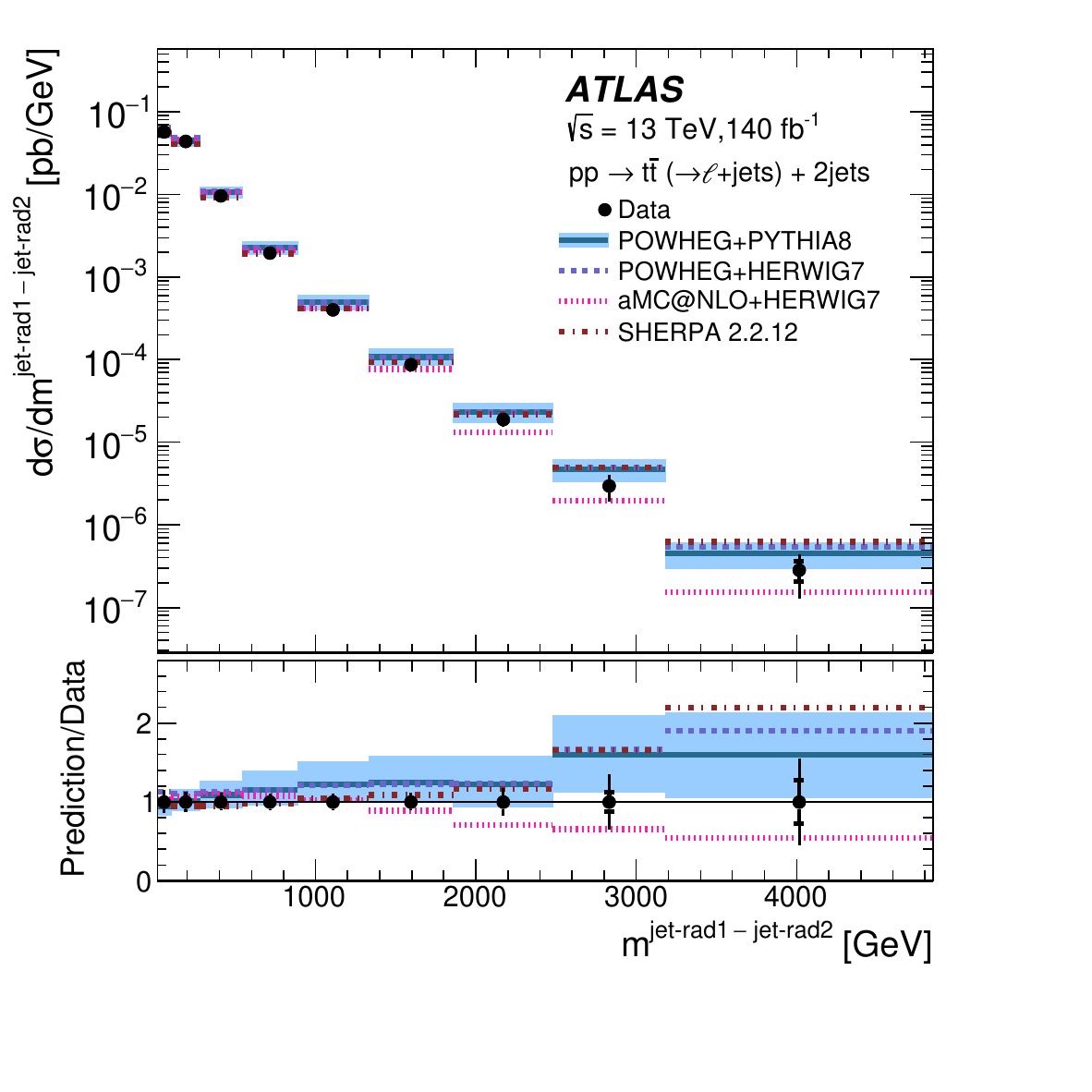}}
\put (3.5,7.2){\small (a)}
\put (11.5,7.2){\small (b)}
\put (3.5,0.2){\small (c)}
\put (11.5,0.2){\small (d)}
\end{picture}
\caption{
The measured differential cross-sections for $\tt$ production in
the \ljets\ decay mode as functions of
(a)~$|\deltaphitlf|$, (b) $|\deltaphithf|$,
(c) $|\deltaphiaf|$ and (d) $\mef$ (dots)
in the $\tt+$2jets channel.
The NLO QCD predictions from \pwp\ (solid lines), \pwh\ (dashed lines),
\amcnloh\ (dotted lines) and \sher~2.2.12 (dot-dashed lines)
normalised to ${\cal O}$(NNLO+NNLL) are also shown.
The inner (outer) error bars represent the statistical uncertainties
(the statistical and systematic uncertainties added in
quadrature). For most of the points, the inner error bars are smaller than
the marker size and, thus, not visible.
The shaded bands represent
the theoretical uncertainty in the \pwp\ prediction. The lower panels of
the figures show the ratios of the predicted and the measured
cross-sections.
}
\label{figres4}
\end{figure}

\begin{figure}[h]
\setlength{\unitlength}{1.0cm}
\begin{picture} (18.0,17.5)
\put (0.0,6.3){\includegraphics[width=8cm]{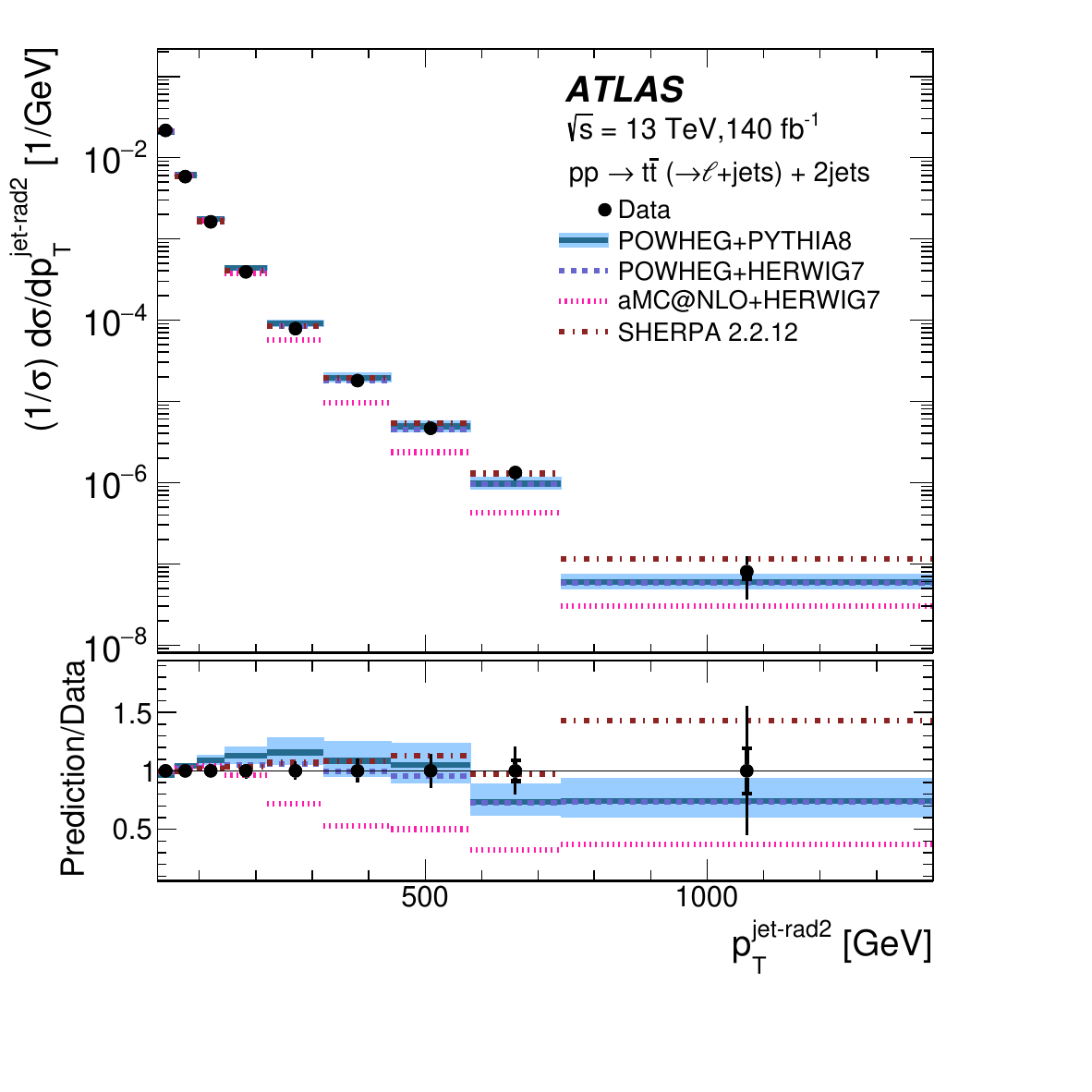}}
\put (8.0,6.3){\includegraphics[width=8cm]{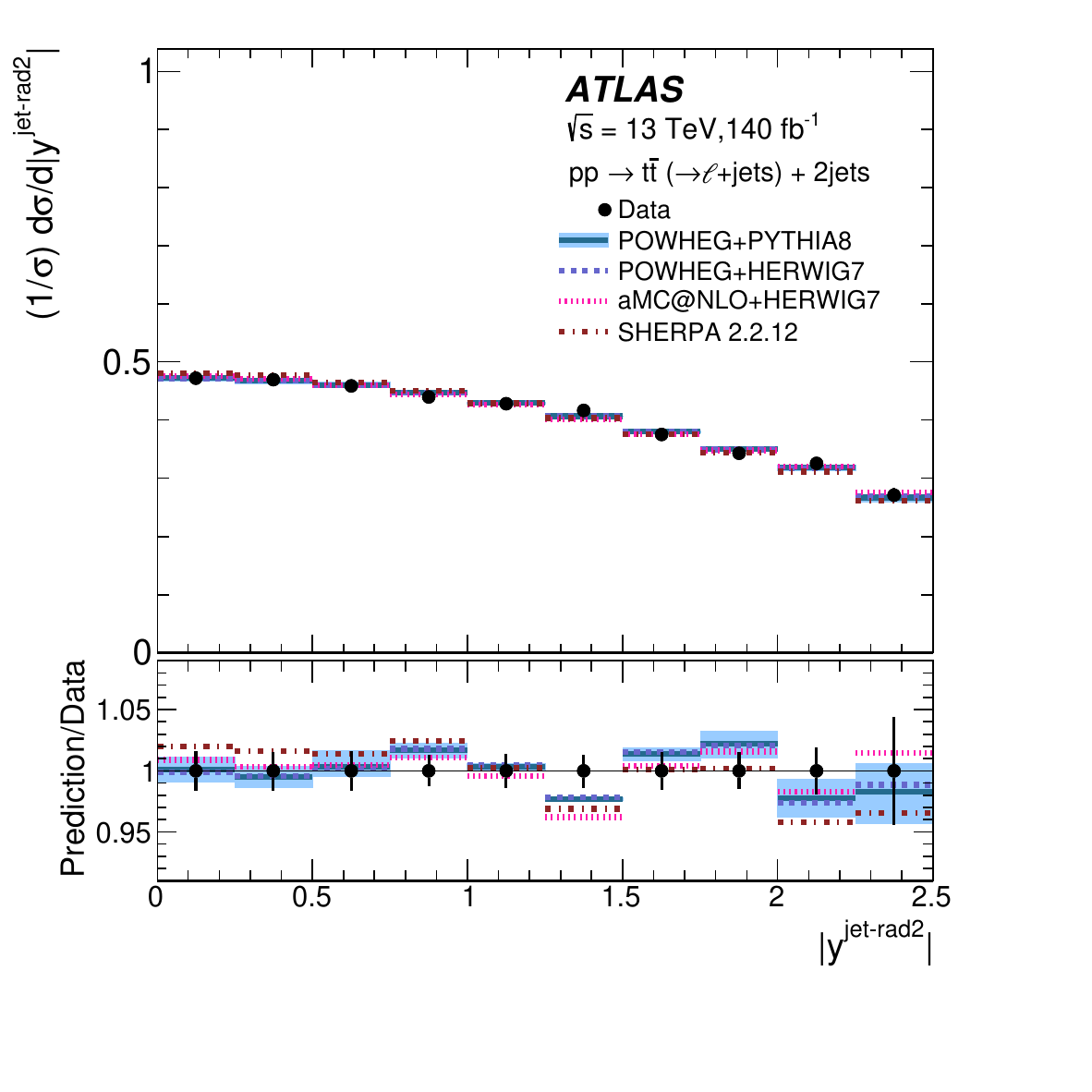}}
\put (0.0,-0.7){\includegraphics[width=8cm]{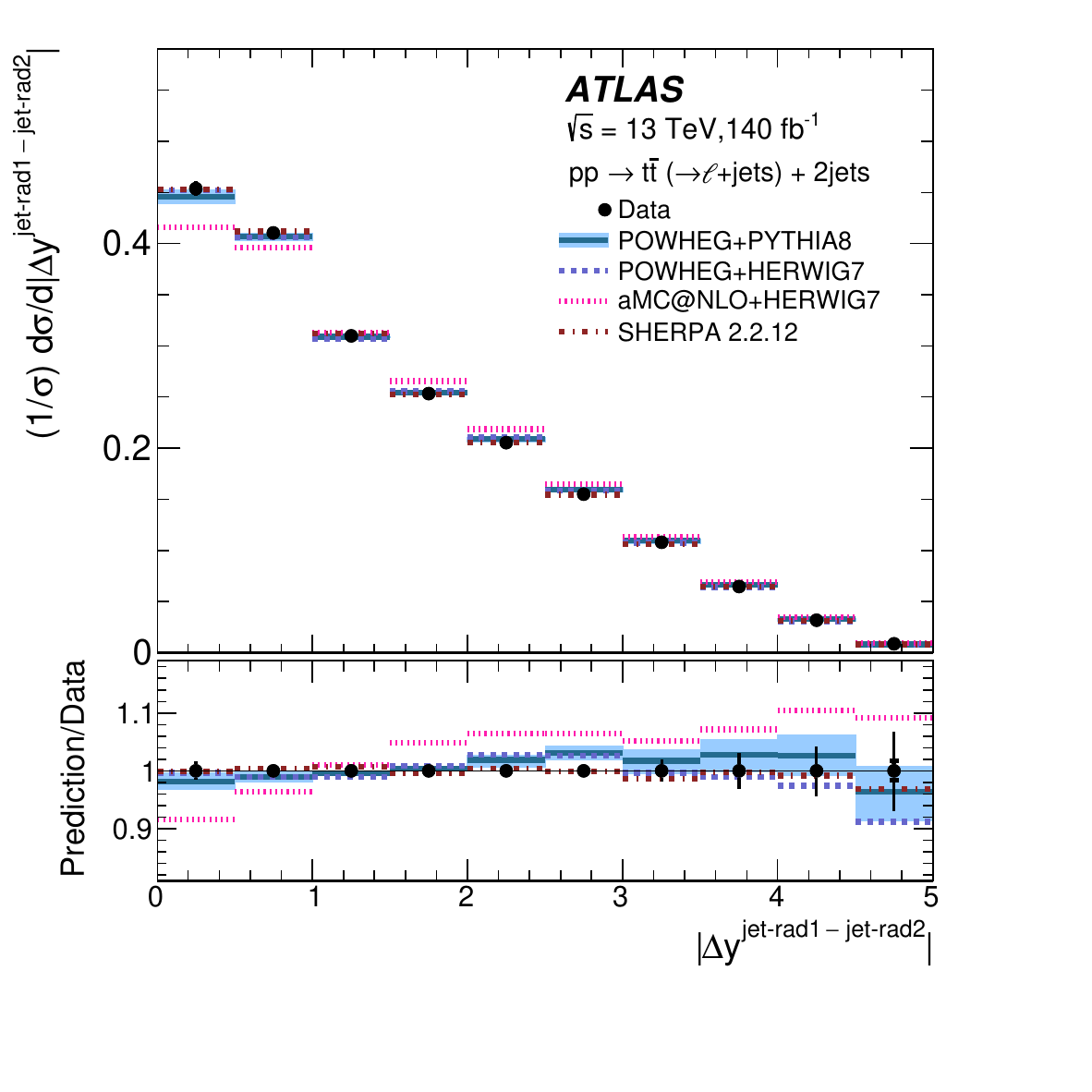}}
\put (8.0,-0.7){\includegraphics[width=8cm]{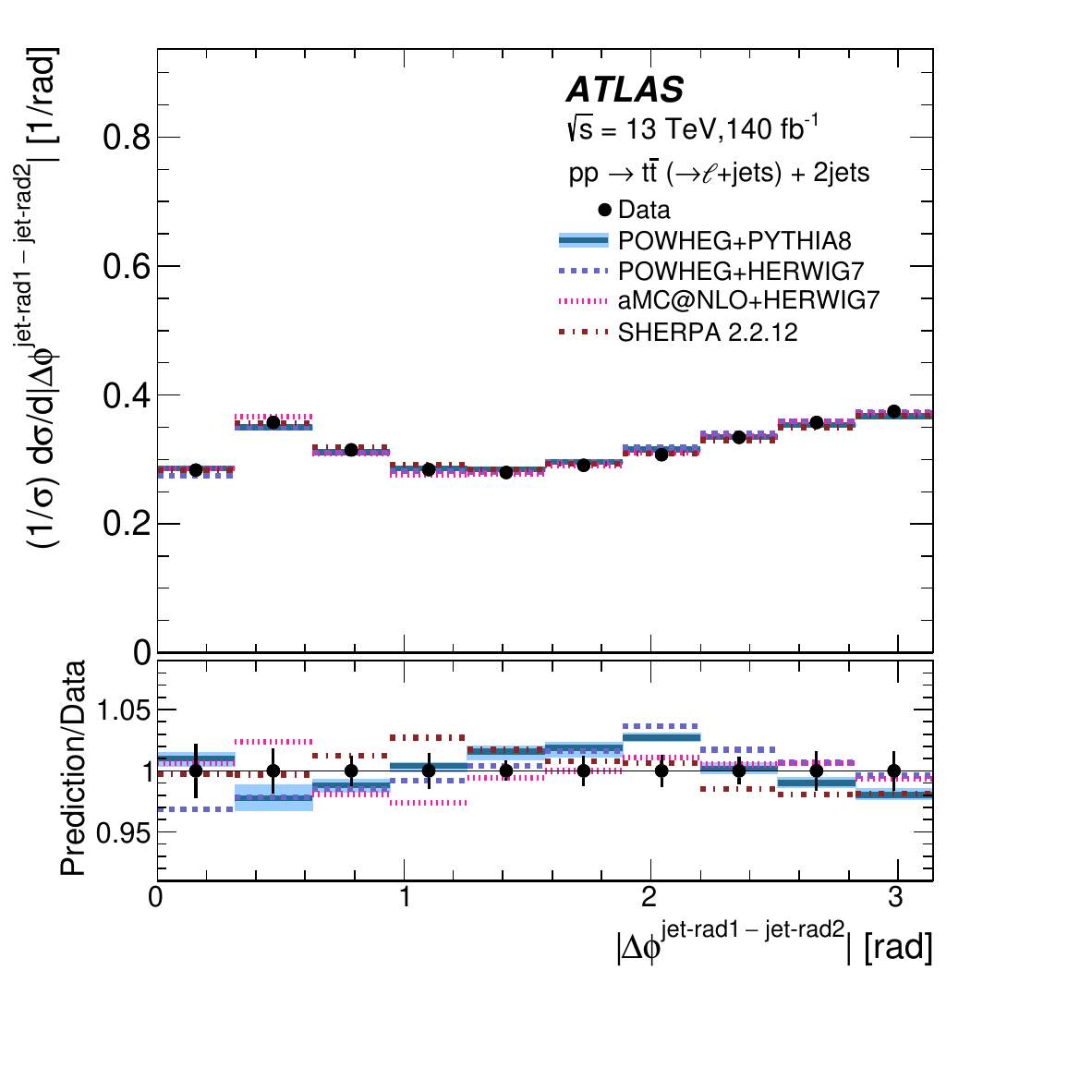}}
\put (3.5,7.2){\small (a)}
\put (11.5,7.2){\small (b)}
\put (3.5,0.2){\small (c)}
\put (11.5,0.2){\small (d)}
\end{picture}
\caption{
The measured normalised differential cross-sections for $\tt$ production in
the \ljets\ decay mode as functions of
(a) $\Ptjetf$, (b) $|\Rapjetf|$,
(c) $|\deltayef|$ and (d) $|\deltaphief|$ (dots)
in the $\tt+$2jets channel.
The NLO QCD predictions from \pwp\ (solid lines), \pwh\ (dashed lines),
\amcnloh\ (dotted lines) and \sher~2.2.12 (dot-dashed lines)
are also shown.
The inner (outer) error bars represent the statistical uncertainties
(the statistical and systematic uncertainties added in
quadrature). For most of the points, the inner error bars are smaller than
the marker size and, thus, not visible.
The shaded bands represent
the theoretical uncertainty in the \pwp\ prediction. The lower panels of
the figures show the ratios of the predicted and the measured
cross-sections.
}
\label{fig_appm3}
\end{figure}

\begin{figure}[h]
\setlength{\unitlength}{1.0cm}
\begin{picture} (18.0,17.5)
\put (0.0,6.3){\includegraphics[width=8cm]{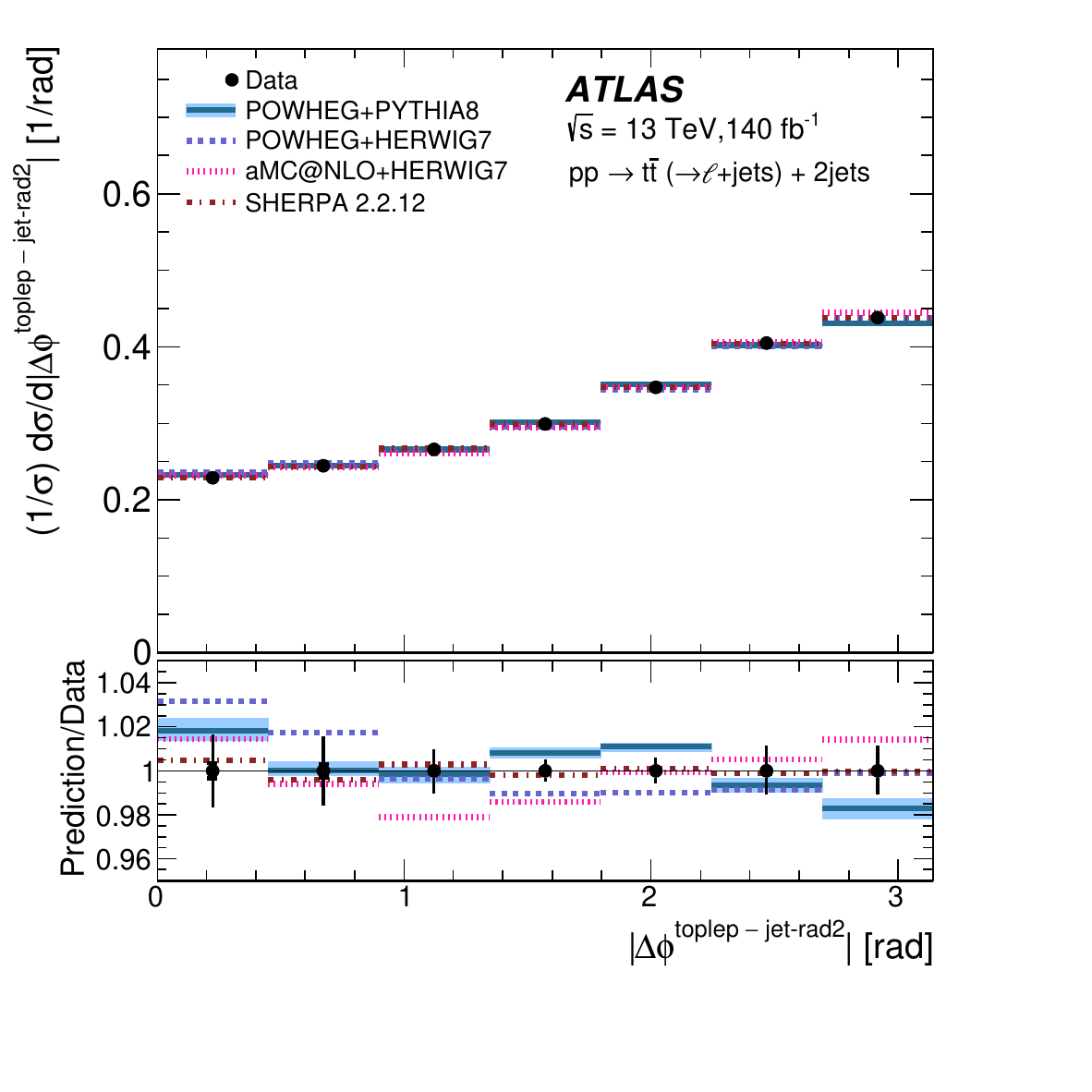}}
\put (8.0,6.3){\includegraphics[width=8cm]{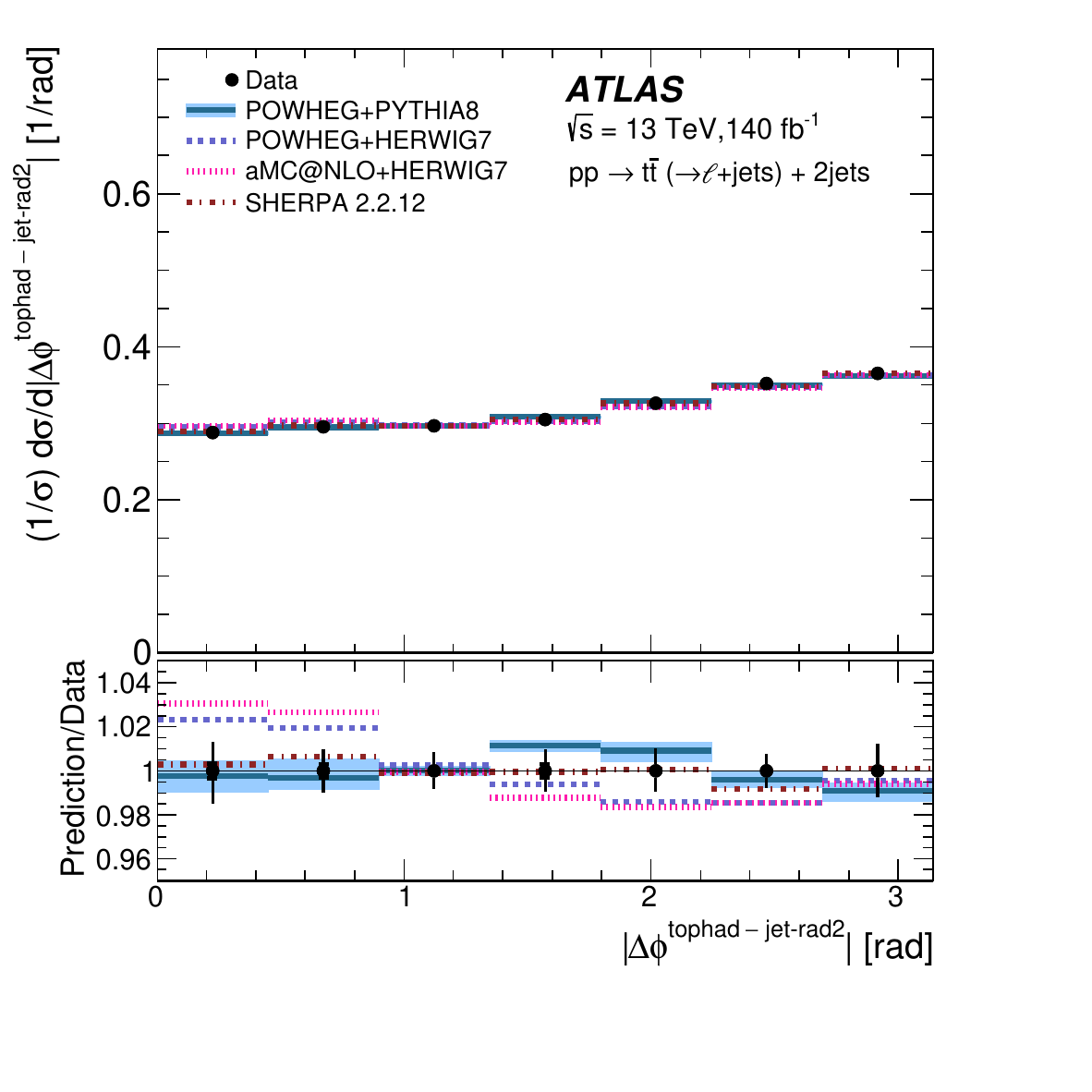}}
\put (0.0,-0.7){\includegraphics[width=8cm]{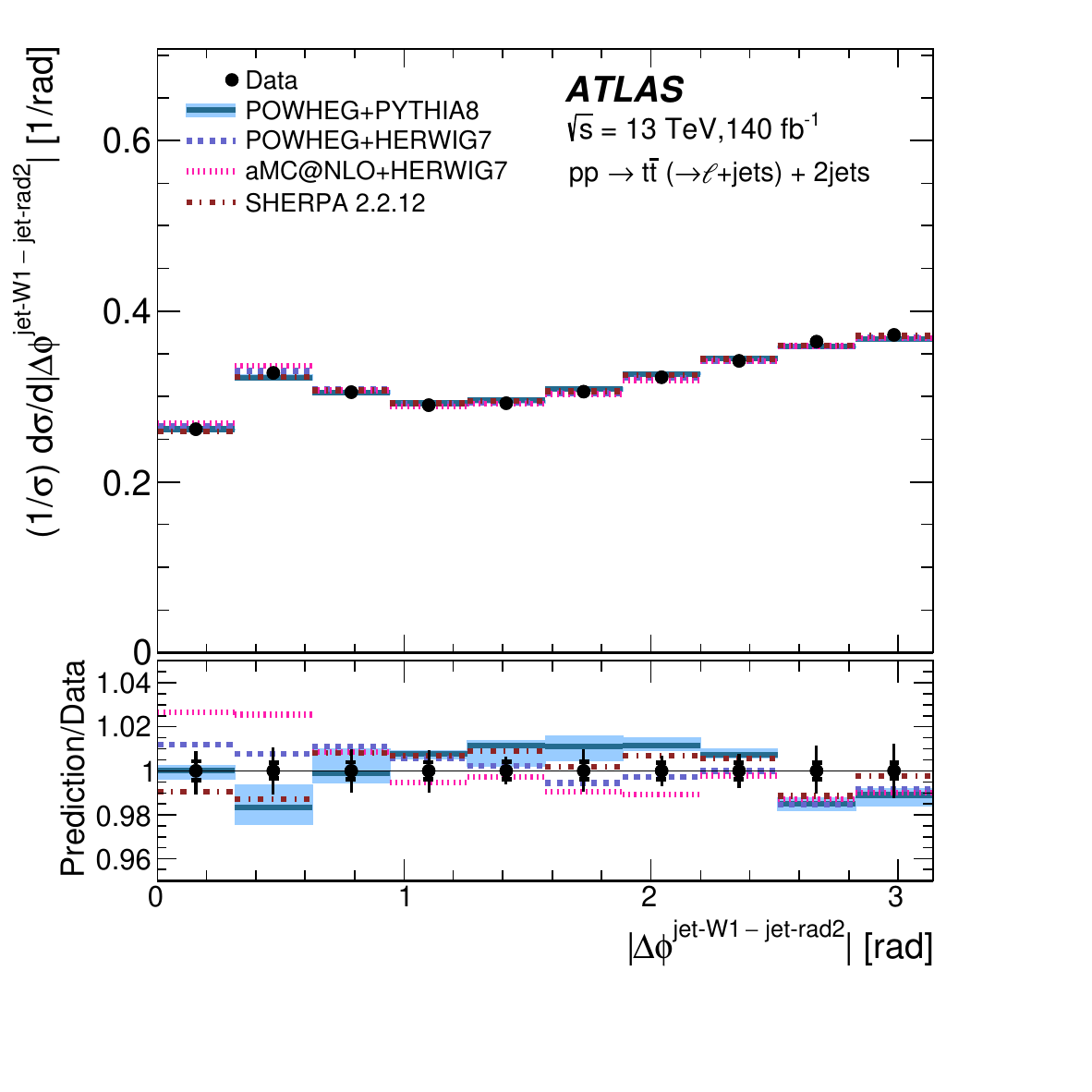}}
\put (8.0,-0.7){\includegraphics[width=8cm]{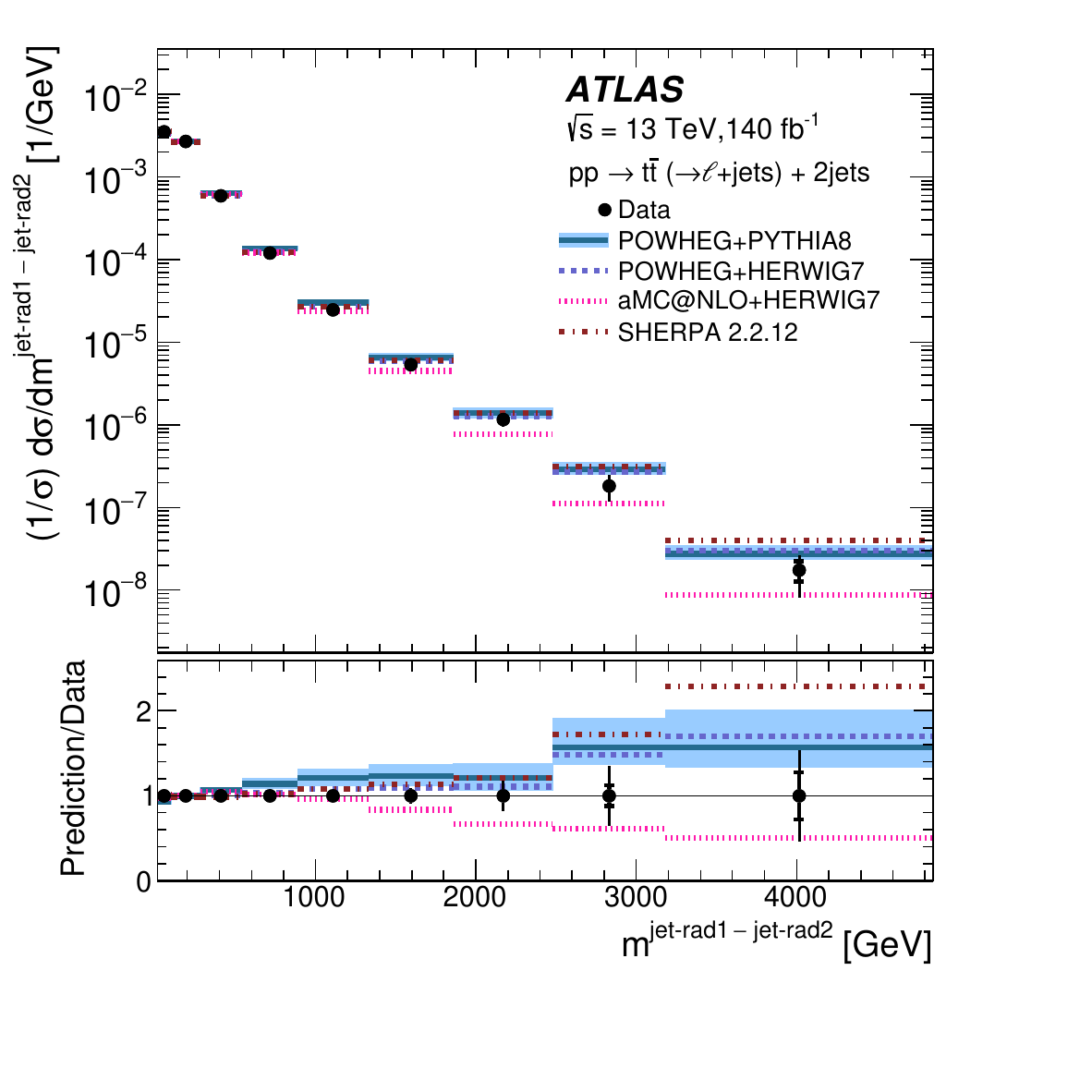}}
\put (3.5,7.2){\small (a)}
\put (11.5,7.2){\small (b)}
\put (3.5,0.2){\small (c)}
\put (11.5,0.2){\small (d)}
\end{picture}
\caption{
The measured normalised differential cross-sections for $\tt$ production in
the \ljets\ decay mode as functions of
(a) $|\deltaphitlf|$, (b) $|\deltaphithf|$,
(c) $|\deltaphiaf|$ and (d) $\mef$ (dots)
in the $\tt+$2jets channel.
The NLO QCD predictions from \pwp\ (solid lines), \pwh\ (dashed lines),
\amcnloh\ (dotted lines) and \sher~2.2.12 (dot-dashed lines)
are also shown.
The inner (outer) error bars represent the statistical uncertainties
(the statistical and systematic uncertainties added in
quadrature). For most of the points, the inner error bars are smaller than
the marker size and, thus, not visible.
The shaded bands represent
the theoretical uncertainty in the \pwp\ prediction. The lower panels of
the figures show the ratios of the predicted and the measured
cross-sections.
}
\label{fig_appm4}
\end{figure}

Figures~\ref{figres3} and \ref{figres4} (Figures~\ref{fig_appm3} and
\ref{fig_appm4})  show the absolute (normalised) differential $\tt$
cross-sections in the \ljets\ decay mode for the $\tt+$2jets channel
as functions of  $\Ptjetf$, $|\Rapjetf|$, $|\deltayef|$,
$|\deltaphief|$, $|\deltaphitlf|$, $|\deltaphithf|$, $|\deltaphiaf|$ and $\mef$. Values
of $\Ptjetf$ and $\mef$ up to $\sim 1.2$~\TeV\ and $\sim 4$~\TeV,
respectively, are measured. The measured cross-sections as functions
of $\Ptjetf$ and $\mef$ decrease by more than five orders of magnitude
within the measured range. The measured cross-section as a function of
$|\Rapjetf|$ decreases as the rapidity increases and has a very
similar shape to the one for $|\Rapjete|$. The measured cross-section
as a function of $|\deltayef|$ decreases rapidly with increasing
$|\deltayef|$ and the $|\deltaphief|$ distribution exhibits peaks at
$0.5$ rad and $\pi$ rad, with approximately the same rate; these
features indicate that the second gluon emission tends to be close in
rapidity to the first gluon emission and that there is a weak dependence on
the difference in $\phi$ between the first and second gluon emissions. The
decrease of the cross-section at $|\deltaphief|\approx 0$ is due to
the fact that since the jets are reconstructed with $R=0.4$, they
cannot be closer than that value in $\phi$ if they have the same
rapidity. The measured cross-sections as functions of
$|\deltaphitlf|$, $|\deltaphithf|$ and $|\deltaphiaf|$ show similar
features to those measured for jet-rad1; therefore, the second gluon emission
also has a higher probability of arising at larger angles from the
leptonically decaying top quark.
 
The NLO QCD predictions describe the measurements as functions of
$|\Rapjetf|$, \break $|\deltayef|$, $|\deltaphief|$,
$|\deltaphitlf|$, $|\deltaphithf|$ and $|\deltaphiaf|$ well, however the
prediction from \amcnloh\ has a different shape than the data
for the normalised $|\deltayef|$ distribution (see \ref{fig_appm3}(c)). The
measurement as a function of $\Ptjetf$ is described adequately by the
predictions of \pwp, \pwh\ and \sher~2.2.12, whereas the prediction
from \amcnloh\ disagrees with the data for $\Ptjetf>200$~\GeV. For
$\mef$, the predictions describe the data only for $\mef<1$~\TeV; for
larger $\mef$ values, the predictions of \pwp, \pwh\ and \sher~2.2.12
(\amcnloh) overestimate (underestimate) the data.
 
\begin{figure}[h]
\setlength{\unitlength}{1.0cm}
\begin{picture} (18.0,17.5)
\put (0.0,6.3){\includegraphics[width=8cm]{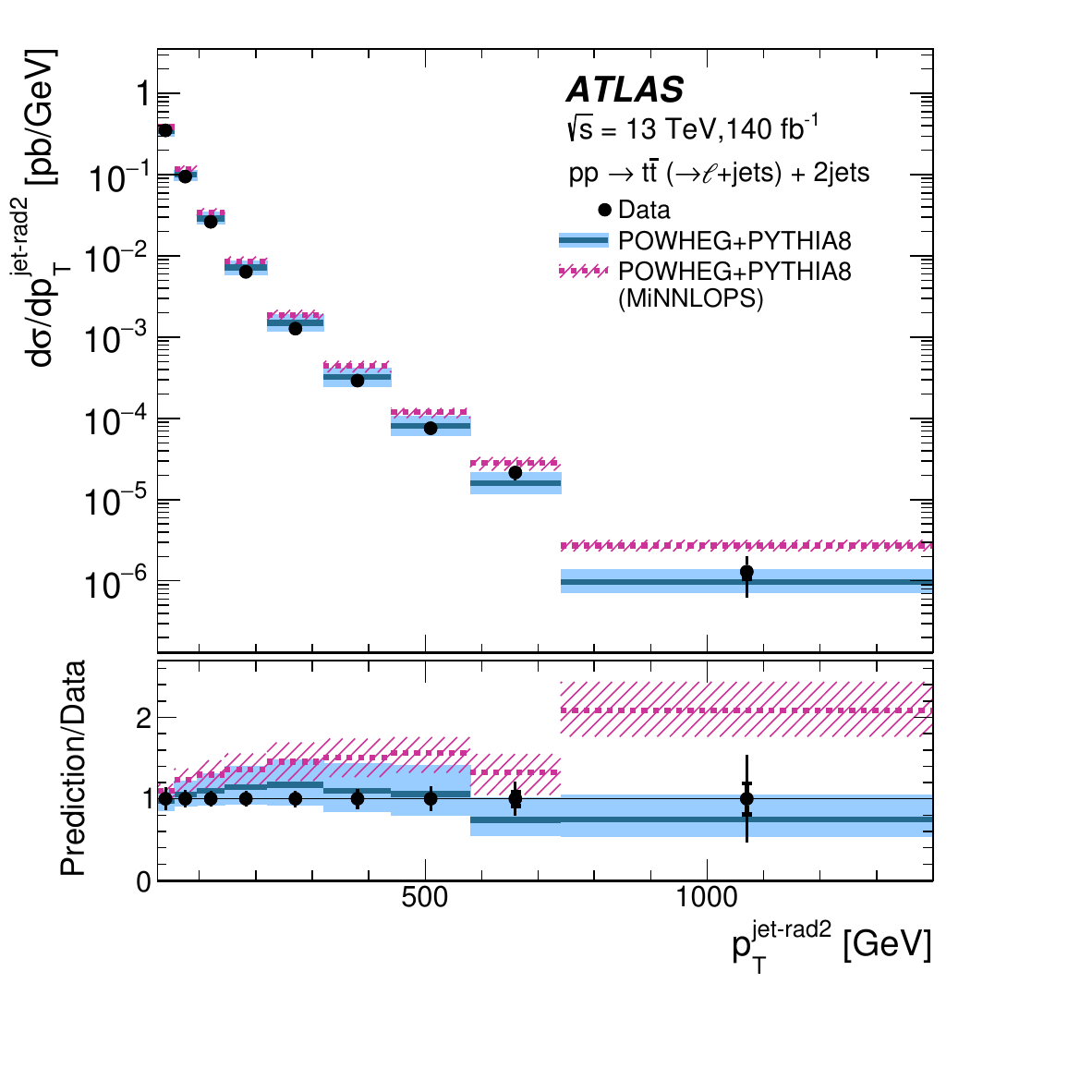}}
\put (8.0,6.3){\includegraphics[width=8cm]{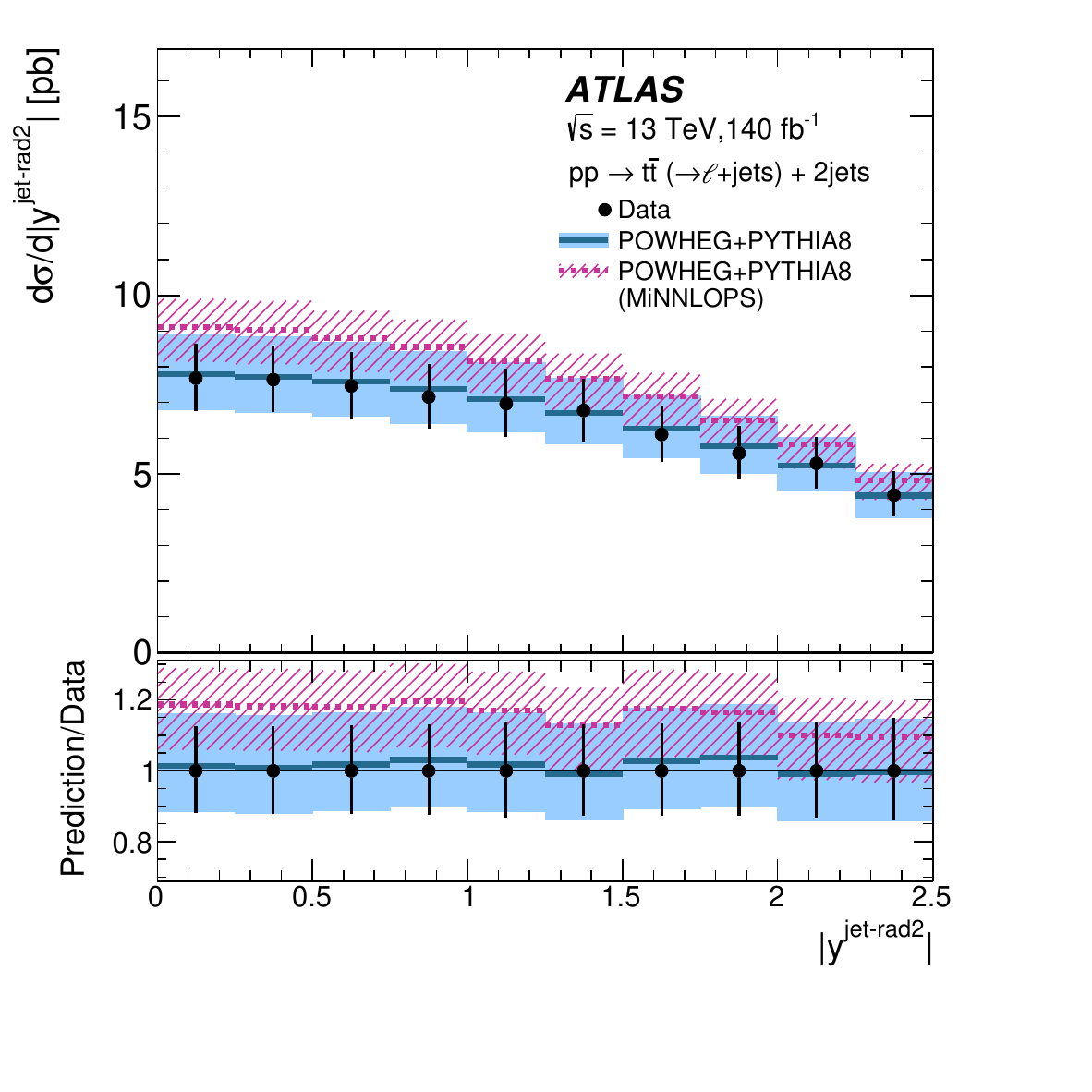}}
\put (0.0,-0.7){\includegraphics[width=8cm]{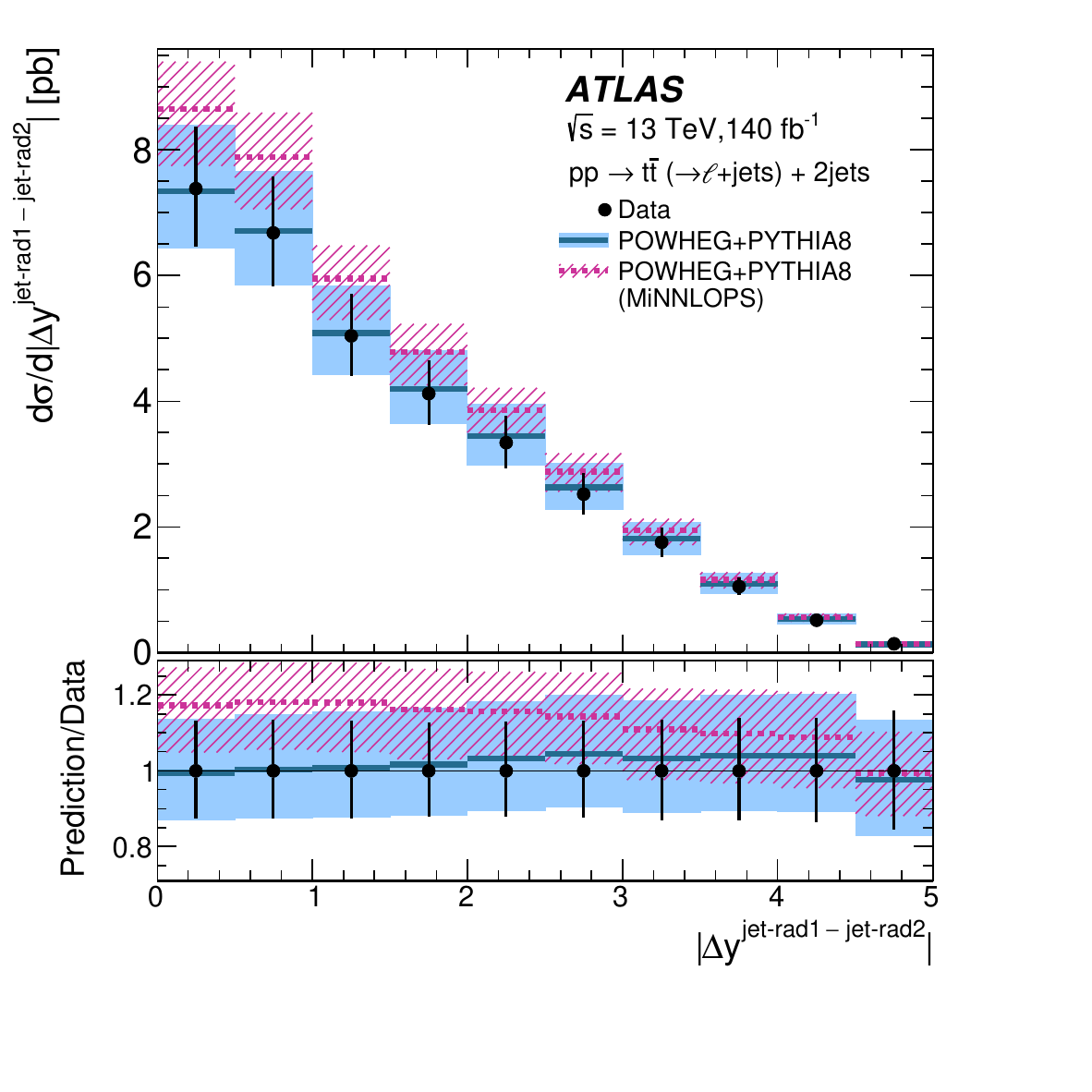}}
\put (8.0,-0.7){\includegraphics[width=8cm]{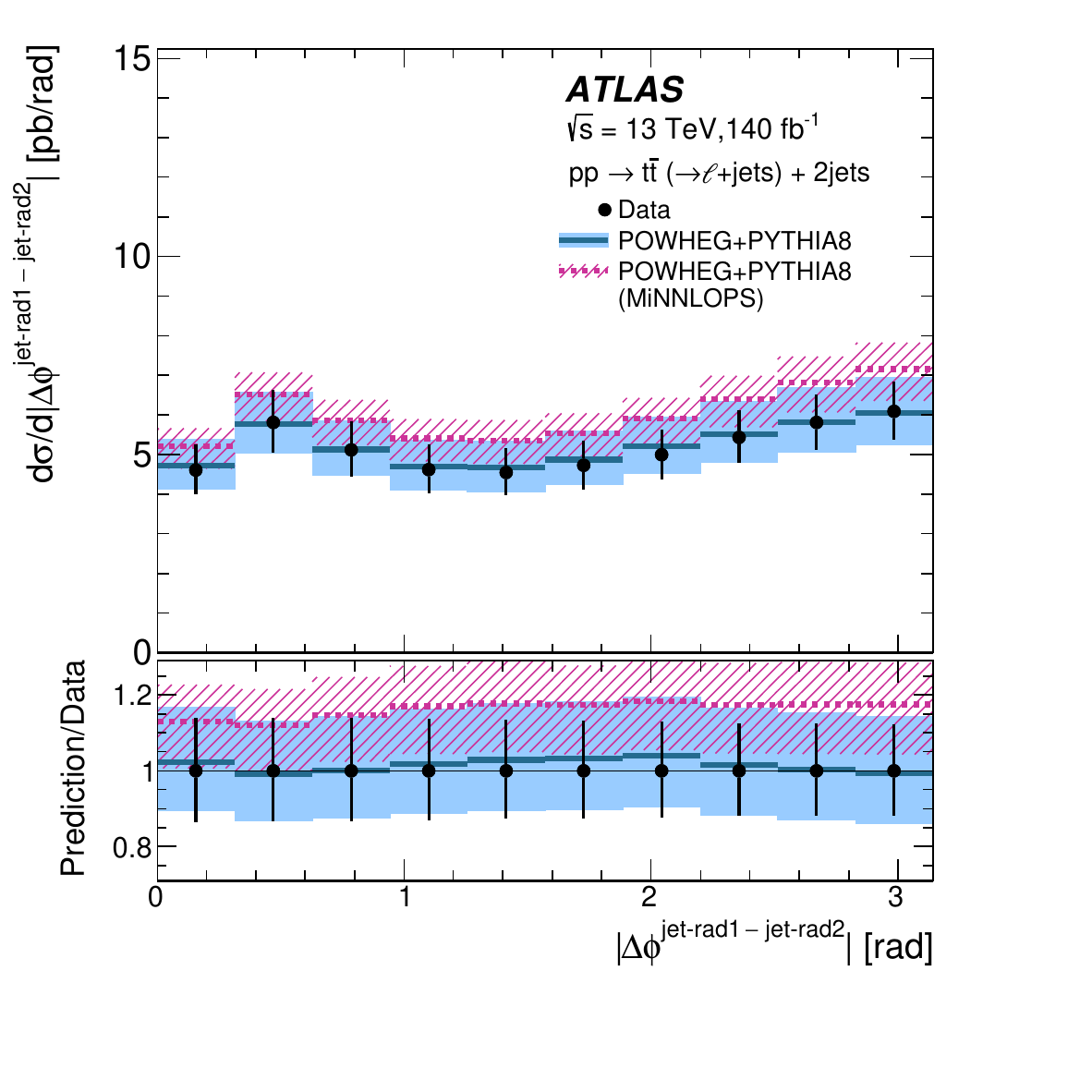}}
\put (3.5,7.2){\small (a)}
\put (11.5,7.2){\small (b)}
\put (3.5,0.2){\small (c)}
\put (11.5,0.2){\small (d)}
\end{picture}
\caption{
The measured differential cross-sections for $\tt$ production in
the \ljets\ decay mode as functions of
(a)~$\Ptjetf$, (b) $|\Rapjetf|$,
(c) $|\deltayef|$ and (d) $|\deltaphief|$ (dots)
in the $\tt+$2jets channel.
The NLO QCD predictions from \pwp\ (solid lines) and the NNLO QCD
predictions from \pwpm\ (dashed lines) are also shown. Both
predictions are normalised to the ${\cal O}$(NNLO+NNLL) total cross-section.
The inner (outer) error bars represent the statistical uncertainties
(the statistical and systematic uncertainties added in
quadrature). For most of the points, the inner error bars are smaller than
the marker size and, thus, not visible.
The shaded (hatched) bands represent
the theoretical uncertainty in the \pwp\ (\pwpm) prediction. The lower panels of
the figures show the ratios of the predicted and the measured
cross-sections.
}
\label{figres7}
\end{figure}

The comparison of the NNLO predictions with the measurements in the
$\tt+$2jets channel is shown in Figures~\ref{figres7} to \ref{fig_appm8}.
The central value of the NNLO predictions for the angular variables in
this channel has a tendency to be above the data, though the \pwpm\
predictions are compatible with the measurements within the
experimental and theoretical uncertainties, except for the normalised
cross-section as a function of $|\deltayef|$, for which the \pwpm\
prediction is below the data for $|\deltayef|>3$.
The description of the data by the NNLO prediction is somewhat worse
than that at NLO for the $\Ptjetf$ observable (see
Figures~\ref{figres7}(a) and \ref{fig_appm7}(a)), especially at
high $\Ptjetf$ values. This effect might be attributed to the fact that these
calculations are only at lowest order in QCD for the second gluon emission,
with higher-order corrections being modelled by the PS, hence
further tuning of the parameters might be necessary.
 
\FloatBarrier
 
\newpage
 
\FloatBarrier
 
\newpage
 
\FloatBarrier
 
\newpage
 
\FloatBarrier
 
\newpage
 
\FloatBarrier
 
\newpage
 
\FloatBarrier
 
\newpage
 
\begin{figure}[h]
\setlength{\unitlength}{1.0cm}
\begin{picture} (18.0,17.5)
\put (0.0,6.3){\includegraphics[width=8cm]{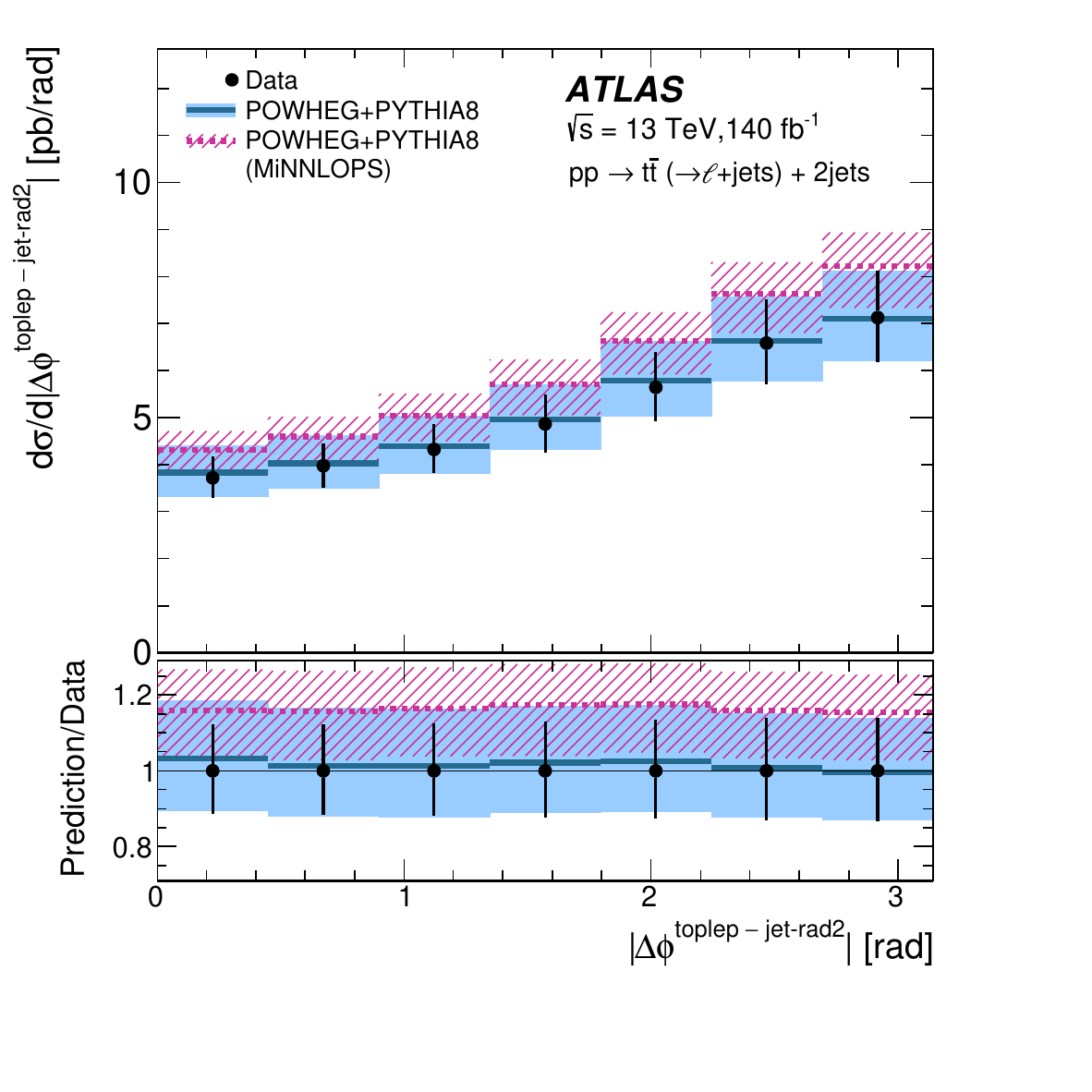}}
\put (8.0,6.3){\includegraphics[width=8cm]{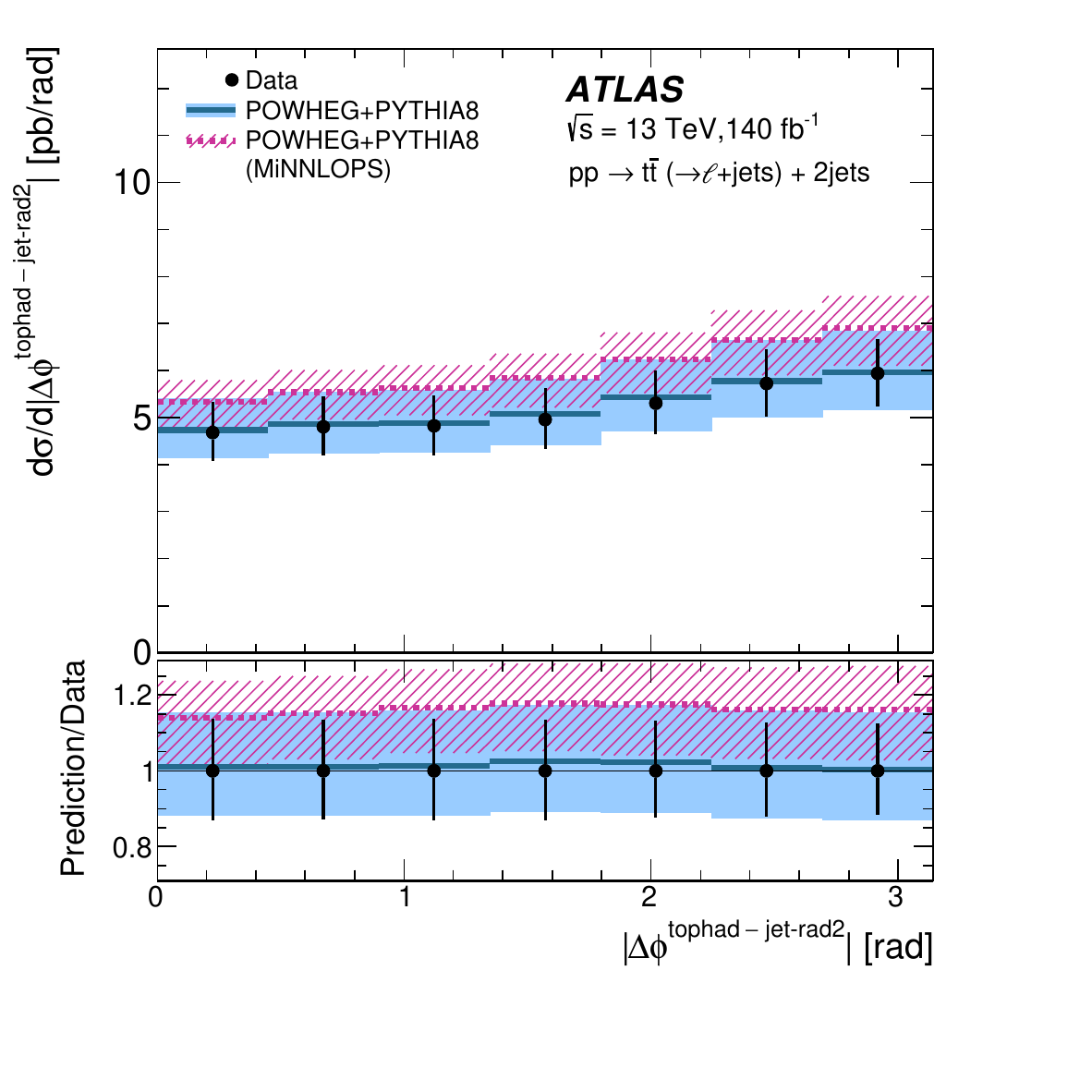}}
\put (0.0,-0.7){\includegraphics[width=8cm]{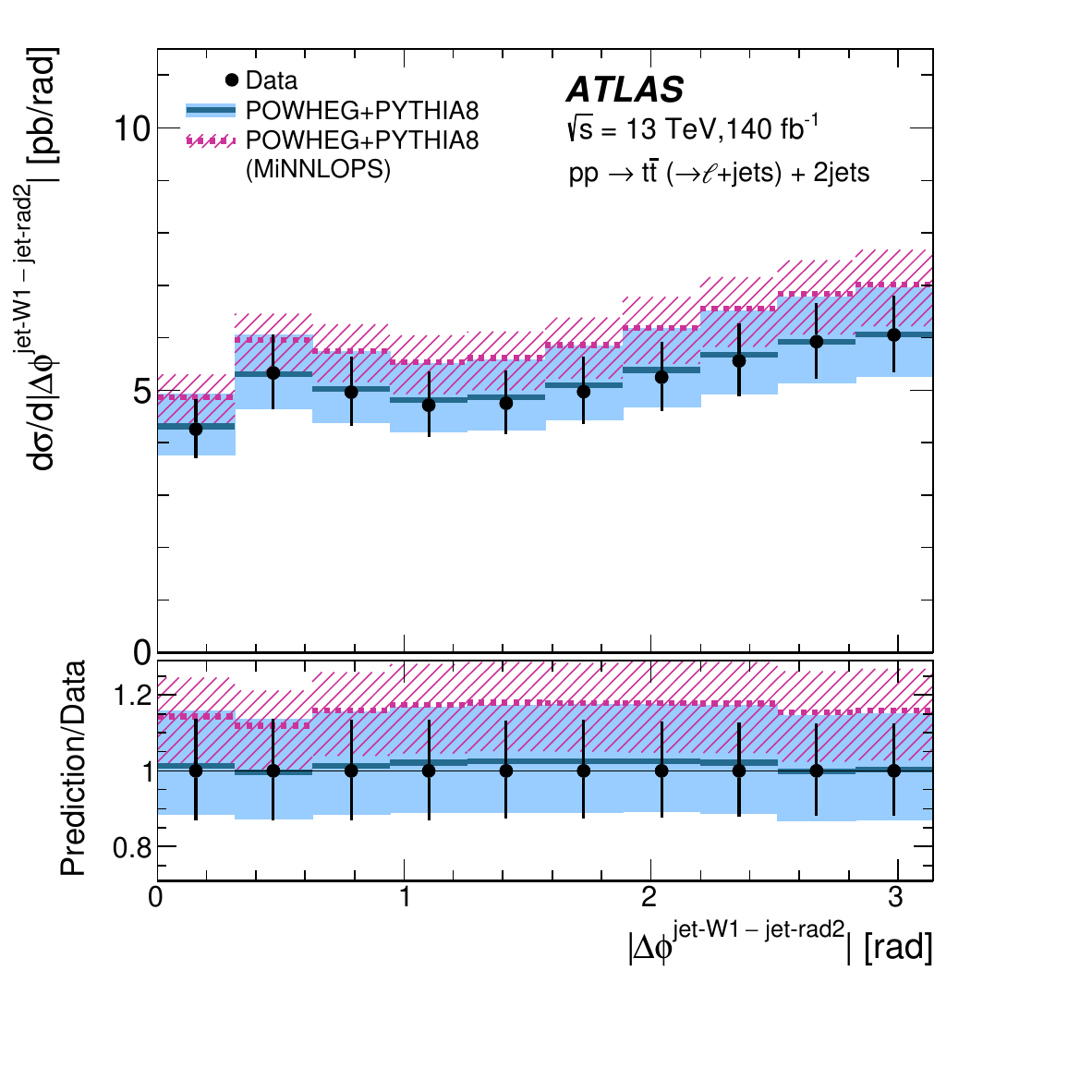}}
\put (8.0,-0.7){\includegraphics[width=8cm]{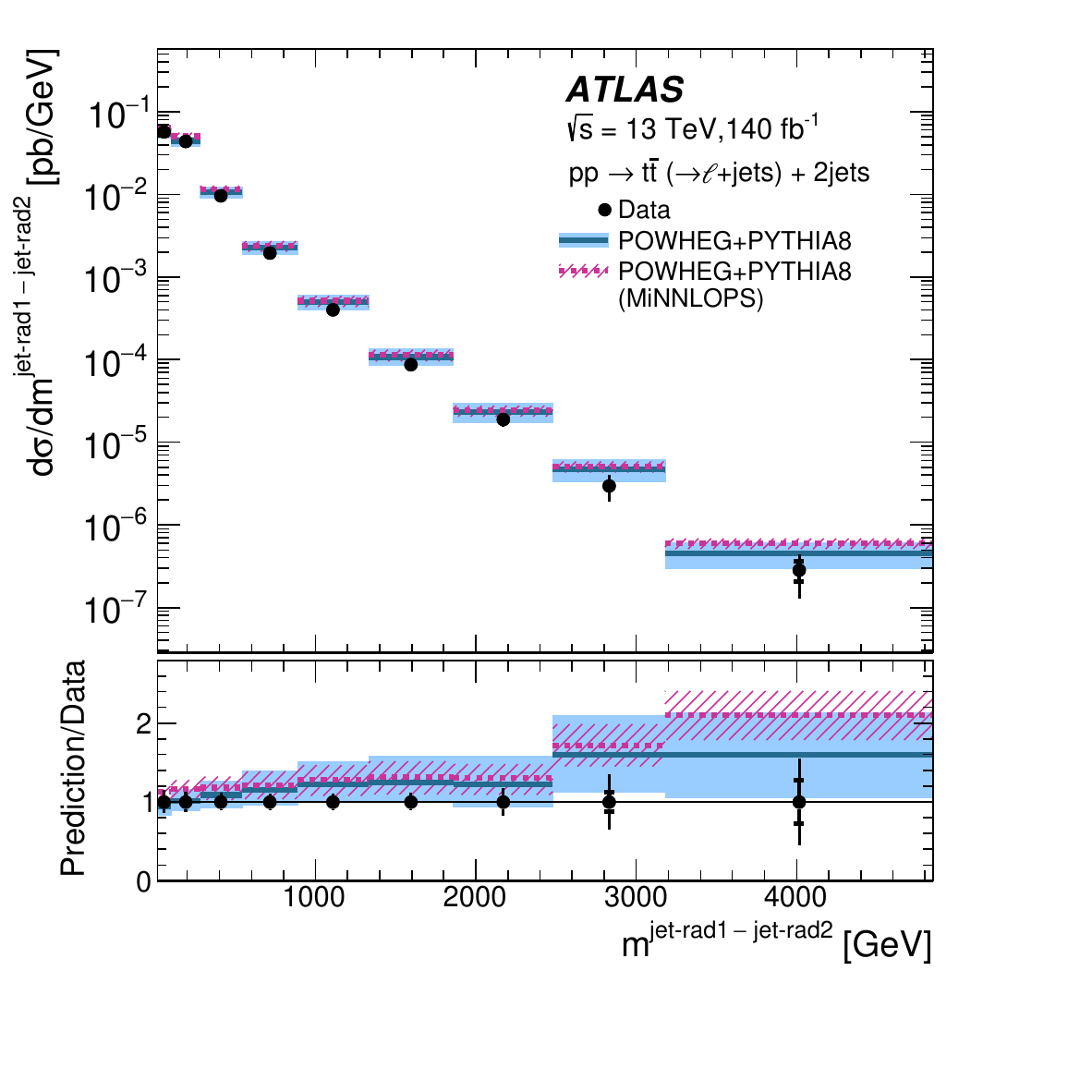}}
\put (3.5,7.2){\small (a)}
\put (11.5,7.2){\small (b)}
\put (3.5,0.2){\small (c)}
\put (11.5,0.2){\small (d)}
\end{picture}
\caption{
The measured differential cross-sections for $\tt$ production in
the \ljets\ decay mode as functions of
(a)~$|\deltaphitlf|$, (b) $|\deltaphithf|$,
(c) $|\deltaphiaf|$ and (d) $\mef$ (dots)
in the $\tt+$2jets channel.
The NLO QCD predictions from \pwp\ (solid lines) and the NNLO QCD
predictions from \pwpm\ (dashed lines) are also shown. Both
predictions are normalised to the ${\cal O}$(NNLO+NNLL) total cross-section.
The inner (outer) error bars represent the statistical uncertainties
(the statistical and systematic uncertainties added in
quadrature). For most of the points, the inner error bars are smaller than
the marker size and, thus, not visible.
The shaded (hatched) bands represent
the theoretical uncertainty in the \pwp\ (\pwpm) prediction. The lower panels of
the figures show the ratios of the predicted and the measured
cross-sections.
}
\end{figure}
 
\FloatBarrier
 
\newpage
 
\begin{figure}[h]
\setlength{\unitlength}{1.0cm}
\begin{picture} (18.0,17.5)
\put (0.0,6.3){\includegraphics[width=8cm]{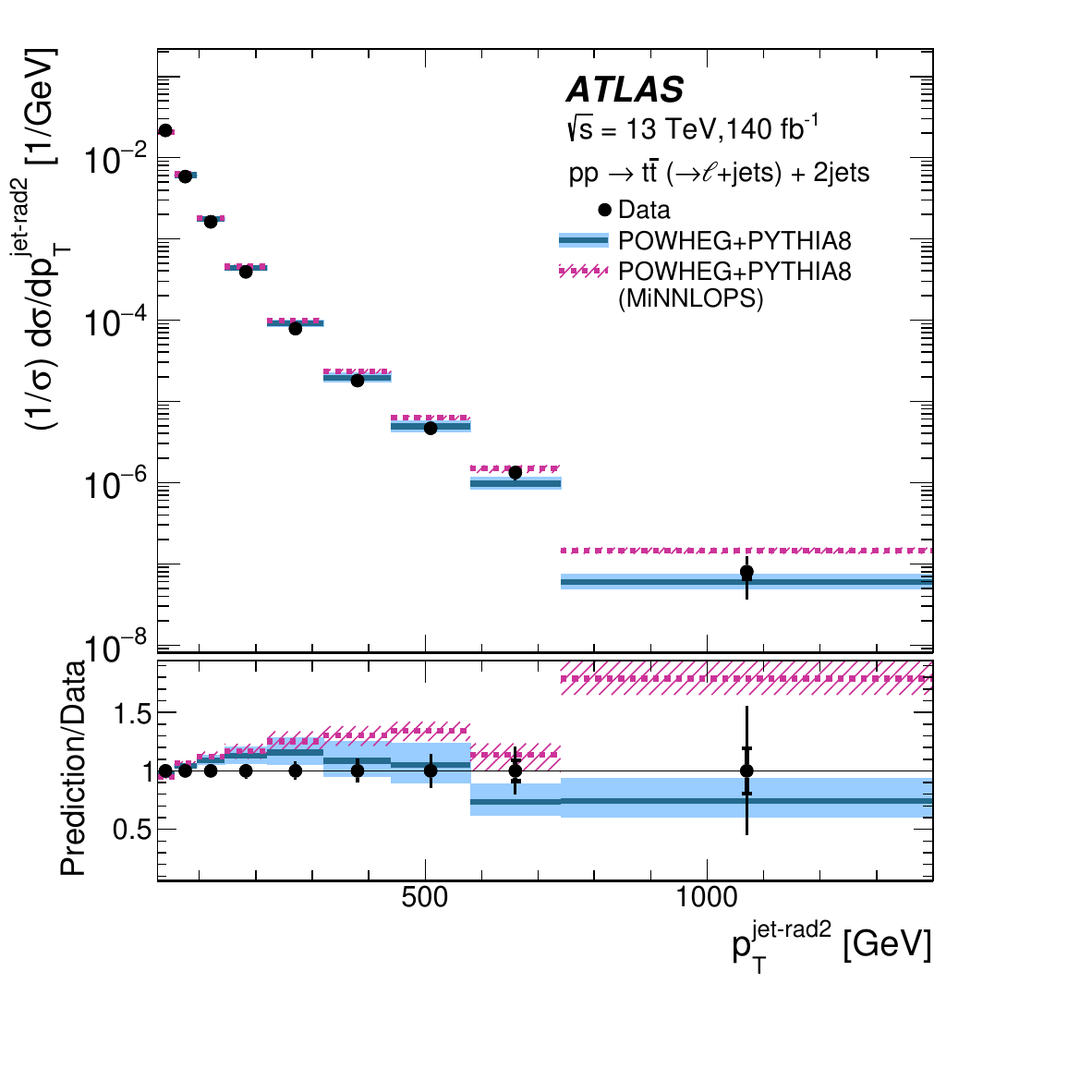}}
\put (8.0,6.3){\includegraphics[width=8cm]{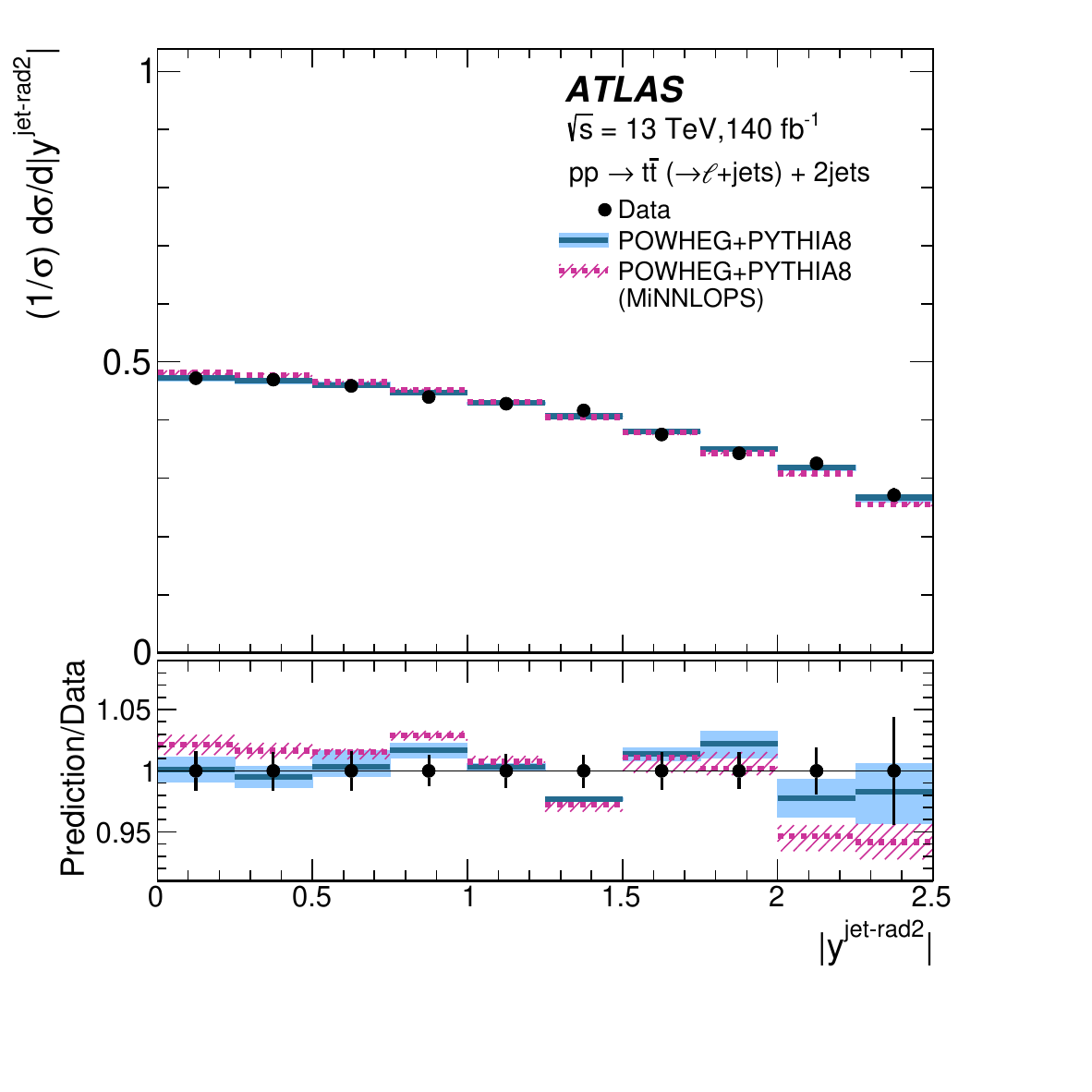}}
\put (0.0,-0.7){\includegraphics[width=8cm]{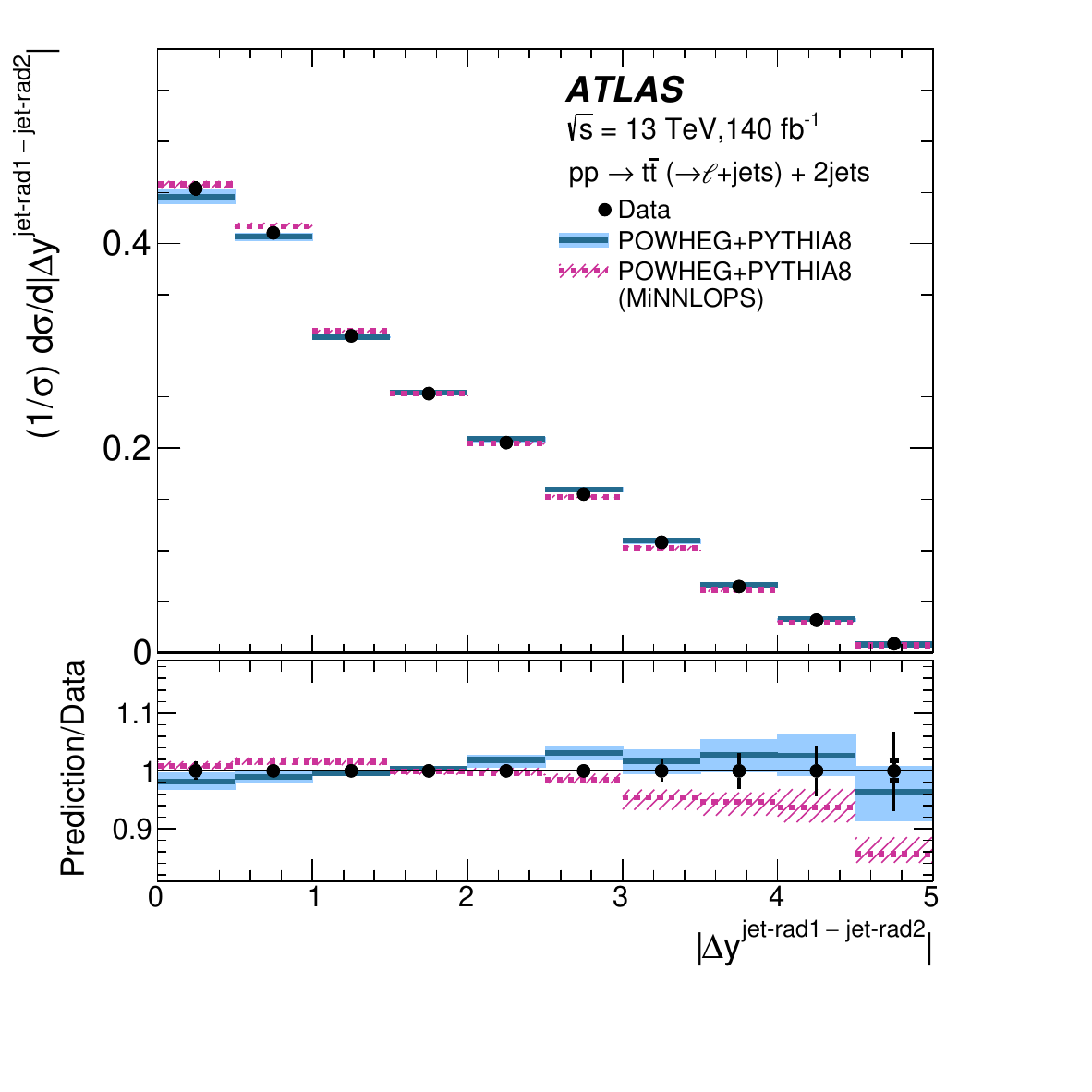}}
\put (8.0,-0.7){\includegraphics[width=8cm]{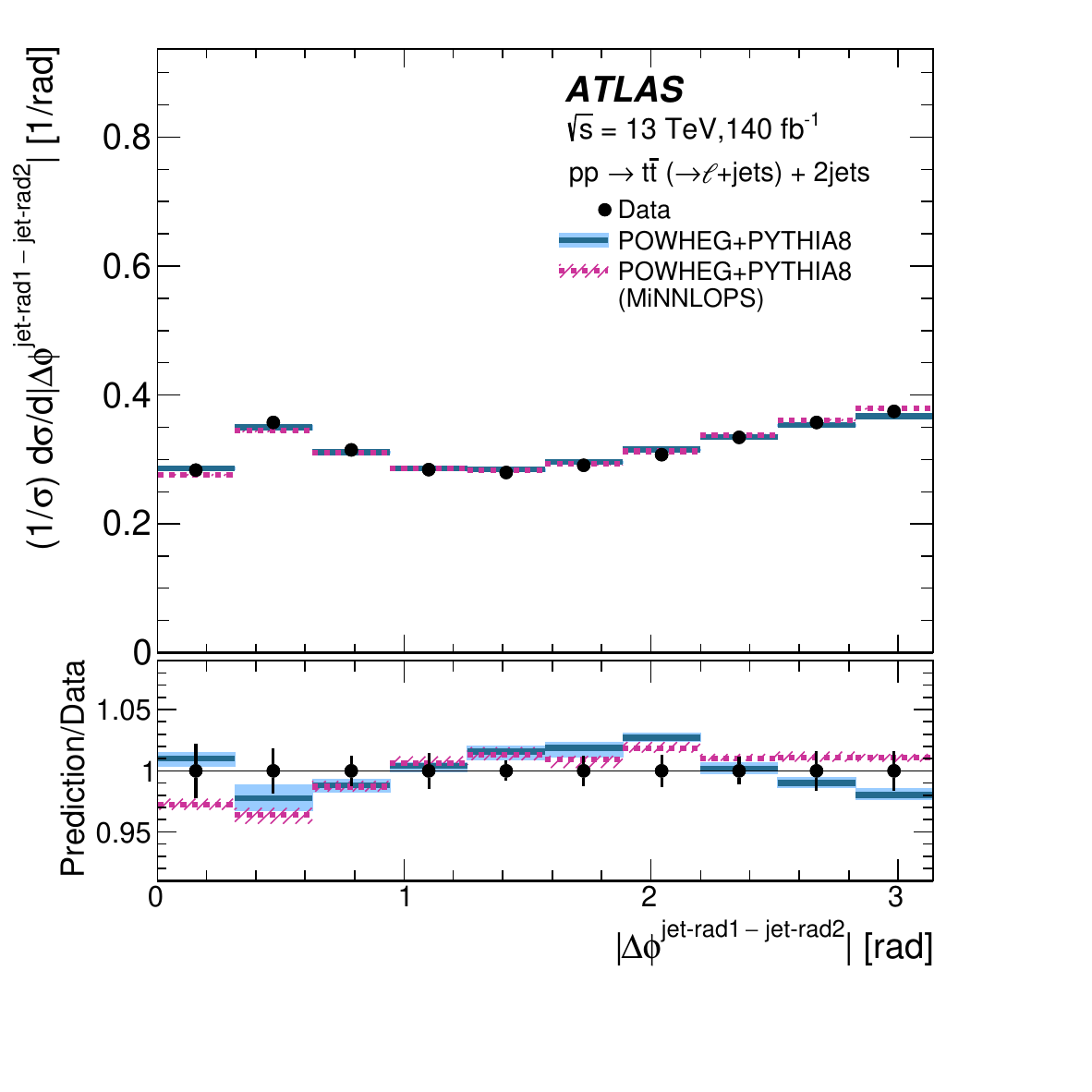}}
\put (3.5,7.2){\small (a)}
\put (11.5,7.2){\small (b)}
\put (3.5,0.2){\small (c)}
\put (11.5,0.2){\small (d)}
\end{picture}
\caption{
The measured normalised differential cross-sections for $\tt$ production in
the \ljets\ decay mode as functions of
(a) $\Ptjetf$, (b) $|\Rapjetf|$,
(c) $|\deltayef|$ and (d) $|\deltaphief|$ (dots)
in the $\tt+$2jets channel.
The NLO QCD predictions from \pwp\ (solid lines) and the NNLO QCD
predictions from \pwpm\ (dashed lines) are also shown.
The inner (outer) error bars represent the statistical uncertainties
(the statistical and systematic uncertainties added in
quadrature). For most of the points, the inner error bars are smaller than
the marker size and, thus, not visible.
The shaded (hatched) bands represent
the theoretical uncertainty in the \pwp\ (\pwpm) prediction. The lower panels of
the figures show the ratios of the predicted and the measured
cross-sections.
}
\label{fig_appm7}
\end{figure}
 
\FloatBarrier
 
\newpage
 
\begin{figure}[h]
\setlength{\unitlength}{1.0cm}
\begin{picture} (18.0,17.5)
\put (0.0,6.3){\includegraphics[width=8cm]{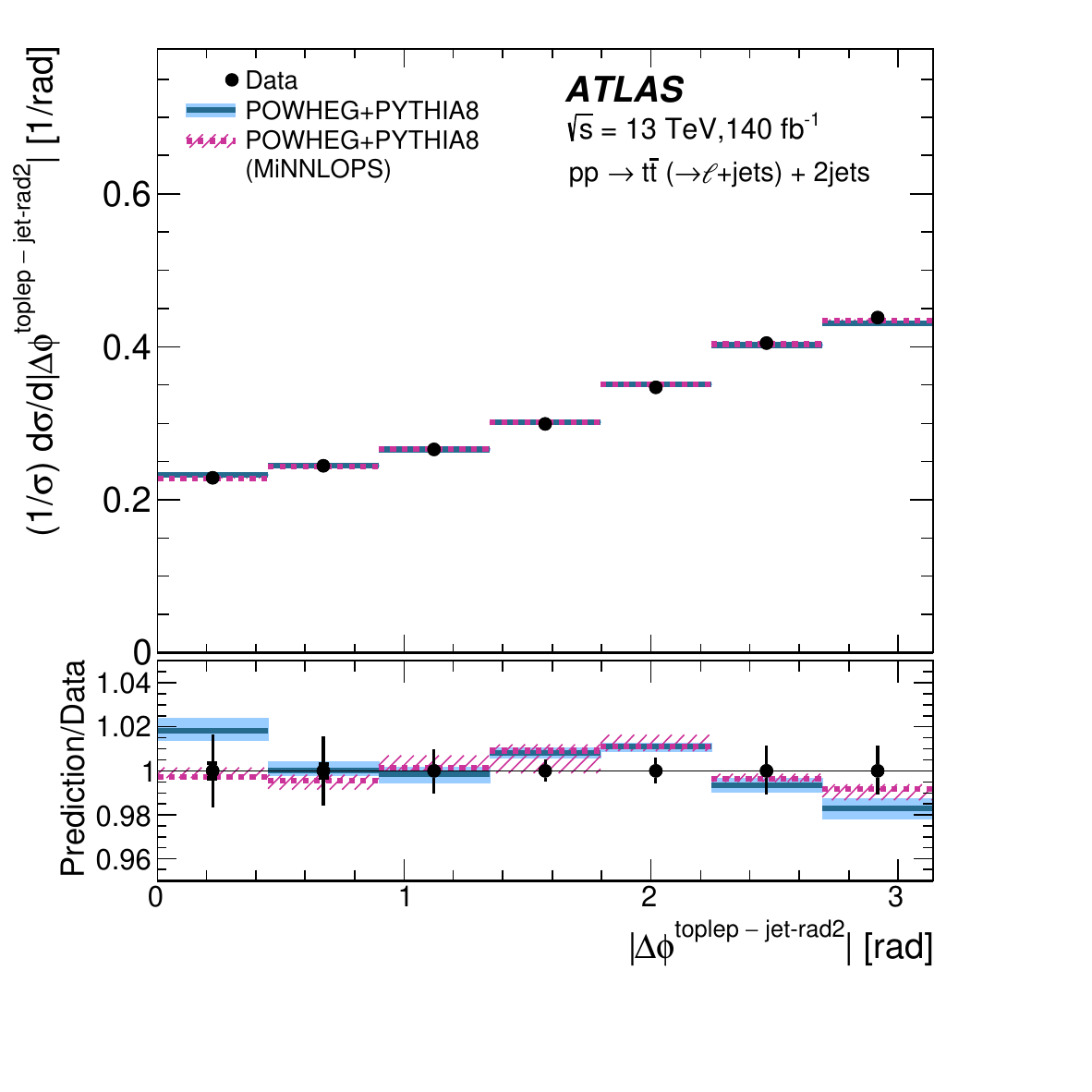}}
\put (8.0,6.3){\includegraphics[width=8cm]{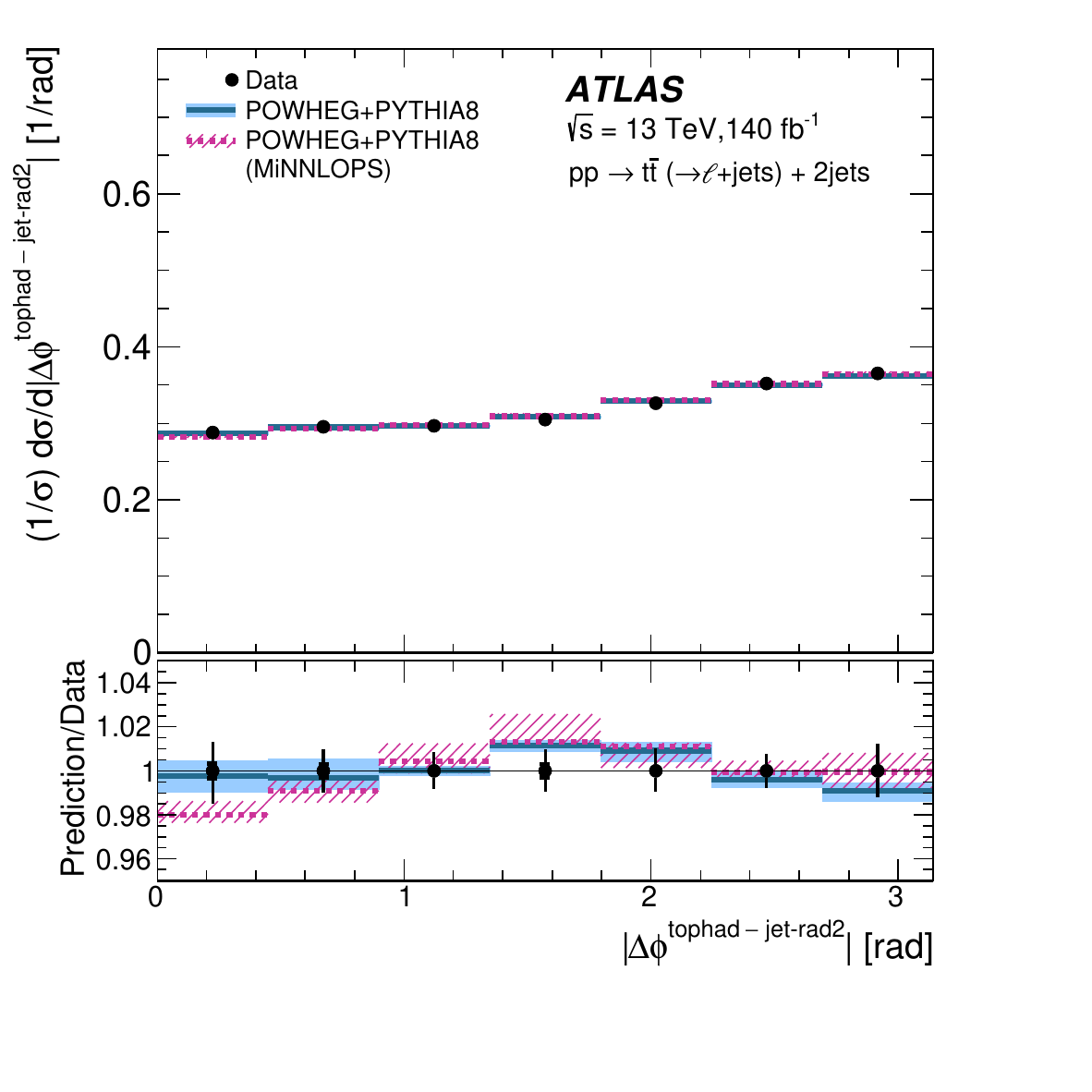}}
\put (0.0,-0.7){\includegraphics[width=8cm]{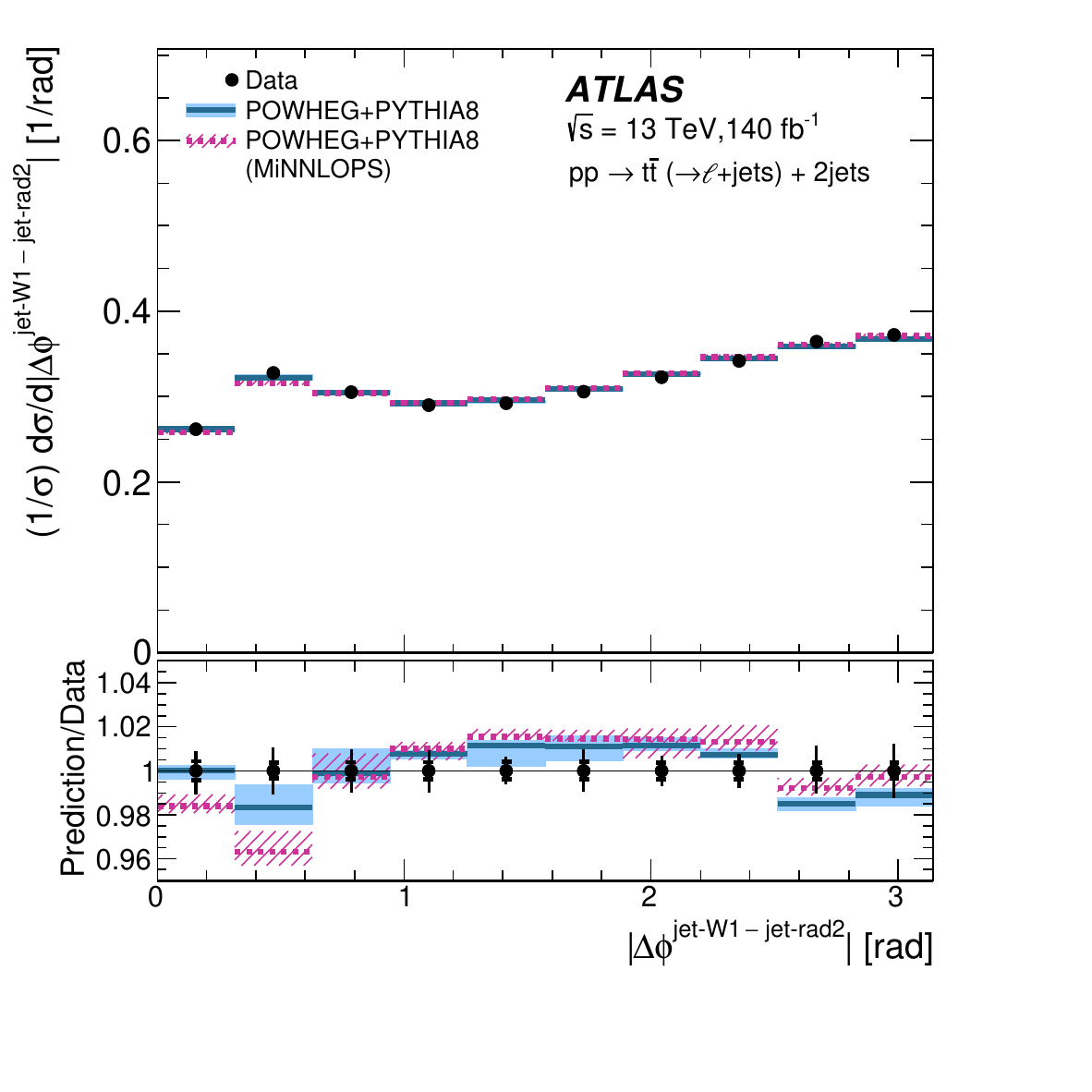}}
\put (8.0,-0.7){\includegraphics[width=8cm]{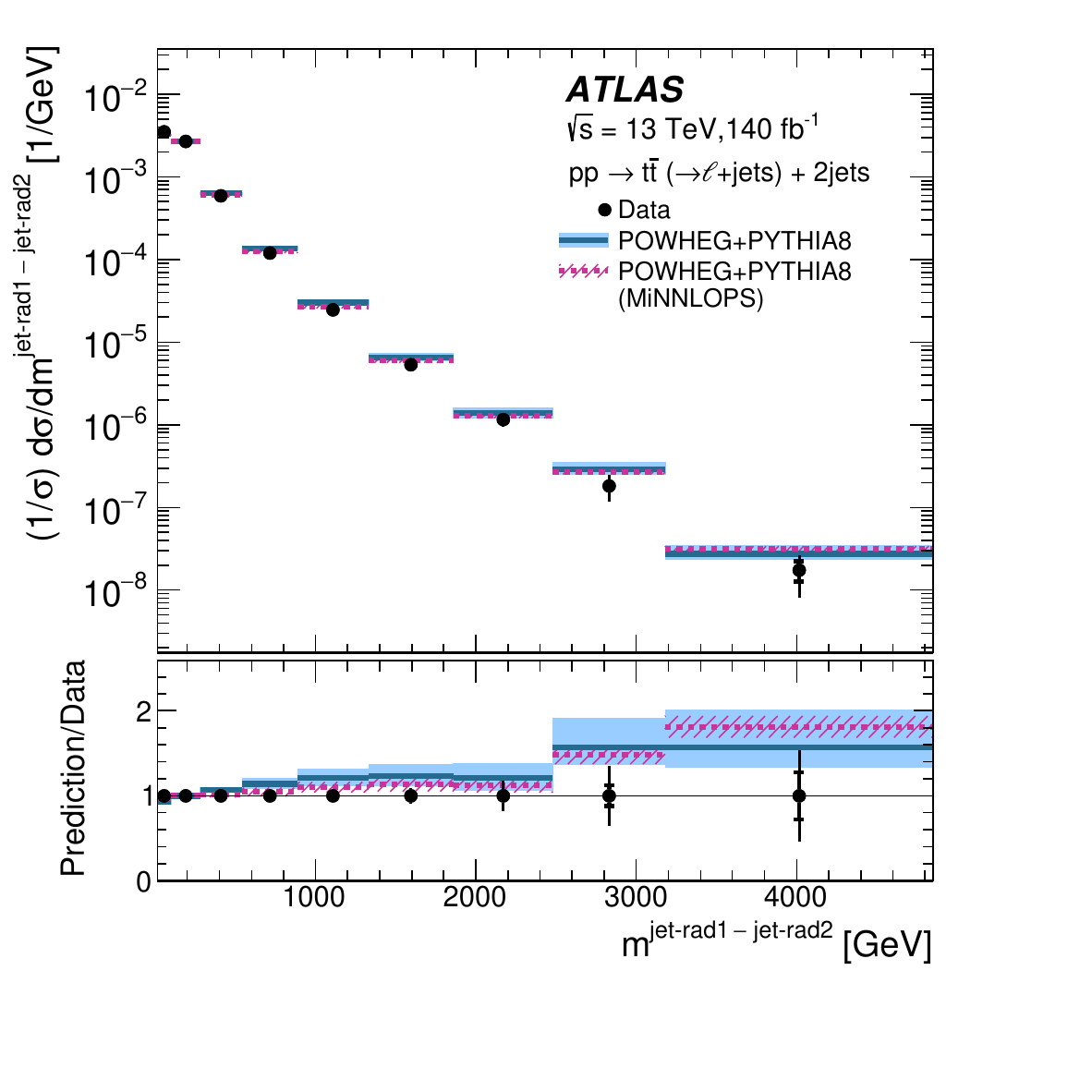}}
\put (3.5,7.2){\small (a)}
\put (11.5,7.2){\small (b)}
\put (3.5,0.2){\small (c)}
\put (11.5,0.2){\small (d)}
\end{picture}
\caption{
The measured normalised differential cross-sections for $\tt$ production in
the \ljets\ decay mode as functions of
(a) $|\deltaphitlf|$, (b) $|\deltaphithf|$,
(c) $|\deltaphiaf|$ and (d) $\mef$ (dots)
in the $\tt+$2jets channel.
The NLO QCD predictions from \pwp\ (solid lines) and the NNLO QCD
predictions from \pwpm\ (dashed lines) are also shown.
The inner (outer) error bars represent the statistical uncertainties
(the statistical and systematic uncertainties added in
quadrature). For most of the points, the inner error bars are smaller than
the marker size and, thus, not visible.
The shaded (hatched) bands represent
the theoretical uncertainty in the \pwp\ (\pwpm) prediction. The lower panels of
the figures show the ratios of the predicted and the measured
cross-sections.
}
\label{fig_appm8}
\end{figure}
 
\FloatBarrier
 
\subsection{$\chi^2$ and $p$-values}
\label{pval}
To quantify the level of agreement between the measured differential
cross-sections and the theoretical predictions, $\chi^2$ values are
evaluated using the following equation~\cite{epj:c82:438}
 
\begin{equation}
\begin{aligned}
\chi^2 = &\sum_{i,k}\left(D_i - T_i(1-\sum_{j}\gamma_{ij}b_{j})\right)C^{-1}_{\text{stat,uncor},ik}\left(D_k - T_k(1-\sum_{j}\gamma_{kj}b_{j})\right)\\
&+ \sum_{i} \log \frac{\delta^{2}_{i,\text{uncor}}T_i^2 + \delta^{2}_{i,\text{stat}}D_i T_i}{\delta^{2}_{i,\text{uncor}}D_i^2 + \delta^{2}_{i,\text{stat}}D_i^2}\\
&+ \sum_{j} b_j^2,
\end{aligned}
\label{eq6}
\end{equation}
known as the Hessian method~\cite{pr:d65:014013}.
 
In Eq.~\ref{eq6}, indices $i$ and $k$ represent different bins in
a differential cross-section and index $j$ represents each source of
uncertainty. The symbol $D$ represents the measured
differential cross-section in a given bin as a function of each
observable. The statistical uncertainties are incorporated in the
covariance matrix $C_{\text{stat,uncor}}$. This matrix also contains
the systematic uncertainty due to non-closure (see
Section~\ref{syst}), which is the only experimental uncertainty
considered uncorrelated between bins. All the other systematic
uncertainties, denoted by the symbol $\gamma$ in Eq.~\ref{eq6},
are considered as correlated between bins and are accounted for by
using nuisance parameters (symbols $b$ in Eq.~\ref{eq6}). The
parameters $b$ are obtained from a minimisation using the
{\textsc{Minuit}}~\cite{cpc:10:343} program.
 
The symbol $T$ represents the theoretical prediction in each bin of
each differential cross-section. The uncertainties that affect the
theoretical predictions (see Section~\ref{thunc}) are taken into
account: the uncertainties arising from $\muR$, $\muF$ and the amount
of initial- and final-state radiation in the PS are
considered as uncorrelated whereas those arising from the PDFs and the
value of $\as$ are considered as correlated bin to bin.
 
The first term in Eq.~\ref{eq6} gives the main contribution to
the $\chi^2$ and the second term is a small bias correction term,
referred to as the {\em log penalty}~\cite{pr:d65:014013}, which
arises because the diagonal term of $C_{\text{stat,uncor}}$ is given
by
$C_{\text{stat,uncor},ii}=\delta^{2}_{i,\text{uncor}}T_i^2+\delta^{2}_{i,\text{stat}}D_iT_i$,
with different weights for the statistical ($\delta_{i,\text{stat}}$)
and uncorrelated ($\delta_{i,\text{uncor}}$) uncertainties. The third
term is the sum in quadrature of the nuisance parameters and gives the
contribution to the $\chi^2$ arising from the correlated sources of
systematic uncertainty.
 
Table~\ref{tab_pvalue_n19} shows the $\chi^2$ and $p$-values for each
differential cross-section and for each theoretical prediction tested
separately. Table~\ref{tab_pvalue_n19_norm} shows the $\chi^2$ and
$p$-values for the normalised differential
cross-sections.\footnote{The normalisation constraint used to derive
the normalised differential cross-sections lowers the number of
degrees of freedom and the rank of the covariance matrix by one unit.}
The $p$-values obtained confirm the qualitative
conclusions discussed in the previous sections. For \sher~2.2.12, the
$p$-values are much larger than for the other predictions because the
theoretical uncertainties for this prediction are larger.
 
\FloatBarrier
 
\newpage
 
\begin{table}[h]
\caption{
$\chi^2$ values, number of degrees of freedom (NDF) and $p$-values for
each observable and each prediction for the absolute differential cross-sections.
}
\setlength{\unitlength}{1.0cm}
\begin{picture} (18.0,12.0)
\put (0.0,4.5){\centerline{\scalebox{0.90}{
\begin{tabular}{l r
r @{.} l r @{.} l
r @{.} l r @{.} l
r @{.} l r @{.} l
r @{.} l r @{.} l
r @{.} l r @{.} l }
\toprule
Prediction & &
\multicolumn{4}{c}{\pwp} &
\multicolumn{4}{c}{\pwh} &
\multicolumn{4}{c}{\amcnloh} &
\multicolumn{4}{c}{\sher~2.2.12} &
\multicolumn{4}{c}{\pwp} \\
& &
\multicolumn{4}{c}{} &
\multicolumn{4}{c}{} &
\multicolumn{4}{c}{} &
\multicolumn{4}{c}{} &
\multicolumn{4}{c}{\textsc{MiNNLOPS}}
\\ \midrule
Observable & NDF &
\multicolumn{2}{c}{$\chi^2$} & \multicolumn{2}{c}{$p$-value} &
\multicolumn{2}{c}{$\chi^2$} & \multicolumn{2}{c}{$p$-value} &
\multicolumn{2}{c}{$\chi^2$} & \multicolumn{2}{c}{$p$-value} &
\multicolumn{2}{c}{$\chi^2$} & \multicolumn{2}{c}{$p$-value} &
\multicolumn{2}{c}{$\chi^2$} & \multicolumn{2}{c}{$p$-value}
\\ \midrule\midrule
$\Ptjeta$ & 10 & 6 & 2 & 0 & 79 & 4 & 1 & 0 & 94 & 2 & 7 & 0 & 99 & 1 & 8 & 1 & 0 & 3 & 4 & 0 & 97 \\ \midrule
$|\Rapjeta|$ & 10 & 1 & 8 & 1 & 0 & 1 & 6 & 1 & 0 & 2 & 3 & 0 & 99 & 1 & 1 & 1 & 0 & 2 & 7 & 0 & 99 \\ \midrule
$\Ptjetb$ & 8 & 2 & 3 & 0 & 97 & 0 & 45 & 1 & 0 & 2 & 8 & 0 & 94 & 0 & 53 & 1 & 0 & 2 & 6 & 0 & 96 \\ \midrule
$|\Rapjetb|$ & 10 & 1 & 9 & 1 & 0 & 1 & 7 & 1 & 0 & 2 & 4 & 0 & 99 & 1 & 1 & 1 & 0 & 2 & 7 & 0 & 99 \\ \midrule
$|\deltayab|$ & 10 & 1 & 6 & 1 & 0 & 2 & 7 & 0 & 99 & 4 & 3 & 0 & 93 & 0 & 90 & 1 & 0 & 6 & 0 & 0 & 81 \\ \midrule
$|\deltaphiab|$ & 10 & 1 & 7 & 1 & 0 & 1 & 6 & 1 & 0 & 2 & 5 & 0 & 99 & 1 & 2 & 1 & 0 & 2 & 1 & 1 & 0 \\ \midrule
$\Ptjete$ & 11 & 7 & 7 & 0 & 74 & 7 & 1 & 0 & 79 & 3 & 3 & 0 & 99 & 2 & 6 & 1 & 0 & 6 & 4 & 0 & 85 \\ \midrule
$|\Rapjete|$ & 10 & 1 & 6 & 1 & 0 & 2 & 8 & 0 & 99 & 3 & 4 & 0 & 97 & 0 & 69 & 1 & 0 & 3 & 0 & 0 & 98 \\ \midrule
$|\deltaphitle|$ & 7 & 1 & 4 & 0 & 99 & 2 & 2 & 0 & 95 & 2 & 9 & 0 & 89 & 0 & 81 & 1 & 0 & 2 & 3 & 0 & 94 \\ \midrule
$|\deltaphithe|$ & 7 & 1 & 4 & 0 & 98 & 2 & 6 & 0 & 92 & 2 & 9 & 0 & 89 & 0 & 85 & 1 & 0 & 2 & 4 & 0 & 93 \\ \midrule
$|\deltaphiae|$ & 10 & 1 & 6 & 1 & 0 & 2 & 9 & 0 & 98 & 3 & 4 & 0 & 97 & 0 & 77 & 1 & 0 & 2 & 7 & 0 & 99 \\ \midrule
$\mtte$ & 8 & 7 & 7 & 0 & 46 & 6 & 5 & 0 & 59 & 4 & 4 & 0 & 81 & 4 & 4 & 0 & 82 & 5 & 5 & 0 & 71 \\ \midrule
$\Ptjetf$ & 9 & 3 & 2 & 0 & 96 & 4 & 0 & 0 & 91 & 37 & 0 & $<0$ & 01 & 1 & 7 & 1 & 0 & 11 & 0 & 0 & 28 \\ \midrule
$|\Rapjetf|$ & 10 & 1 & 4 & 1 & 0 & 3 & 9 & 0 & 95 & 2 & 4 & 0 & 99 & 0 & 20 & 1 & 0 & 5 & 3 & 0 & 87 \\ \midrule
$|\deltayef|$ & 10 & 1 & 6 & 1 & 0 & 4 & 0 & 0 & 95 & 4 & 0 & 0 & 95 & 0 & 13 & 1 & 0 & 5 & 4 & 0 & 86 \\ \midrule
$|\deltaphief|$ & 10 & 1 & 4 & 1 & 0 & 4 & 0 & 0 & 95 & 2 & 4 & 0 & 99 & 0 & 25 & 1 & 0 & 5 & 2 & 0 & 88 \\ \midrule
$|\deltaphitlf|$ & 7 & 1 & 3 & 0 & 99 & 3 & 2 & 0 & 86 & 2 & 0 & 0 & 96 & 0 & 46 & 1 & 0 & 4 & 1 & 0 & 77 \\ \midrule
$|\deltaphithf|$ & 7 & 1 & 2 & 0 & 99 & 3 & 2 & 0 & 87 & 2 & 1 & 0 & 96 & 0 & 46 & 1 & 0 & 4 & 1 & 0 & 77 \\ \midrule
$|\deltaphiaf|$ & 10 & 1 & 3 & 1 & 0 & 3 & 8 & 0 & 95 & 2 & 4 & 0 & 99 & 0 & 22 & 1 & 0 & 5 & 1 & 0 & 88 \\ \midrule
$\mef$ & 9 & 6 & 2 & 0 & 72 & 6 & 7 & 0 & 66 & 2 & 6 & 0 & 98 & 4 & 2 & 0 & 90 & 8 & 8 & 0 & 45 \\ \midrule
\end{tabular}
}}}
\end{picture}
\label{tab_pvalue_n19}
\end{table}
 
\FloatBarrier
 
\newpage
 
\begin{table}[h]
\caption{
$\chi^2$ values, number of degrees of freedom (NDF) and $p$-values for
each observable and each prediction for the normalised differential cross-sections.
}
\setlength{\unitlength}{1.0cm}
\begin{picture} (18.0,12.0)
\put (0.0,4.5){\centerline{\scalebox{0.90}{
\begin{tabular}{l r
r @{.} l r @{.} l
r @{.} l r @{.} l
r @{.} l r @{.} l
r @{.} l r @{.} l
r @{.} l r @{.} l }
\toprule
Prediction & &
\multicolumn{4}{c}{\pwp} &
\multicolumn{4}{c}{\pwh} &
\multicolumn{4}{c}{\amcnloh} &
\multicolumn{4}{c}{\sher~2.2.12} &
\multicolumn{4}{c}{\pwp} \\
& &
\multicolumn{4}{c}{} &
\multicolumn{4}{c}{} &
\multicolumn{4}{c}{} &
\multicolumn{4}{c}{} &
\multicolumn{4}{c}{\textsc{MiNNLOPS}}
\\ \midrule
Observable & NDF &
\multicolumn{2}{c}{$\chi^2$} & \multicolumn{2}{c}{$p$-value} &
\multicolumn{2}{c}{$\chi^2$} & \multicolumn{2}{c}{$p$-value} &
\multicolumn{2}{c}{$\chi^2$} & \multicolumn{2}{c}{$p$-value} &
\multicolumn{2}{c}{$\chi^2$} & \multicolumn{2}{c}{$p$-value} &
\multicolumn{2}{c}{$\chi^2$} & \multicolumn{2}{c}{$p$-value}
\\ \midrule\midrule
$\Ptjeta$ & 9 & 12 & 0 & 0 & 21 & 16 & 0 & 0 & 074 & 13 & 0 & 0 & 19 & 19 & 0 & 0 & 024 & 14 & 0 & 0 & 13 \\ \midrule
$|\Rapjeta|$ & 9 & 63 & 0 & $<0$ & 01 & 76 & 0 & $<0$ & 01 & 65 & 0 & $<0$ & 01 & 56 & 0 & $<0$ & 01 & 13 & 0 & 0 & 17 \\ \midrule
$\Ptjetb$ & 7 & 7 & 9 & 0 & 34 & 9 & 8 & 0 & 20 & 8 & 0 & 0 & 33 & 13 & 0 & 0 & 064 & 8 & 2 & 0 & 31 \\ \midrule
$|\Rapjetb|$ & 9 & 134 & 0 & $<0$ & 01 & 220 & 0 & $<0$ & 01 & 219 & 0 & $<0$ & 01 & 103 & 0 & $<0$ & 01 & 25 & 0 & $<0$ & 01 \\ \midrule
$|\deltayab|$ & 9 & 7 & 0 & 0 & 63 & 8 & 2 & 0 & 52 & 7 & 9 & 0 & 54 & 6 & 1 & 0 & 73 & 5 & 2 & 0 & 81 \\ \midrule
$|\deltaphiab|$ & 9 & 9 & 5 & 0 & 39 & 12 & 0 & 0 & 21 & 13 & 0 & 0 & 17 & 11 & 0 & 0 & 25 & 8 & 7 & 0 & 47 \\ \midrule
$\Ptjete$ & 10 & 9 & 4 & 0 & 50 & 9 & 5 & 0 & 48 & 8 & 5 & 0 & 58 & 10 & 3 & 0 & 42 & 12 & 0 & 0 & 29 \\ \midrule
$|\Rapjete|$ & 9 & 49 & 0 & $<0$ & 01 & 80 & 0 & $<0$ & 01 & 68 & 0 & $<0$ & 01 & 42 & 0 & $<0$ & 01 & 21 & 0 & 0 & 011 \\ \midrule
$|\deltaphitle|$ & 6 & 2 & 8 & 0 & 83 & 3 & 0 & 0 & 80 & 2 & 7 & 0 & 85 & 2 & 9 & 0 & 82 & 2 & 5 & 0 & 87 \\ \midrule
$|\deltaphithe|$ & 6 & 2 & 6 & 0 & 86 & 2 & 9 & 0 & 82 & 2 & 2 & 0 & 90 & 2 & 6 & 0 & 85 & 2 & 1 & 0 & 91 \\ \midrule
$|\deltaphiae|$ & 9 & 7 & 7 & 0 & 56 & 11 & 0 & 0 & 27 & 6 & 3 & 0 & 71 & 8 & 6 & 0 & 47 & 5 & 6 & 0 & 78 \\ \midrule
$\mtte$ & 7 & 9 & 1 & 0 & 24 & 9 & 8 & 0 & 20 & 7 & 9 & 0 & 35 & 7 & 8 & 0 & 35 & 8 & 2 & 0 & 31 \\ \midrule
$\Ptjetf$ & 8 & 12 & 0 & 0 & 13 & 13 & 0 & 0 & 11 & 11 & 0 & 0 & 22 & 10 & 5 & 0 & 23 & 14 & 0 & 0 & 072 \\ \midrule
$|\Rapjetf|$ & 9 & 24 & 0 & $<0$ & 01 & 45 & 0 & $<0$ & 01 & 38 & 0 & $<0$ & 01 & 30 & 0 & $<0$ & 01 & 16 & 0 & 0 & 069 \\ \midrule
$|\deltayef|$ & 9 & 5 & 6 & 0 & 78 & 15 & 0 & 0 & 090 & 7 & 7 & 0 & 57 & 5 & 9 & 0 & 75 & 6 & 4 & 0 & 70 \\ \midrule
$|\deltaphief|$ & 9 & 25 & 0 & $<0$ & 01 & 69 & 0 & $<0$ & 01 & 33 & 0 & $<0$ & 01 & 15 & 0 & 0 & 084 & 26 & 0 & $<0$ & 01 \\ \midrule
$|\deltaphitlf|$ & 6 & 12 & 0 & 0 & 072 & 15 & 0 & 0 & 018 & 12 & 0 & 0 & 062 & 13 & 0 & 0 & 036 & 8 & 5 & 0 & 21 \\ \midrule
$|\deltaphithf|$ & 6 & 4 & 7 & 0 & 59 & 5 & 5 & 0 & 49 & 4 & 8 & 0 & 57 & 4 & 6 & 0 & 59 & 2 & 6 & 0 & 86 \\ \midrule
$|\deltaphiaf|$ & 9 & 12 & 0 & 0 & 23 & 22 & 0 & $<0$ & 01 & 18 & 0 & 0 & 040 & 14 & 0 & 0 & 13 & 11 & 0 & 0 & 26 \\ \midrule
$\mef$ & 8 & 14 & 0 & 0 & 094 & 16 & 0 & 0 & 042 & 12 & 0 & 0 & 15 & 10 & 0 & 0 & 26 & 14 & 0 & 0 & 085 \\ \midrule
\end{tabular}
}}}
\end{picture}
\label{tab_pvalue_n19_norm}
\end{table}
 
\FloatBarrier
 
\newpage
 
\vspace{1cm}
 
\section{Summary and Conclusions}
Measurements are presented of differential cross-sections at particle
level for $\tt$ production in $pp$ collisions at $\sqrt s=13$~\TeV\
using the ATLAS detector at the LHC with an integrated luminosity
of $140$~\fb1\ of 2015--2018 data. The $\tt$ events are selected in
the \ljets\ final state. The measurements presented benefit from the large
number of events available from the full Run 2 sample and a reduction of
the systematic uncertainties compared to previous analyses.
 
Absolute and normalised differential cross-sections as functions of
jet observables, including jet transverse momenta and rapidities,
angular correlations and invariant masses, are measured in three channels:
$\tt$ inclusive with $N^\mathrm{jets}\geq 4$, $\tt+$1jet with
$N^\mathrm{jets}\geq 5$ and $\tt+$2jets with $N^\mathrm{jets}\geq
6$. The observables, either constructed from the jets forming the
$\tt$ system or arising from hard QCD radiation, are measured
to characterise the kinematics, topology and
dynamics of the $\tt$ system; they also prove to be sensitive to
initial- and final-state QCD radiation effects, represented by the
production of additional jets. Angular correlations between the top
quarks and the additional jets, as well as between the jets, are
measured for the first time.
 
The measured cross-section as a function of $\Ptjeta$ has a harder
spectrum than that of $\Ptjetb$. The measured cross-sections as
functions of $|\Rapjeta|$ and $|\Rapjetb|$ have very similar shape and
normalisation. The measured cross-section as a function of
$|\deltayab|$ ($|\deltaphiab|$) exhibits a peak around $|\deltayab|=0$
($|\deltaphiab|=1$~rad). These observables characterise the kinematics of
the two light jets forming the hadronically decaying top quark.
 
Values of $\mtte$ ($\mef$) up to $\sim 5.3$~\TeV\ ($\sim 4$~\TeV) are
measured. The measured cross-sections as functions of $\mtte$ and $\mef$
decrease by more than five orders of magnitude within the measured
range. These observables characterise the dynamics of the first and
second gluon emissions.
 
From the event topology point of view, differences between the shapes of the
measured cross-sections as functions of the rapidities and angular
correlations for the $\tt+$1jet and $\tt+$2jets
channels and those measured in the
$\tt$ inclusive channel indicate that the first and
second gluon emissions tend to be more isotropic in rapidity than the jets
forming the hadronically decaying top quark. This is supported by the
measurements as functions of $|\deltaphitle|$, $|\deltaphithe|$,
$|\deltaphitlf|$ and $|\deltaphithf|$,
which show that the first and second gluon emissions have a
higher probability of arising at larger angles from the leptonically decaying top quark
and that there are more events with a gluon emission back-to-back with
the jets forming the leptonically decaying top.
In addition, the measured features indicate that the second gluon
emission tends to be close in rapidity to the first gluon emission and that
there is a weak dependence on the difference in $\phi$ between the
first and second gluon emissions.
 
Next-to-leading-order QCD predictions normalised to ${\cal
O}$(NNLO+NNLL) describe well the shape and normalisation of the
measured absolute differential cross-sections as functions of
angular observables, but the transverse momentum and invariant mass
observables are described well only at low values.
For $\Ptjeta$, $\Ptjetb$ and $\mef$ ($\Ptjete$ and $\Ptjetf$),
the predictions from \pwh\ (\sher~2.2.12) are closest to the data,
whereas the prediction from \amcnloh\ gives the best description of
the measured cross-section as a function of $\mtte$.
The same conclusions hold for the normalised differential cross-sections.
 
Recently, predictions for $\tt$ production at
NNLO QCD have become available and the
interface with the PS provided by \pyt\ via the
{\textsc{MiNNLOPS}} matching scheme allows the computation of
differential cross-sections for jet observables such as those presented
here. The NNLO predictions are compared with the measured cross-sections
for the first time and provide a good description of the angular
observables, though there is a tendency for the NNLO predictions to be
below the data for $|\deltayab|>3$. The description of the data by the
NNLO prediction is somewhat worse than that at NLO for $\Ptjetf$.
For the transverse momentum and
invariant mass observables, the NNLO QCD predictions provide a
significant improvement in the description of the $\Ptjeta$,
$\Ptjetb$, $\Ptjete$, $\mtte$ and $\mef$ observables relative to
that of the NLO predictions.
 
The $\chi^2$ and $p$-values calculated for each differential
cross-section and each theoretical prediction confirm the qualitative
conclusions.
 
\section*{Acknowledgements}
\addcontentsline{toc}{section}{Acknowledgements}


We thank CERN for the very successful operation of the LHC and its injectors, as well as the support staff at
CERN and at our institutions worldwide without whom ATLAS could not be operated efficiently.
 
The crucial computing support from all WLCG partners is acknowledged gratefully, in particular from CERN, the ATLAS Tier-1 facilities at TRIUMF/SFU (Canada), NDGF (Denmark, Norway, Sweden), CC-IN2P3 (France), KIT/GridKA (Germany), INFN-CNAF (Italy), NL-T1 (Netherlands), PIC (Spain), RAL (UK) and BNL (USA), the Tier-2 facilities worldwide and large non-WLCG resource providers. Major contributors of computing resources are listed in Ref.~\cite{ATL-SOFT-PUB-2023-001}.
 
We gratefully acknowledge the support of ANPCyT, Argentina; YerPhI, Armenia; ARC, Australia; BMWFW and FWF, Austria; ANAS, Azerbaijan; CNPq and FAPESP, Brazil; NSERC, NRC and CFI, Canada; CERN; ANID, Chile; CAS, MOST and NSFC, China; Minciencias, Colombia; MEYS CR, Czech Republic; DNRF and DNSRC, Denmark; IN2P3-CNRS and CEA-DRF/IRFU, France; SRNSFG, Georgia; BMBF, HGF and MPG, Germany; GSRI, Greece; RGC and Hong Kong SAR, China; ISF and Benoziyo Center, Israel; INFN, Italy; MEXT and JSPS, Japan; CNRST, Morocco; NWO, Netherlands; RCN, Norway; MEiN, Poland; FCT, Portugal; MNE/IFA, Romania; MESTD, Serbia; MSSR, Slovakia; ARIS and MVZI, Slovenia; DSI/NRF, South Africa; MICINN, Spain; SRC and Wallenberg Foundation, Sweden; SERI, SNSF and Cantons of Bern and Geneva, Switzerland; NSTC, Taipei; TENMAK, T\"urkiye; STFC/UKRI, United Kingdom; DOE and NSF, United States of America.
 
Individual groups and members have received support from BCKDF, CANARIE, CRC and DRAC, Canada; CERN-CZ, PRIMUS 21/SCI/017 and UNCE SCI/013, Czech Republic; COST, ERC, ERDF, Horizon 2020, ICSC-NextGenerationEU and Marie Sk{\l}odowska-Curie Actions, European Union; Investissements d'Avenir Labex, Investissements d'Avenir Idex and ANR, France; DFG and AvH Foundation, Germany; Herakleitos, Thales and Aristeia programmes co-financed by EU-ESF and the Greek NSRF, Greece; BSF-NSF and MINERVA, Israel; Norwegian Financial Mechanism 2014-2021, Norway; NCN and NAWA, Poland; La Caixa Banking Foundation, CERCA Programme Generalitat de Catalunya and PROMETEO and GenT Programmes Generalitat Valenciana, Spain; G\"{o}ran Gustafssons Stiftelse, Sweden; The Royal Society and Leverhulme Trust, United Kingdom.
 
In addition, individual members wish to acknowledge support from CERN: European Organization for Nuclear Research (CERN PJAS); Chile: Agencia Nacional de Investigaci\'on y Desarrollo (FONDECYT 1190886, FONDECYT 1210400, FONDECYT 1230812, FONDECYT 1230987); China: National Natural Science Foundation of China (NSFC - 12175119, NSFC 12275265, NSFC-12075060); Czech Republic: PRIMUS Research Programme (PRIMUS/21/SCI/017); EU: H2020 European Research Council (ERC - 101002463); European Union: European Research Council (ERC - 948254), Horizon 2020 Framework Programme (MUCCA - CHIST-ERA-19-XAI-00), European Union, Future Artificial Intelligence Research (FAIR-NextGenerationEU PE00000013), Italian Center for High Performance Computing, Big Data and Quantum Computing (ICSC, NextGenerationEU), Marie Sklodowska-Curie Actions (EU H2020 MSC IF GRANT NO 101033496); France: Agence Nationale de la Recherche (ANR-20-CE31-0013, ANR-21-CE31-0013, ANR-21-CE31-0022, ANR-22-EDIR-0002), Investissements d'Avenir Idex (ANR-11-LABX-0012), Investissements d'Avenir Labex (ANR-11-LABX-0012); Germany: Baden-Württemberg Stiftung (BW Stiftung-Postdoc Eliteprogramme), Deutsche Forschungsgemeinschaft (DFG - 469666862, DFG - CR 312/5-1); Italy: Istituto Nazionale di Fisica Nucleare (FELLINI G.A. n. 754496, ICSC, NextGenerationEU); Japan: Japan Society for the Promotion of Science (JSPS KAKENHI JP21H05085, JSPS KAKENHI JP22H01227, JSPS KAKENHI JP22H04944, JSPS KAKENHI JP22KK0227); Netherlands: Netherlands Organisation for Scientific Research (NWO Veni 2020 - VI.Veni.202.179); Norway: Research Council of Norway (RCN-314472); Poland: Polish National Agency for Academic Exchange (PPN/PPO/2020/1/00002/U/00001), Polish National Science Centre (NCN 2021/42/E/ST2/00350, NCN OPUS nr 2022/47/B/ST2/03059, NCN UMO-2019/34/E/ST2/00393, UMO-2020/37/B/ST2/01043, UMO-2021/40/C/ST2/00187); Slovenia: Slovenian Research Agency (ARIS grant J1-3010); Spain: BBVA Foundation (LEO22-1-603), Generalitat Valenciana (Artemisa, FEDER, IDIFEDER/2018/048), La Caixa Banking Foundation (LCF/BQ/PI20/11760025), Ministry of Science and Innovation (MCIN \& NextGenEU PCI2022-135018-2, MICIN \& FEDER PID2021-125273NB, RYC2019-028510-I, RYC2020-030254-I, RYC2021-031273-I, RYC2022-038164-I), PROMETEO and GenT Programmes Generalitat Valenciana (CIDEGENT/2019/023, CIDEGENT/2019/027); Sweden: Swedish Research Council (VR 2018-00482, VR 2022-03845, VR 2022-04683, VR grant 2021-03651), Knut and Alice Wallenberg Foundation (KAW 2017.0100, KAW 2018.0157, KAW 2018.0458, KAW 2019.0447); Switzerland: Swiss National Science Foundation (SNSF - PCEFP2\_194658); United Kingdom: Leverhulme Trust (Leverhulme Trust RPG-2020-004); United States of America: U.S. Department of Energy (ECA DE-AC02-76SF00515), Neubauer Family Foundation.


\printbibliography
\addcontentsline{toc}{section}{References}
 
\clearpage
 
\begin{flushleft}
\hypersetup{urlcolor=black}
{\Large The ATLAS Collaboration}

\bigskip

\AtlasOrcid[0000-0002-6665-4934]{G.~Aad}$^\textrm{\scriptsize 102}$,
\AtlasOrcid[0000-0002-5888-2734]{B.~Abbott}$^\textrm{\scriptsize 120}$,
\AtlasOrcid[0000-0002-1002-1652]{K.~Abeling}$^\textrm{\scriptsize 55}$,
\AtlasOrcid[0000-0001-5763-2760]{N.J.~Abicht}$^\textrm{\scriptsize 49}$,
\AtlasOrcid[0000-0002-8496-9294]{S.H.~Abidi}$^\textrm{\scriptsize 29}$,
\AtlasOrcid[0000-0002-9987-2292]{A.~Aboulhorma}$^\textrm{\scriptsize 35e}$,
\AtlasOrcid[0000-0001-5329-6640]{H.~Abramowicz}$^\textrm{\scriptsize 151}$,
\AtlasOrcid[0000-0002-1599-2896]{H.~Abreu}$^\textrm{\scriptsize 150}$,
\AtlasOrcid[0000-0003-0403-3697]{Y.~Abulaiti}$^\textrm{\scriptsize 117}$,
\AtlasOrcid[0000-0002-8588-9157]{B.S.~Acharya}$^\textrm{\scriptsize 69a,69b,m}$
% ...
\bigskip
\\

% ...
$^{29}$Physics Department, Brookhaven National Laboratory, Upton NY; United States of America.\\
% ...
$^{35}$$^{(a)}$Facult\'e des Sciences Ain Chock, Universit\'e Hassan II de Casablanca;$^{(b)}$Facult\'{e} des Sciences, Universit\'{e} Ibn-Tofail, K\'{e}nitra;$^{(c)}$Facult\'e des Sciences Semlalia, Universit\'e Cadi Ayyad, LPHEA-Marrakech;$^{(d)}$LPMR, Facult\'e des Sciences, Universit\'e Mohamed Premier, Oujda;$^{(e)}$Facult\'e des sciences, Universit\'e Mohammed V, Rabat;$^{(f)}$Institute of Applied Physics, Mohammed VI Polytechnic University, Ben Guerir; Morocco.\\
% ...
$^{49}$Fakult\"{a}t Physik , Technische Universit{\"a}t Dortmund, Dortmund; Germany.\\
% ...
$^{55}$II. Physikalisches Institut, Georg-August-Universit\"{a}t G\"ottingen, G\"ottingen; Germany.\\
% ...
$^{69}$$^{(a)}$INFN Gruppo Collegato di Udine, Sezione di Trieste, Udine;$^{(b)}$ICTP, Trieste;$^{(c)}$Dipartimento Politecnico di Ingegneria e Architettura, Universit\`a di Udine, Udine; Italy.\\
% ...
$^{102}$CPPM, Aix-Marseille Universit\'e, CNRS/IN2P3, Marseille; France.\\
% ...
$^{117}$Department of Physics, New York University, New York NY; United States of America.\\
% ...
$^{120}$Homer L. Dodge Department of Physics and Astronomy, University of Oklahoma, Norman OK; United States of America.\\
% ...
$^{150}$Department of Physics, Technion, Israel Institute of Technology, Haifa; Israel.\\
$^{151}$Raymond and Beverly Sackler School of Physics and Astronomy, Tel Aviv University, Tel Aviv; Israel.\\
% ...
$^{a}$ Also Affiliated with an institute covered by a cooperation agreement with CERN.\\
$^{b}$ Also at An-Najah National University, Nablus; Palestine.\\
$^{c}$ Also at Borough of Manhattan Community College, City University of New York, New York NY; United States of America.\\
$^{d}$ Also at Center for High Energy Physics, Peking University; China.\\
$^{e}$ Also at Center for Interdisciplinary Research and Innovation (CIRI-AUTH), Thessaloniki; Greece.\\
$^{f}$ Also at Centro Studi e Ricerche Enrico Fermi; Italy.\\
% ...
$^{m}$ Also at Department of Physics, King's College London, London; United Kingdom.\\
% ...

\end{flushleft}


\clearpage

\end{document}